\documentclass{aa}  
\usepackage{graphicx}
\usepackage{txfonts}
\usepackage{natbib}

\bibpunct[]{(}{)}{;}{a}{}{,}
\begin{document}
   \title{Detection of amino acetonitrile in Sgr~B2(N)
   \thanks{Based on observations carried out with the IRAM Plateau de Bure 
   Interferometer, the IRAM 30m telescope, the Australia Telescope Compact 
   Array, and the NRAO Very Large Array. IRAM is supported by INSU/CNRS 
   (France), MPG (Germany) and IGN 
   (Spain). The Australia Telescope Compact Array is part of the Australia 
   Telescope which is funded by the Commonwealth of Australia 
   for operation as a National Facility managed by CSIRO. The National Radio 
   Astronomy Observatory is a facility of the National Science Foundation 
   operated under cooperative agreement by Associated Universities, Inc.}}


   \author{A. Belloche\inst{1}
          \and
          K.~M. Menten\inst{1}
          \and
          C. Comito\inst{1}
          \and
          H.~S.~P. M{\"uller}\inst{1,2}
          \and
          P. Schilke\inst{1}
          \and 
          J. Ott\inst{3,4,5}
          \and
          S. Thorwirth\inst{1}
          \and
          C. Hieret\inst{1}
          }

   \offprints{A. Belloche}

   \institute{Max-Planck Institut f\"ur Radioastronomie, Auf dem H\"ugel 69,
              D-53121 Bonn, Germany\\
              \email{belloche,kmenten,ccomito,schilke,sthorwirth,chieret@mpifr-bonn.mpg.de}
         \and
             I. Physikalisches Institut, Universit{\"a}t zu K{\"o}ln, 
             Z{\"u}lpicher Str. 77, D-50937 K{\"o}ln, Germany\\
             \email{hspm@ph1.uni-koeln.de}
         \and
             National Radio Astronomy Observatory, 520 Edgemont Road, 
             Charlottesville, VA 22903-2475, USA\\
             \email{jott@nrao.edu}
         \and
             California Institute of Technology, 1200 E. California Blvd., 
             Caltech Astronomy, 105-24, Pasadena, CA 91125-2400, USA
         \and
             CSIRO Australia Telescope National Facility, Cnr Vimiera $\&$ 
             Pembroke Roads, Marsfield NSW 2122, Australia
             }

   \date{Received 6 December 2007; accepted 16 January 2008}

 
  \abstract
   {Amino acids are building blocks of proteins and therefore key ingredients
   for the origin of life. The simplest amino acid, glycine
   (NH$_2$CH$_2$COOH), has long been searched for in the interstellar medium
   but has not been unambiguously detected so far. At the same time, 
   more and more complex molecules have been newly found toward the prolific 
   Galactic center source Sagittarius B2.}
   {Since the search for glycine has turned out to be extremely difficult, we
   aimed at detecting a chemically related species (possibly a direct 
   precursor), amino acetonitrile (NH$_2$CH$_2$CN).}
   {With the IRAM 30m telescope we carried out a complete line survey of the 
   hot core regions Sgr~B2(N) and (M) in the  3\,mm range, plus partial 
   surveys at 2 and 1.3\,mm. We analyzed our 30m line
   survey in the LTE approximation and modeled the emission of all known
   molecules simultaneously. We identified spectral features at the 
   frequencies predicted for amino acetonitrile lines having intensities 
   compatible with a unique rotation temperature. We also used the Very Large
   Array to look for cold, extended emission from amino acetonitrile.} 
   {We detected amino acetonitrile in Sgr~B2(N) in our 30m telescope line 
   survey and conducted confirmatory observations of selected lines with the
   IRAM Plateau de Bure and the Australia Telescope Compact Array 
   interferometers. The emission arises from a known hot core, the Large 
   Molecule Heimat, and is compact with a source diameter of 2$\arcsec$ 
   (0.08 pc). We derived a column density of $2.8 \times 10^{16}$ cm$^{-2}$,
   a temperature of 100 K, and a linewidth of 7 km~s$^{-1}$. Based on the 
   simultaneously observed continuum emission, we calculated a density of 
   $1.7 \times 10^8$ cm$^{-3}$, a mass of 2340 M$_\odot$, and an amino 
   acetonitrile fractional abundance of $2.2 \times 10^{-9}$. The high 
   abundance and 
   temperature may indicate that amino acetonitrile is formed by grain surface 
   chemistry. We did not detect any hot, compact amino acetonitrile emission 
   toward Sgr~B2(M) or any cold, extended emission toward Sgr~B2, with 
   column-density upper limits of $6 \times 10^{15}$ and $3 \times 10^{12-14}$ 
   cm$^{-2}$, respectively.}
   {Based on our amino acetonitrile detection toward Sgr~B2(N) and a comparison
   to the pair methylcyanide/acetic acid both detected in this source, we 
   suggest that the column density of both glycine conformers in Sgr~B2(N) is 
   well below the best upper limits published recently by other authors, 
   and probably below the confusion limit in the 1-3\,mm range.}


   \maketitle
%

\section{Introduction -- general methods}
\label{s:intro}

Among the still growing list of complex molecules found in the 
interstellar medium, so-called ``bio''molecules garner special attention. 
In particular, the quest for interstellar amino acids, building blocks 
of proteins, has engaged radio and millimeter wavelength astronomers 
for a long time. Numerous published and unpublished searches have been made 
for interstellar glycine, the simplest amino acid 
\citep{Brown79,Hollis80,Berulis85,Combes96,Ceccarelli00,Hollis03,Jones07,Cunningham07}.
Its recent ``detection'' claimed by \citet{Kuan03} has been persuasively 
rebutted by \citet{Snyder05}. Since the early days of molecular radio 
astronomy, Sagittarius B2 has been a favorite target in searches for complex
molecules in space.

\subsection{The target: Sagittarius B2}
\label{ss:sgrb2intro}
\object{Sagittarius B2} (hereafter Sgr~B2 for short) is a very massive 
(several million 
solar masses) and extremely active region of high-mass star formation at a 
projected distance of $\sim 100$ pc from the Galactic center. Its distance 
from the Sun is assumed to be the same as the Galactic center distance, $R_0$.
\citet{Reid93}, reviewing various methods to determine $R_0$, arrived at a 
``best estimate'' of 8.0 $\pm$ 0.5 kpc, a value that we adopt in this article. 
It is supported by recent modeling of trajectories of stars orbiting the 
central black hole, which yields 7.94 $\pm$ 0.42 kpc \citep{Eisenhauer03}. 

There are two major centers of activity, \object{Sgr~B2(M)} and 
\object{Sgr~B2(N)} separated by 
$\sim 2$ pc. In each of them, recent star formation manifests itself in a 
multitude of H{\sc ii} regions of many sizes, from hypercompact to compact 
\citep[][]{Gaume95}, and there is abundant material to form 
new stars evident by massive sources of molecular line and submillimeter 
continuum emission from dust \citep{Lis91, Lis93}.

\subsubsection{Sgr B2 as part of the Central Molecular Zone}
\label{sss:sgrb2cmz}
Some of the first detections of interstellar organic molecules (at 
cm-wavelengths!) were made toward Sgr~B2 
\citep[see][ for a historical perspective]{Menten04}. The low intrinsic line 
strengths make these cm lines unlikely candidates for detection. However, the 
situation is helped, first, by the fact that many of the transitions in 
question may have inverted levels \citep{Menten04} and amplify background 
radio continuum emission which is very intense at cm wavelengths 
\citep{Hollis07}. Second, the spatial distributions of many 
species are characterized by spatially extended emission covering areas 
beyond Sgr~B2 itself, filling single dish telescope beams, thus producing
appreciable intensity even when observed with low spatial resolution 
\citep[][]{Cummins86,Jones08}. 
This emission is characterized by low rotation temperatures, favoring 
lower frequency lines. Recent identifications of  ``new'' species include 
glycolaldehyde CH$_2$OHCHO \citep[][]{Hollis00,Hollis01}, ethylene glycol 
HOCH$_2$CH$_2$OH \citep[][]{Hollis02}, and vinyl alcohol CH$_2$CHOH 
\citep[][]{Turner01}.

Sgr B2 and its surroundings are part of the Central Molecular Zone (CMZ) of 
our Galaxy, a $\sim \pm 0\fdg3$ latitude wide band stretching around the 
Galactic center from longitude $l \sim +1\fdg6$ to $-1\fdg1$ 
\citep[see, e.g.,][]{Morris96}. The CMZ contains spatially extended emission 
of many complex organic molecules 
\citep{Minh92,Dahmen97,Menten04,RequenaTorres06}.

\subsubsection{The Large Molecule Heimat}
\label{sss:introlmh}
Near Sgr~B2(N), there is a hot, dense compact source  that has 
a mm-wavelength line density second to no other known object. This source, 
for which \citet{Snyder94} coined the name ``Large Molecule Heimat'' (LMH),
is characterized by very high densities ($> 10^7$~cm$^{-3}$) 
and gas temperatures ($> 100$~K). In recent years arcsecond resolution
interferometry with the BIMA array  has resulted in the detection and
imaging of increasingly complex organic species toward the LMH, such as
vinyl cyanide CH$_2$CHCN, methyl formate HCOOCH$_3$, and ethyl cyanide 
CH$_3$CH$_2$CN \citep[][]{Miao95,Miao97}, formamide NH$_2$CHO, isocyanic 
acid HNCO, and methyl formate HCOOCH$_3$ \citep[][]{Kuan96a}, acetic acid 
CH$_3$COOH \citep[][]{Mehringer97b,Remijan02}, formic acid HCOOH 
\citep[][]{Liu01}, and acetone (CH$_3$)$_2$CO \citep[][]{Snyder02}. All the 
interferometric observations are consistent with a compact ($< $few arcsec
diameter) source that had already been identified as the source of 
high-density-tracing non-metastable ammonia line emission by \citet{Vogel87} 
and thermal methanol emission by \citet[][, their source ``i'']{Mehringer97a}. 
The LMH also hosts a powerful H$_2$O maser region \citep[][]{Reid88}, which
provides evidence that it is very young (see Sect.~\ref{ss:sgrb2n}).

\subsection{The complex spectra of complex molecules}
Complex molecules in general have large partition functions, in particular for 
the elevated temperatures ($> 100$ K) in molecular hot cores, dense and compact
cloud condensations internally heated by a deeply embedded, young high-mass 
(proto)stellar object. Therefore, most individual spectral lines are weak and 
might easily get hidden in the ``line forest'' found toward these frequently 
extremely line-rich sources. To a large part, this forest consists of 
rotational lines, many of them presently unidentifiable, from within 
relatively low-lying vibrational states of molecules. Most of the
candidate molecules from which these lines originate are known to exist in 
these sources, but laboratory spectroscopy is presently lacking for lines from 
the states in question. At this point in the game, unequivocally identifying a 
species in a spectrum of a hot core covering a wide spectral range requires 
the following steps: 
as described in detail in Sect.~\ref{ss:modeling30m}, assuming Local 
Thermodynamic Equilibrium (LTE) (which applies at the high densities in hot 
cores) a  model spectrum is calculated for an assumed rotation temperature, 
column density, line width and other parameters. This predicts lines of a 
given intensity at all the known frequencies. Then at least two conditions 
have to be fulfilled:  
(i) All predicted lines should have a counterpart in the observed spectrum 
with the right intensity and width -- no single line should be missing. 
(ii) Follow-up observations with interferometers have to prove whether all 
lines from the candidate species are emitted from the same spot. Given the 
chemical variety in hot core regions, this is a powerful constraint. Moreover, 
interferometer images tend to have less line confusion, since many lines that 
are blended in larger beam single-dish spectra arise from different locations 
or are emitted by an extended region that is spatially filtered out.
Using an interferometer for aiding molecule identifications was 
pioneered by L. Snyder and collaborators who (mostly) used the 
Berkeley-Illinois-Maryland-Array (BIMA) to clearly identify a number of 
species  in the Sgr~B2(N) Large Molecule Heimat (see Sect.~\ref{sss:introlmh}).

We carried out a complete line survey of the hot core regions Sgr~B2(N) and 
(M) with the IRAM 30m telescope at 3\,mm, along with partial surveys at 
2 and 1.3\,mm. 
One of the overall goals of our survey was to better characterize the 
molecular content of both regions. It also allows searches for ``new'' species 
once we have identified the lines emitted by known molecules (including 
vibrationally and torsionally excited states). In 
particular, many complex molecules have enough lines in the covered frequency 
ranges to apply criterion (i) above. Once a species fulfils this criterion, 
interferometer measurements of selected lines can be made to check criterion 
(ii).

\subsection{Amino acetonitrile}
One of our target molecules was amino acetonitrile (NH$_2$CH$_2$CN), 
a molecule chemically related to glycine. Whether it is a precursor to the 
latter is under debate (see Sect.~\ref{ss:precursor}).
Not many astronomical searches for amino acetonitrile have been reported 
in the literature. In his dissertation, \citet[][]{Storey76} reported 
searches for the $J_{K_a,K_c} = 2_{11}-2_{12}$ and $1_{01}-0_{00}$ transitions 
at 1350.5 and 9071.7 MHz, respectively with the Parkes 64\,m telescope. On 
afterthought, the only chance of success for their observations 
would have been if amino acetonitrile existed on large spatial scales,
similar to the molecules described in Sect.~\ref{sss:sgrb2cmz} (see 
Sect.~\ref{ss:aanvla} for further limits on extended amino acetonitrile 
emission). Recently, \citet{Wirstroem07} reported unsuccessful searches of a 
number of mm-wavelength transitions of amino acetonitrile toward a number of 
hot cores.

Here, we report our detection of warm compact emission from
amino acetonitrile in Sgr~B2(N) with the IRAM 30m telescope, the Plateau de 
Bure Interferometer (PdBI) and the Australia Telescope Compact Array (ATCA),
and upper limits on cold, spatially extended emission from amino acetonitrile 
that we obtained with the NRAO Very Large Array (VLA).
Section~\ref{s:obs} summarizes the observational details. We present our
results in Sect.~\ref{s:results}. Implications in terms of interstellar 
chemistry are discussed in Sect.~\ref{s:discussion}. 
Our conclusions are summarized in Sect.~\ref{s:concl}.

\section{Observations and data reduction}
\label{s:obs}

\subsection{Single-dish observations and data reduction}
\label{ss:30m}

We carried out millimeter line observations with the IRAM 30m telescope on
Pico Veleta, Spain, in January 2004, September 2004 and January 2005. We used
four SIS heterodyne receivers simultaneously, two in the 3\,mm window connected
to the autocorrelation spectrometer VESPA and two in the 1.3\,mm window with
filter banks as backends. A few selected frequency ranges were also observed
with one SIS receiver at 2\,mm in January 2004. The channel spacing and
bandwidth were 0.313 and 420~MHz for each receiver at 3 and 2\,mm,
and 1 and 512~MHz for each receiver at 1.3\,mm, respectively. The 
observations were done in single-sideband mode with sideband rejections of 
\hbox{$\sim 1$--3~$\%$} at 3\,mm, $\sim 5$--7~$\%$ at 2\,mm, and 
$\sim 5$--8~$\%$ 
at 1.3\,mm. The half-power beamwidths can be computed with the equation HPBW 
($\arcsec$) = $\frac{2460}{\nu(\mathrm{GHz})}$. The forward efficiencies 
$F_{\mathrm{eff}}$ were 0.95 at 3\,mm, 0.93 at 2\,mm, and 0.91 at 1.3\,mm, 
respectively. The main-beam efficiencies were computed using the Ruze function
$B_{\mathrm{eff}} = 1.2 \epsilon \,\, e^{-(4 \pi R \sigma / \lambda)^2}$, with
$\epsilon = 0.69$, $R \sigma = 0.07$, and $\lambda$ the wavelength in mm (see
the IRAM 30m telescope system summary on http:$//$www.iram.es). 
The system temperatures ranged from 96 to 600 K at 3\,mm, from 220 to 720 K at 
2\,mm (except at 176 GHz where they ranged from 2400 to 3000 K), and from 280 
to 1200~K at 1.3\,mm. The telescope pointing was checked
every $\sim 1.5$ hours on Mercury, Mars, 1757$-$240 or G10.62, and found to be
accurate to 2--3$\arcsec$ (rms). The telescope focus was optimized on Mercury,
Jupiter, Mars or G34.3+0.2 every $\sim 1.5$--3 hours.
The observations were performed toward both sources
Sgr~B2(N) ($\alpha_{\mathrm{J2000}}$=17$^\mathrm{h}$47$^\mathrm{m}$20$\fs$0,
$\delta_{\mathrm{J2000}}$=$-28^\circ$22$\arcmin$19.0$\arcsec$, 
$V_{\mathrm{lsr}}$ = 64 km~s$^{-1}$) and Sgr~B2(M)
($\alpha_{\mathrm{J2000}}$=17$^\mathrm{h}$47$^\mathrm{m}$20$\fs$4, 
$\delta_{\mathrm{J2000}}$=$-28^\circ$23$\arcmin$07.0$\arcsec$,
\hbox{$V_{\mathrm{lsr}}$ = 62 km~s$^{-1}$)} in
position-switching mode with a reference position offset by 
($\Delta\alpha$,$\Delta\delta$)=($-752\arcsec$,$+342\arcsec$) with respect
to the former. The emission toward this reference position was found to be
weak: $T_{\mathrm{a}}^\star$($^{12}$CO~1$-$0) $\sim$ 2 K,
$T_{\mathrm{a}}^\star$(CS~2$-$1) $\la$ 0.05 K,
$T_{\mathrm{a}}^\star$($^{12}$CO~2$-$1) $\sim$ 1.5 K,
$T_{\mathrm{a}}^\star$($^{13}$CO~2$-$1) $\la$ 0.1 K,
and it is negligible for higher excitation lines and/or complex species.

We observed the full 3\,mm window between 80 and 116~GHz toward both sources.
The step between two adjacent tuning frequencies
was 395~MHz, which yielded an overlap of 50~MHz.
The autocorrelator VESPA produces artificial spikes with a width of 3--5
channels at the junction between subbands (typically 2 or 3 spikes per
spectrum). To get rid of these artefacts, half of the integration time at each
tuning frequency was spent with the backend shifted by 50 MHz, so that we
could, without any loss of information, systematically remove
in each spectrum 5 channels at each of the 6 junctions between subbands that
were possibly affected by this phenomenon. At 2\,mm, we observed at only 8
selected frequencies, and removed the artificial spikes in the same way as at 
3\,mm. At 1\,mm, we covered the frequency ranges 201.8 to 204.6 GHz and 205.0 
to 217.7 GHz, plus a number of selected spots at higher frequency. For 
each individual
spectrum, we removed a 0$^{\mathrm{th}}$-order (constant) baseline by selecting
a group of channels which seemed to be free of emission or absorption. However,
many spectra are full of lines, especially at 1.3\,mm where we reached the 
confusion limit, and we may have overestimated the level of the baseline for 
some of them.

In $T_{\mathrm{a}}^\star$ scale, the rms noise level achieved towards 
Sgr~B2(N) is about 15--20 mK below 100 GHz, 20--30 mK between 100 and 114.5 
GHz, and about 50 mK between 114.5 and 116~GHz. At 1.3\,mm, we reached the 
confusion limit for most of the spectra. The data were reduced with the
CLASS software, which is part of the GILDAS software package 
(see http://www.iram.fr/IRAMFR/GILDAS).

\subsection{Interferometric observations with the PdBI}
\label{ss:pdbi}

We observed Sgr~B2(N) with the PdBI for 4.7 hours on February 
7$^{\mathrm{th}}$, 2006 with 6 antennas in the high-resolution $A$ 
configuration (E24E68E04N46W27N29). The coordinates of the phase center were
$\alpha_{\mathrm{J2000}} = 17^{\mathrm{h}}47^{\mathrm{m}}20\fs00$, 
$\delta_{\mathrm{J2000}} = -28^\circ22\arcmin19.0\arcsec$. The 3 and 1.2\,mm 
receivers were tuned to 81.982 and 245.380~GHz, respectively, in single side 
band mode. At 3\,mm, there were two spectral windows centered at 81.736 and 
82.228~GHz with a bandwidth of 80 MHz and a channel separation of 0.313~MHz, 
and two continuum windows of 320 MHz bandwidth centered at 81.852 and 
82.112~GHz. The atmospheric phase stability was good for the 3\,mm band but 
bad for the 1.2\,mm band. Therefore we do not analyze the 1.2\,mm data. The 
system temperatures were typically 150--220~K at 3\,mm in the lower sideband. 
The (naturally-weighted) synthesized half-power beam width was 
$3.4\arcsec \times 0.8\arcsec$ with P.A. 10$^\circ$, and the primary beam was 
$\sim 61\arcsec$ FWHM.  The correlator bandpass was calibrated on the quasar 
3C~273. Phase calibration was determined on the nearby sources NRAO~530 
and 1622$-$297. The time-dependent amplitude calibration was done on 
1622$-$297, NRAO~530, 1334$-$127, and 1749$+$096, 
while the absolute flux density scale was derived from MWC~349. The absolute 
calibration uncertainty is estimated to be $\sim$ 15$\%$. The data were 
calibrated and imaged using the GILDAS software. The continuum emission was
estimated on line-free portions of the bands and removed in the \textit{uv} 
plane. The deconvolution was performed with the CLEAN method 
\citep[][]{Clark80}.

\subsection{Interferometric observations with the ATCA}
\label{ss:atca}

We observed Sgr~B2(N) with the ATCA on May 17$^{\mathrm{th}}$, 2006 in the 
hybrid H\,214 configuration for 6 hours, for 7 hours on July 
30$^{\mathrm{th}}$, 2006 in the H\,168 configuration, and for 6 hours in the 
compact H\,75 configuration on September 25$^{\mathrm{th}}$, 2006. 
The coordinates of the phase center were
$\alpha_{\mathrm{J2000}} = 17^{\mathrm{h}}47^{\mathrm{m}}20\fs00$, 
$\delta_{\mathrm{J2000}} = -28^\circ22\arcmin19.0\arcsec$.
The 3\,mm receiver 
was alternately tuned to three frequency pairs of 90.550 and 93.200~GHz, 
90.779 and 93.200~GHz, and 99.978 and 97.378~GHz in single side band mode, 
where only the first frequency of each pair was in spectral line mode with 
32~MHz bandwidth and 128 channels. The second frequency of each pair was
configured for continuum observations with 128~MHz bandwidth
each. The system temperatures were typically 60 K. The
(naturally-weighted) synthesized half-power beam width was 
$2.8\arcsec \times 1.9\arcsec$ with P.A. $72^\circ$ for H\,214,
$3.9\arcsec \times 1.9\arcsec$ with P.A. $-86^\circ$ for H\,168, and
$6.9\arcsec \times 5.4\arcsec$ with P.A. $-73^\circ$ for H\,75. The 
combination of H\,214 and H\,168 yields a synthesized half-power beam width of 
$3.0\arcsec \times 1.9\arcsec$ with P.A. $82^\circ$, and the combination of 
all three configurations a synthesized half-power beam width of 
$3.4\arcsec \times 2.3\arcsec$ with P.A. $83^\circ$.
The primary beam was $\sim 2.4\arcmin$ FWHM. The correlator bandpass was
calibrated on PKS~1253$-$055. The phase and gain calibration was
determined on the nearby source PKS~1759$-$39. The absolute flux
density scale was derived from Uranus. The absolute calibration
uncertainty is estimated to be $\sim$ 20$\%$. The data were
calibrated, continuum subtracted, imaged, and deconvolved using the
software package MIRIAD \citep*[][]{Sault95}.

Our ATCA data are affected by two problems. First, the tuning frequency used
in May was not updated for the new observatory velocity in July and September. 
As a result, the observed bands were shifted by +12 and +16 MHz in rest 
frequency in the H\,168 and H\,75 configurations with respect to the H\,214 
configuration. Second, we suspect a technical problem with the tuning at 99 GHz
in May and July because we do not detect any line in the H\,214 and H\,168 
configurations while we easily detect two lines in the H\,75 configuration: 
one unidentified line and one line from CH$_3$CH$_3$CO, v$_{\mathrm{t}}$=1 
according to our line survey with the IRAM 30m telescope. Comparing this band 
to the two other bands where we detect every line in each configuration 
(albeit with different intensities due to variable spatial filtering), we 
consider it to be very 
unlikely that the two lines detected at 99 GHz in the H\,75 configuration are 
completely filtered out in the H\,214 and H\,168 configurations. Since the 
amino acetonitrile transition is shifted out of the H\,75 band at 99 GHz (due 
to the variation of the observatory velocity), we do not analyze this dataset 
in the present article. 

\subsection{Interferometric observations with the VLA}
\label{ss:vla}

We used the NRAO Very Large Array to search for the $1_{01}-0_{00}$ multiplet 
of amino acetonitrile at 9071.208 MHz and examine the possibility of cold 
extended emission from this molecule. The VLA data were taken over a 1.5~h 
interval on 
February 13$^{\mathrm{th}}$, 2003 when the array was in its lowest-resolution 
($D$) configuration. Three $\sim 20$ minute long scans of the following 
position in Sgr~B2 were alternated with scans of the phase calibrator 
NRAO~530. For absolute flux density calibration, 3C286 was observed. Our phase 
center in Sgr~B2 was at 
$\alpha_{\mathrm{J2000}}$=17$^\mathrm{h}$47$^\mathrm{m}$20$\fs$00,
$\delta_{\mathrm{J2000}}$=$-28^\circ$22$\arcmin$51.0$\arcsec$. 
This is $32\arcsec$ South and $16\arcsec$ North of our 30m telescope pointing 
positions for Sgr~B2(N) and (M), respectively.

Our observations were done in spectral line mode with one intermediate 
frequency (IF) band split into 32 channels, each of which had a width of 
0.1953 MHz, corresponding to 6.46~km~s$^{-1}$. The usable central 72$\%$ 
frequency range of the IF bandwidth, 4.49 MHz, covered all the multiplet's 7 
hyperfine structure (hfs) components\footnote{The $1_{01}-0_{00}$ multiplet 
consists of 9 hyperfine components with significant intensity spanning the 
frequency range from 9070.233 to 9073.867~MHz. Twice two components have 
basically the same frequency (difference of about 0.1 kHz) so that under 
favorable conditions there are 7 hyperfine lines observable. Full 
spectroscopic information for amino acetonitrile can be found in the CDMS 
(entry 56507).}.
This frequency range corresponds to a total velocity coverage of 
148~km~s$^{-1}$. The center velocity was set to  V$_{\mathrm{lsr}}$ = 65 
km~s$^{-1}$. A 4.49 MHz bandwidth ``pseudo continuum'' database (the so-called 
``channel 0'') was created by averaging the central 23 channels. The $uv$-data 
were calibrated using the NRAO's Astronomical Imaging Processing System 
(AIPS). Several 
iterations of self calibration delivered a high quality continuum image. Using 
UVLIN, the average of selected regions of the line $uv$-database were 
subtracted channel by channel from the latter to remove the continuum level. 
To calibrate the spectral line data, the phase and amplitude corrections 
determined by the initial calibration, as well as by the self calibration were 
transferred to the line database and applied to them successively, producing a 
23 channel database which was imaged channel by channel using natural 
weighting. The synthesized beam width of the images is 
$20.5\arcsec \times 7.0\arcsec$ FWHM with a position angle of $-6.3^\circ$ 
East of North\footnote{The calibrated and deconvolved data cubes and images
(line and continuum) obtained with the PdBI, the ATCA, and the VLA are 
available in FITS format at the CDS via anonymous ftp to cdsarc.u-strasbg.fr 
(130.79.128.5) or via http://cdsweb.u-strasbg.fr/cgi-bin/qcat?J/A+A/.}.

\section{Identification of amino acetonitrile}
\label{s:results}

\subsection{Amino acetonitrile frequencies}
\label{ss:aanfreq}

\begin{table}
\centering
\caption{Spectroscopic parameters$^a$ (MHz) of amino acetonitrile.}
\label{t:spectro_param}
\begin{tabular}{lr@{}l}
\hline\hline
Parameter                        &\multicolumn{2}{c}{Value} \\
\hline
$A$                       & 30\,246&.456\,1\,(71)   \\
$B$                       &  4\,761&.061\,02\,(84)  \\
$C$                       &  4\,310&.751\,23\,(76)  \\
$D_K \times 10^3$        &     676&.62\,(99)       \\
$D_{JK} \times 10^3$     &   $-$55&.298\,6\,(69)   \\
$D_J \times 10^3$        &       3&.068\,53\,(68)  \\
$d_1 \times 10^6$        &  $-$671&.60\,(40)       \\
$d_2 \times 10^6$        &   $-$28&.893\,(106)     \\
$H_K \times 10^6$        &      30&.$^b$           \\
$H_{KJ} \times 10^6$     &    $-$2&.686\,1\,(268)  \\
$H_{JK} \times 10^9$     &  $-$120&.1\,(72)        \\
$H_J \times 10^9$        &       9&.593\,(276)     \\
$h_1 \times 10^9$        &       2&.989\,(225)     \\
\hline
\end{tabular}
 \begin{list}{}{}
  \item[$(a)$] Watson's $S$-reduction was employed in the representation $I^r$.
The dimensionless weighted rms of the fit is 0.51, therefore, the numbers in
parentheses may be considered as two standard deviations in units
of the least significant figures.
  \item [$(b)$] Assumed value (see Sect.~\ref{ss:aanfreq}).
 \end{list}
\end{table}

In the course of the present investigation an amino acetonitrile entry (tag: 
56507) has been prepared for the catalog of the Cologne Database for Molecular 
Spectroscopy \citep[CDMS, see][]{Mueller01,Mueller05}.
The laboratory transition frequencies were summarized by \citet{Bogey90}. 
Their work included microwave transitions reported without $^{14}$N 
quadrupole splitting by \citet{MacDonald72}, \citet{Pickett73}, as well as 
microwave transitions reported with quadrupole splitting by \citet{Brown77}; 
the latter data were used with the reported splitting. Line fitting and 
prediction of transition frequencies was done with the SPFIT/SPCAT suite of 
programs \citep{Pickett91} using a Watson type Hamiltonian
in the $S$ reduction in the representation $I^r$ \citep[see, e.g.,][]{Gordy84}.

The set of spectroscopic parameters reported by \citet*{Bogey90} 
included terms of up to decic order ($S_K$), rather unusual for an apparently 
rigid and fairly heavy molecule, and we found the higher order parameters to 
be surprisingly  large. Moreover, the off-diagonal sextic distortion parameter 
$h_3$ was larger in magnitude than $h_2$, and $h_1$ was not even used in the 
fit; the importance of these parameters is reversed to what is more commonly 
found. Therefore, we performed a trial fit with the octic and decic 
parameters as well as $h_2$ and $h_3$ omitted and the sextic term $H_K$ fixed 
to a value that was estimated from  $A/D_K \approx D_K/H_K$. 
Subsequently, we found that inclusion of $h_1$ improved the quality of the 
fit. All transitions but five having $\Delta K_a \ge 1$ and $K''_a \ge 3$ 
could be reproduced well. Two of these had $K_a = 3 - 2$ and deviated 
$\sim3.5$~MHz from the predicted frequencies which was only twice the 
predicted uncertainty. The inclusion of these transitions in the fit caused 
relatively small changes in $A$ and $D_K$; changes in the remaining parameters 
were within the uncertainties. Therefore, it is likely that the assignments of 
these two transitions are correct. Effects on the predicted $\Delta K_a = 0$ 
transition frequencies are negligible. The remaining three transitions are 
considered to be mis-assignments as their inclusion would require many more 
higher order distortion parameters with apparently unphysical values and a 
poorer quality of the fit. Hence, those transitions were omitted from our fit.
The resulting spectroscopic parameters are given in 
Table~\ref{t:spectro_param}. The rotational partition function at 75 and 150~K 
is 4403 and 12460, respectively.

\subsection{Modeling of the 30m line survey}
\label{ss:modeling30m}

The average line density above 3$\sigma$ in our 30m survey is about 100 and 25 
features per GHz for Sgr~B2(N) and (M), respectively, translating into about 
3700 and 950 lines over the whole 80--116 GHz band. To identify a new molecule 
in such a line forest and reduce the risk of mis-assignments, it is essential 
to model first the emission of all known molecules, including vibrationally 
and torsionally excited states, and their isotopologues. We used the 
XCLASS software \citep[see][ and references therein]{Comito05}
to model the emission and absorption lines in the LTE approximation. These
calculations take into account the beam dilution, the line opacity, and the 
line blending. The molecular spectroscopic parameters are taken from our line 
catalog which contains all entries from the CDMS catalog 
\citep[][]{Mueller01,Mueller05} and from the molecular 
spectroscopic database of the Jet Propulsion Laboratory 
\citep[JPL, see][]{Pickett98}, plus additional ``private'' entries.

Each molecule is modeled separately with the following set of input 
parameters: source size, rotational temperature, column density, velocity 
linewidth, velocity offset with respect to the systemic velocity
of the source, and a flag indicating if it is an emission or absorption 
component. 
For some of the molecules, it was necessary to include several velocity 
components to reproduce the observed spectra. The velocity components in 
emission are supposed to be non-interacting, i.e. the intensities add up 
linearly. The radiative transfer is computed in the following way:
first the emission line spectrum is calculated, and then the absorption lines,
using the full (lines + continuum) emission spectrum as background to absorb
against.
The vibrationally and/or torsionally excited states of some molecules
were modeled separately from the ground state. The input parameters were 
varied until a good fit to the data was obtained for each molecule. The whole 
spectrum including all the identified molecules was then computed at once, and 
the parameters for each molecule were adjusted again when necessary. The 
quality of the fit was checked by eye over the whole frequency coverage of the 
line survey. We favored our eye-checking method against an automated fitting 
because the high occurence of line blending and the uncertainty in the 
baseline removal would in many cases make an automated fitting procedure fail.

The detailed results of this modeling will be published in a forthcoming 
article describing the complete survey (Belloche et al., \textit{in prep}). So 
far, we have identified 51 different molecules, 60 isotopologues, and 41 
vibrationally/torsionally excited
states in Sgr~B2(N), which represent about 60$\%$ of the lines detected above
the 3$\sigma$ level. In Sgr~B2(M), the corresponding numbers are 41, 50, 20, 
and 50$\%$, respectively.

\subsection{Detection of amino acetonitrile with the 30m telescope}
\label{ss:aan30m}

\addtocounter{table}{1}
\newcounter{apptable}
\setcounter{apptable}{\value{table}}
\onltab{\value{apptable}}{
\begin{table*}
 \centering
 \caption{
 Transitions of amino acetonitrile observed with the IRAM 30m telescope toward Sgr~B2(N).
The horizontal lines mark discontinuities in the observed frequency coverage.
 All lines with S$\mu^2$ $<$ 20 D$^2$ are not listed, since they are expected to be much too weak to be detectable with our sensitivity.
 }
 \label{t:aan30m}
 \vspace*{0.0ex}
 \begin{tabular}{rlrcrrrl}
 \hline\hline
 \multicolumn{1}{c}{N$^a$} & \multicolumn{1}{c}{Transition$^b$} & \multicolumn{1}{c}{Frequency} & \multicolumn{1}{c}{Unc.$^c$} & \multicolumn{1}{c}{E$_\mathrm{l}$$^d$} & \multicolumn{1}{c}{S$\mu^2$} &  \multicolumn{1}{c}{$\sigma^e$} &  \multicolumn{1}{c}{Comments} \\ 
  & & \multicolumn{1}{c}{\scriptsize (MHz)} & \multicolumn{1}{c}{\scriptsize (kHz)} & \multicolumn{1}{c}{\scriptsize (K)} & \multicolumn{1}{c}{\scriptsize (D$^2$)} &  \multicolumn{1}{c}{\scriptsize (mK)} &  \\ 
 \multicolumn{1}{c}{(1)} & \multicolumn{1}{c}{(2)} & \multicolumn{1}{c}{(3)} & \multicolumn{1}{c}{(4)} & \multicolumn{1}{c}{(5)} & \multicolumn{1}{c}{(6)} & \multicolumn{1}{c}{(7)} & \multicolumn{1}{c}{(8)} \\ 
 \hline
   1 &  9$_{ 0, 9}$ -  8$_{ 0, 8}$ &   80947.479 &     7 &   16 &          60 &   33 & \textbf{Detected} \\ 
   2 &  9$_{ 2, 8}$ -  8$_{ 2, 7}$ &   81535.184 &     6 &   21 &          57 &   18 & Strong HC$^{13}$CCN, v=0 \\ 
   3 &  9$_{ 5, 5}$ -  8$_{ 5, 4}$$^\star$ &   81700.966 &     6 &   47 &          41 &   13 & \textbf{Group detected}, partial blend with U-line \\ 
   5 &  9$_{ 6, 3}$ -  8$_{ 6, 2}$$^\star$ &   81702.498 &     5 &   60 &          33 &   13 & \textbf{Group detected}, partial blend with U-line \\ 
   7 &  9$_{ 4, 6}$ -  8$_{ 4, 5}$$^\star$ &   81709.838 &     6 &   35 &          48 &   13 & \textbf{Group detected} \\ 
   9 &  9$_{ 7, 3}$ -  8$_{ 7, 2}$ &   81709.848 &     6 &   76 &          24 &   13 & \textbf{Group detected} \\ 
  10 &  9$_{ 4, 5}$ -  8$_{ 4, 4}$ &   81710.098 &     6 &   35 &          48 &   13 & \textbf{Group detected} \\ 
  11 &  9$_{ 3, 7}$ -  8$_{ 3, 6}$ &   81733.892 &     6 &   27 &          53 &   13 & \textbf{Detected}, blend with CH$_3$OCH$_3$ and HCC$^{13}$CN, v$_6$=1 \\ 
  12 &  9$_{ 3, 6}$ -  8$_{ 3, 5}$ &   81756.174 &     6 &   27 &          53 &   13 & \textbf{Detected}, blend with U-line \\ 
  13 &  9$_{ 2, 7}$ -  8$_{ 2, 6}$ &   82224.644 &     7 &   21 &          57 &   19 & \textbf{Detected}, uncertain baseline \\ 
  14 &  9$_{ 1, 8}$ -  8$_{ 1, 7}$ &   83480.894 &     8 &   17 &          59 &   17 & Blend with C$_2$H$_3$CN, v=0 \\ 
  15 & 10$_{ 1,10}$ -  9$_{ 1, 9}$ &   88240.541 &     8 &   20 &          66 &   19 & Strong HNCO, v=0 and HN$^{13}$CO, v=0 \\ 
  16 & 10$_{ 0,10}$ -  9$_{ 0, 9}$ &   89770.285 &     7 &   19 &          66 &   18 & Blend with U-line \\ 
  17 & 10$_{ 2, 9}$ -  9$_{ 2, 8}$ &   90561.332 &     6 &   25 &          64 &   20 & \textbf{Detected}, blend with weak C$_2$H$_5$CN, v$_{13}$=1/v$_{21}$=1 \\ 
  18 & 10$_{ 6, 4}$ -  9$_{ 6, 3}$$^\star$ &   90783.538 &     6 &   64 &          43 &   14 & \textbf{Group detected}, partial blend with CH$_2$(OH)CHO and U-line \\ 
  20 & 10$_{ 5, 6}$ -  9$_{ 5, 5}$$^\star$ &   90784.281 &     6 &   50 &          50 &   14 & \textbf{Group detected}, partial blend with CH$_2$(OH)CHO and U-line \\ 
  22 & 10$_{ 7, 3}$ -  9$_{ 7, 2}$$^\star$ &   90790.259 &     6 &   80 &          34 &   14 & \textbf{Group detected}, blend with U-line \\ 
  24 & 10$_{ 4, 7}$ -  9$_{ 4, 6}$ &   90798.685 &     6 &   39 &          56 &   14 & \textbf{Group detected}, blend with U-line \\ 
  25 & 10$_{ 4, 6}$ -  9$_{ 4, 5}$ &   90799.249 &     6 &   39 &          56 &   14 & \textbf{Group detected}, blend with U-line \\ 
  26 & 10$_{ 8, 2}$ -  9$_{ 8, 1}$$^\star$ &   90801.896 &     7 &   98 &          24 &   14 & Blend with U-line and HC$^{13}$CCN, v$_7$=1 \\ 
  28 & 10$_{ 3, 8}$ -  9$_{ 3, 7}$ &   90829.945 &     6 &   31 &          60 &   14 & \textbf{Detected}, blend with U-line also in M? \\ 
  29 & 10$_{ 3, 7}$ -  9$_{ 3, 6}$ &   90868.038 &     6 &   31 &          60 &   14 & \textbf{Detected}, partial blend with U-line \\ 
  30 & 10$_{ 2, 8}$ -  9$_{ 2, 7}$ &   91496.108 &     8 &   25 &          64 &   24 & \textbf{Detected}, partial blend with CH$_3$CN, v$_4$=1 and U-line \\ 
  31 & 10$_{ 1, 9}$ -  9$_{ 1, 8}$ &   92700.172 &     8 &   21 &          66 &   28 & Blend with U-line and CH$_3$OCH$_3$ \\ 
  32 & 11$_{ 1,11}$ - 10$_{ 1,10}$ &   97015.224 &     8 &   25 &          72 &   21 & \textbf{Detected}, partial blend with C$_2$H$_5$OH and CH$_3$OCHO \\ 
  33 & 11$_{ 0,11}$ - 10$_{ 0,10}$ &   98548.363 &     8 &   24 &          73 &   18 & Blend with C$_2$H$_5$CN, v=0 \\ 
  34 & 11$_{ 2,10}$ - 10$_{ 2, 9}$ &   99577.063 &     7 &   29 &          71 &   19 & Blend with CH$_3$OCHO, v$_{\mathrm{t}}$=1 \\ 
  35 & 11$_{ 6, 6}$ - 10$_{ 6, 5}$$^\star$ &   99865.516 &     6 &   68 &          51 &   14 & Strong blend with C$_2$H$_5$OH, CCS \\ 
  37 & 11$_{ 5, 7}$ - 10$_{ 5, 6}$$^\star$ &   99869.306 &     6 &   55 &          58 &   14 & Strong CH$_3$OCHO, v$_{\mathrm{t}}$=1 \\ 
  39 & 11$_{ 7, 4}$ - 10$_{ 7, 3}$$^\star$ &   99871.151 &     6 &   84 &          43 &   14 & Strong CH$_3$OCHO, v$_{\mathrm{t}}$=1 \\ 
  41 & 11$_{ 8, 3}$ - 10$_{ 8, 2}$$^\star$ &   99882.826 &     7 &  103 &          34 &   14 & Strong blend with C$_2$H$_5$CN, v$_{13}$=1/v$_{21}$=1 \\ 
  43 & 11$_{ 4, 8}$ - 10$_{ 4, 7}$ &   99890.599 &     6 &   44 &          63 &   14 & Strong blend with HC$^{13}$CCN, v$_7$=1 and HC$^{13}$CCN, v$_6$=1  \\ 
  44 & 11$_{ 4, 7}$ - 10$_{ 4, 6}$ &   99891.725 &     6 &   44 &          63 &   14 & Strong blend with HC$^{13}$CCN, v$_6$=1 and NH$_2$CN \\ 
  45 & 11$_{ 9, 2}$ - 10$_{ 9, 1}$$^\star$ &   99898.969 &     8 &  124 &          24 &   14 & Strong HCC$^{13}$CN, v$_6$=1 \\ 
  47 & 11$_{ 3, 9}$ - 10$_{ 3, 8}$ &   99928.886 &     6 &   35 &          68 &   14 & \textbf{Detected}, partial blend with NH$_2$CN and U-line \\ 
  48 & 11$_{ 3, 8}$ - 10$_{ 3, 7}$ &   99990.567 &     7 &   35 &          68 &   14 & \textbf{Detected} \\ 
  49 & 11$_{ 2, 9}$ - 10$_{ 2, 8}$ &  100800.876 &     8 &   29 &          71 &   20 & \textbf{Detected}, partial blend with CH$_3$CH$_3$CO, v=0 and U-line \\ 
  50 & 11$_{ 1,10}$ - 10$_{ 1, 9}$ &  101899.795 &     8 &   26 &          72 &   34 & \textbf{Detected}, uncertain baseline \\ 
  51 & 12$_{ 1,12}$ - 11$_{ 1,11}$ &  105777.991 &     8 &   29 &          79 &   43 & \textbf{Detected}, blend with c-C$_2$H$_4$O and C$_2$H$_5$CN, v=0 \\ 
  52 & 12$_{ 0,12}$ - 11$_{ 0,11}$ &  107283.142 &     8 &   29 &          80 &   24 & \textbf{Detected}, blend with C$_2$H$_5$OH and U-line \\ 
  53 & 12$_{ 2,11}$ - 11$_{ 2,10}$ &  108581.408 &     7 &   34 &          77 &   20 & \textbf{Detected}, weak blend with C$_2$H$_5$OH \\ 
  54 & 12$_{ 6, 7}$ - 11$_{ 6, 6}$$^\star$ &  108948.523 &     6 &   73 &          60 &   29 & Blend with C$_2$H$_5$CN, v=0 and C$_2$H$_5$OH \\ 
  56 & 12$_{ 7, 5}$ - 11$_{ 7, 4}$$^\star$ &  108952.574 &     6 &   89 &          53 &   29 & Blend with C$_2$H$_5$OH \\ 
  58 & 12$_{ 5, 8}$ - 11$_{ 5, 7}$$^\star$ &  108956.206 &     6 &   60 &          66 &   29 & \textbf{Group detected}, blend with C$_2$H$_5$OH \\ 
  60 & 12$_{ 8, 4}$ - 11$_{ 8, 3}$$^\star$ &  108963.964 &     7 &  108 &          44 &   29 & Strong blend with HC$^{13}$CCN, v$_7$=1 \\ 
  62 & 12$_{ 9, 3}$ - 11$_{ 9, 2}$$^\star$ &  108980.660 &     8 &  128 &          35 &   29 & Strong blend with HCC$^{13}$CN, v$_6$=1 \\ 
  64 & 12$_{ 4, 9}$ - 11$_{ 4, 8}$ &  108985.821 &     6 &   48 &          71 &   29 & Blend with HCC$^{13}$CN, v$_7$=1 \\ 
  65 & 12$_{ 4, 8}$ - 11$_{ 4, 7}$ &  108987.928 &     6 &   48 &          71 &   29 & Blend with HCC$^{13}$CN, v$_7$=1 \\ 
  66 & 12$_{10, 2}$ - 11$_{10, 1}$$^\star$ &  109001.603 &    10 &  152 &          24 &   29 & Weak \\ 
  68 & 12$_{ 3,10}$ - 11$_{ 3, 9}$ &  109030.225 &     6 &   40 &          75 &   29 & \textbf{Detected}, partial blend with HC$_3$N, v$_4$=1, C$_2$H$_5$OH, and U-line \\ 
  69 & 12$_{ 3, 9}$ - 11$_{ 3, 8}$ &  109125.734 &     7 &   40 &          75 &   29 & Strong HC$^{13}$CCN, v$_7$=1 \\ 
 \hline
 \end{tabular}
 \begin{list}{}{}
 \item[$(a)$]{Numbering of the observed transitions with S$\mu^2$ $>$ 20 D$^2$.}
 \item[$(b)$]{Transitions marked with a $^\star$ are double with a frequency difference less than 0.1 MHz. The quantum numbers of the second one are not shown.}
 \item[$(c)$]{Frequency uncertainty.}
 \item[$(d)$]{Lower energy level in temperature units (E$_\mathrm{l}/$k$_\mathrm{B}$).}
 \item[$(e)$]{Calculated rms noise level in T$_{\mathrm{mb}}$ scale.}
 \end{list}
 \end{table*}
\begin{table*}
 \centering
 \addtocounter{table}{-1}
 \caption{
(continued)
 }
 \label{t:aan30m}
 \begin{tabular}{rlrcrrrl}
 \hline\hline
 \multicolumn{1}{c}{N$^a$} & \multicolumn{1}{c}{Transition$^b$} & \multicolumn{1}{c}{Frequency} & \multicolumn{1}{c}{Unc.$^c$} & \multicolumn{1}{c}{E$_\mathrm{l}$$^d$} & \multicolumn{1}{c}{S$\mu^2$} &  \multicolumn{1}{c}{$\sigma^e$} &  \multicolumn{1}{c}{Comments} \\ 
  & & \multicolumn{1}{c}{\scriptsize (MHz)} & \multicolumn{1}{c}{\scriptsize (kHz)} & \multicolumn{1}{c}{\scriptsize (K)} & \multicolumn{1}{c}{\scriptsize (D$^2$)} &  \multicolumn{1}{c}{\scriptsize (mK)} &  \\ 
 \multicolumn{1}{c}{(1)} & \multicolumn{1}{c}{(2)} & \multicolumn{1}{c}{(3)} & \multicolumn{1}{c}{(4)} & \multicolumn{1}{c}{(5)} & \multicolumn{1}{c}{(6)} & \multicolumn{1}{c}{(7)} & \multicolumn{1}{c}{(8)} \\ 
 \hline
  70 & 12$_{ 2,10}$ - 11$_{ 2, 9}$ &  110136.314 &     8 &   34 &          77 &   24 & Blend with $^{13}$CH$_3$OH, v=0 \\ 
  71 & 12$_{ 1,11}$ - 11$_{ 1,10}$ &  111076.901 &     8 &   31 &          79 &   25 & \textbf{Detected}, slightly shifted? \\ 
  72 & 13$_{ 1,13}$ - 12$_{ 1,12}$ &  114528.654 &     8 &   34 &          86 &   37 & \textbf{Detected}, partial blend with U-line \\ 
  73 & 13$_{ 0,13}$ - 12$_{ 0,12}$ &  115977.853 &    50 &   34 &          86 &   79 & Blend with CH$_3$$^{13}$CH$_2$CN, v=0, CH$_3$CH$_3$CO, and C$_2$H$_5$CN, v=0 \\ 
\hline
  74 & 15$_{ 7, 9}$ - 14$_{ 7, 8}$$^\star$ &  136200.478 &     6 &  106 &          78 &   28 & Strong HC$^{13}$CCN, v$_7$=1 \\ 
  76 & 15$_{ 6,10}$ - 14$_{ 6, 9}$$^\star$ &  136204.641 &     5 &   90 &          84 &   28 & Strong HC$^{13}$CCN, v$_7$=1 \\ 
  78 & 15$_{ 8, 7}$ - 14$_{ 8, 6}$$^\star$ &  136208.805 &     7 &  125 &          71 &   28 & Strong HC$^{13}$CCN, v$_7$=1 and HC$^{13}$CCN, v$_6$=1 \\ 
  80 & 15$_{ 9, 6}$ - 14$_{ 9, 5}$$^\star$ &  136225.653 &     8 &  145 &          64 &   28 & Strong HCC$^{13}$CN, v$_6$=1 and HCC$^{13}$CN, v$_7$=1 \\ 
  82 & 15$_{ 5,11}$ - 14$_{ 5,10}$ &  136229.823 &     5 &   77 &          89 &   28 & Blend with HCC$^{13}$CN, v$_7$=1 \\ 
  83 & 15$_{ 5,10}$ - 14$_{ 5, 9}$ &  136230.008 &     5 &   77 &          89 &   28 & Blend with HCC$^{13}$CN, v$_7$=1 \\ 
  84 & 15$_{10, 5}$ - 14$_{10, 4}$$^\star$ &  136248.969 &    10 &  169 &          55 &   28 & \textbf{Group detected}, blend with U-line \\ 
  86 & 15$_{11, 4}$ - 14$_{11, 3}$$^\star$ &  136277.600 &    13 &  195 &          46 &   28 & Strong HC$_3$N, v$_4$=1 and CH$_3$OCHO \\ 
  88 & 15$_{ 4,12}$ - 14$_{ 4,11}$ &  136293.271 &     6 &   65 &          93 &   28 & Strong CH$_3$$^{13}$CH$_2$CN \\ 
  89 & 15$_{ 4,11}$ - 14$_{ 4,10}$ &  136303.599 &     6 &   65 &          93 &   28 & \textbf{Detected}, blend with a(CH$_2$OH)$_2$ and CH$_3$C$_3$N \\ 
  90 & 15$_{12, 3}$ - 14$_{12, 2}$$^\star$ &  136310.849 &    16 &  223 &          36 &   28 & Weak, blend with CH$_3$C$_3$N \\ 
  92 & 15$_{ 3,13}$ - 14$_{ 3,12}$ &  136341.155 &     6 &   57 &          96 &   28 & \textbf{Detected}, partial blend with U-line also in M \\ 
  93 & 15$_{13, 2}$ - 14$_{13, 1}$$^\star$ &  136348.271 &    21 &  253 &          25 &   28 & Weak, blend with U-line \\ 
\hline
  95 & 16$_{ 7, 9}$ - 15$_{ 7, 8}$$^\star$ &  145284.487 &    30 &  113 &          86 &   25 & Strong HC$^{13}$CCN, v$_7$=1 \\ 
  97 & 16$_{ 8, 8}$ - 15$_{ 8, 7}$$^\star$ &  145290.958 &    30 &  131 &          80 &   25 & Blend with HC$^{13}$CCN, v$_6$=1 \\ 
  99 & 16$_{ 6,11}$ - 15$_{ 6,10}$$^\star$ &  145292.688 &    30 &   97 &          91 &   25 & Blend with HC$^{13}$CCN, v$_6$=1 \\ 
 101 & 16$_{ 9, 7}$ - 15$_{ 9, 6}$$^\star$ &  145307.254 &    30 &  152 &          73 &   25 & Strong C$_2$H$_5$CN, v$_{13}$=1/v$_{21}$=1, HCC$^{13}$CN, v$_7$=1, C$_2$H$_3$CN, v$_{15}$=1 \\ 
 103 & 16$_{ 5,12}$ - 15$_{ 5,11}$ &  145325.871 &    30 &   83 &          96 &   25 & \textbf{Group detected}, uncertain baseline, partial blend with  \\ 
 & & & & & & & C$_2$H$_5$CN, v$_{13}$=1/v$_{21}$=1 \\ 
 104 & 16$_{ 5,11}$ - 15$_{ 5,10}$ &  145326.209 &    30 &   83 &          96 &   25 & \textbf{Group detected}, uncertain baseline, partial blend with  \\ 
 & & & & & & & C$_2$H$_5$CN, v$_{13}$=1/v$_{21}$=1 \\ 
 105 & 16$_{10, 6}$ - 15$_{10, 5}$$^\star$ &  145330.985 &    40 &  175 &          65 &   25 & \textbf{Group detected}, uncertain baseline \\ 
 107 & 16$_{11, 5}$ - 15$_{11, 4}$$^\star$ &  145360.684 &    30 &  201 &          56 &   25 & Strong HC$_3$N, v$_4$=1 \\ 
 109 & 16$_{12, 4}$ - 15$_{12, 3}$$^\star$ &  145395.485 &    60 &  229 &          46 &   25 & Blend with C$_3$H$_7$CN \\ 
 111 & 16$_{ 4,13}$ - 15$_{ 4,12}$ &  145403.421 &    30 &   72 &         100 &   25 & Blend with U-line or wing of C$_2$H$_5$CN, v=0 \\ 
 112 & 16$_{ 4,12}$ - 15$_{ 4,11}$ &  145419.704 &    30 &   72 &         100 &   25 & Strong C$_2$H$_5$CN, v=0 \\ 
 113 & 16$_{13, 3}$ - 15$_{13, 2}$$^\star$ &  145434.928 &    60 &  260 &          36 &   25 & Weak, blend with U-line \\ 
 115 & 16$_{ 3,14}$ - 15$_{ 3,13}$ &  145443.850 &    30 &   63 &         103 &   25 & \textbf{Detected}, blend with C$_2$H$_5$CN, v=0 and U-line \\ 
 116 & 16$_{14, 2}$ - 15$_{14, 1}$$^\star$ &  145478.462 &    60 &  293 &          25 &   25 & Weak, strong O$^{13}$CS \\ 
\hline
 118 & 16$_{ 1,15}$ - 15$_{ 1,14}$ &  147495.789 &     6 &   55 &         106 &   31 & \textbf{Detected}, partial blend with H$_3$C$^{13}$CN, v$_8$=1 \\ 
 119 & 16$_{ 2,14}$ - 15$_{ 2,13}$ &  147675.839 &    30 &   58 &         105 &   31 & Blend with H$_3$C$^{13}$CN, v$_8$=1, U-line, and CH$_3$OCHO \\ 
\hline
 120 & 17$_{ 7,10}$ - 16$_{ 7, 9}$$^\star$ &  154369.232 &    40 &  120 &          94 &  112 & Strong HC$^{13}$CCN, v$_7$=1 and HC$^{13}$CCN, v$_6$=1 \\ 
 122 & 17$_{ 8, 9}$ - 16$_{ 8, 8}$$^\star$ &  154373.384 &    40 &  138 &          88 &  112 & Strong HC$^{13}$CCN, v$_6$=1 \\ 
 124 & 17$_{ 6,12}$ - 16$_{ 6,11}$$^\star$ &  154382.222 &    40 &  104 &          99 &  112 & Strong HCC$^{13}$CN, v$_6$=1 and HCC$^{13}$CN, v$_7$=1 \\ 
 126 & 17$_{ 9, 8}$ - 16$_{ 9, 7}$$^\star$ &  154388.904 &    40 &  159 &          81 &  112 & Strong HCC$^{13}$CN, v$_7$=1 \\ 
 128 & 17$_{10, 7}$ - 16$_{10, 6}$$^\star$ &  154412.758 &    50 &  182 &          74 &  112 & Strong HNCO, v=0 \\ 
 130 & 17$_{ 5,13}$ - 16$_{ 5,12}$ &  154424.604 &    40 &   90 &         103 &  112 & Strong CH$_3$OH, v=0 \\ 
 131 & 17$_{ 5,12}$ - 16$_{ 5,11}$ &  154425.216 &    40 &   90 &         103 &  112 & Strong CH$_3$OH, v=0 \\ 
 132 & 17$_{11, 6}$ - 16$_{11, 5}$$^\star$ &  154443.330 &    60 &  208 &          66 &  112 & Strong HC$_3$N, v$_4$=1 and CH$_3$$^{13}$CH$_2$CN \\ 
 134 & 17$_{12, 5}$ - 16$_{12, 4}$$^\star$ &  154479.566 &    15 &  236 &          57 &  112 & Strong C$_2$H$_5$CN, v=0 \\ 
 136 & 17$_{ 4,14}$ - 16$_{ 4,13}$ &  154517.470 &     5 &   79 &         107 &  112 & Blend with C$_2$H$_5$CN, v$_{13}$=1/v$_{21}$=1 \\ 
 137 & 17$_{13, 4}$ - 16$_{13, 3}$$^\star$ &  154520.861 &    19 &  267 &          47 &  112 & Blend with C$_2$H$_5$CN, v$_{13}$=1/v$_{21}$=1 \\ 
 139 & 17$_{ 4,13}$ - 16$_{ 4,12}$ &  154542.406 &     5 &   79 &         107 &  112 & \textbf{Group detected}, blend with U-line \\ 
 140 & 17$_{ 3,15}$ - 16$_{ 3,14}$ &  154544.046 &     5 &   70 &         109 &  112 & \textbf{Group detected}, blend with U-line \\ 
 141 & 17$_{14, 3}$ - 16$_{14, 2}$$^\star$ &  154566.773 &    25 &  300 &          36 &  112 & Weak, strong U-line \\ 
 143 & 17$_{15, 2}$ - 16$_{15, 1}$$^\star$ &  154617.004 &    33 &  335 &          25 &  112 & Weak, strong U-line and C$_2$H$_5$CN, v$_{13}$=1/v$_{21}$=1 \\ 
\hline
 145 & 18$_{ 7,12}$ - 17$_{ 7,11}$$^\star$ &  163454.794 &     5 &  127 &         101 &   38 & \textbf{Group detected}, partial blend with HC$^{13}$CCN, v$_6$=1  \\ 
 & & & & & & & and HCC$^{13}$CN, v$_6$=1 \\ 
 147 & 18$_{ 8,10}$ - 17$_{ 8, 9}$$^\star$ &  163456.136 &     6 &  146 &          96 &   38 & \textbf{Group detected}, partial blend with HC$^{13}$CCN, v$_6$=1  \\ 
 & & & & & & & and HCC$^{13}$CN, v$_6$=1 \\ 
 149 & 18$_{ 9, 9}$ - 17$_{ 9, 8}$$^\star$ &  163470.472 &     8 &  166 &          90 &   38 & \textbf{Group detected}, partial blend with HCC$^{13}$CN,v$_7$=1 \\ 
 151 & 18$_{ 6,13}$ - 17$_{ 6,12}$$^\star$ &  163473.305 &     5 &  111 &         106 &   38 & \textbf{Group detected}, partial blend with HCC$^{13}$CN,v$_7$=1 \\ 
 153 & 18$_{10, 8}$ - 17$_{10, 7}$$^\star$ &  163494.265 &     9 &  190 &          83 &   38 & Strong CH$_3$$^{13}$CH$_2$CN \\ 
 \hline
 \end{tabular}
 \end{table*}
\begin{table*}
 \centering
 \addtocounter{table}{-1}
 \caption{
(continued)
 }
 \label{t:aan30m}
 \begin{tabular}{rlrcrrrl}
 \hline\hline
 \multicolumn{1}{c}{N$^a$} & \multicolumn{1}{c}{Transition$^b$} & \multicolumn{1}{c}{Frequency} & \multicolumn{1}{c}{Unc.$^c$} & \multicolumn{1}{c}{E$_\mathrm{l}$$^d$} & \multicolumn{1}{c}{S$\mu^2$} &  \multicolumn{1}{c}{$\sigma^e$} &  \multicolumn{1}{c}{Comments} \\ 
  & & \multicolumn{1}{c}{\scriptsize (MHz)} & \multicolumn{1}{c}{\scriptsize (kHz)} & \multicolumn{1}{c}{\scriptsize (K)} & \multicolumn{1}{c}{\scriptsize (D$^2$)} &  \multicolumn{1}{c}{\scriptsize (mK)} &  \\ 
 \multicolumn{1}{c}{(1)} & \multicolumn{1}{c}{(2)} & \multicolumn{1}{c}{(3)} & \multicolumn{1}{c}{(4)} & \multicolumn{1}{c}{(5)} & \multicolumn{1}{c}{(6)} & \multicolumn{1}{c}{(7)} & \multicolumn{1}{c}{(8)} \\ 
 \hline
 155 & 18$_{11, 7}$ - 17$_{11, 6}$$^\star$ &  163525.533 &    11 &  216 &          75 &   38 & \textbf{Group detected}, blend with HC$_3$N, v$_4$=1 \\ 
 157 & 18$_{ 5,14}$ - 17$_{ 5,13}$ &  163526.183 &     4 &   97 &         110 &   38 & \textbf{Group detected}, blend with HC$_3$N, v$_4$=1 \\ 
 158 & 18$_{ 5,13}$ - 17$_{ 5,12}$ &  163527.171 &     4 &   97 &         110 &   38 & \textbf{Group detected}, blend with HC$_3$N, v$_4$=1 \\ 
 159 & 18$_{12, 6}$ - 17$_{12, 5}$$^\star$ &  163563.084 &    14 &  244 &          66 &   38 & Blend with C$_2$H$_3$CN, v$_{15}$=2 and SO$_2$, v=0 \\ 
 161 & 18$_{13, 5}$ - 17$_{13, 4}$$^\star$ &  163606.161 &    18 &  274 &          57 &   38 & Strong SO$_2$, v=0 \\ 
 163 & 18$_{ 4,15}$ - 17$_{ 4,14}$ &  163635.326 &     5 &   86 &         114 &   38 & \textbf{Detected}, partial blend with C$_3$H$_7$CN \\ 
 164 & 18$_{ 3,16}$ - 17$_{ 3,15}$ &  163640.468 &     5 &   78 &         116 &   38 & \textbf{Detected}, partial blend with C$_3$H$_7$CN \\ 
 165 & 18$_{14, 4}$ - 17$_{14, 3}$$^\star$ &  163654.261 &    24 &  307 &          47 &   38 & Weak, blend with H$^{13}$CONH$_2$, v=0 \\ 
 167 & 18$_{ 4,14}$ - 17$_{ 4,13}$ &  163672.524 &     5 &   86 &         114 &   38 & Strong C$_2$H$_5$CN, v$_{13}$=1/v$_{21}$=1 \\ 
 168 & 18$_{15, 3}$ - 17$_{15, 2}$$^\star$ &  163707.031 &    32 &  343 &          37 &   38 & Weak, blend with CH$_2$(OH)CHO \\ 
\hline
 170 & 18$_{ 2,16}$ - 17$_{ 2,15}$ &  166463.884 &    30 &   72 &         118 &   66 & Blend with $^{13}$CH$_2$CHCN \\ 
\hline
 171 & 19$_{ 8,12}$ - 18$_{ 8,11}$$^\star$ &  172539.195 &    40 &  153 &         104 &   44 & Strong HCC$^{13}$CN, v$_6$=1 \\ 
 173 & 19$_{ 7,13}$ - 18$_{ 7,12}$$^\star$ &  172541.207 &    40 &  135 &         109 &   44 & Strong HCC$^{13}$CN, v$_6$=1 \\ 
 175 & 19$_{ 9,10}$ - 18$_{ 9, 9}$$^\star$ &  172552.010 &    40 &  174 &          98 &   44 & Strong HCC$^{13}$CN, v$_7$=1 \\ 
 177 & 19$_{ 6,14}$ - 18$_{ 6,13}$$^\star$ &  172566.092 &    50 &  119 &         114 &   44 & \textbf{Group detected}, partial blend with U-line and HCC$^{13}$CN, v$_7$=1 \\ 
 179 & 19$_{10, 9}$ - 18$_{10, 8}$$^\star$ &  172575.485 &    50 &  198 &          91 &   44 & Strong U-line \\ 
 181 & 19$_{11, 8}$ - 18$_{11, 7}$$^\star$ &  172607.260 &    60 &  223 &          84 &   44 & Strong HC$_3$N, v$_4$=1 \\ 
 183 & 19$_{ 5,15}$ - 18$_{ 5,14}$ &  172630.750 &    30 &  105 &         117 &   44 & Strong U-line and t-HCOOH \\ 
 184 & 19$_{ 5,14}$ - 18$_{ 5,13}$ &  172632.382 &    30 &  105 &         117 &   44 & Strong U-line and t-HCOOH \\ 
 185 & 19$_{12, 7}$ - 18$_{12, 6}$$^\star$ &  172645.994 &    13 &  252 &          76 &   44 & Blend with t-HCOOH, H$^{13}$CN, and U-line \\ 
 187 & 19$_{13, 6}$ - 18$_{13, 5}$$^\star$ &  172690.745 &    17 &  282 &          67 &   44 & Strong H$^{13}$CN and CH$_3$OCHO \\ 
 189 & 19$_{ 3,17}$ - 18$_{ 3,16}$ &  172731.699 &    30 &   86 &         123 &   44 & Blend with U-line and C$_2$H$_3$CN, v=0 \\ 
 190 & 19$_{14, 5}$ - 18$_{14, 4}$$^\star$ &  172740.945 &    23 &  315 &          58 &   44 & Strong C$_2$H$_3$CN, v=0 \\ 
 192 & 19$_{ 4,16}$ - 18$_{ 4,15}$ &  172756.821 &    30 &   94 &         121 &   44 & Baseline problem? \\ 
 193 & 19$_{15, 4}$ - 18$_{15, 3}$$^\star$ &  172796.183 &    30 &  351 &          48 &   44 & Weak, strong HCC$^{13}$CN, v$_7$=1 \\ 
 195 & 19$_{ 4,15}$ - 18$_{ 4,14}$ &  172811.041 &    30 &   94 &         121 &   44 & Strong C$_2$H$_5$CN, v=0 \\ 
 196 & 19$_{16, 3}$ - 18$_{16, 2}$$^\star$ &  172856.161 &    40 &  388 &          37 &   44 & Weak, strong HC$_3$N, v=0 and HC$_3$N, v$_5$=1$/$v$_7$=3 \\ 
\hline
 198 & 20$_{ 0,20}$ - 19$_{ 0,19}$ &  176174.096 &    30 &   81 &         132 &  365 & Noisy, partial blend with HNCO, v$_5$=1 and U-line \\ 
\hline
 199 & 23$_{ 0,23}$ - 22$_{ 0,22}$ &  201875.876 &    10 &  108 &         152 &  138 & Strong $^{13}$CH$_3$CH$_2$CN, v=0 \\ 
 200 & 22$_{ 2,20}$ - 21$_{ 2,19}$ &  203812.341 &    10 &  107 &         145 &  364 & Strong C$_2$H$_3$CN, v=0 \\ 
 201 & 23$_{ 2,22}$ - 22$_{ 2,21}$ &  206652.454 &     9 &  115 &         151 &  106 & Blend with C$_2$H$_5$CN, v=0 \\ 
 202 & 23$_{ 8,16}$ - 22$_{ 8,15}$$^\star$ &  208875.040 &     8 &  189 &         134 &  160 & Blend with HCC$^{13}$CN, v$_7$=1, CH$_3$CH$_3$CO, v$_{\mathrm{t}}$=1, and CH$_3$OCHO \\ 
 204 & 23$_{ 9,14}$ - 22$_{ 9,13}$$^\star$ &  208877.860 &     9 &  210 &         129 &  160 & Blend with HCC$^{13}$CN, v$_7$=1, CH$_3$CH$_3$CO, v$_{\mathrm{t}}$=1, and CH$_3$OCHO \\ 
 206 & 23$_{ 7,17}$ - 22$_{ 7,16}$$^\star$ &  208896.163 &     7 &  171 &         139 &  160 & Strong C$_2$H$_3$CN, v=0 \\ 
 208 & 23$_{10,13}$ - 22$_{10,12}$$^\star$ &  208897.241 &    10 &  233 &         124 &  160 & Strong C$_2$H$_3$CN, v=0 \\ 
 210 & 23$_{11,12}$ - 22$_{11,11}$$^\star$ &  208929.064 &    10 &  259 &         118 &  160 & Strong C$_2$H$_3$CN, v=0 \\ 
 212 & 23$_{ 6,18}$ - 22$_{ 6,17}$ &  208955.555 &     6 &  155 &         142 &  160 & Blend with C$_2$H$_3$CN, v=0 and U-line \\ 
 213 & 23$_{ 6,17}$ - 22$_{ 6,16}$ &  208955.797 &     6 &  155 &         142 &  160 & Blend with C$_2$H$_3$CN, v=0 and U-line \\ 
 214 & 23$_{12,11}$ - 22$_{12,10}$$^\star$ &  208970.868 &    11 &  287 &         111 &  160 & Strong C$_2$H$_3$CN, v=0 \\ 
 216 & 23$_{ 3,21}$ - 22$_{ 3,20}$ &  209015.508 &     7 &  121 &         150 &  160 & Strong C$_2$H$_5$CN, v=0 and C$_2$H$_3$CN, v=0 \\ 
 217 & 23$_{13,10}$ - 22$_{13, 9}$$^\star$ &  209021.101 &    13 &  318 &         104 &  160 & Strong C$_2$H$_3$CN, v=0 \\ 
 219 & 23$_{14, 9}$ - 22$_{14, 8}$$^\star$ &  209078.736 &    16 &  351 &          96 &  160 & Strong C$_2$H$_3$CN, v=0 \\ 
 221 & 23$_{ 5,19}$ - 22$_{ 5,18}$ &  209081.032 &     6 &  141 &         146 &  160 & Strong C$_2$H$_3$CN, v=0 \\ 
 222 & 23$_{ 5,18}$ - 22$_{ 5,17}$ &  209090.124 &     6 &  141 &         146 &  160 & Strong C$_2$H$_3$CN, v=0 \\ 
 223 & 23$_{15, 8}$ - 22$_{15, 7}$$^\star$ &  209143.066 &    23 &  386 &          88 &  160 & Weak, strong HC$^{13}$CCN, v$_7$=1 \\ 
 225 & 23$_{16, 7}$ - 22$_{16, 6}$$^\star$ &  209213.584 &    33 &  424 &          79 &   58 & Weak, strong H$_2$CS and C$_2$H$_3$CN, v$_{15}$=1 \\ 
 227 & 23$_{ 4,20}$ - 22$_{ 4,19}$ &  209272.189 &     6 &  130 &         148 &   58 & \textbf{Detected}, blend CH$_3$CH$_3$CO, v=0 \\ 
 228 & 23$_{17, 6}$ - 22$_{17, 5}$$^\star$ &  209289.914 &    46 &  464 &          69 &   58 & Weak, blend with C$_2$H$_5$CN, v$_{13}$=1/v$_{21}$=1 and C$_2$H$_3$CN, v$_{15}$=1 \\ 
 230 & 23$_{18, 5}$ - 22$_{18, 4}$$^\star$ &  209371.766 &    64 &  507 &          59 &   58 & Weak, blend with C$_2$H$_5$OH \\ 
 232 & 23$_{19, 4}$ - 22$_{19, 3}$$^\star$ &  209458.910 &    86 &  552 &          49 &   58 & Weak, strong C$_2$H$_3$CN, v$_{15}$=2 and C$_2$H$_5$CN, v=0 \\ 
 234 & 23$_{ 4,19}$ - 22$_{ 4,18}$ &  209473.790 &     7 &  130 &         148 &   58 & Strong C$_2$H$_3$CN, v$_{15}$=2 and CH$_3$OCHO \\ 
 235 & 23$_{20, 3}$ - 22$_{20, 2}$$^\star$ &  209551.156 &   114 &  599 &          37 &   58 & Weak, strong C$_2$H$_3$CN, v$_{11}$=1 \\ 
 237 & 23$_{ 1,22}$ - 22$_{ 1,21}$ &  209629.913 &     9 &  113 &         152 &   45 & \textbf{Detected}, blend with HC$^{13}$CCN, v$_7$=2 and HCC$^{13}$CN, v$_7$=2 \\ 
 238 & 23$_{21, 2}$ - 22$_{21, 1}$$^\star$ &  209648.342 &   146 &  649 &          25 &   45 & Weak, blend with U-line and C$_2$H$_3$CN, v$_{11}$=1 \\ 
 240 & 24$_{ 1,24}$ - 23$_{ 1,23}$ &  210072.793 &    12 &  118 &         159 &   45 & Blend with C$_2$H$_5$CN, v$_{13}$=1/v$_{21}$=1 and $^{13}$CH$_3$CH$_2$CN, v=0 \\ 
 241 & 24$_{ 0,24}$ - 23$_{ 0,23}$ &  210448.044 &    12 &  118 &         159 &   64 & Strong CH$_3$OCHO \\ 
 242 & 23$_{ 3,20}$ - 22$_{ 3,19}$ &  211099.150 &    12 &  122 &         150 &   33 & Blend with U-lines \\ 
 \hline
 \end{tabular}
 \end{table*}
\begin{table*}
 \centering
 \addtocounter{table}{-1}
 \caption{
(continued)
 }
 \label{t:aan30m}
 \begin{tabular}{rlrcrrrl}
 \hline\hline
 \multicolumn{1}{c}{N$^a$} & \multicolumn{1}{c}{Transition$^b$} & \multicolumn{1}{c}{Frequency} & \multicolumn{1}{c}{Unc.$^c$} & \multicolumn{1}{c}{E$_\mathrm{l}$$^d$} & \multicolumn{1}{c}{S$\mu^2$} &  \multicolumn{1}{c}{$\sigma^e$} &  \multicolumn{1}{c}{Comments} \\ 
  & & \multicolumn{1}{c}{\scriptsize (MHz)} & \multicolumn{1}{c}{\scriptsize (kHz)} & \multicolumn{1}{c}{\scriptsize (K)} & \multicolumn{1}{c}{\scriptsize (D$^2$)} &  \multicolumn{1}{c}{\scriptsize (mK)} &  \\ 
 \multicolumn{1}{c}{(1)} & \multicolumn{1}{c}{(2)} & \multicolumn{1}{c}{(3)} & \multicolumn{1}{c}{(4)} & \multicolumn{1}{c}{(5)} & \multicolumn{1}{c}{(6)} & \multicolumn{1}{c}{(7)} & \multicolumn{1}{c}{(8)} \\ 
 \hline
 243 & 23$_{ 2,21}$ - 22$_{ 2,20}$ &  213074.653 &    70 &  117 &         152 &   48 & Strong SO$_2$, v=0 \\ 
 244 & 24$_{ 2,23}$ - 23$_{ 2,22}$ &  215466.138 &    10 &  124 &         158 &   74 & Strong C$_2$H$_5$CN, v$_{13}$=1/v$_{21}$=1 \\ 
\hline
 245 & 24$_{ 3,21}$ - 23$_{ 3,20}$ &  220537.064 &    14 &  132 &         157 &   98 & Strong CH$_3$CN, v$_8$=0 \\ 
\hline
 246 & 25$_{ 1,24}$ - 24$_{ 1,23}$ &  226957.428 &    40 &  134 &         165 &   96 & Strong CN absorption \\ 
 247 & 25$_{ 9,16}$ - 24$_{ 9,15}$$^\star$ &  227040.487 &    50 &  230 &         145 &   96 & \textbf{Group detected}, partial blend with CN absorption and  \\ 
 & & & & & & & CH$_3$CH$_3$CO, v$_{\mathrm{t}}$=1 \\ 
 249 & 25$_{ 8,18}$ - 24$_{ 8,17}$$^\star$ &  227045.287 &    50 &  210 &         149 &   96 & \textbf{Group detected}, partial blend with CN absorption and  \\ 
 & & & & & & & CH$_3$CH$_3$CO, v$_{\mathrm{t}}$=1 \\ 
 251 & 25$_{10,15}$ - 24$_{10,14}$$^\star$ &  227055.944 &    50 &  254 &         139 &   96 & \textbf{Group detected}, partial blend with CN absorption \\ 
 253 & 25$_{ 7,19}$ - 24$_{ 7,18}$$^\star$ &  227079.847 &    50 &  191 &         153 &   96 & \textbf{Group detected}, blend with CH$_2$CH$^{13}$CN and CH$_3$OH, v=0 \\ 
 255 & 25$_{11,14}$ - 24$_{11,13}$$^\star$ &  227086.424 &    50 &  280 &         134 &   96 & Blend with CH$_2$CH$^{13}$CN and CH$_3$OH, v=0 \\ 
 257 & 25$_{ 3,23}$ - 24$_{ 3,22}$ &  227088.938 &    40 &  142 &         164 &   96 & Blend with CH$_2$CH$^{13}$CN and CH$_3$OH, v=0 \\ 
 258 & 25$_{12,13}$ - 24$_{12,12}$$^\star$ &  227128.728 &    60 &  308 &         128 &   96 & Blend with C$_2$H$_5$CN, v=0 and  C$_2$H$_3$CN, v=0 \\ 
\hline
 260 & 25$_{17, 8}$ - 24$_{17, 7}$$^\star$ &  227467.235 &    45 &  485 &          89 &   85 & Weak, blend with CH$_2$(OH)CHO, CH$_2$CH$^{13}$CN and  \\ 
 & & & & & & & t-C$_2$H$_5$OCHO \\ 
 262 & 25$_{ 4,22}$ - 24$_{ 4,21}$ &  227539.318 &    40 &  151 &         162 &   85 & Strong HCONH$_2$, v$_{12}$=1 and C$_2$H$_5$CN, v=0 \\ 
 263 & 25$_{18, 7}$ - 24$_{18, 6}$$^\star$ &  227555.295 &    63 &  528 &          80 &   85 & Weak, strong CH$_3$OCHO \\ 
 265 & 26$_{ 0,26}$ - 25$_{ 0,25}$ &  227601.595 &    16 &  138 &         172 &   85 & Strong HCONH$_2$, v=0 \\ 
 266 & 25$_{19, 6}$ - 24$_{19, 5}$$^\star$ &  227649.239 &    85 &  572 &          70 &   85 & Weak, blend with CH$_3$OCH$_3$ \\ 
 268 & 25$_{20, 5}$ - 24$_{20, 4}$$^\star$ &  227748.834 &   113 &  620 &          60 &   85 & Weak, blend with $^{13}$CH$_3$CH$_2$CN, v=0 \\ 
 270 & 25$_{21, 4}$ - 24$_{21, 3}$$^\star$ &  227853.885 &   147 &  669 &          49 &   85 & Weak, strong HC$^{13}$CCN, v$_7$=2 and HCC$^{13}$CN, v$_7$=2 \\ 
 272 & 25$_{ 4,21}$ - 24$_{ 4,20}$ &  227892.614 &    60 &  151 &         162 &   85 & Strong HCC$^{13}$CN, v$_7$=2, C$_2$H$_5$OH, and HC$^{13}$CCN, v$_7$=2 \\ 
\hline
 273 & 25$_{ 2,23}$ - 24$_{ 2,22}$ &  231485.527 &    50 &  138 &         165 &   40 & \textbf{Detected}, blend with U-line? \\ 
\hline
 274 & 26$_{ 1,25}$ - 25$_{ 1,24}$ &  235562.532 &    50 &  145 &         172 &  131 & Strong C$_2$H$_3$CN, v=0 \\ 
 275 & 27$_{ 1,27}$ - 26$_{ 1,26}$ &  235964.814 &    19 &  149 &         179 &  131 & Strong $^{13}$CH$_3$OH, v=0 \\ 
\hline
 276 & 26$_{ 3,24}$ - 25$_{ 3,23}$ &  236103.949 &    40 &  153 &         170 &   37 & Blend with HCC$^{13}$CN, v$_7$=1 and C$_2$H$_5$CN, v$_{13}$=1/v$_{21}$=1 \\ 
 277 & 26$_{ 9,17}$ - 25$_{ 9,16}$$^\star$ &  236121.689 &    50 &  241 &         152 &   37 & Blend with $^{13}$CH$_3$CH$_2$CN \\ 
 279 & 26$_{ 8,19}$ - 25$_{ 8,18}$$^\star$ &  236131.044 &    50 &  220 &         156 &   37 & Blend with CH$_2$$^{13}$CHCN \\ 
 281 & 26$_{10,16}$ - 25$_{10,15}$$^\star$ &  236134.730 &    50 &  265 &         147 &   37 & Blend with CH$_2$$^{13}$CHCN \\ 
 283 & 26$_{11,15}$ - 25$_{11,14}$$^\star$ &  236164.129 &    60 &  291 &         142 &   37 & Strong C$_2$H$_3$CN, v$_{11}$=1 and $^{13}$CH$_3$CH$_2$CN, v=0 \\ 
 285 & 26$_{ 7,20}$ - 25$_{ 7,19}$$^\star$ &  236173.437 &    60 &  202 &         160 &   37 & Blend with C$_2$H$_3$CN, v$_{11}$=1, CH$_2$$^{13}$CHCN, and  \\ 
 & & & & & & & $^{13}$CH$_3$CH$_2$CN, v=0 \\ 
 287 & 27$_{ 0,27}$ - 26$_{ 0,26}$ &  236182.602 &    18 &  149 &         179 &   37 & Blend with CH$_2$$^{13}$CHCN, $^{13}$CH$_3$CH$_2$CN, v=0, and HC$_3$N, v$_4$=1 \\ 
 288 & 26$_{12,14}$ - 25$_{12,13}$$^\star$ &  236206.427 &    19 &  319 &         136 &   37 & Strong SO$_2$, v=0 \\ 
 290 & 26$_{13,13}$ - 25$_{13,12}$$^\star$ &  236259.339 &    21 &  349 &         130 &   37 & Blend with C$_2$H$_5$CN, v$_{13}$=1/v$_{21}$=1 and t-C$_2$H$_5$OCHO \\ 
 292 & 26$_{ 6,21}$ - 25$_{ 6,20}$ &  236269.491 &    60 &  186 &         163 &   37 & \textbf{Group detected}, partial blend with t-C$_2$H$_5$OCHO and U-line \\ 
 293 & 26$_{ 6,20}$ - 25$_{ 6,19}$ &  236270.459 &    60 &  186 &         163 &   37 & \textbf{Group detected}, partial blend with t-C$_2$H$_5$OCHO and U-line \\ 
 294 & 26$_{14,12}$ - 25$_{14,11}$$^\star$ &  236321.411 &    23 &  382 &         123 &   37 & Weak, blend with C$_2$H$_5$OH and CH$_3$$^{13}$CH$_2$CN, v=0 \\ 
 296 & 26$_{15,11}$ - 25$_{15,10}$$^\star$ &  236391.638 &    27 &  418 &         115 &   37 & Weak, blend with CH$_2$$^{13}$CHCN \\ 
 298 & 26$_{ 5,22}$ - 25$_{ 5,21}$ &  236454.502 &    40 &  173 &         166 &   37 & Blend with SO, v=0 and $^{13}$CH$_3$CH$_2$CN, v=0, baseline problem? \\ 
 299 & 26$_{16,10}$ - 25$_{16, 9}$$^\star$ &  236469.303 &    35 &  456 &         107 &   37 & Weak, blend with CH$_2$CH$^{13}$CN \\ 
 301 & 26$_{ 5,21}$ - 25$_{ 5,20}$ &  236481.643 &    40 &  173 &         166 &   37 & Blend with CH$_2$$^{13}$CHCN \\ 
 302 & 26$_{17, 9}$ - 25$_{17, 8}$$^\star$ &  236553.897 &    80 &  496 &          99 &   37 & Weak, blend with t-C$_2$H$_5$OCHO, $^{13}$CH$_3$CH$_2$CN, v=0, and  \\ 
 & & & & & & & CH$_3$COOH, v$_{\mathrm{t}}$=0 \\ 
\hline
 304 & 27$_{ 1,26}$ - 26$_{ 1,25}$ &  244135.969 &    14 &  156 &         178 &   46 & Blend with C$_2$H$_5$CN, v$_{13}$=1/v$_{21}$=1 \\ 
 305 & 28$_{ 1,28}$ - 27$_{ 1,27}$ &  244585.841 &    21 &  161 &         186 &   39 & Strong CH$_3$OCHO and HCC$^{13}$CN, v=0 \\ 
 306 & 28$_{ 0,28}$ - 27$_{ 0,27}$ &  244765.968 &    21 &  160 &         186 &   39 & \textbf{Detected}, blend with CH$_3$$^{13}$CH$_2$CN, v=0 and U-line \\ 
 307 & 27$_{ 3,25}$ - 26$_{ 3,24}$ &  245102.984 &    11 &  164 &         177 &   72 & Blend with $^{13}$CH$_3$CH$_2$CN, v=0 \\ 
 308 & 27$_{ 9,18}$ - 26$_{ 9,17}$$^\star$ &  245202.855 &    16 &  253 &         159 &   72 & Blend with $^{13}$CH$_3$CH$_2$CN, v=0 and U-line? \\ 
 310 & 27$_{10,17}$ - 26$_{10,16}$$^\star$ &  245213.055 &    19 &  276 &         155 &   72 & Strong CH$_3$OH, v=0 \\ 
 312 & 27$_{ 8,20}$ - 26$_{ 8,19}$$^\star$ &  245217.230 &    13 &  232 &         164 &   72 & Strong CH$_3$OH, v=0 \\ 
 314 & 27$_{11,16}$ - 26$_{11,15}$$^\star$ &  245241.163 &    21 &  302 &         150 &   72 & Blend with C$_2$H$_5$$^{13}$CN, v=0 and C$_2$H$_3$CN, v$_{15}$=1 \\ 
 316 & 27$_{ 7,21}$ - 26$_{ 7,20}$$^\star$ &  245268.168 &    11 &  213 &         167 &   72 & Blend with HC$_3$N, v$_4$=1 \\ 
 318 & 27$_{12,15}$ - 26$_{12,14}$$^\star$ &  245283.214 &    24 &  330 &         144 &   72 & Blend with C$_2$H$_5$$^{13}$CN, v=0 \\ 
 320 & 27$_{13,14}$ - 26$_{13,13}$$^\star$ &  245336.715 &    26 &  361 &         138 &   72 & Strong SO$_2$, v=0 \\ 
 322 & 27$_{ 6,22}$ - 26$_{ 6,21}$ &  245378.722 &    10 &  197 &         170 &   72 & \textbf{Group detected}, blend with $^{13}$CH$_3$CH$_2$CN, v=0? \\ 
 323 & 27$_{ 6,21}$ - 26$_{ 6,20}$ &  245380.146 &    10 &  197 &         170 &   72 & \textbf{Group detected}, blend with $^{13}$CH$_3$CH$_2$CN, v=0? \\ 
 \hline
 \end{tabular}
 \end{table*}
\begin{table*}
 \centering
 \addtocounter{table}{-1}
 \caption{
(continued)
 }
 \label{t:aan30m}
 \begin{tabular}{rlrcrrrl}
 \hline\hline
 \multicolumn{1}{c}{N$^a$} & \multicolumn{1}{c}{Transition$^b$} & \multicolumn{1}{c}{Frequency} & \multicolumn{1}{c}{Unc.$^c$} & \multicolumn{1}{c}{E$_\mathrm{l}$$^d$} & \multicolumn{1}{c}{S$\mu^2$} &  \multicolumn{1}{c}{$\sigma^e$} &  \multicolumn{1}{c}{Comments} \\ 
  & & \multicolumn{1}{c}{\scriptsize (MHz)} & \multicolumn{1}{c}{\scriptsize (kHz)} & \multicolumn{1}{c}{\scriptsize (K)} & \multicolumn{1}{c}{\scriptsize (D$^2$)} &  \multicolumn{1}{c}{\scriptsize (mK)} &  \\ 
 \multicolumn{1}{c}{(1)} & \multicolumn{1}{c}{(2)} & \multicolumn{1}{c}{(3)} & \multicolumn{1}{c}{(4)} & \multicolumn{1}{c}{(5)} & \multicolumn{1}{c}{(6)} & \multicolumn{1}{c}{(7)} & \multicolumn{1}{c}{(8)} \\ 
 \hline
 324 & 27$_{14,13}$ - 26$_{14,12}$$^\star$ &  245400.026 &    29 &  394 &         131 &   72 & Blend with U-line, CH$_3$CH$_3$CO, v=0 and $^{13}$CH$_3$CH$_2$CN, v=0? \\ 
 326 & 27$_{15,12}$ - 26$_{15,11}$$^\star$ &  245472.024 &    33 &  429 &         124 &   72 & Weak, strong C$_2$H$_5$CN, v$_{13}$=1/v$_{21}$=1 and C$_2$H$_3$CN, v$_{11}$=1 \\ 
 328 & 27$_{16,11}$ - 26$_{16,10}$$^\star$ &  245551.911 &    40 &  467 &         116 &   53 & Weak, blend with SO$_2$, v=0 \\ 
 330 & 27$_{ 5,23}$ - 26$_{ 5,22}$ &  245585.766 &    10 &  184 &         173 &   53 & Strong CH$_2$$^{13}$CHCN and HC$_3$N, v=0 \\ 
 331 & 27$_{ 5,22}$ - 26$_{ 5,21}$ &  245623.711 &    10 &  184 &         173 &   53 & Strong HC$_3$N, v$_5$=1$/$v$_7$=3 and CH$_2$$^{13}$CHCN \\ 
 332 & 27$_{17,10}$ - 26$_{17, 9}$$^\star$ &  245639.099 &    51 &  507 &         108 &   53 & Weak, blend with C$_2$H$_3$CN, v$_{15}$=2 \\ 
 334 & 27$_{18, 9}$ - 26$_{18, 8}$$^\star$ &  245733.144 &    67 &  550 &         100 &   53 & Weak, strong HNCO, v$_5$=1 \\ 
 336 & 27$_{ 4,24}$ - 26$_{ 4,23}$ &  245803.548 &    10 &  173 &         175 &   53 & Blend with CH$_2$CH$^{13}$CN \\ 
 337 & 27$_{19, 8}$ - 26$_{19, 7}$$^\star$ &  245833.697 &    88 &  595 &          91 &   53 & Weak, strong U-line, partial blend with CH$_2$CH$^{13}$CN and  \\ 
 & & & & & & & C$_2$H$_3$CN, v$_{11}$=1$/$v$_{15}$=1 \\ 
 339 & 27$_{20, 7}$ - 26$_{20, 6}$$^\star$ &  245940.476 &   116 &  642 &          81 &   53 & Weak, strong U-line and CH$_3$OCHO, v$_{\mathrm{t}}$=1 \\ 
 341 & 27$_{21, 6}$ - 26$_{21, 5}$$^\star$ &  246053.247 &   150 &  692 &          71 &   53 & Weak, strong CH$_3$OCHO \\ 
\hline
 343 & 28$_{ 3,26}$ - 27$_{ 3,25}$ &  254085.051 &    12 &  176 &         184 &   32 & Blend with CH$_3$CH$_3$CO, v=0 and C$_2$H$_5$CN, v=0 \\ 
 344 & 28$_{ 9,19}$ - 27$_{ 9,18}$$^\star$ &  254283.930 &    19 &  264 &         167 &   32 & Strong SO$_2$, v=0 and $^{13}$CH$_3$OH, v=0 \\ 
 346 & 28$_{10,18}$ - 27$_{10,17}$$^\star$ &  254290.939 &    22 &  288 &         162 &   32 & Strong SO$_2$, v=0, $^{13}$CH$_3$OH, v=0, $^{13}$CH$_2$CHCN,  \\ 
 & & & & & & & $^{13}$CH$_3$CH$_2$CN, v=0, and H$^{13}$CONH$_2$, v$_{12}$=1 \\ 
 348 & 28$_{ 8,21}$ - 27$_{ 8,20}$$^\star$ &  254303.873 &    15 &  244 &         171 &   32 & Blend with $^{13}$CH$_3$CH$_2$CN, v=0 and C$_2$H$_5$CN, v=0 \\ 
 350 & 28$_{11,17}$ - 27$_{11,16}$$^\star$ &  254317.460 &    26 &  314 &         157 &   32 & Strong C$_2$H$_5$CN, v=0 and $^{13}$CH$_3$OH, v=0 \\ 
 352 & 28$_{12,16}$ - 27$_{12,15}$$^\star$ &  254359.074 &    29 &  342 &         152 &   32 & Blend with H$^{13}$CONH$_2$, v=0 \\ 
 354 & 28$_{ 7,22}$ - 27$_{ 7,21}$$^\star$ &  254364.160 &    12 &  225 &         174 &   32 & Blend with H$^{13}$CONH$_2$, v=0 \\ 
 356 & 28$_{13,15}$ - 27$_{13,14}$$^\star$ &  254413.004 &    32 &  372 &         146 &   32 & Strong C$_2$H$_5$CN, v=0, H$^{13}$CONH$_2$, v=0, and CH$_3$OH, v=0 \\ 
\hline
 358 & 28$_{22, 6}$ - 27$_{22, 5}$$^\star$ &  255273.081 &   196 &  755 &          71 &  217 & Weak, strong $^{13}$CH$_3$OH, v=0 and CH$_2$CH$^{13}$CN \\ 
 360 & 28$_{23, 5}$ - 27$_{23, 4}$$^\star$ &  255401.564 &   245 &  809 &          60 &  217 & Weak, blend with CH$_3$CH$_3$CO, v$_{\mathrm{t}}$=1 \\ 
 362 & 28$_{24, 4}$ - 27$_{24, 3}$$^\star$ &  255535.733 &   304 &  866 &          49 &  217 & Weak, blend with SO$_2$, v=0 and U-line? \\ 
 364 & 28$_{ 4,24}$ - 27$_{ 4,23}$ &  255674.369 &    18 &  185 &         182 &  217 & Strong HC$_3$N, v$_7$=1 \\ 
 365 & 28$_{25, 3}$ - 27$_{25, 2}$$^\star$ &  255675.432 &   372 &  925 &          38 &  217 & Strong HC$_3$N, v$_7$=1 \\ 
\hline
 367 & 28$_{ 2,26}$ - 27$_{ 2,25}$ &  258775.885 &    19 &  172 &         185 & 1609 & Weak, blend with CH$_3$OH, v=0 \\ 
\hline
 368 & 29$_{ 9,20}$ - 28$_{ 9,19}$$^\star$ &  263364.923 &    22 &  277 &         174 &   74 & \textbf{Group detected}, baseline problem?, blend with U-line \\ 
 370 & 29$_{10,19}$ - 28$_{10,18}$$^\star$ &  263368.355 &    26 &  300 &         170 &   74 & \textbf{Group detected}, baseline problem?, blend with U-line \\ 
 372 & 29$_{ 8,22}$ - 28$_{ 8,21}$$^\star$ &  263390.982 &    17 &  256 &         178 &   74 & Blend with HNCO, v$_4$=1 and CH$_3$OCH$_3$ \\ 
 374 & 29$_{11,18}$ - 28$_{11,17}$$^\star$ &  263393.008 &    31 &  326 &         165 &   74 & Blend with HNCO, v$_4$=1 and CH$_3$OCH$_3$ \\ 
 376 & 29$_{12,17}$ - 28$_{12,16}$$^\star$ &  263433.971 &    35 &  354 &         160 &   74 & Strong HC$_3$N, v$_4$=1, HNCO, v$_5$=1, and HNCO, v=0 \\ 
 378 & 29$_{ 7,23}$ - 28$_{ 7,22}$$^\star$ &  263461.436 &    14 &  237 &         181 &   74 & Blend with HNCO, v=0, CH$_3$CH$_3$CO, v=0 and C$_2$H$_5$OH \\ 
 380 & 29$_{13,16}$ - 28$_{13,15}$$^\star$ &  263488.164 &    40 &  385 &         154 &   74 & Blend with C$_2$H$_5$$^{13}$CN, v=0 \\ 
 382 & 29$_{14,15}$ - 28$_{14,14}$$^\star$ &  263553.566 &    44 &  418 &         148 &   74 & Strong SO$_2$, v=0, HCONH$_2$, v$_{12}$=1, and CH$_2$(OH)CHO \\ 
 384 & 29$_{ 6,24}$ - 28$_{ 6,23}$ &  263604.573 &    12 &  221 &         184 &   74 & \textbf{Group detected}, baseline problem?, partial blend with  \\ 
 & & & & & & & CH$_3$CH$_3$CO, v$_{\mathrm{t}}$=1 and CH$_3$OCH$_3$ \\ 
 385 & 29$_{ 6,23}$ - 28$_{ 6,22}$ &  263607.689 &    12 &  221 &         184 &   74 & \textbf{Group detected}, baseline problem?, partial blend with  \\ 
 & & & & & & & CH$_3$CH$_3$CO, v$_{\mathrm{t}}$=1 and CH$_3$OCH$_3$ \\ 
 386 & 29$_{15,14}$ - 28$_{15,13}$$^\star$ &  263628.792 &    50 &  453 &         141 &   74 & Strong CH$_3$OCH$_3$ \\ 
 388 & 29$_{16,13}$ - 28$_{16,12}$$^\star$ &  263712.865 &    57 &  491 &         134 &  108 & Weak, strong HCC$^{13}$CN, v$_7$=1 and HNCO, v$_6$=1 \\ 
 390 & 29$_{17,12}$ - 28$_{17,11}$$^\star$ &  263805.068 &    67 &  531 &         126 &  108 & Weak, strong HC$_3$N, v$_5$=1$/$v$_7$=3 and C$_2$H$_5$CN, v=0 \\ 
 392 & 29$_{ 5,25}$ - 28$_{ 5,24}$ &  263857.842 &    12 &  208 &         187 &  108 & Blend with HCONH$_2$, v=0 and C$_2$H$_5$$^{13}$CN, v=0 \\ 
 393 & 29$_{18,11}$ - 28$_{18,10}$$^\star$ &  263904.858 &    81 &  574 &         118 &  108 & Weak, blend with SO$_2$, v=0, C$_2$H$_5$CN, v$_{13}$=1/v$_{21}$=1, and U-line? \\ 
 395 & 29$_{ 5,24}$ - 28$_{ 5,23}$ &  263928.994 &    13 &  208 &         187 &  108 & Blend with C$_2$H$_5$CN, v$_{13}$=1/v$_{21}$=1 and C$_2$H$_5$CN, v=0 \\ 
 396 & 29$_{19,10}$ - 28$_{19, 9}$$^\star$ &  264011.816 &   101 &  619 &         110 &  108 & Weak, strong CH$_3$$^{13}$CH$_2$CN, v=0 \\ 
 398 & 29$_{ 4,26}$ - 28$_{ 4,25}$ &  264055.836 &    13 &  197 &         189 &  108 & \textbf{Detected}, partial blend with C$_2$H$_5$CN, v=0 and  \\ 
 & & & & & & & CH$_3$CH$_3$CO, v=0 \\ 
 \hline
 \end{tabular}
 \end{table*}

}

\addtocounter{figure}{1}
\newcounter{appfig}
\setcounter{appfig}{\value{figure}}
\onlfig{\value{appfig}}{
\begin{figure*}
 
\centerline{\resizebox{0.85\hsize}{!}{\includegraphics[angle=270]{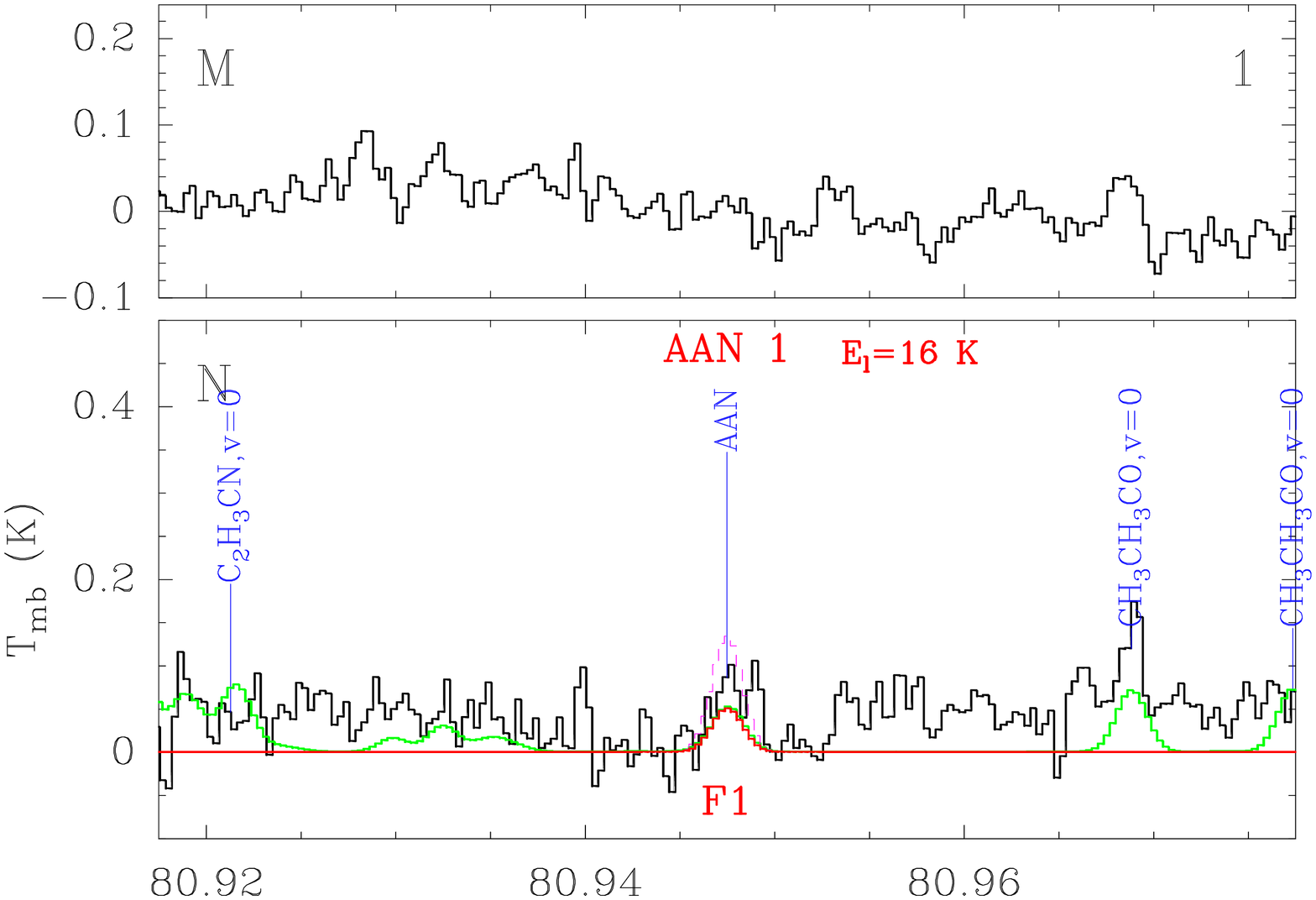}\includegraphics[angle=270]{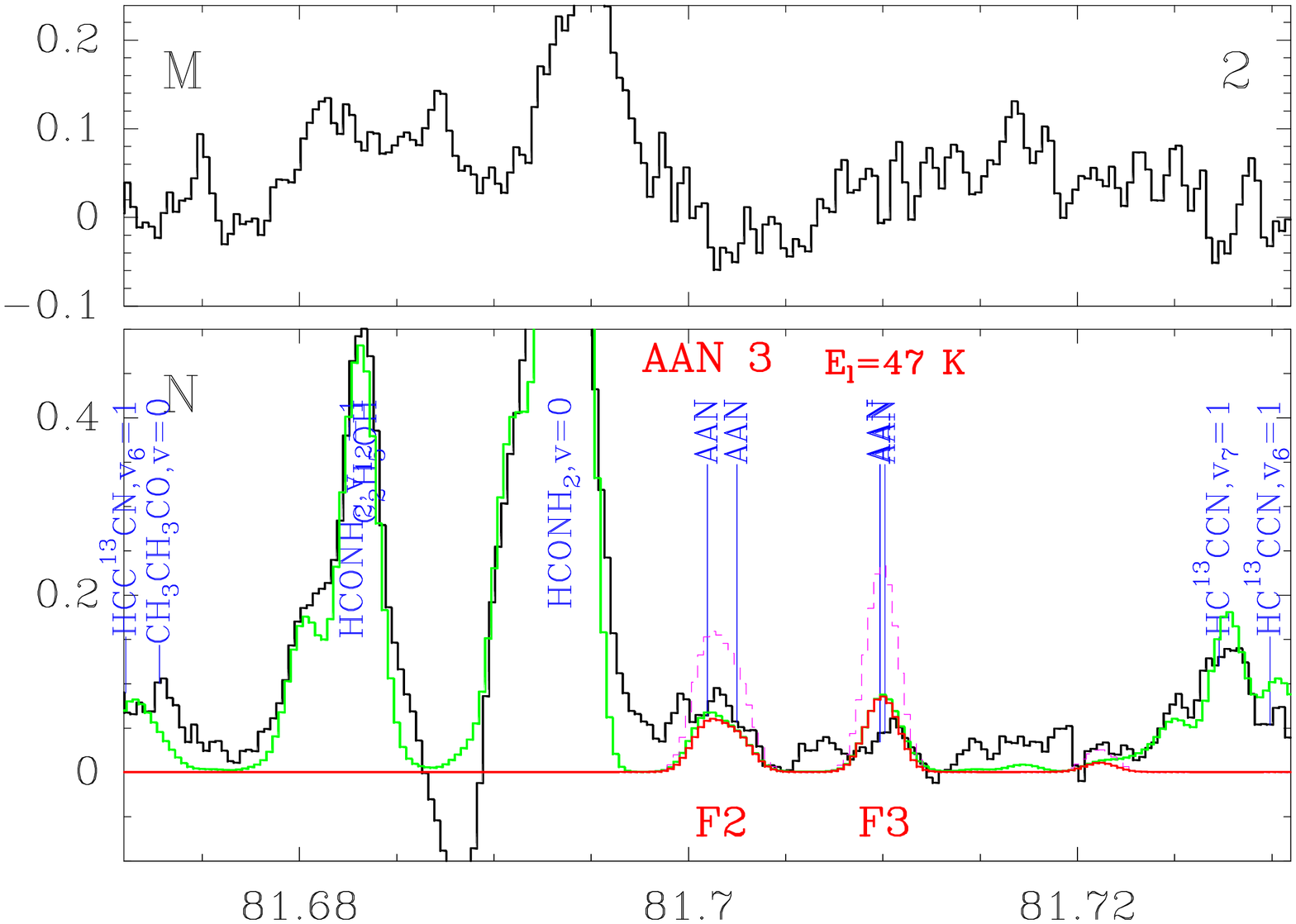}}}
\vspace*{-0.4ex}
\centerline{\resizebox{0.85\hsize}{!}{\includegraphics[angle=270]{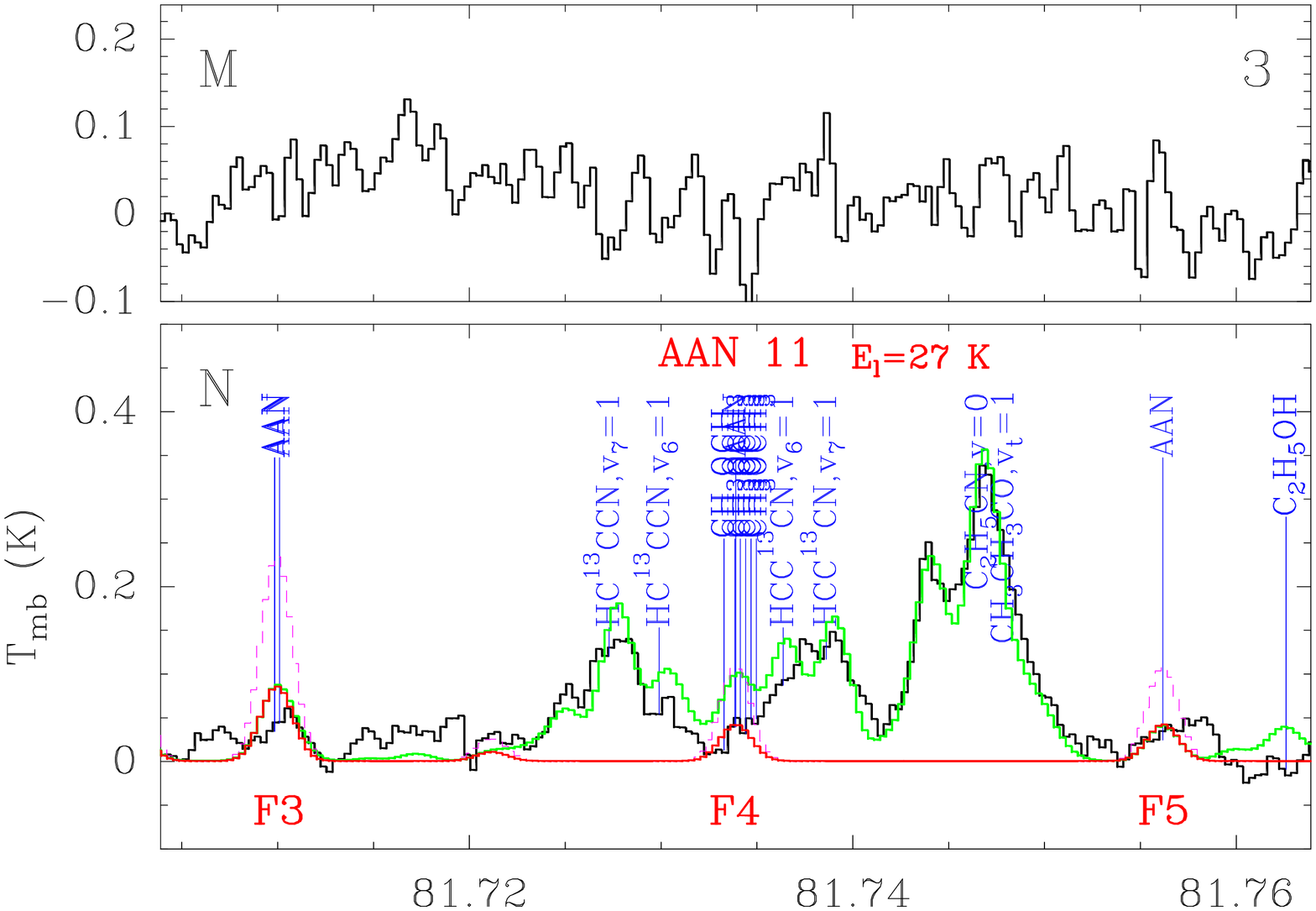}\includegraphics[angle=270]{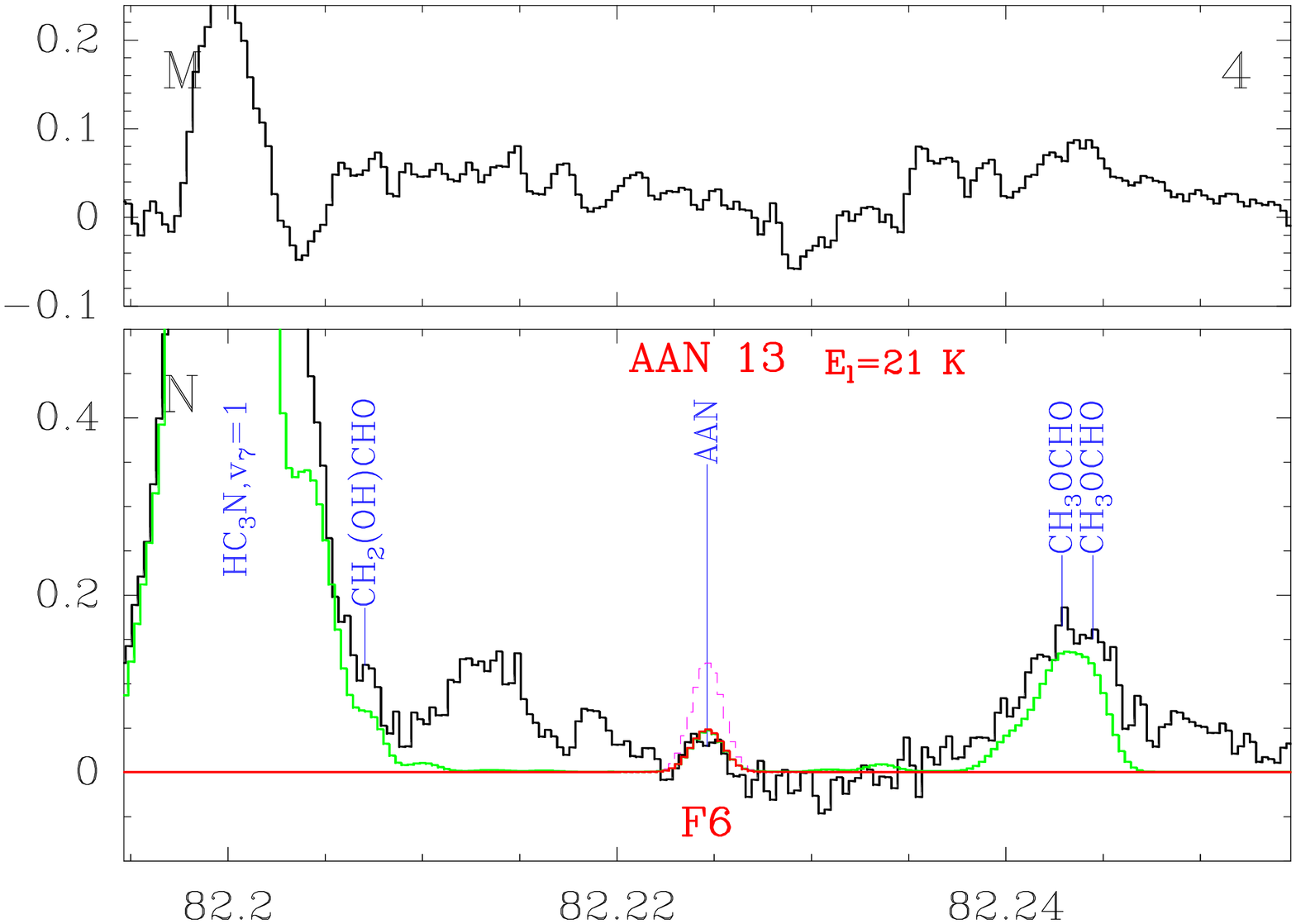}}}
\vspace*{-0.4ex}
\centerline{\resizebox{0.85\hsize}{!}{\includegraphics[angle=270]{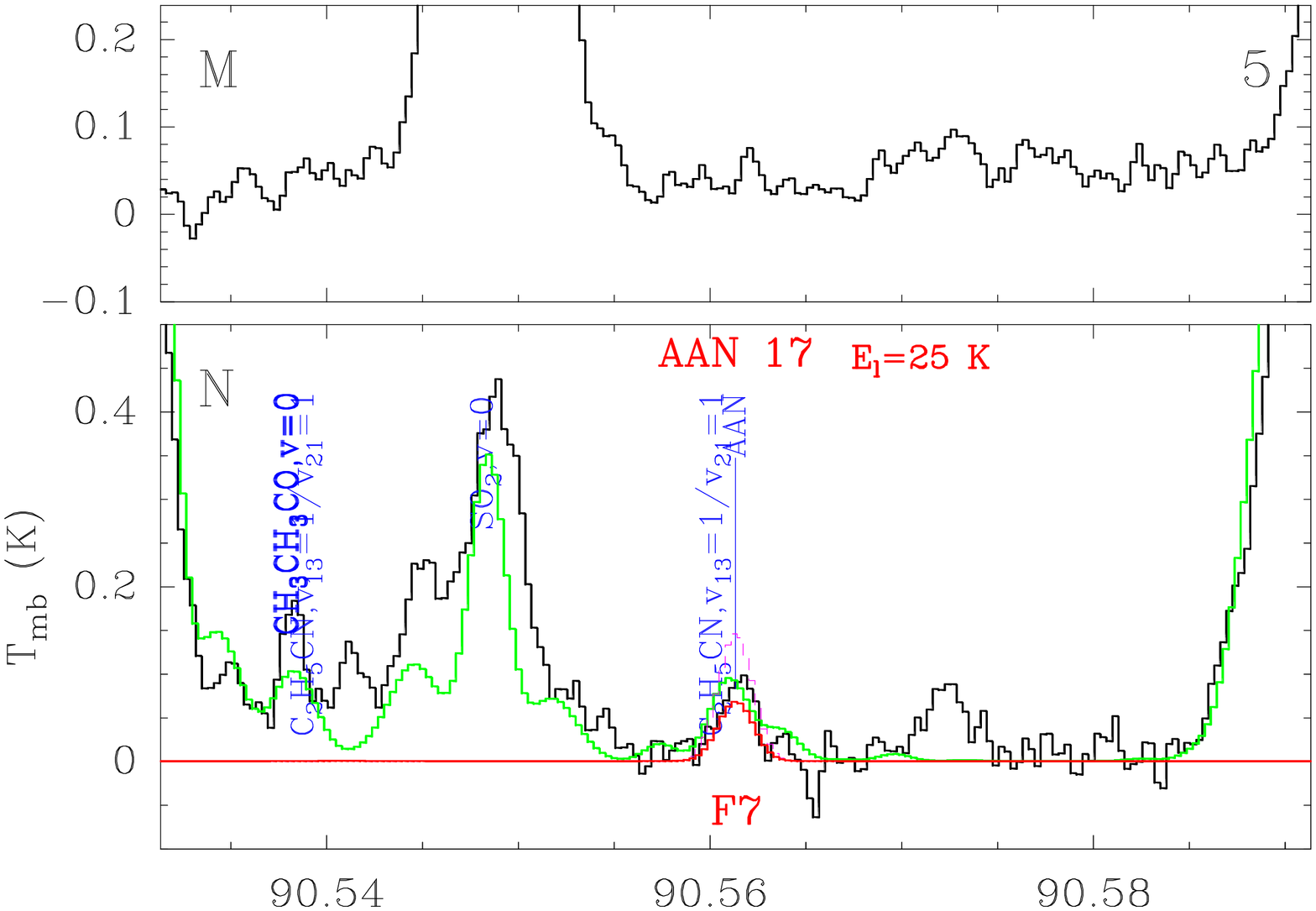}\includegraphics[angle=270]{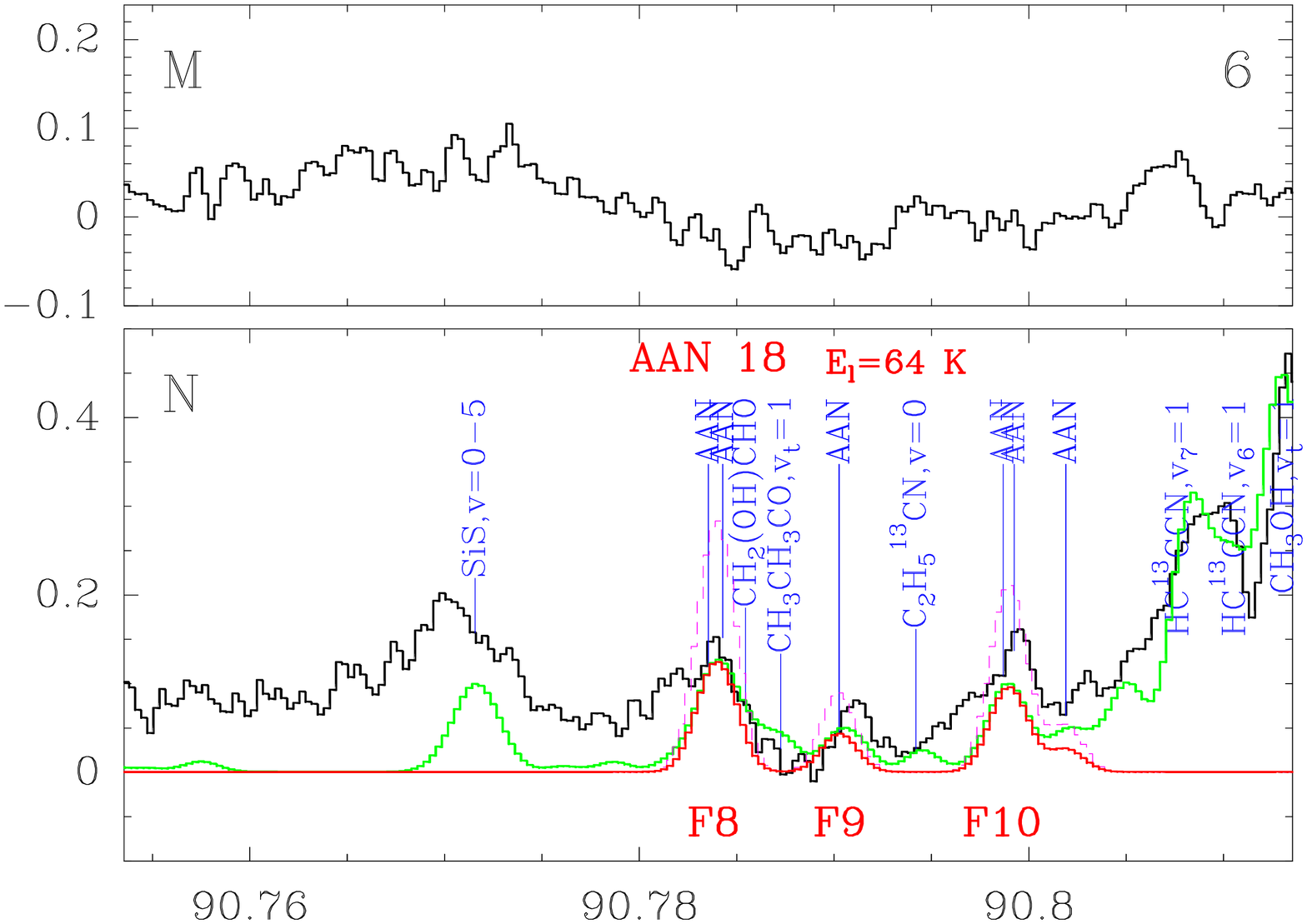}}}
\vspace*{-0.4ex}
\centerline{\resizebox{0.85\hsize}{!}{\includegraphics[angle=270]{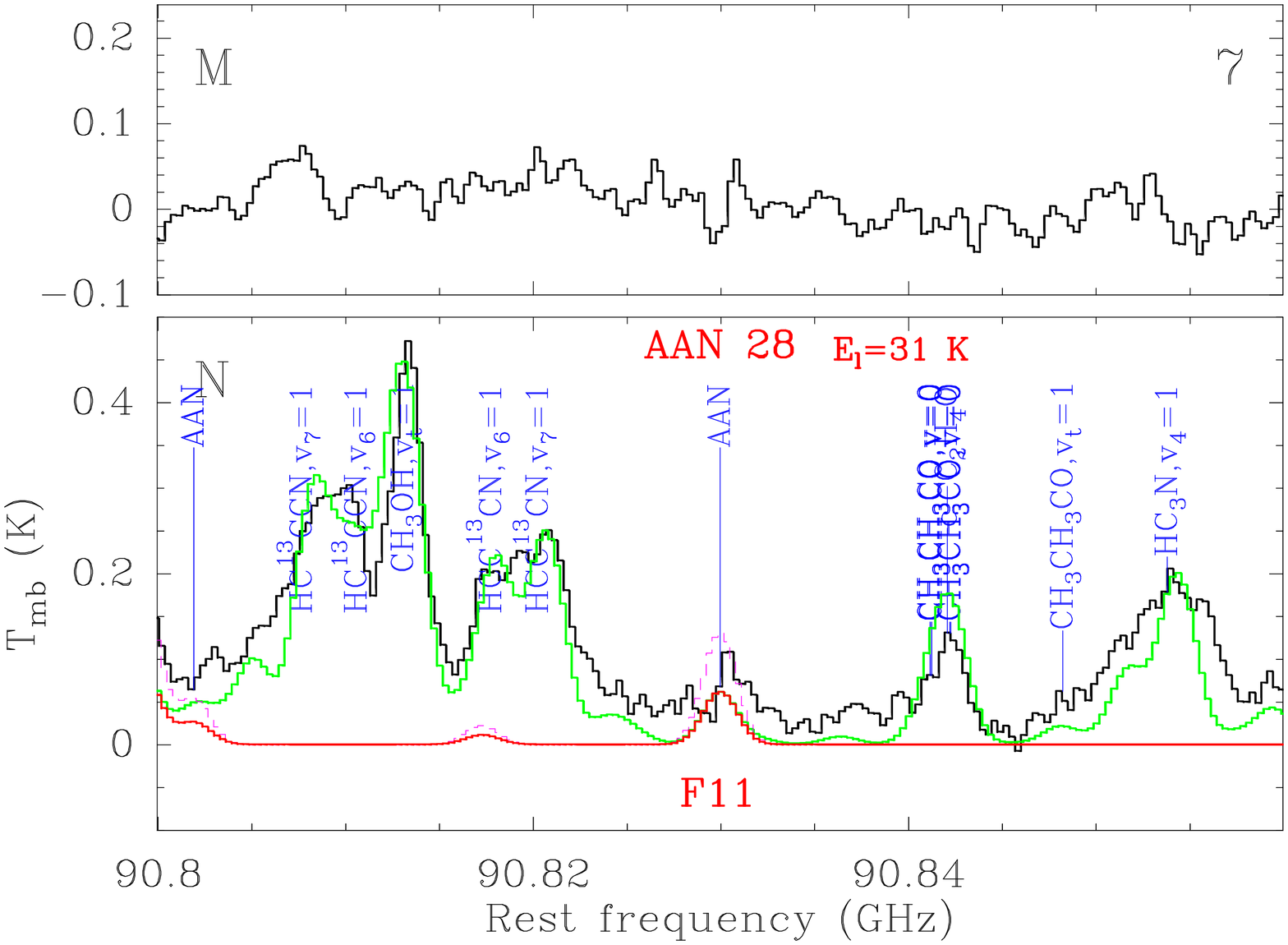}\includegraphics[angle=270]{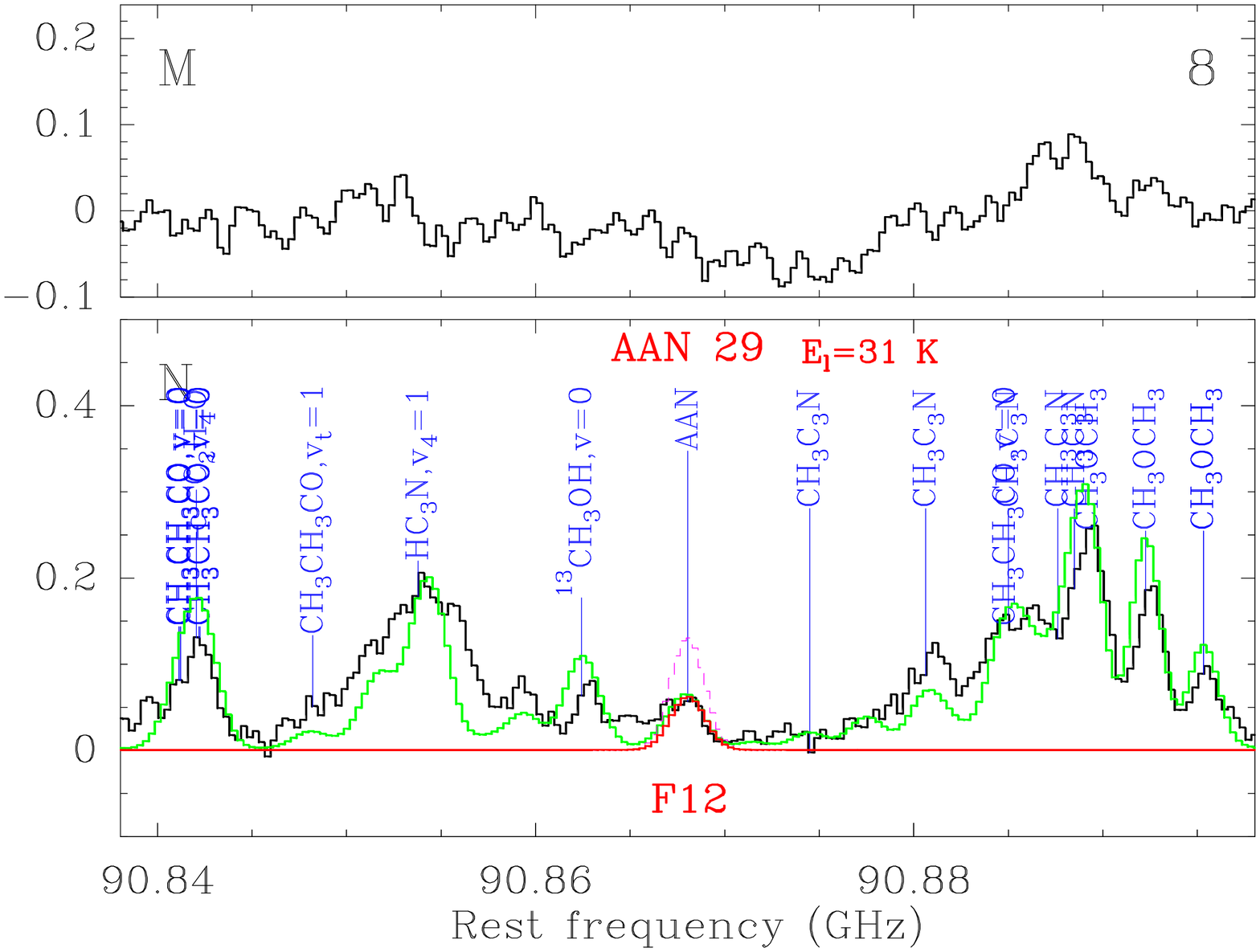}}}
\vspace*{-0.4ex}
\caption{
Transitions of amino acetonitrile (AAN) detected with the IRAM 30m telescope.
Each panel consists of two plots and is labeled in black in the upper right corner.
The lower plot shows in black the spectrum obtained toward Sgr~B2(N) in main-beam temperature scale (K), while the upper plot shows the spectrum toward Sgr~B2(M). The rest frequency axis is labeled in GHz. The systemic velocities assumed for Sgr~B2(N) and (M) are 64 and 62 km~s$^{-1}$, respectively.
The lines identified in the Sgr~B2(N) spectrum are labeled in blue. The top red label indicates the AAN transition centered in each plot (numbered like in col.~1 of Table~\ref{t:detectaan30m}), along with the energy of its lower level in K ($E_l/k_{\mathrm{B}}$).
The other AAN lines are labeled in blue only.
The bottom red label is the feature number (see col.~8 of Table~\ref{t:detectaan30m}).
The green spectrum shows our LTE model containing all identified molecules, including AAN.
The LTE synthetic spectrum of AAN alone is overlaid in red, and its opacity in dashed violet.
All observed lines which have no counterpart in the green spectrum are still unidentified in Sgr~B2(N).
}
\label{f:detectaan30m}
\end{figure*}
\begin{figure*}

\centerline{\resizebox{0.85\hsize}{!}{\includegraphics[angle=270]{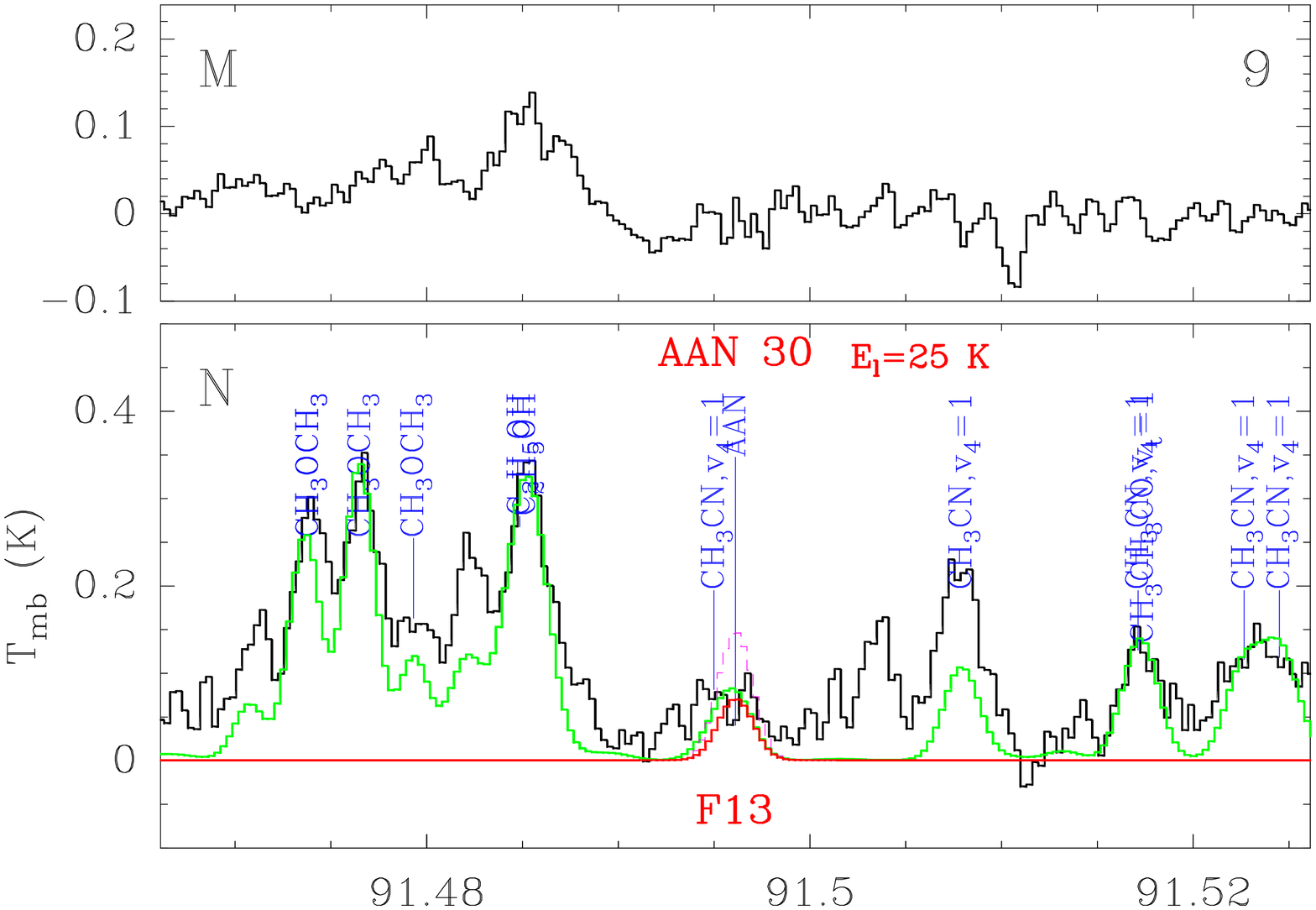}\includegraphics[angle=270]{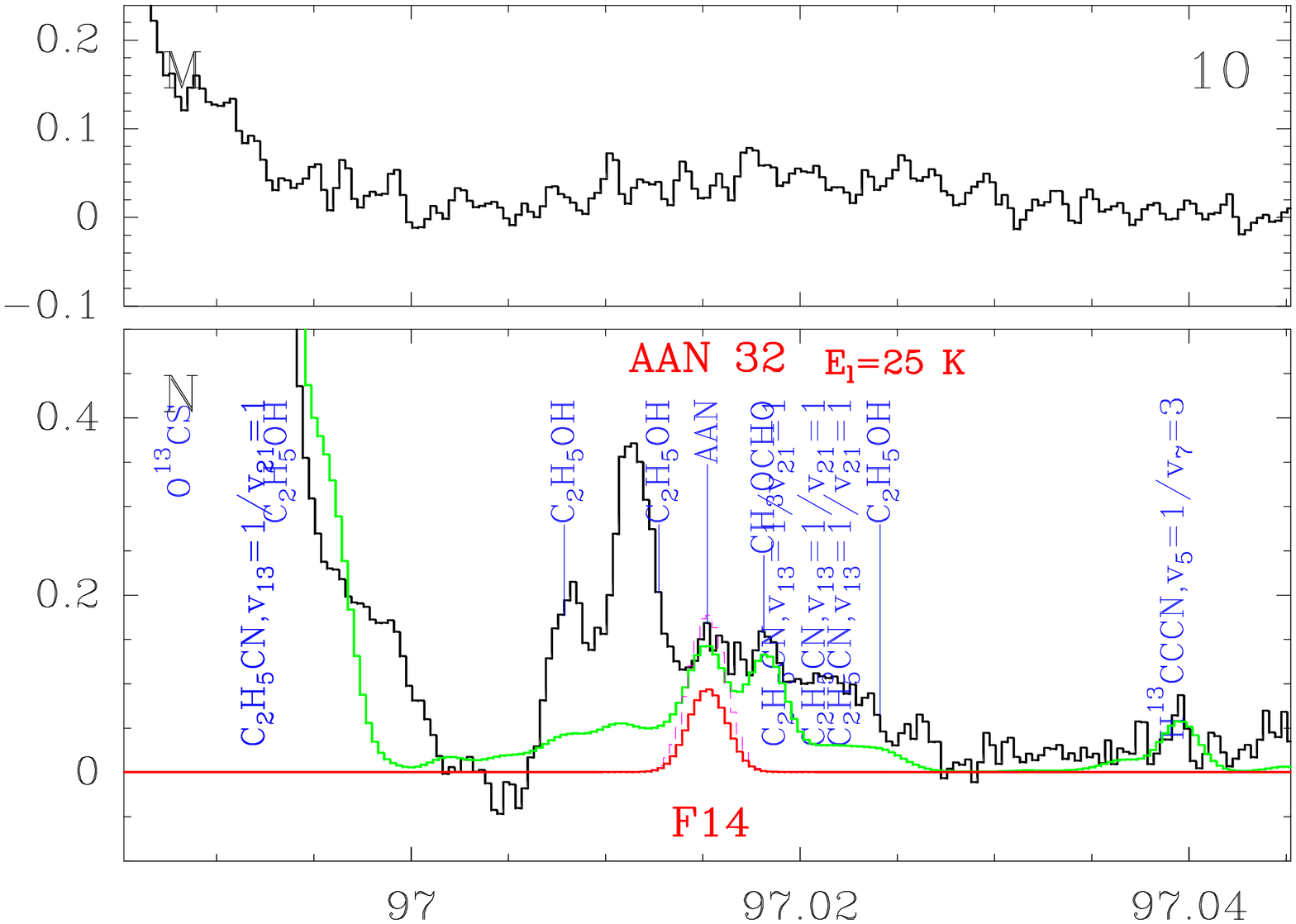}}}
\vspace*{-0.4ex}
\centerline{\resizebox{0.85\hsize}{!}{\includegraphics[angle=270]{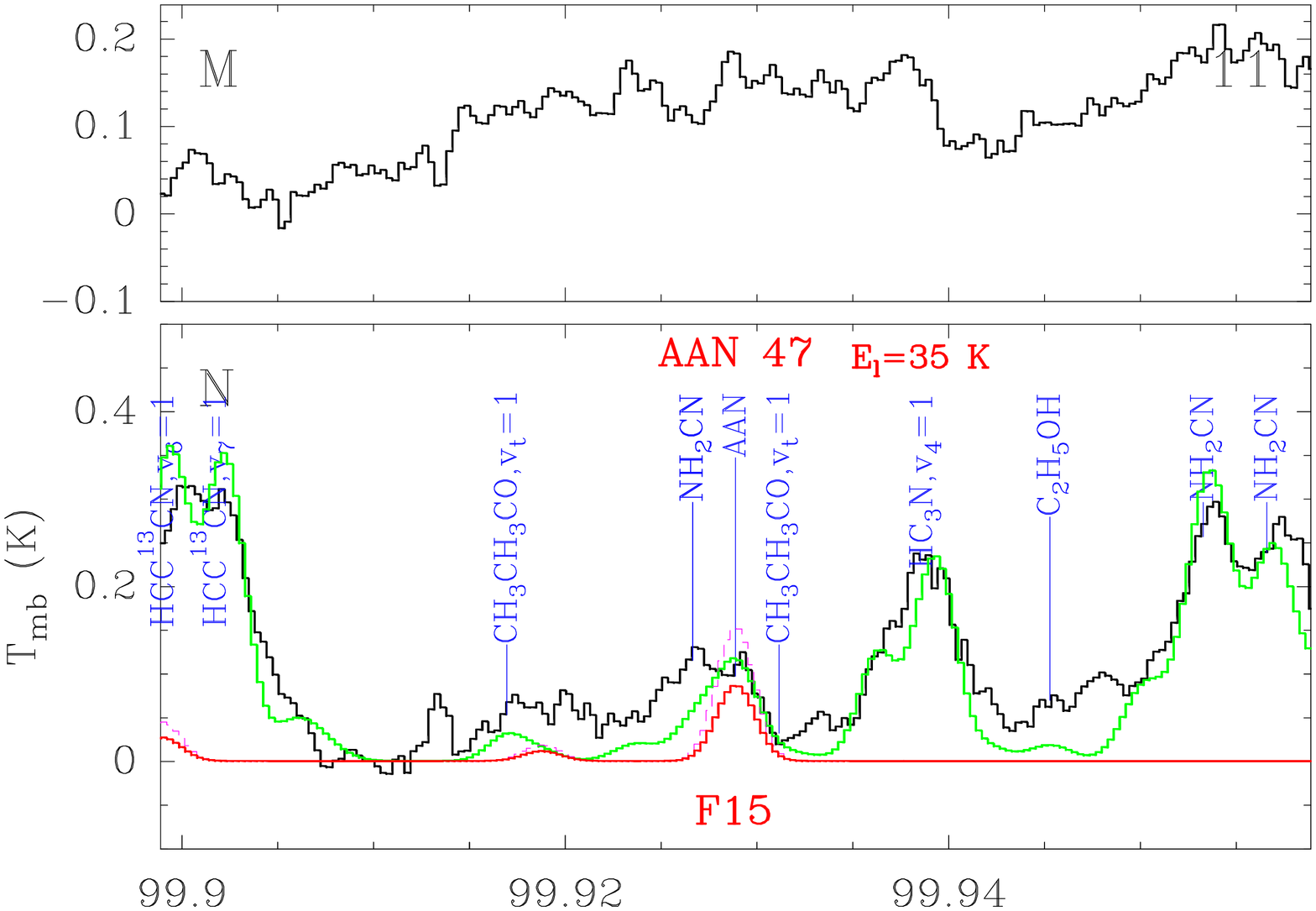}\includegraphics[angle=270]{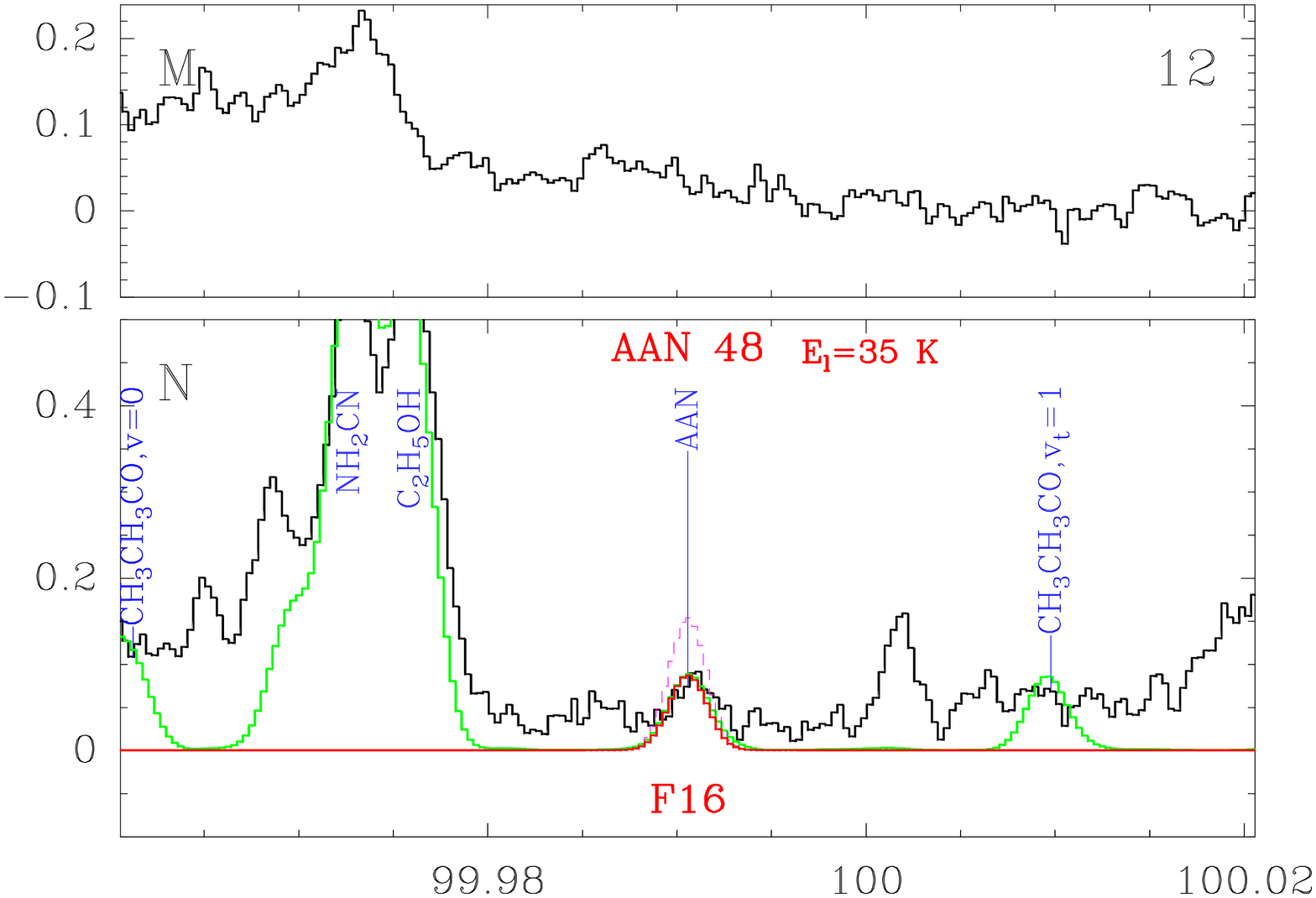}}}
\vspace*{-0.4ex}
\centerline{\resizebox{0.85\hsize}{!}{\includegraphics[angle=270]{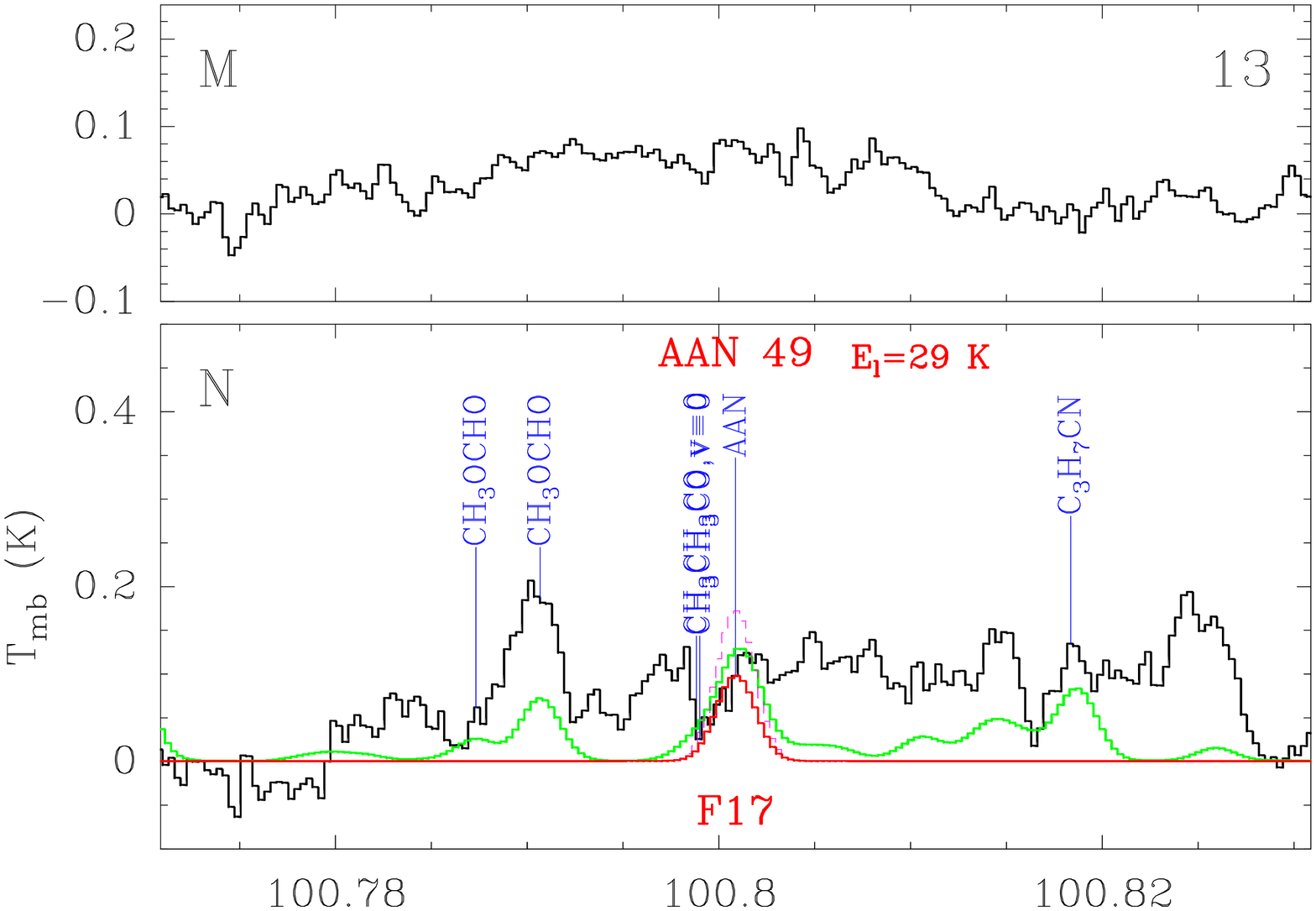}\includegraphics[angle=270]{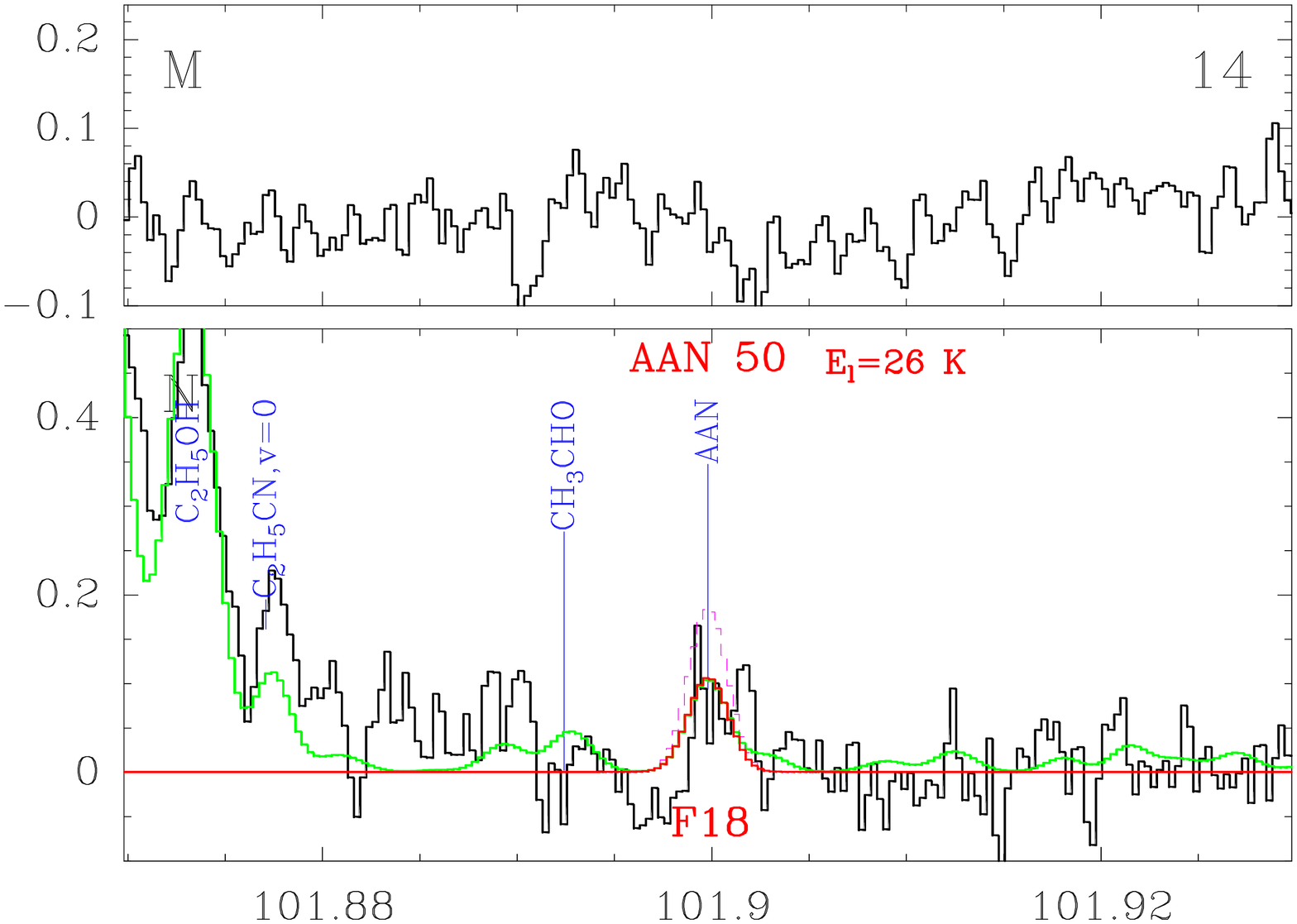}}}
\vspace*{-0.4ex}
\centerline{\resizebox{0.85\hsize}{!}{\includegraphics[angle=270]{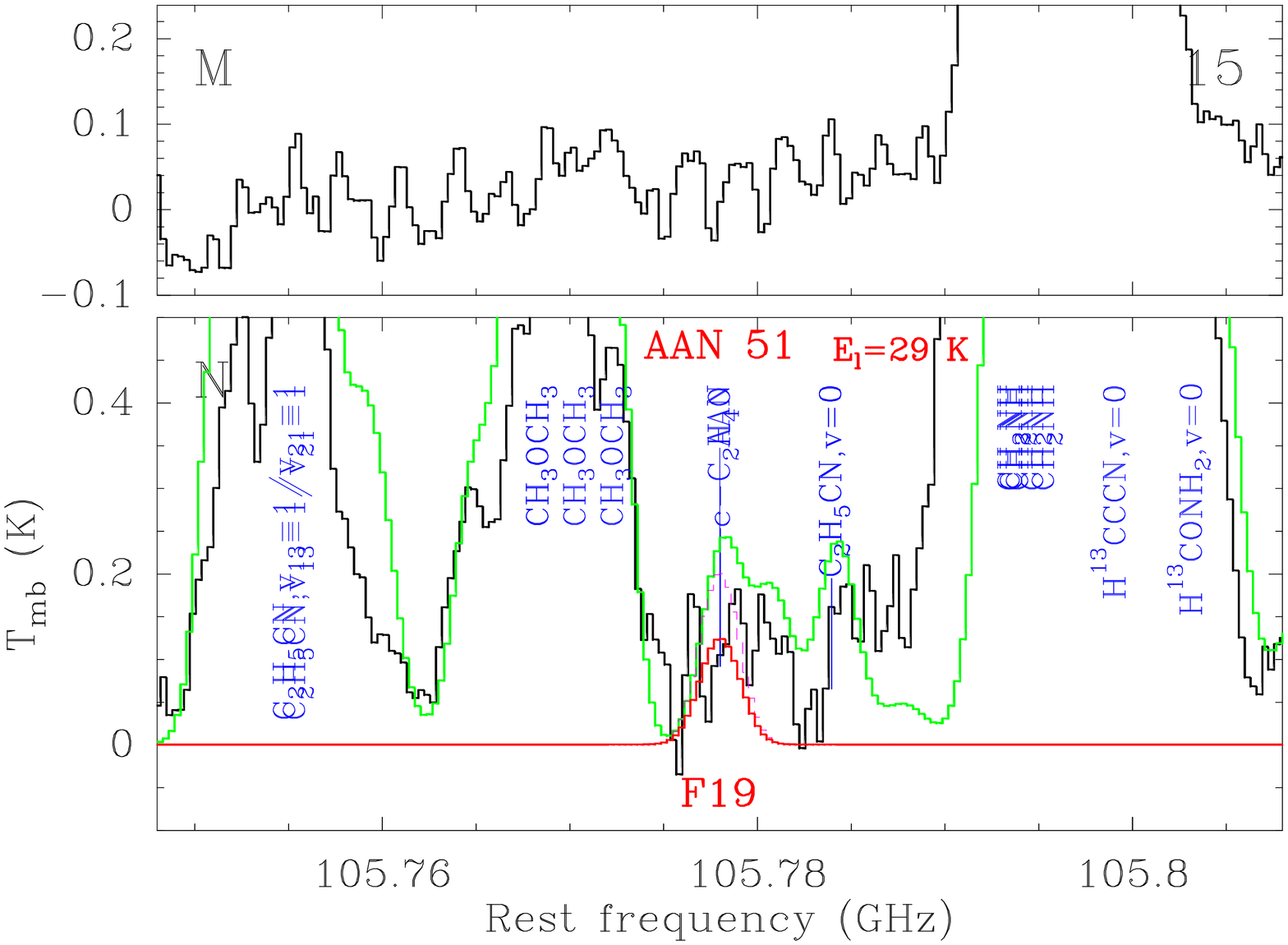}\includegraphics[angle=270]{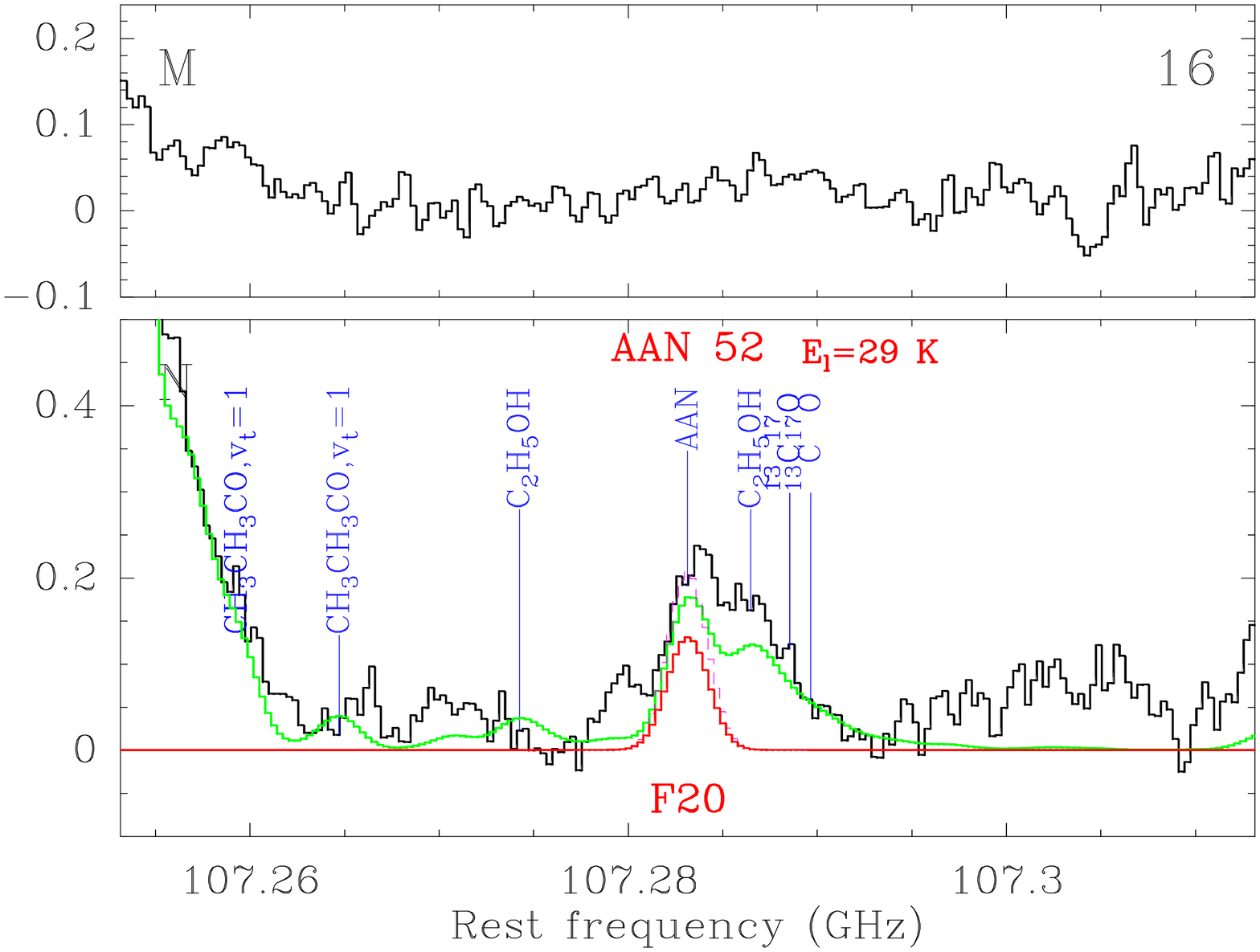}}}
\vspace*{-0.4ex}
\addtocounter{figure}{-1}
\caption{
(continued)
}
\label{f:detectaan30m}
\end{figure*}
\begin{figure*}

\centerline{\resizebox{0.85\hsize}{!}{\includegraphics[angle=270]{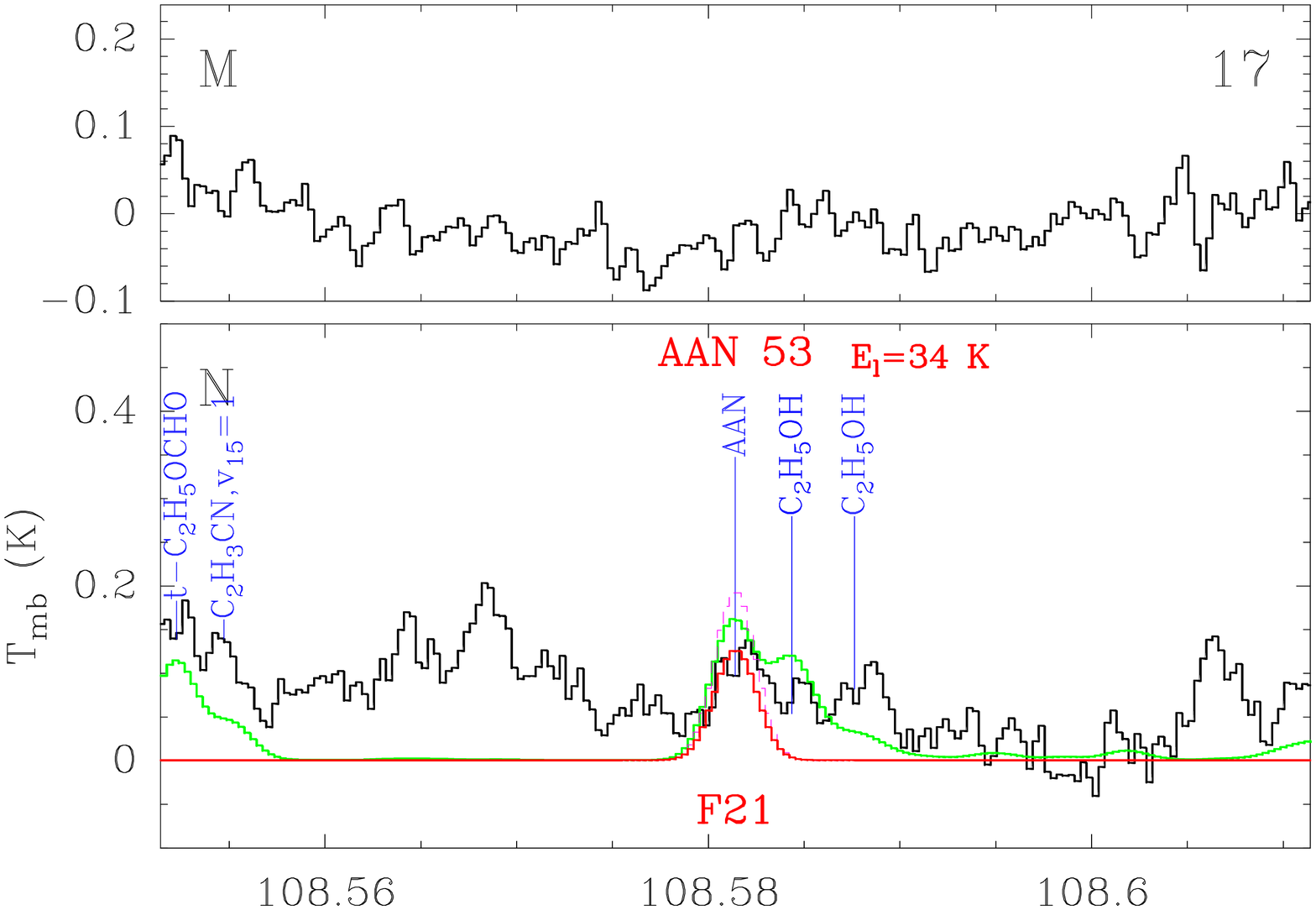}\includegraphics[angle=270]{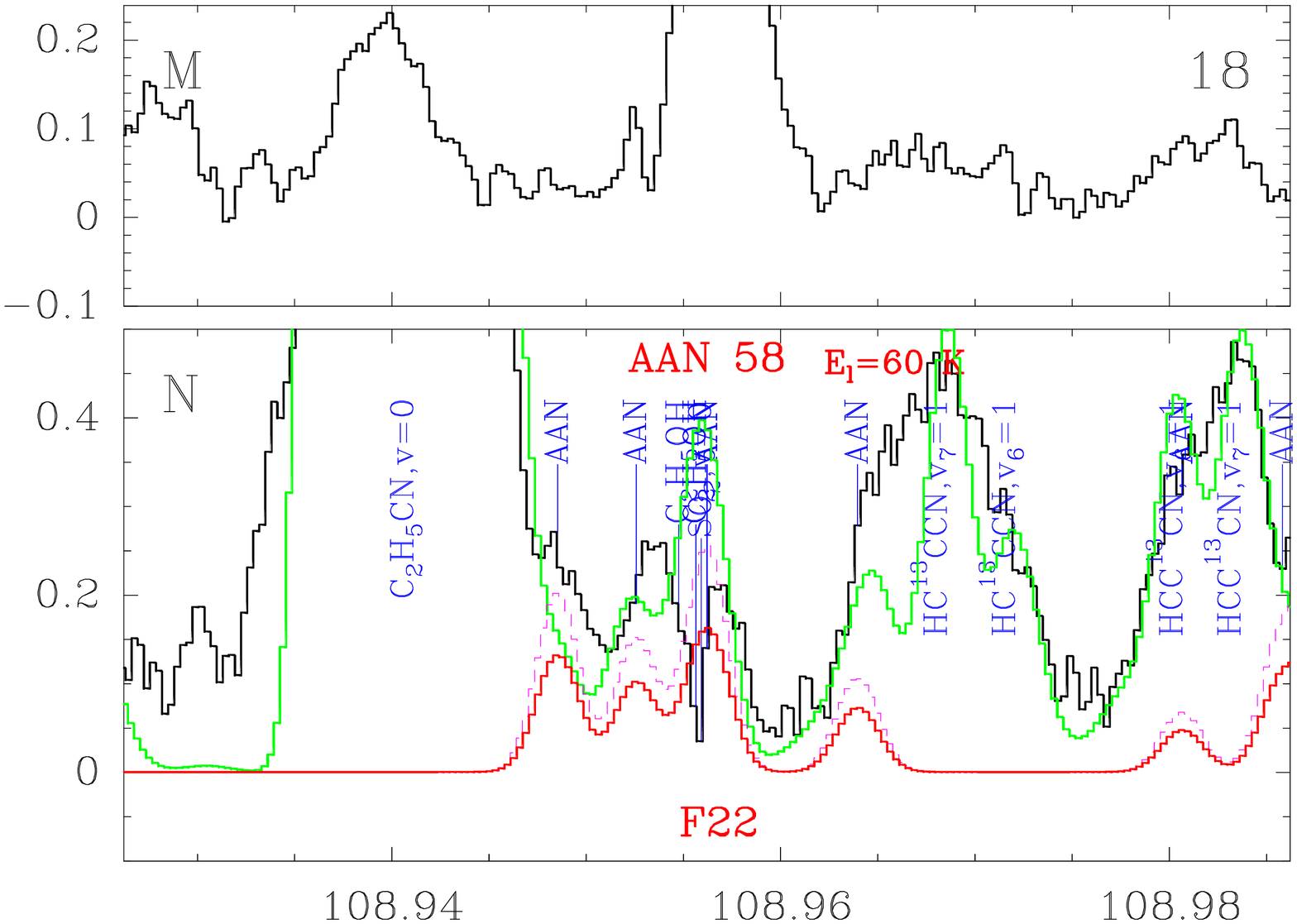}}}
\vspace*{-0.4ex}
\centerline{\resizebox{0.85\hsize}{!}{\includegraphics[angle=270]{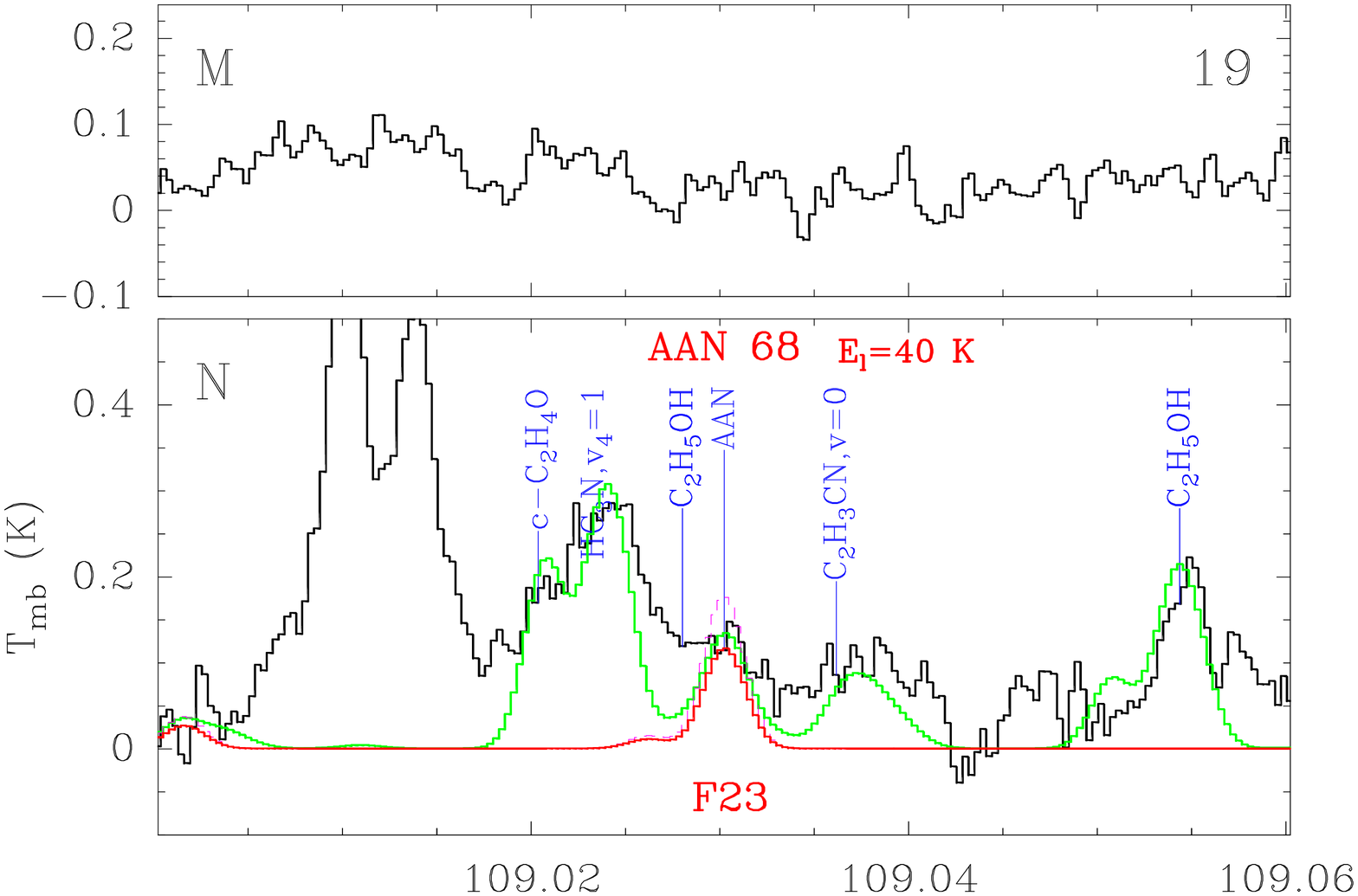}\includegraphics[angle=270]{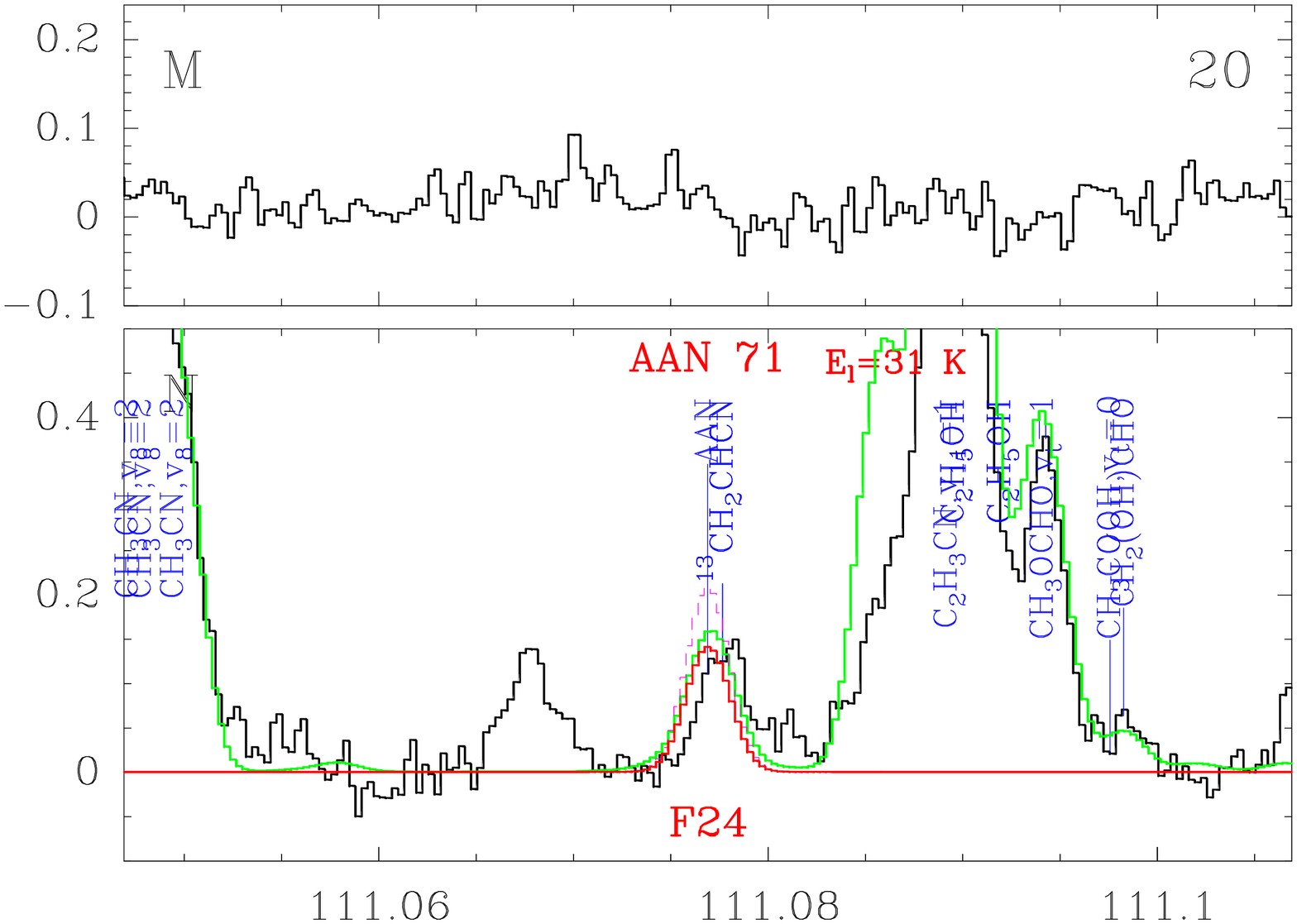}}}
\vspace*{-0.4ex}
\centerline{\resizebox{0.85\hsize}{!}{\includegraphics[angle=270]{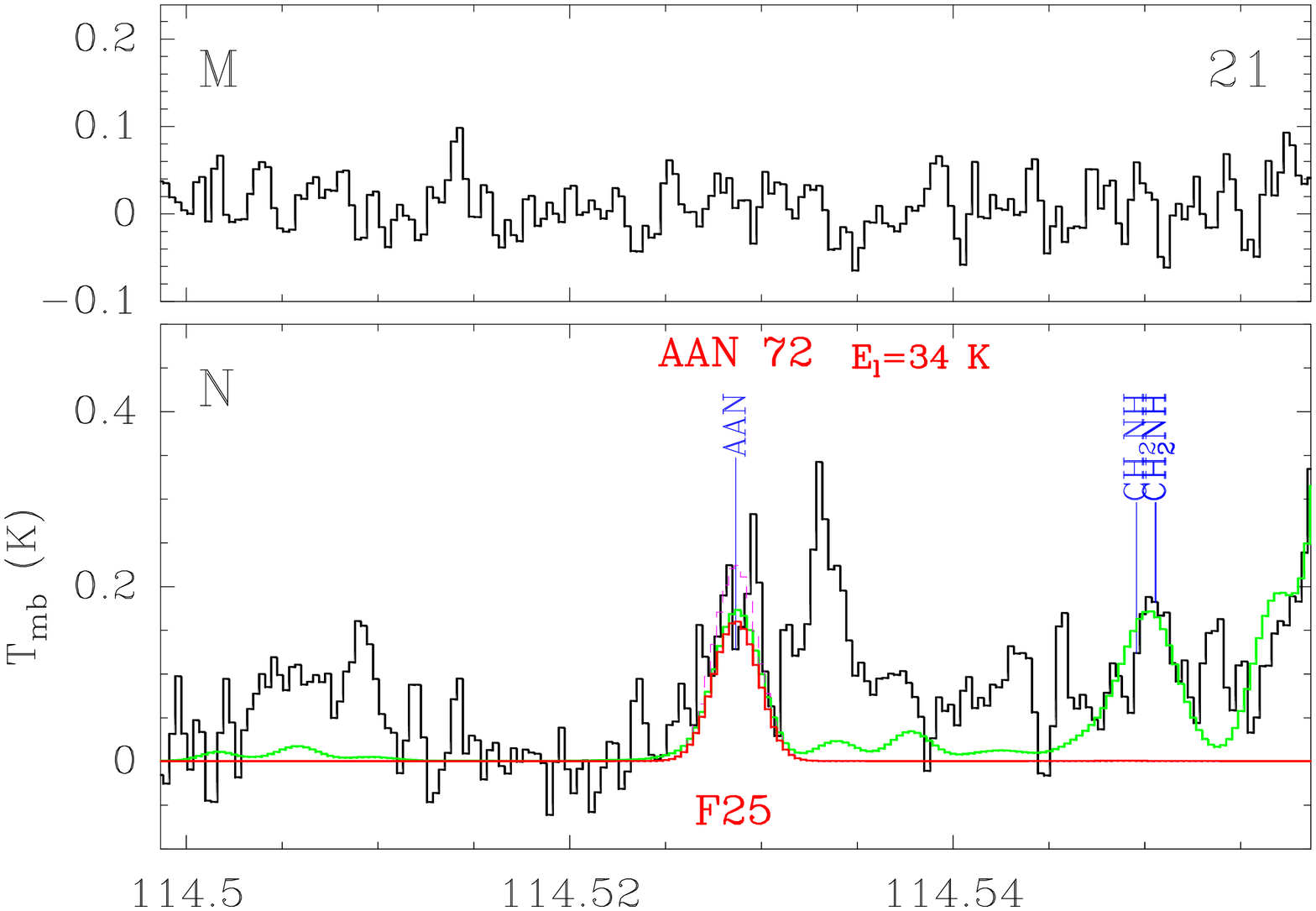}\includegraphics[angle=270]{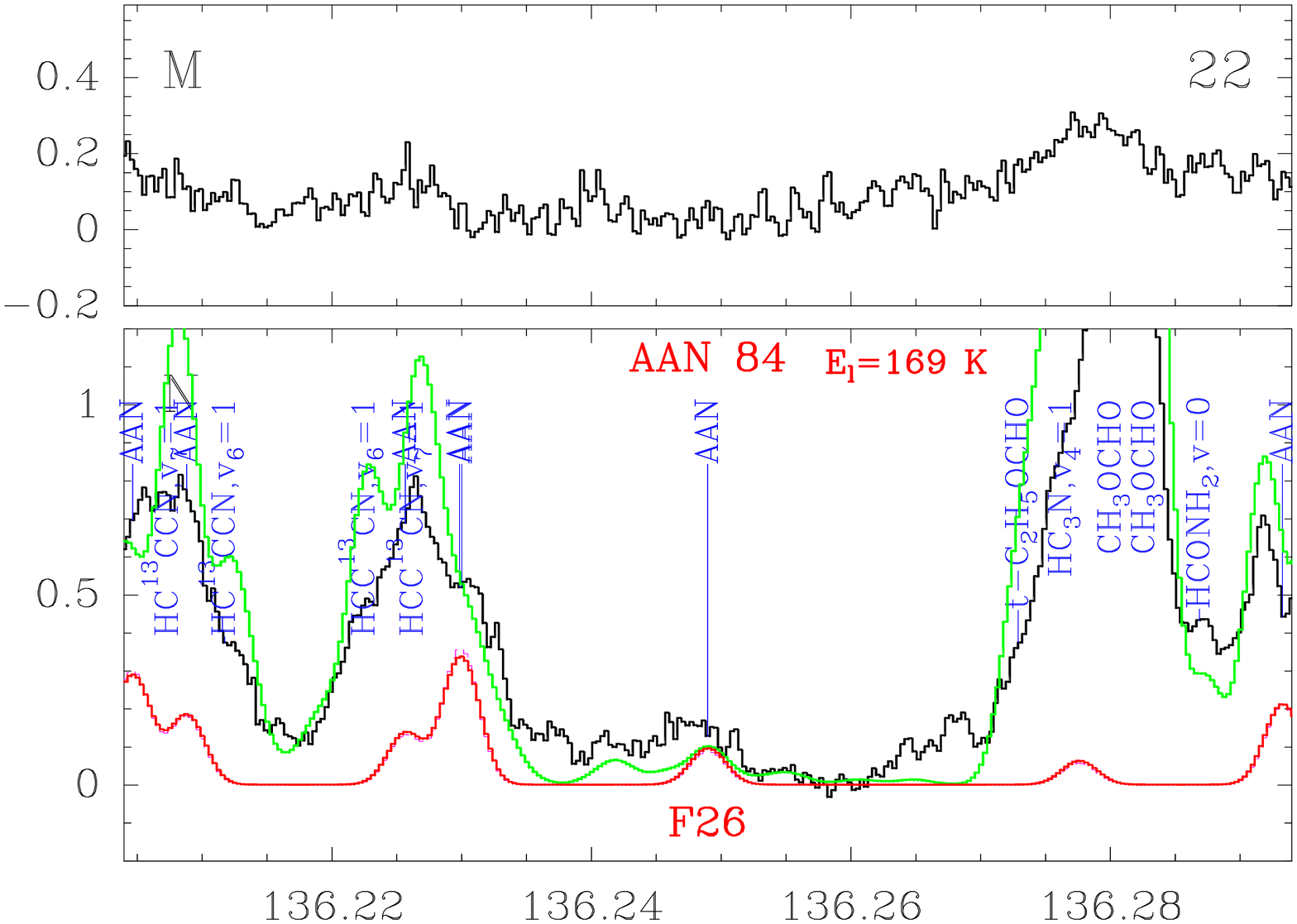}}}
\vspace*{-0.4ex}
\centerline{\resizebox{0.85\hsize}{!}{\includegraphics[angle=270]{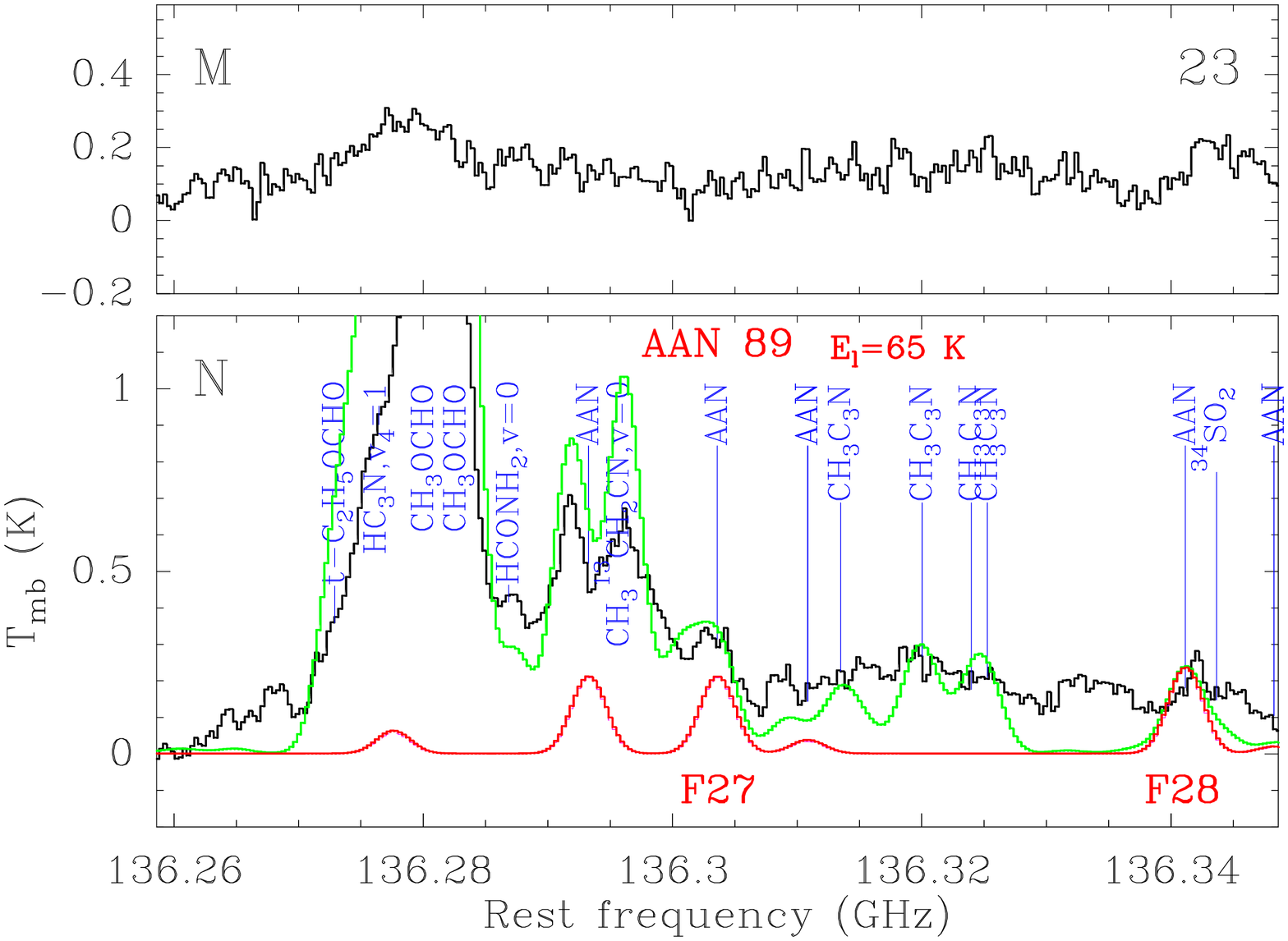}\includegraphics[angle=270]{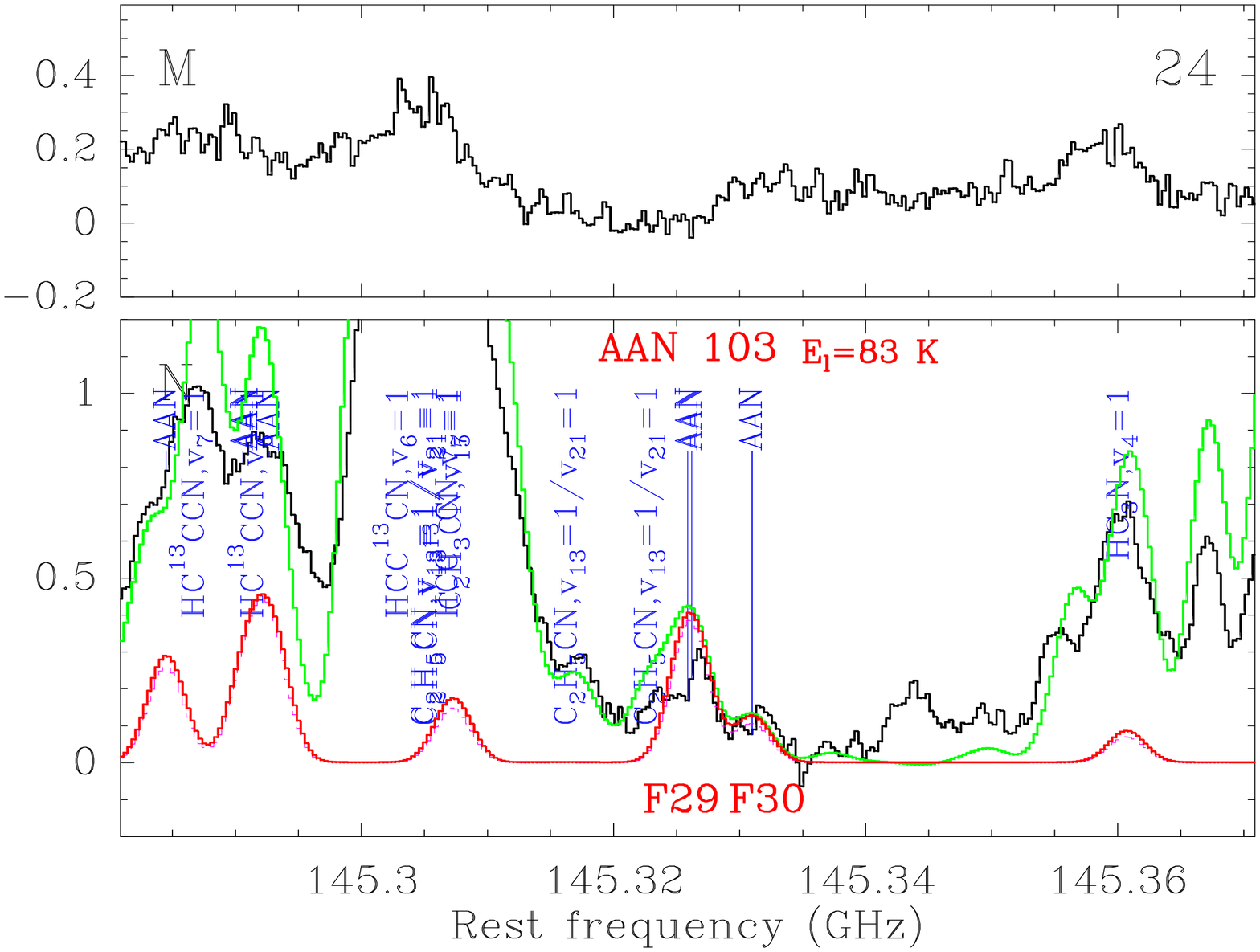}}}
\vspace*{-0.4ex}
\addtocounter{figure}{-1}
\caption{
(continued)
}
\label{f:detectaan30m}
\end{figure*}
\begin{figure*}

\centerline{\resizebox{0.85\hsize}{!}{\includegraphics[angle=270]{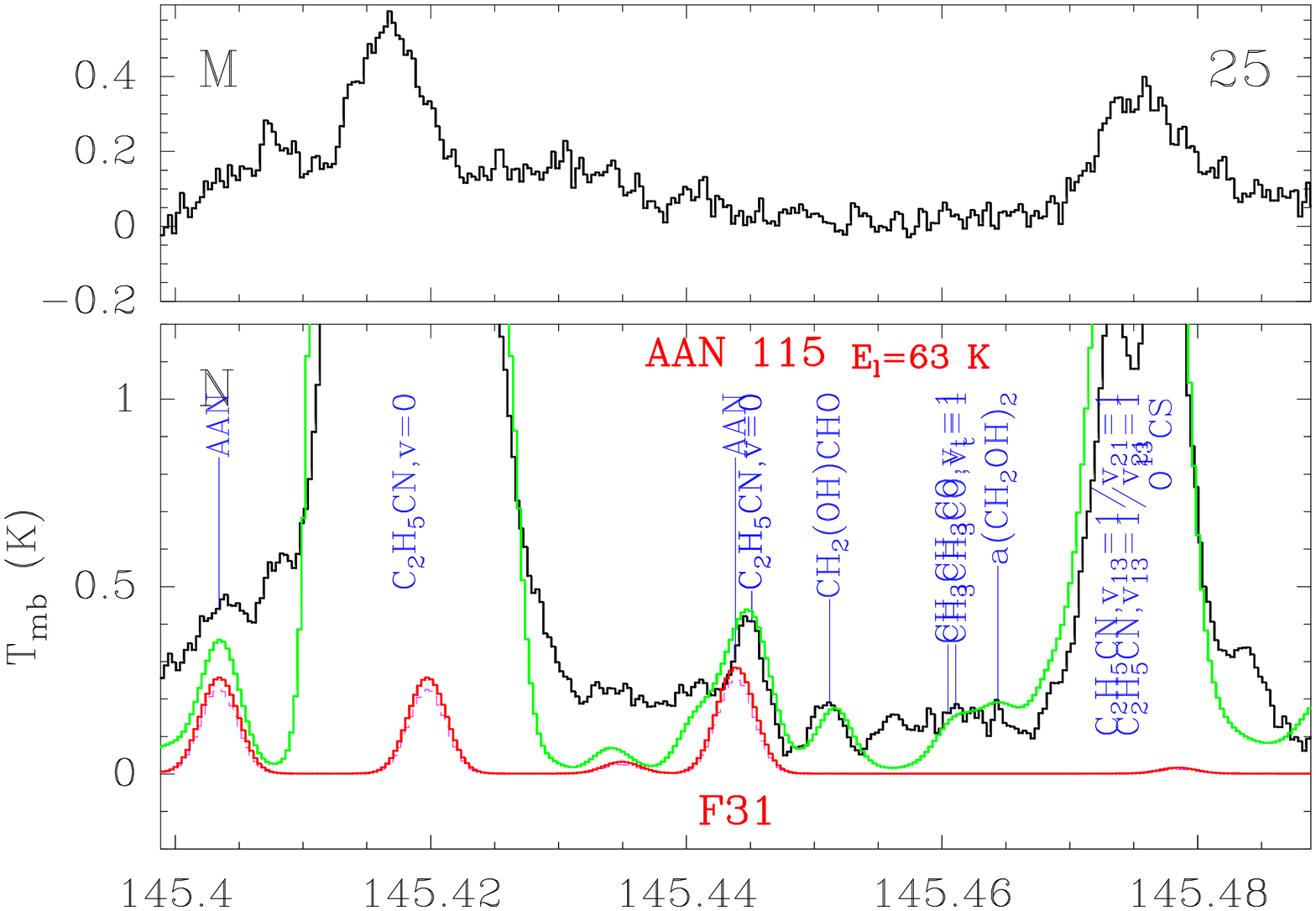}\includegraphics[angle=270]{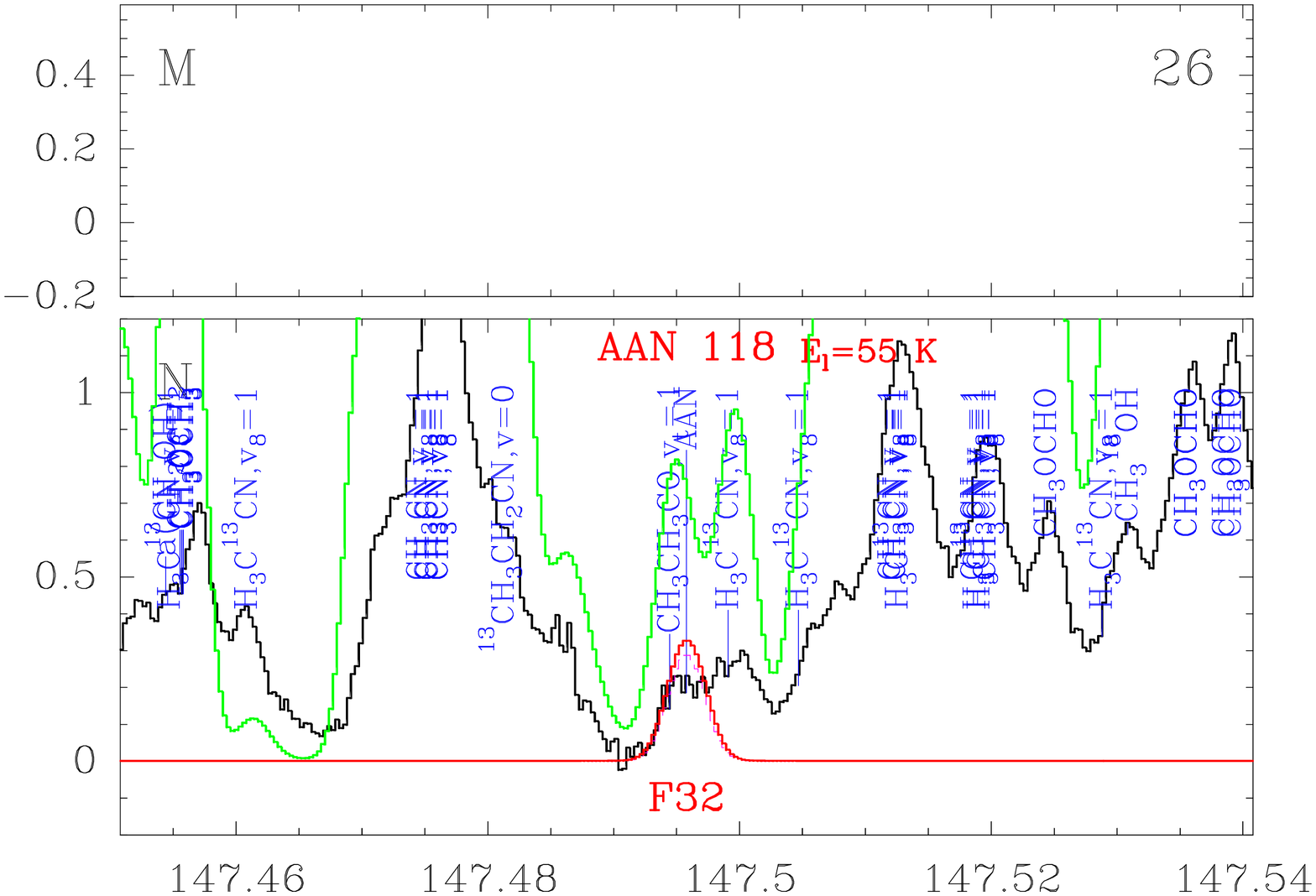}}}
\vspace*{-0.4ex}
\centerline{\resizebox{0.85\hsize}{!}{\includegraphics[angle=270]{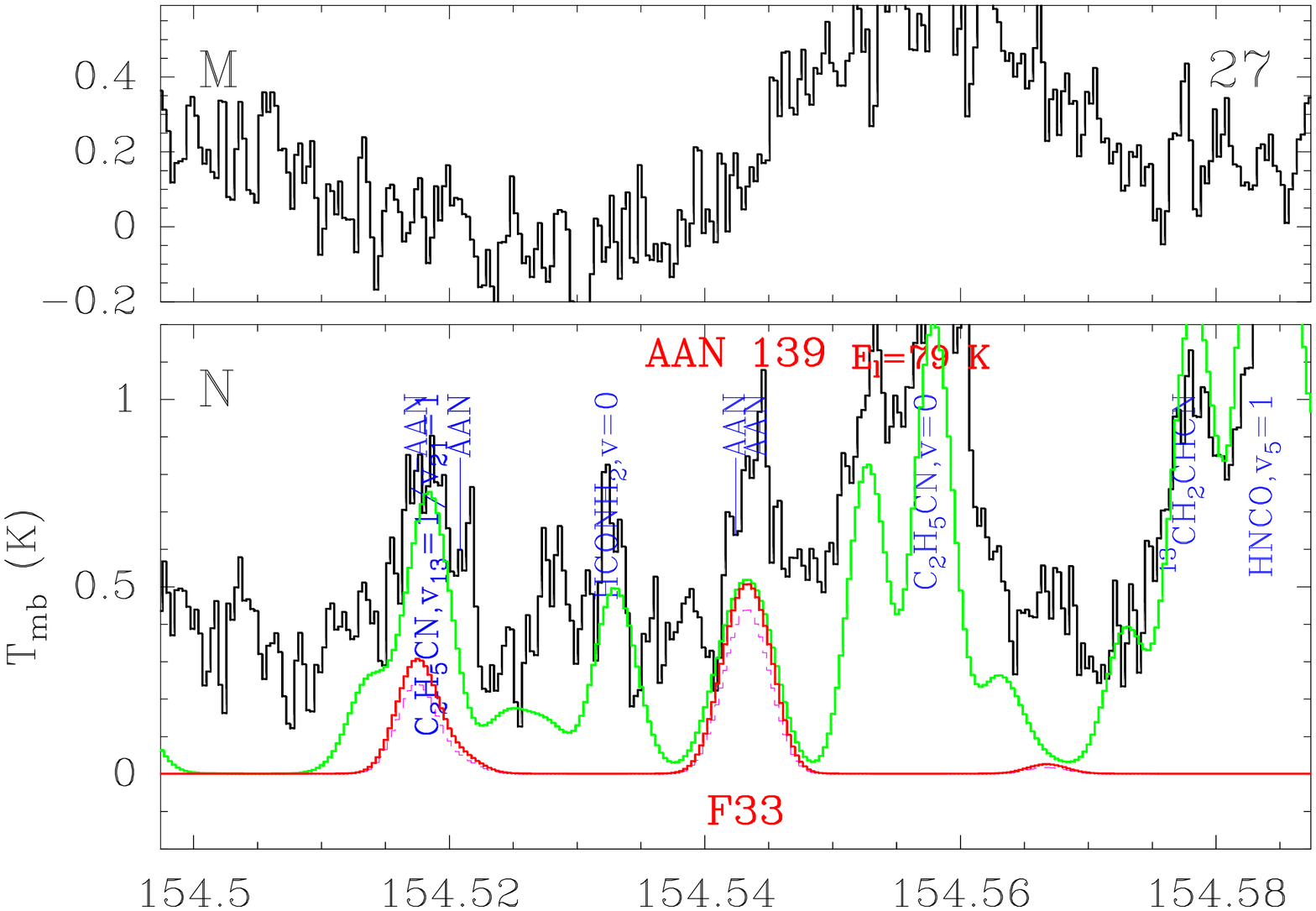}\includegraphics[angle=270]{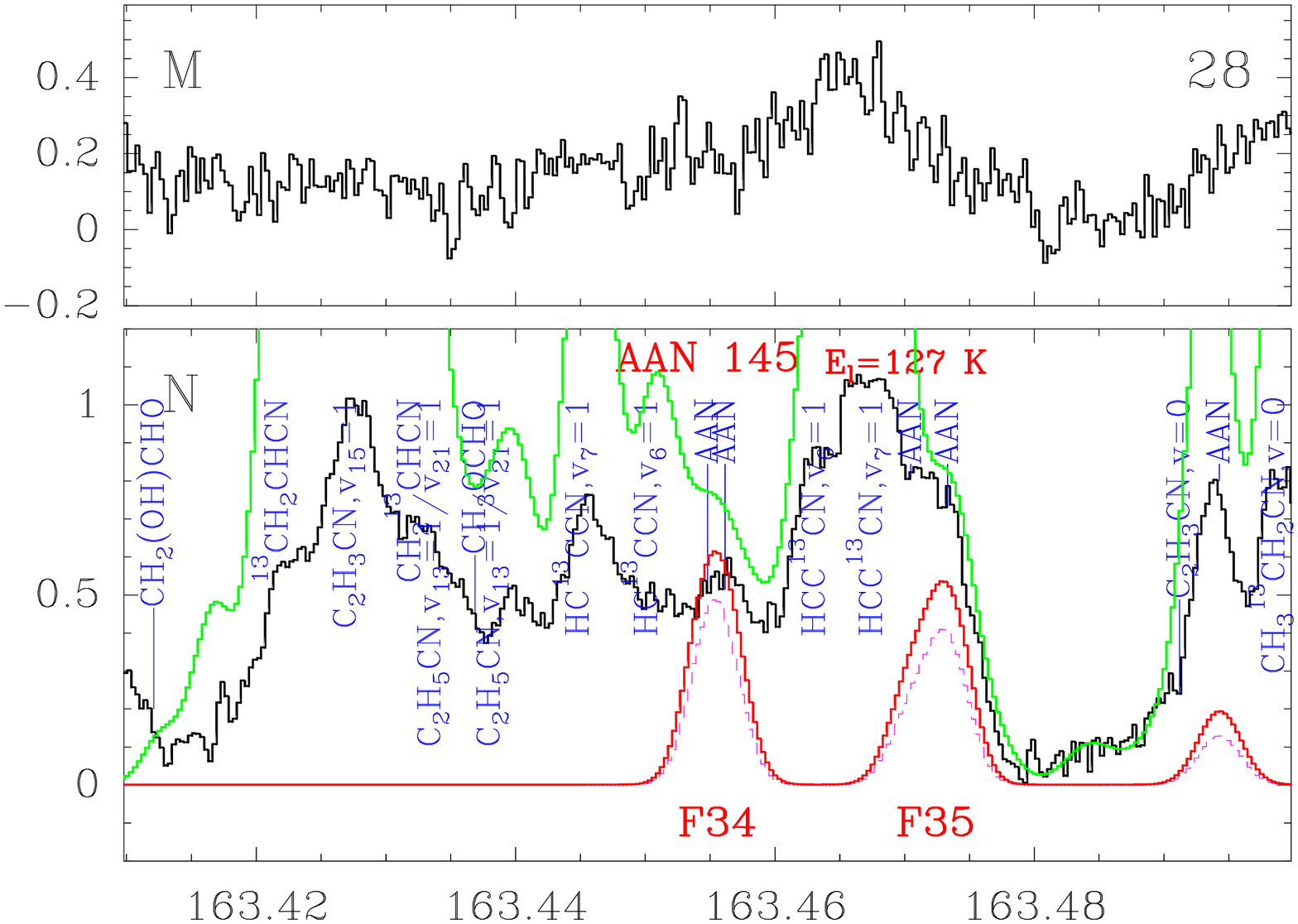}}}
\vspace*{-0.4ex}
\centerline{\resizebox{0.85\hsize}{!}{\includegraphics[angle=270]{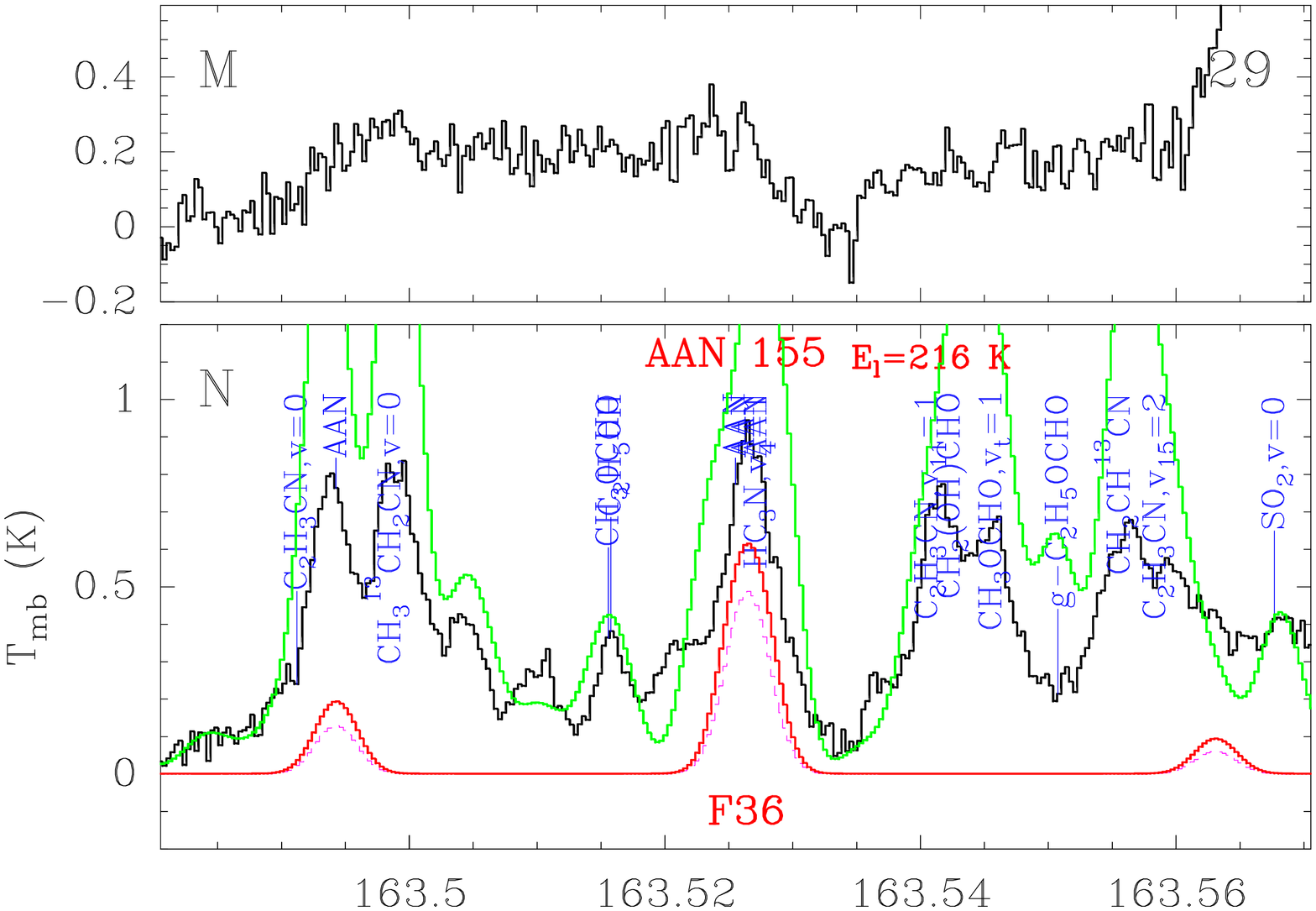}\includegraphics[angle=270]{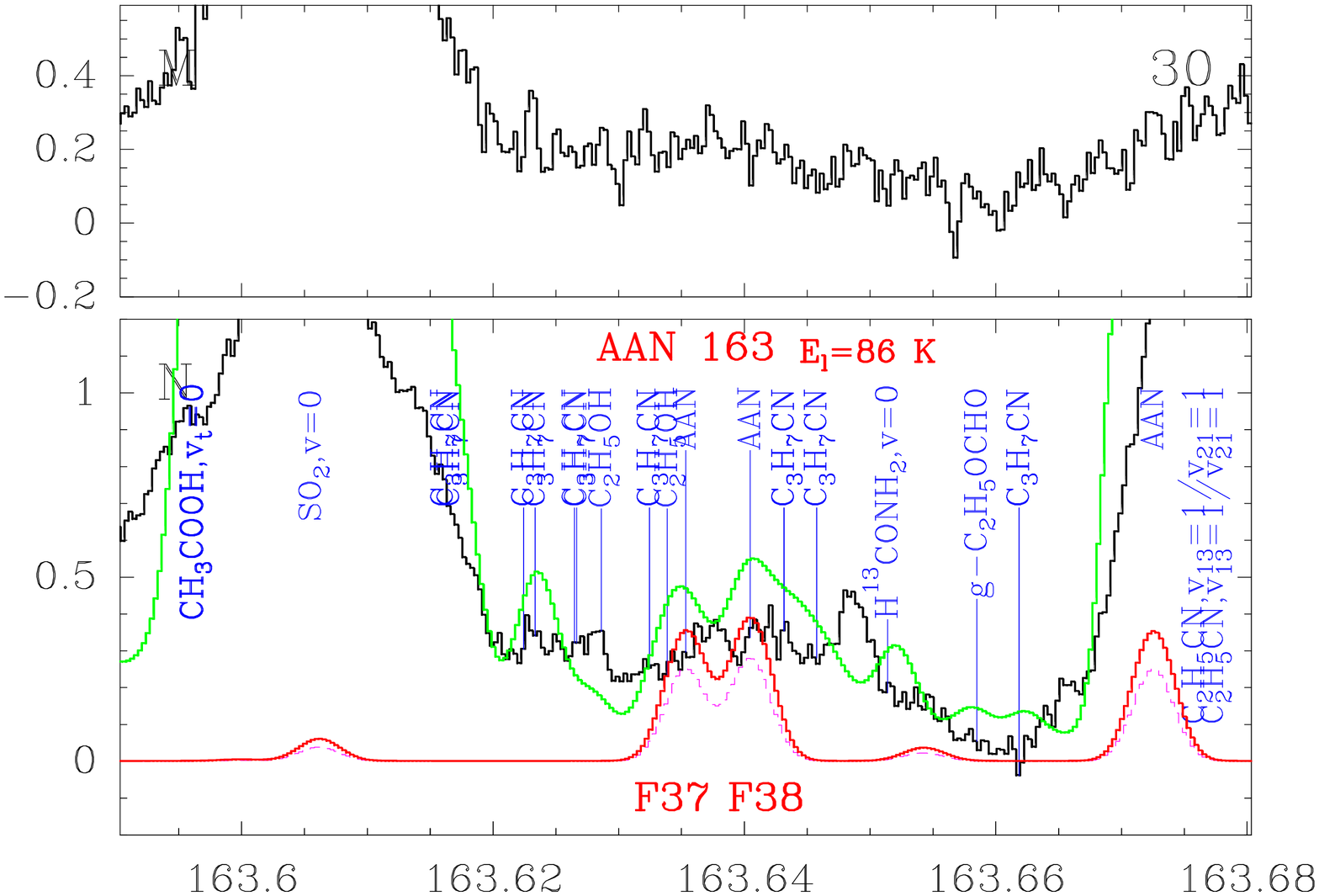}}}
\vspace*{-0.4ex}
\centerline{\resizebox{0.85\hsize}{!}{\includegraphics[angle=270]{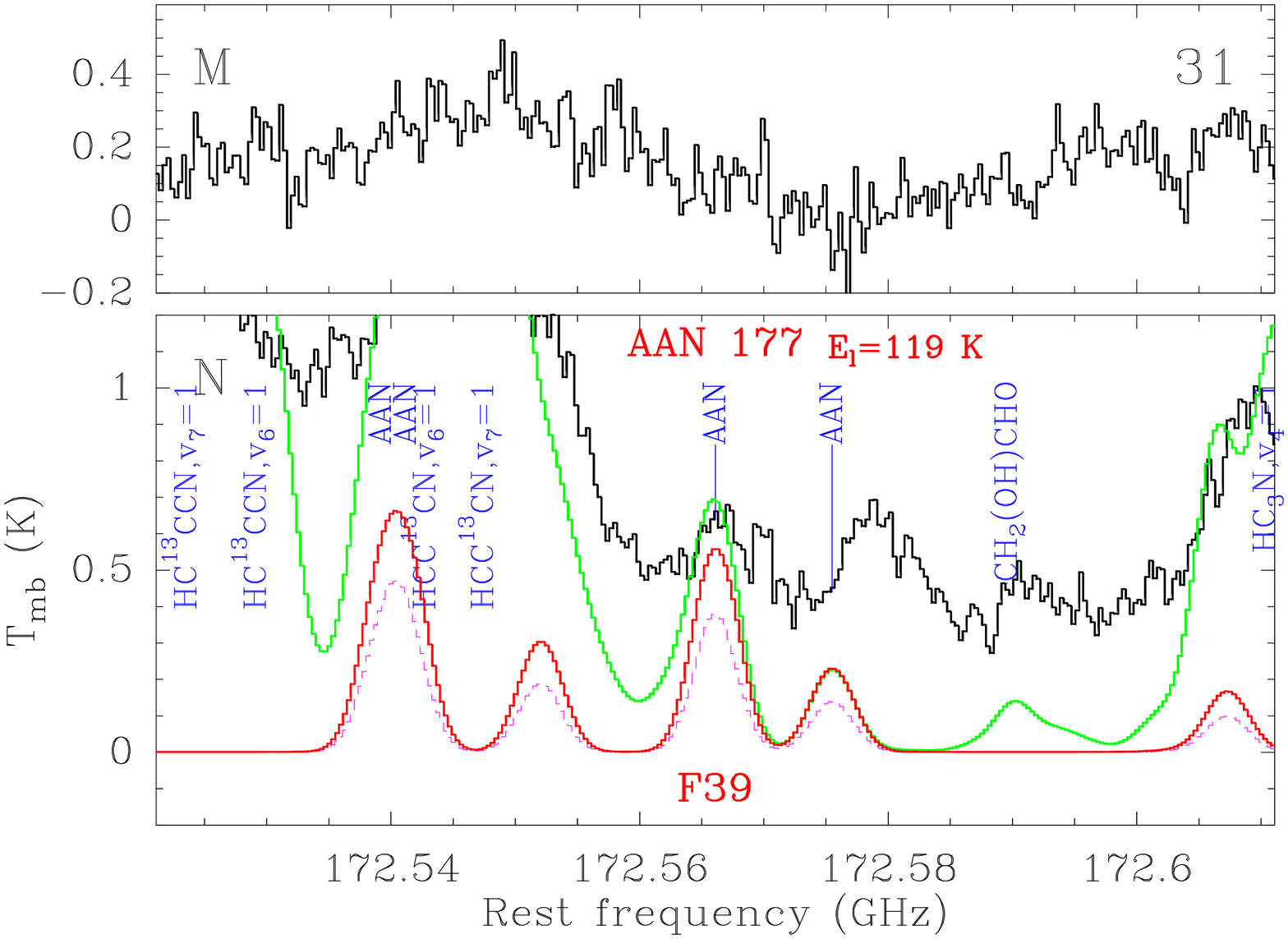}\includegraphics[angle=270]{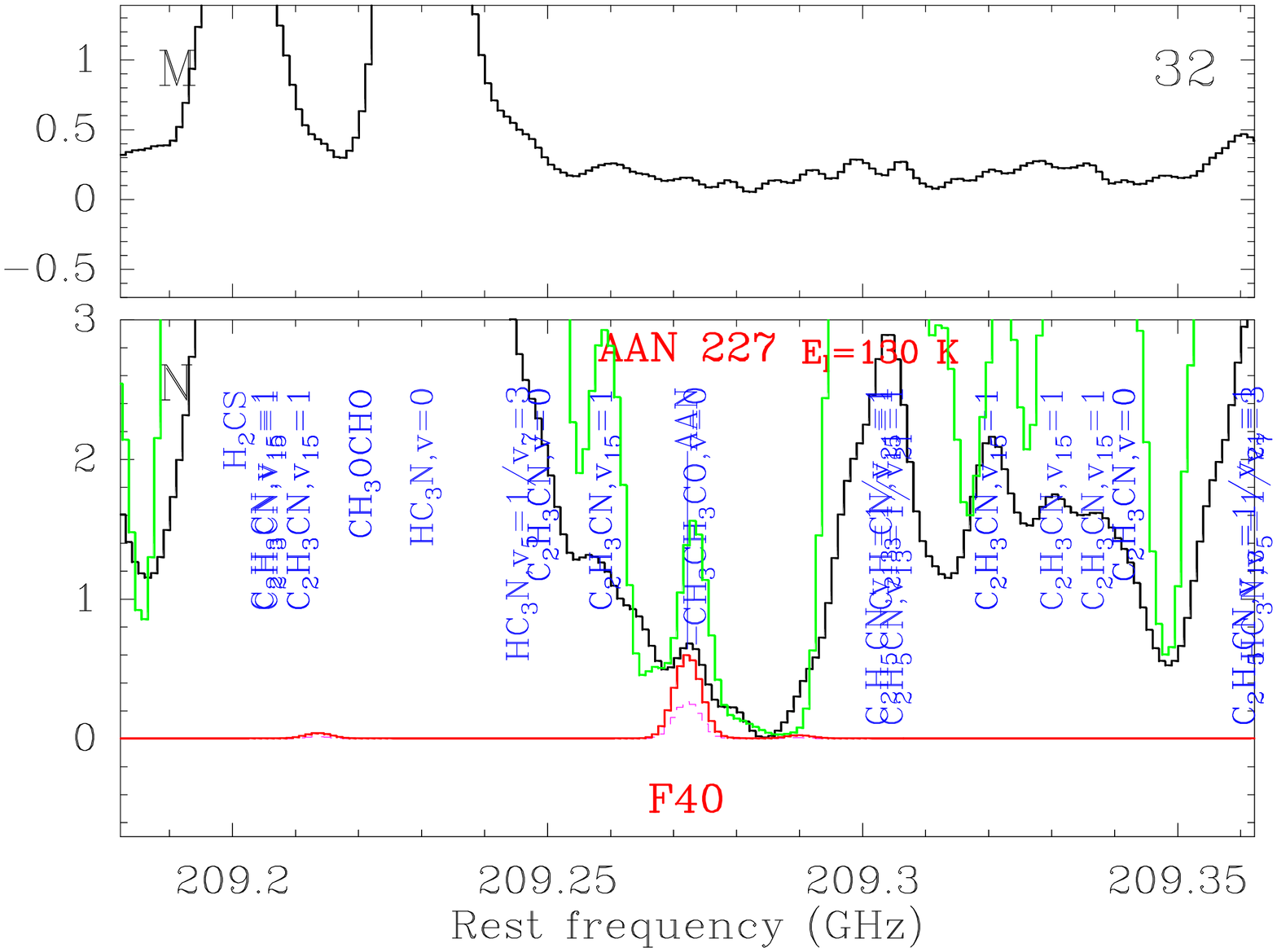}}}
\vspace*{-0.4ex}
\addtocounter{figure}{-1}
\caption{
(continued)
}
\label{f:detectaan30m}
\end{figure*}
\begin{figure*}

\centerline{\resizebox{0.85\hsize}{!}{\includegraphics[angle=270]{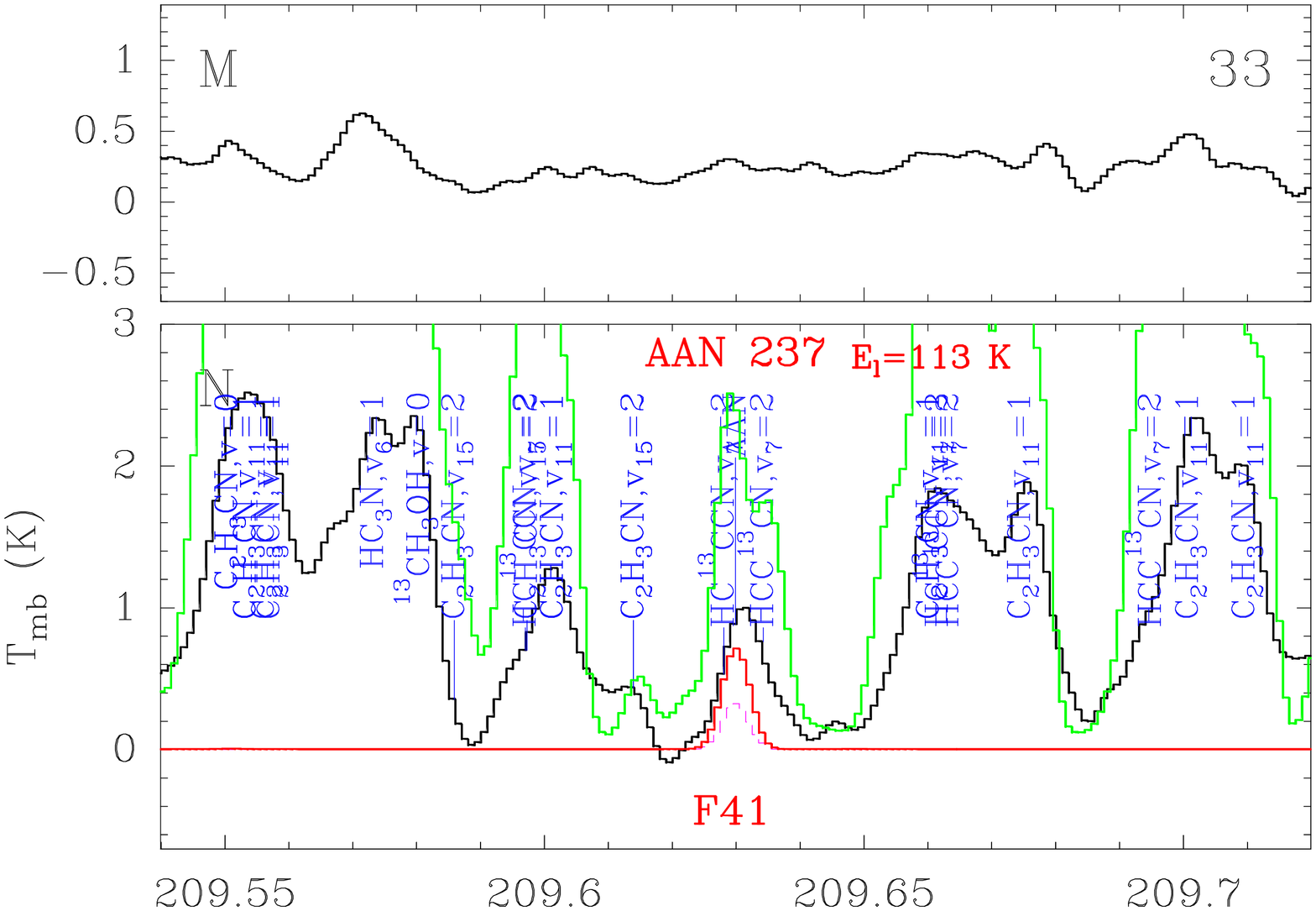}\includegraphics[angle=270]{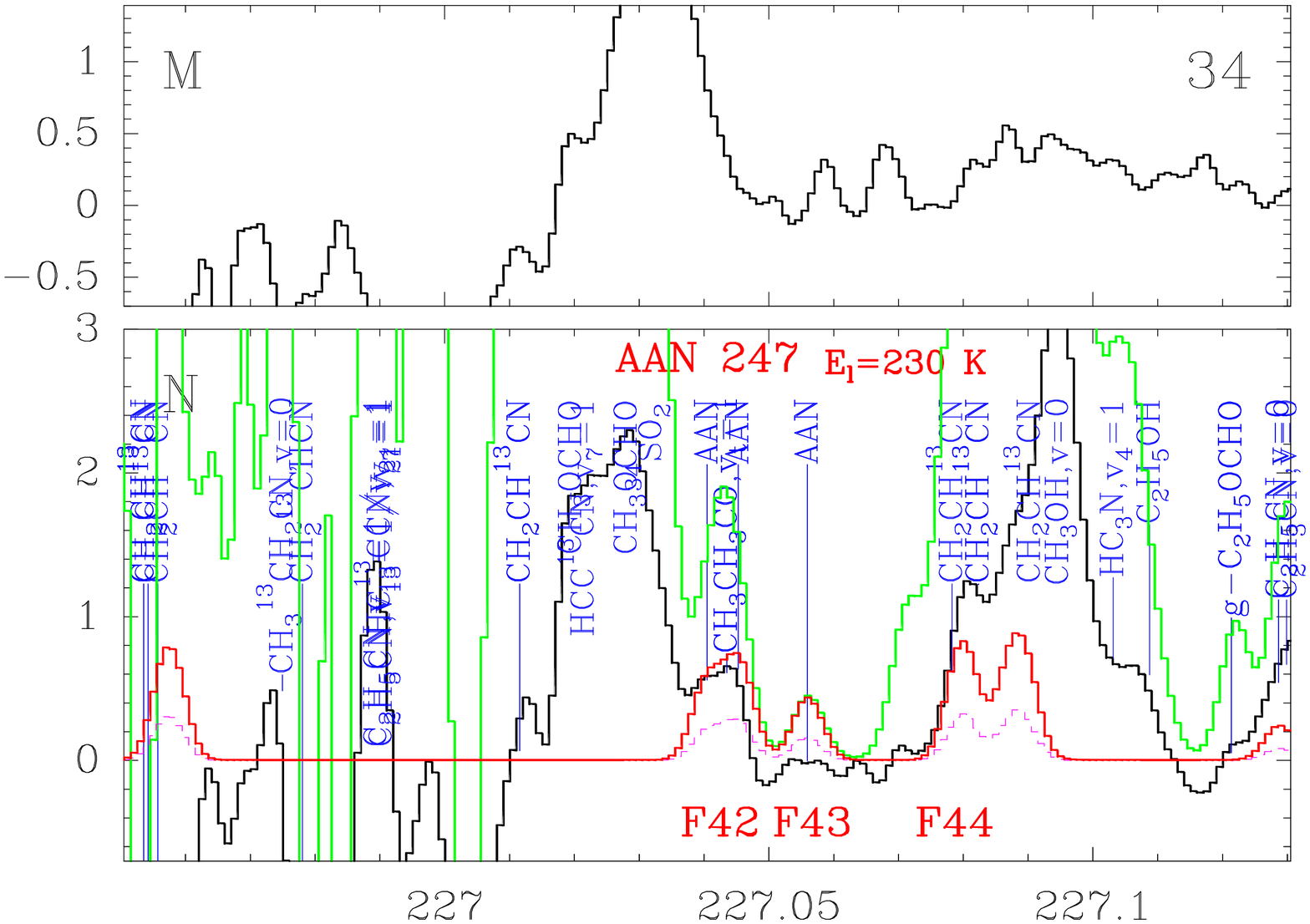}}}
\vspace*{-0.4ex}
\centerline{\resizebox{0.85\hsize}{!}{\includegraphics[angle=270]{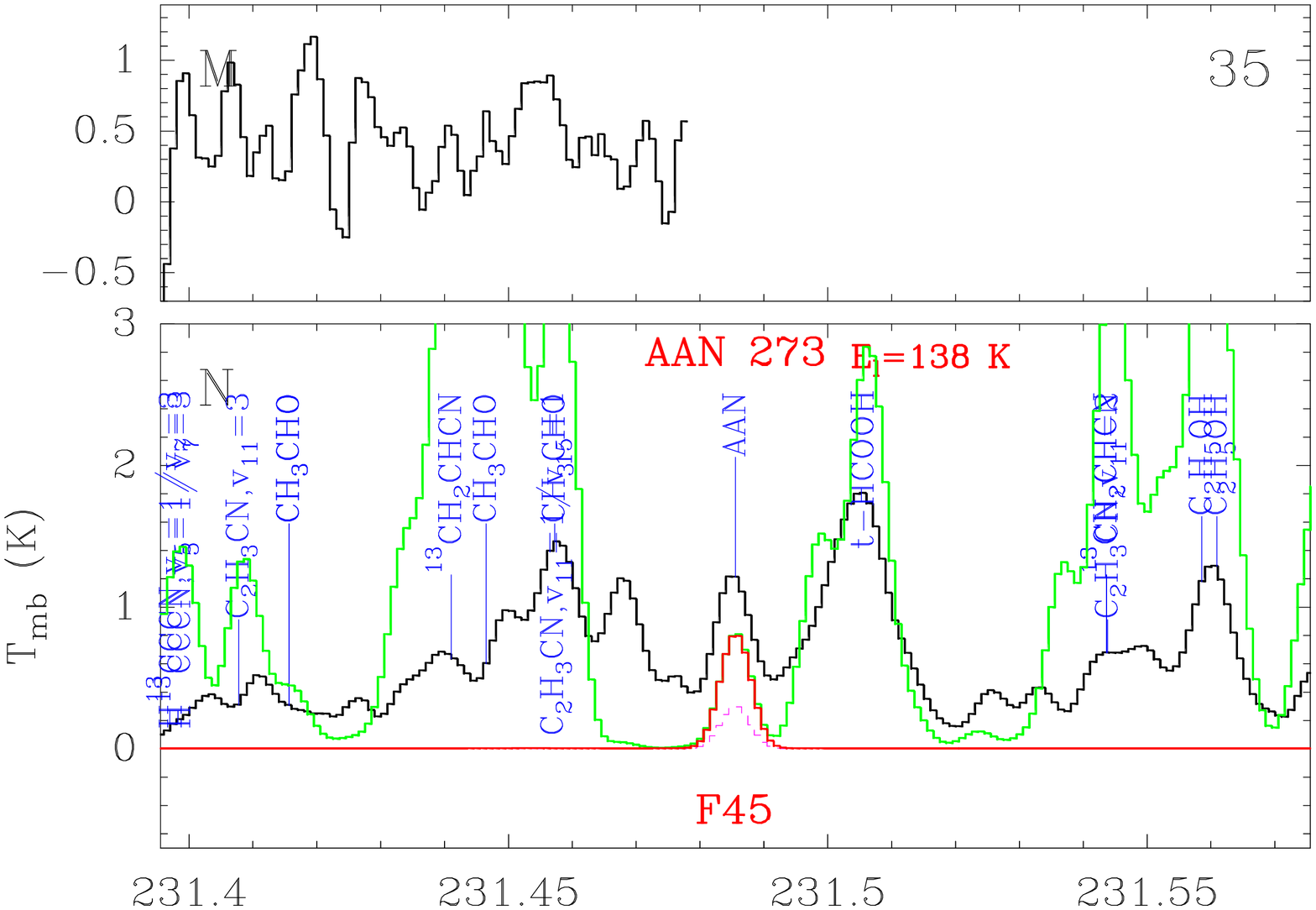}\includegraphics[angle=270]{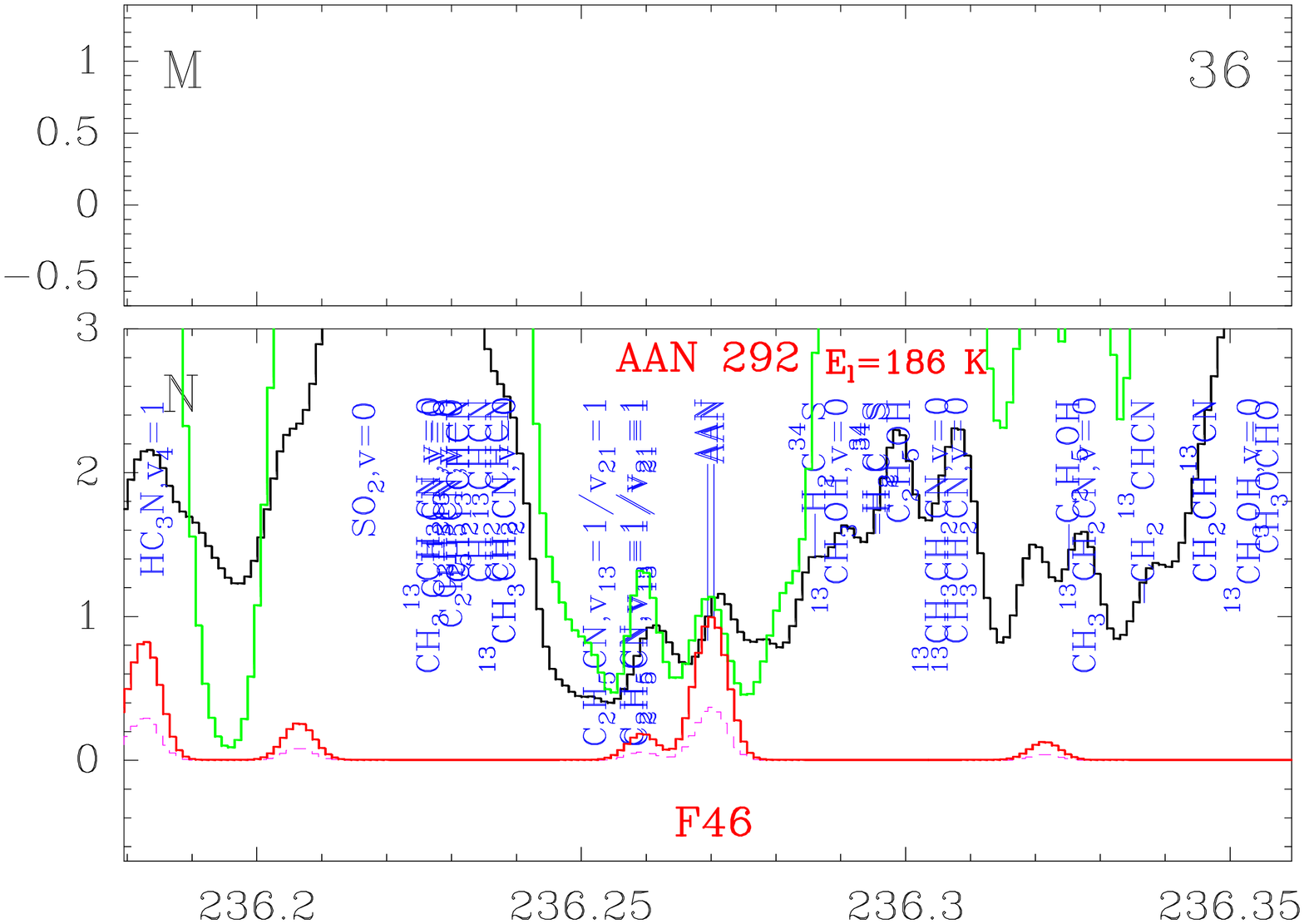}}}
\vspace*{-0.4ex}
\centerline{\resizebox{0.85\hsize}{!}{\includegraphics[angle=270]{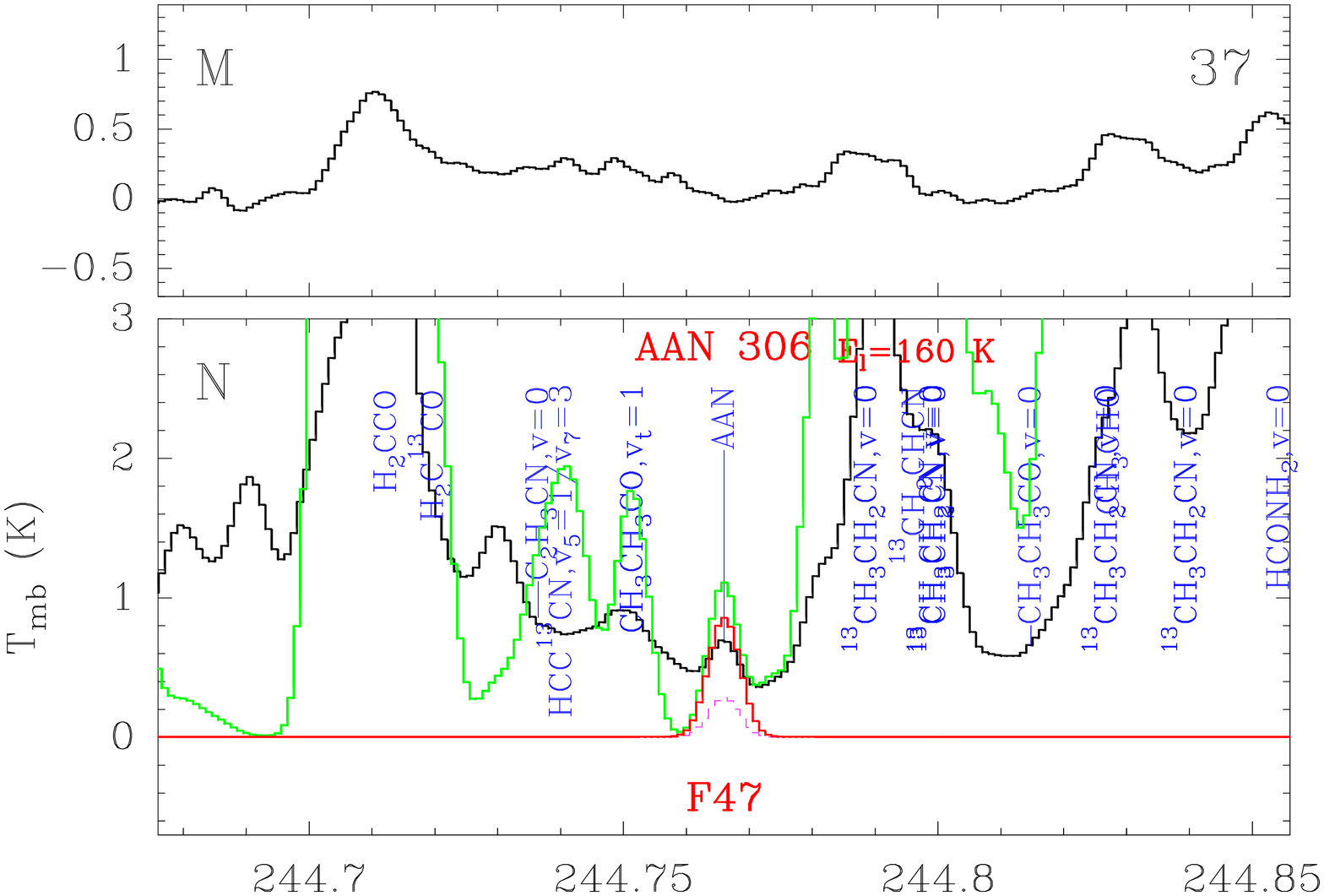}\includegraphics[angle=270]{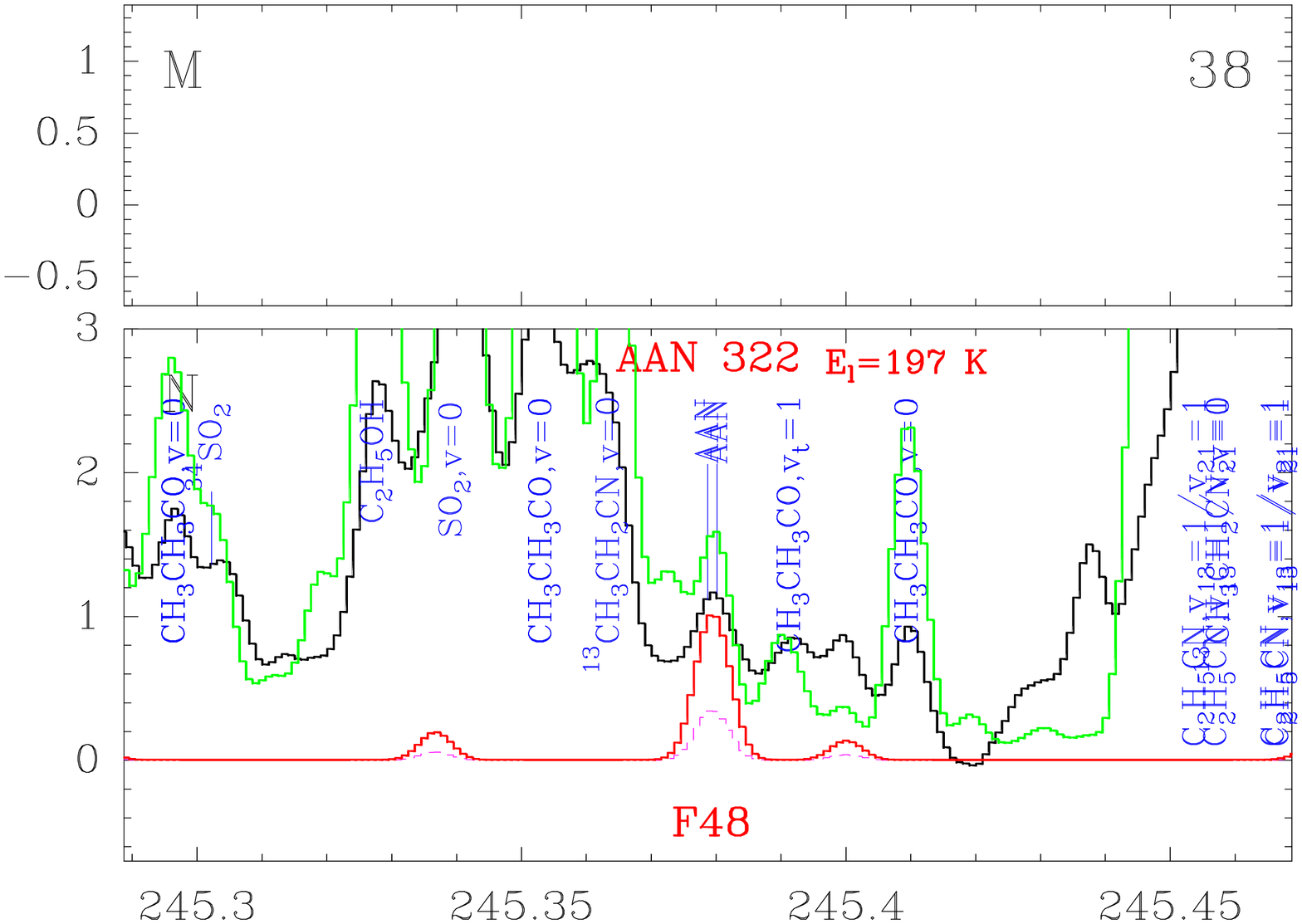}}}
\vspace*{-0.4ex}
\centerline{\resizebox{0.85\hsize}{!}{\includegraphics[angle=270]{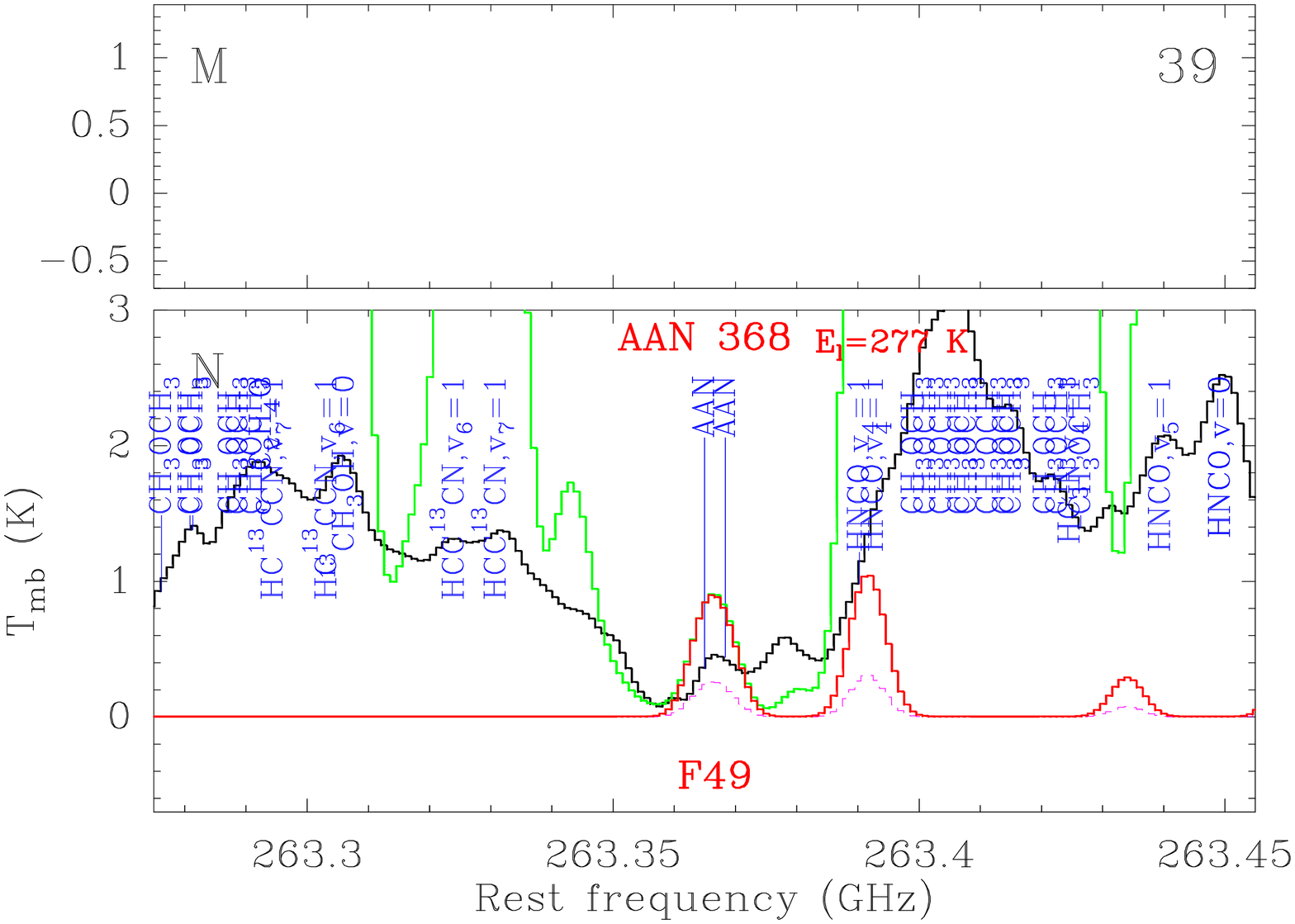}\includegraphics[angle=270]{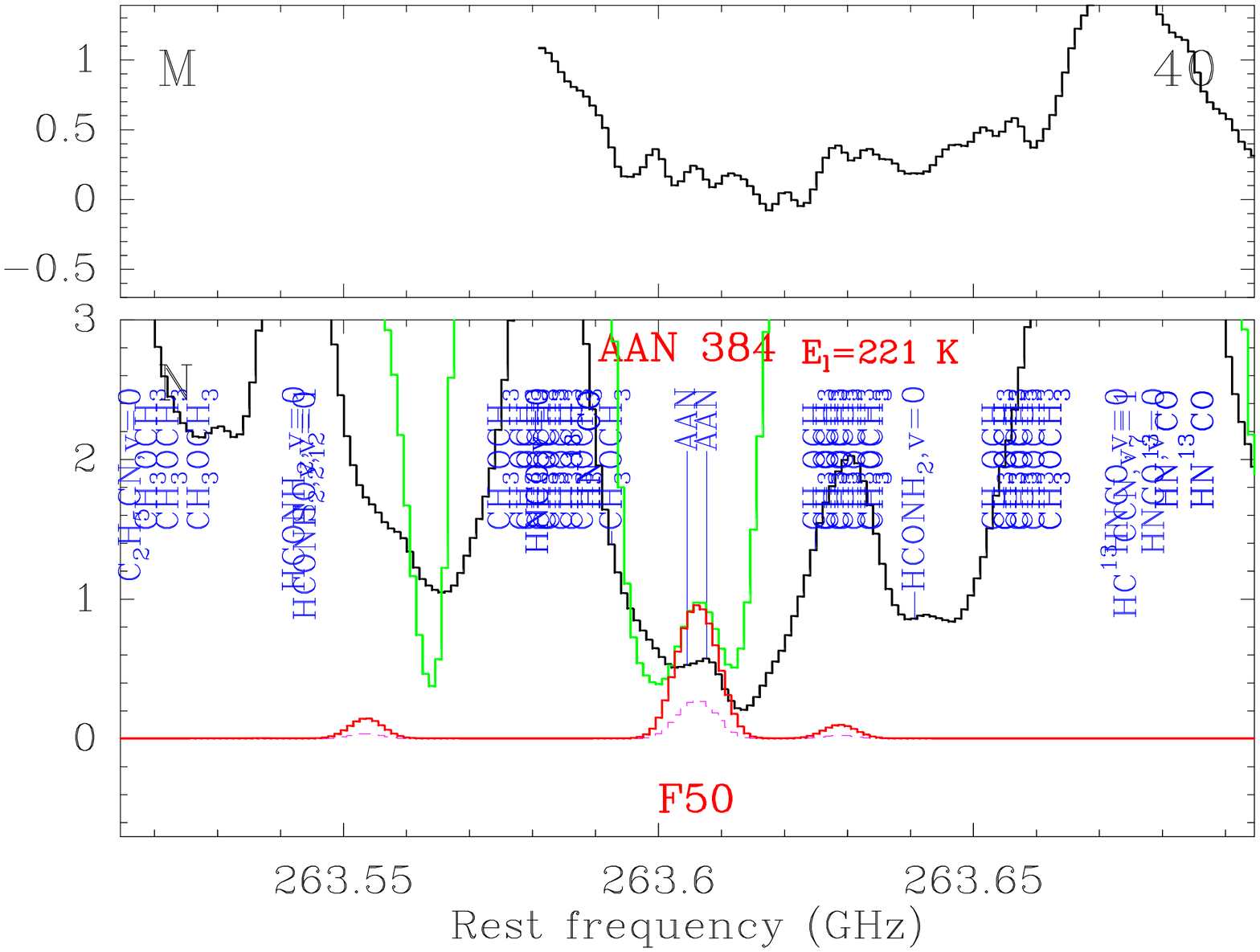}}}
\vspace*{-0.4ex}
\addtocounter{figure}{-1}
\caption{
(continued)
}
\label{f:detectaan30m}
\end{figure*}
\begin{figure*}

\centerline{\resizebox{0.425\hsize}{!}{\includegraphics[angle=270]{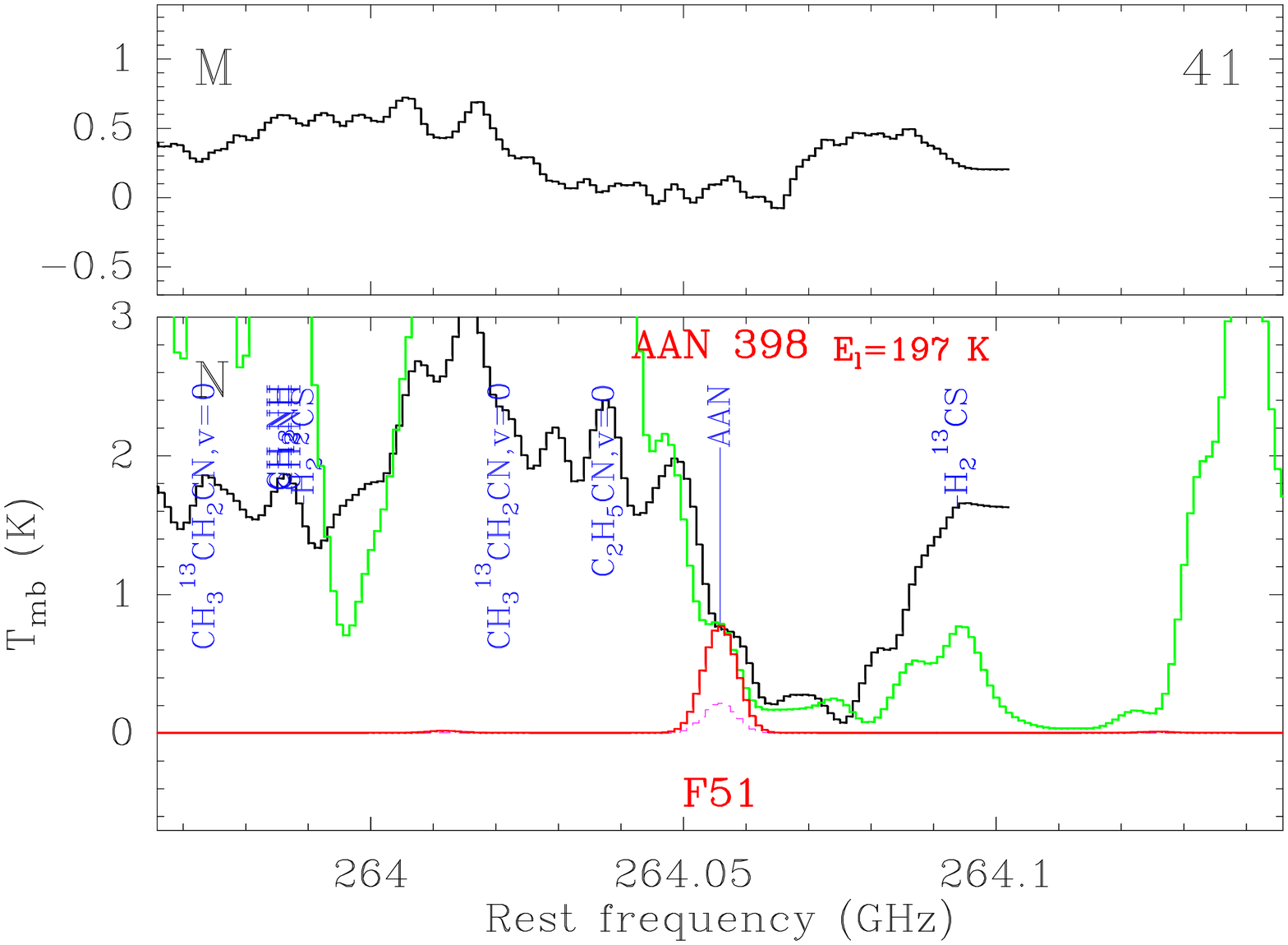}}}
\addtocounter{figure}{-1}
\caption{
(continued)
}
\label{f:detectaan30m}
\end{figure*}

}

\begin{table*}
 {\centering
 \caption{
 Transitions of amino acetonitrile detected toward Sgr~B2(N) with the IRAM 30m telescope.
}
 \label{t:detectaan30m}
 \vspace*{-1.0ex}
 \begin{tabular}{rlrcrrrcrrrrl}
 \hline\hline
 \multicolumn{1}{c}{N$^a$} & \multicolumn{1}{c}{Transition} & \multicolumn{1}{c}{\hspace*{-2ex} Frequency} & \multicolumn{1}{c}{\hspace*{-3ex} Unc.$^b$} & \multicolumn{1}{c}{\hspace*{-3ex} E$_\mathrm{l}^c$} & \multicolumn{1}{c}{S$\mu^2$} & \multicolumn{1}{c}{\hspace*{-2ex} $\sigma^d$} & \multicolumn{1}{c}{F$^e$} & \multicolumn{1}{c}{$\tau^f$} & \multicolumn{1}{c}{I$_{\mathrm{obs}}^g$} & \multicolumn{1}{c}{\hspace*{-2ex} I$_{\mathrm{AAN}}^g$} & \multicolumn{1}{c}{\hspace*{-2ex} I$_{\mathrm{all}}^g$} & \multicolumn{1}{c}{Comments} \\ 
  & & \multicolumn{1}{c}{\hspace*{-2ex} \scriptsize (MHz)} & \multicolumn{1}{c}{\hspace*{-3ex} \scriptsize (kHz)} & \multicolumn{1}{c}{\hspace*{-3ex} \scriptsize (K)} & \multicolumn{1}{c}{\scriptsize (D$^2$)} & \multicolumn{1}{c}{\hspace*{-2ex} \scriptsize (mK)} & & & \multicolumn{1}{c}{\scriptsize (K~km$/$s)} & \multicolumn{1}{c}{\hspace*{-2ex} \scriptsize (K~km$/$s)} & \multicolumn{1}{c}{\hspace*{-2ex} \scriptsize (K~km$/$s)} & \\ 
 \multicolumn{1}{c}{(1)} & \multicolumn{1}{c}{(2)} & \multicolumn{1}{c}{\hspace*{-2ex} (3)} & \multicolumn{1}{c}{\hspace*{-3ex} (4)} & \multicolumn{1}{c}{\hspace*{-3ex} (5)} & \multicolumn{1}{c}{(6)} & \multicolumn{1}{c}{\hspace*{-2ex} (7)} & \multicolumn{1}{c}{(8)} & \multicolumn{1}{c}{(9)} & \multicolumn{1}{c}{(10)} & \multicolumn{1}{c}{(11)} & \multicolumn{1}{c}{\hspace*{-2ex} (12)} & \multicolumn{1}{c}{\hspace*{-2ex} (13)} \\ 
 \hline
   1 &  9$_{ 0, 9}$ -  8$_{ 0, 8}$ & \hspace*{-2ex}   80947.479 & \hspace*{-3ex}    7 & \hspace*{-3ex}   16 &          60 & \hspace*{-2ex}    33 &    1 & 0.13 &        0.65(16) & \hspace*{-2ex}        0.38 & \hspace*{-2ex}        0.42 & no blend \\ 
   3 &  9$_{ 5, 5}$ -  8$_{ 5, 4}$ & \hspace*{-2ex}   81700.966 & \hspace*{-3ex}    6 & \hspace*{-3ex}   47 &          41 & \hspace*{-2ex}    13 &    2 & 0.16 &        0.92(07) & \hspace*{-2ex}        0.67 & \hspace*{-2ex}        0.75 &  partial blend with U-line \\ 
   4 &  9$_{ 5, 4}$ -  8$_{ 5, 3}$ & \hspace*{-2ex}   81700.967 & \hspace*{-3ex}    6 & \hspace*{-3ex}   47 &          41 & \hspace*{-2ex}    13 &    2 & - & - & \hspace*{-2ex} - & \hspace*{-2ex} - & - \\ 
   5 &  9$_{ 6, 3}$ -  8$_{ 6, 2}$ & \hspace*{-2ex}   81702.498 & \hspace*{-3ex}    5 & \hspace*{-3ex}   60 &          33 & \hspace*{-2ex}    13 &    2 & - & - & \hspace*{-2ex} - & \hspace*{-2ex} - & - \\ 
   6 &  9$_{ 6, 4}$ -  8$_{ 6, 3}$ & \hspace*{-2ex}   81702.498 & \hspace*{-3ex}    5 & \hspace*{-3ex}   60 &          33 & \hspace*{-2ex}    13 &    2 & - & - & \hspace*{-2ex} - & \hspace*{-2ex} - & - \\ 
   7 &  9$_{ 4, 6}$ -  8$_{ 4, 5}$ & \hspace*{-2ex}   81709.838 & \hspace*{-3ex}    6 & \hspace*{-3ex}   35 &          48 & \hspace*{-2ex}    13 &    3 & 0.23 &        0.39(06) & \hspace*{-2ex}        0.66 & \hspace*{-2ex}        0.73 & no blend \\ 
   8 &  9$_{ 7, 2}$ -  8$_{ 7, 1}$ & \hspace*{-2ex}   81709.848 & \hspace*{-3ex}    6 & \hspace*{-3ex}   76 &          24 & \hspace*{-2ex}    13 &    3 & - & - & \hspace*{-2ex} - & \hspace*{-2ex} - & - \\ 
   9 &  9$_{ 7, 3}$ -  8$_{ 7, 2}$ & \hspace*{-2ex}   81709.848 & \hspace*{-3ex}    6 & \hspace*{-3ex}   76 &          24 & \hspace*{-2ex}    13 &    3 & - & - & \hspace*{-2ex} - & \hspace*{-2ex} - & - \\ 
  10 &  9$_{ 4, 5}$ -  8$_{ 4, 4}$ & \hspace*{-2ex}   81710.098 & \hspace*{-3ex}    6 & \hspace*{-3ex}   35 &          48 & \hspace*{-2ex}    13 &    3 & - & - & \hspace*{-2ex} - & \hspace*{-2ex} - & - \\ 
  11 &  9$_{ 3, 7}$ -  8$_{ 3, 6}$ & \hspace*{-2ex}   81733.892 & \hspace*{-3ex}    6 & \hspace*{-3ex}   27 &          53 & \hspace*{-2ex}    13 &    4 & 0.11 &        0.50(06) & \hspace*{-2ex}        0.32 & \hspace*{-2ex}        1.46 &  blend with CH$_3$OCH$_3$ and \\ 
 & & & & & & & & & & & &  HCC$^{13}$CN, v$_6$=1 \\ 
  12 &  9$_{ 3, 6}$ -  8$_{ 3, 5}$ & \hspace*{-2ex}   81756.174 & \hspace*{-3ex}    6 & \hspace*{-3ex}   27 &          53 & \hspace*{-2ex}    13 &    5 & 0.11 &        0.39(06) & \hspace*{-2ex}        0.32 & \hspace*{-2ex}        0.32 &  blend with U-line \\ 
  13 &  9$_{ 2, 7}$ -  8$_{ 2, 6}$ & \hspace*{-2ex}   82224.644 & \hspace*{-3ex}    7 & \hspace*{-3ex}   21 &          57 & \hspace*{-2ex}    19 &    6 & 0.12 &        0.19(08) & \hspace*{-2ex}        0.36 & \hspace*{-2ex}        0.35 &  uncertain baseline \\ 
  17 & 10$_{ 2, 9}$ -  9$_{ 2, 8}$ & \hspace*{-2ex}   90561.332 & \hspace*{-3ex}    6 & \hspace*{-3ex}   25 &          64 & \hspace*{-2ex}    20 &    7 & 0.14 &        0.64(09) & \hspace*{-2ex}        0.52 & \hspace*{-2ex}        1.01 &  blend with weak \\ 
 & & & & & & & & & & & &  C$_2$H$_5$CN, v$_{13}$=1/v$_{21}$=1 \\ 
  18 & 10$_{ 6, 4}$ -  9$_{ 6, 3}$ & \hspace*{-2ex}   90783.538 & \hspace*{-3ex}    6 & \hspace*{-3ex}   64 &          43 & \hspace*{-2ex}    14 &    8 & 0.28 &        1.54(06) & \hspace*{-2ex}        1.05 & \hspace*{-2ex}        1.40 &  partial blend with CH$_2$(OH)CHO and \\ 
 & & & & & & & & & & & &  U-line \\ 
  19 & 10$_{ 6, 5}$ -  9$_{ 6, 4}$ & \hspace*{-2ex}   90783.538 & \hspace*{-3ex}    6 & \hspace*{-3ex}   64 &          43 & \hspace*{-2ex}    14 &    8 & - & - & \hspace*{-2ex} - & \hspace*{-2ex} - & - \\ 
  20 & 10$_{ 5, 6}$ -  9$_{ 5, 5}$ & \hspace*{-2ex}   90784.281 & \hspace*{-3ex}    6 & \hspace*{-3ex}   50 &          50 & \hspace*{-2ex}    14 &    8 & - & - & \hspace*{-2ex} - & \hspace*{-2ex} - & - \\ 
  21 & 10$_{ 5, 5}$ -  9$_{ 5, 4}$ & \hspace*{-2ex}   90784.285 & \hspace*{-3ex}    6 & \hspace*{-3ex}   50 &          50 & \hspace*{-2ex}    14 &    8 & - & - & \hspace*{-2ex} - & \hspace*{-2ex} - & - \\ 
  22 & 10$_{ 7, 3}$ -  9$_{ 7, 2}$ & \hspace*{-2ex}   90790.259 & \hspace*{-3ex}    6 & \hspace*{-3ex}   80 &          34 & \hspace*{-2ex}    14 &    9 & 0.09 &        0.51(06) & \hspace*{-2ex}        0.33 & \hspace*{-2ex}        0.56 &  blend with U-line \\ 
  23 & 10$_{ 7, 4}$ -  9$_{ 7, 3}$ & \hspace*{-2ex}   90790.259 & \hspace*{-3ex}    6 & \hspace*{-3ex}   80 &          34 & \hspace*{-2ex}    14 &    9 & - & - & \hspace*{-2ex} - & \hspace*{-2ex} - & - \\ 
  24 & 10$_{ 4, 7}$ -  9$_{ 4, 6}$ & \hspace*{-2ex}   90798.685 & \hspace*{-3ex}    6 & \hspace*{-3ex}   39 &          56 & \hspace*{-2ex}    14 &   10 & 0.21 &        1.42(06) & \hspace*{-2ex}        0.81 & \hspace*{-2ex}        0.95 &  blend with U-line \\ 
  25 & 10$_{ 4, 6}$ -  9$_{ 4, 5}$ & \hspace*{-2ex}   90799.249 & \hspace*{-3ex}    6 & \hspace*{-3ex}   39 &          56 & \hspace*{-2ex}    14 &   10 & - & - & \hspace*{-2ex} - & \hspace*{-2ex} - & - \\ 
  28 & 10$_{ 3, 8}$ -  9$_{ 3, 7}$ & \hspace*{-2ex}   90829.945 & \hspace*{-3ex}    6 & \hspace*{-3ex}   31 &          60 & \hspace*{-2ex}    14 &   11 & 0.13 &        0.84(06) & \hspace*{-2ex}        0.47 & \hspace*{-2ex}        0.51 &  blend with U-line also in M? \\ 
  29 & 10$_{ 3, 7}$ -  9$_{ 3, 6}$ & \hspace*{-2ex}   90868.038 & \hspace*{-3ex}    6 & \hspace*{-3ex}   31 &          60 & \hspace*{-2ex}    14 &   12 & 0.13 &        0.49(06) & \hspace*{-2ex}        0.47 & \hspace*{-2ex}        0.57 &  partial blend with U-line \\ 
  30 & 10$_{ 2, 8}$ -  9$_{ 2, 7}$ & \hspace*{-2ex}   91496.108 & \hspace*{-3ex}    8 & \hspace*{-3ex}   25 &          64 & \hspace*{-2ex}    24 &   13 & 0.15 &        0.86(11) & \hspace*{-2ex}        0.53 & \hspace*{-2ex}        0.71 &  partial blend with CH$_3$CN, v$_4$=1 and \\ 
 & & & & & & & & & & & &  U-line \\ 
  32 & 11$_{ 1,11}$ - 10$_{ 1,10}$ & \hspace*{-2ex}   97015.224 & \hspace*{-3ex}    8 & \hspace*{-3ex}   25 &          72 & \hspace*{-2ex}    21 &   14 & 0.18 &        2.05(09) & \hspace*{-2ex}        0.71 & \hspace*{-2ex}        1.78 &  partial blend with C$_2$H$_5$OH and \\ 
 & & & & & & & & & & & &  CH$_3$OCHO \\ 
  47 & 11$_{ 3, 9}$ - 10$_{ 3, 8}$ & \hspace*{-2ex}   99928.886 & \hspace*{-3ex}    6 & \hspace*{-3ex}   35 &          68 & \hspace*{-2ex}    14 &   15 & 0.15 &        1.31(06) & \hspace*{-2ex}        0.66 & \hspace*{-2ex}        1.24 &  partial blend with NH$_2$CN and U-line \\ 
  48 & 11$_{ 3, 8}$ - 10$_{ 3, 7}$ & \hspace*{-2ex}   99990.567 & \hspace*{-3ex}    7 & \hspace*{-3ex}   35 &          68 & \hspace*{-2ex}    14 &   16 & 0.15 &        0.80(06) & \hspace*{-2ex}        0.66 & \hspace*{-2ex}        0.74 & no blend \\ 
  49 & 11$_{ 2, 9}$ - 10$_{ 2, 8}$ & \hspace*{-2ex}  100800.876 & \hspace*{-3ex}    8 & \hspace*{-3ex}   29 &          71 & \hspace*{-2ex}    20 &   17 & 0.17 &        1.38(08) & \hspace*{-2ex}        0.75 & \hspace*{-2ex}        1.25 &  partial blend with CH$_3$CH$_3$CO, v=0 \\ 
 & & & & & & & & & & & &  and U-line \\ 
  50 & 11$_{ 1,10}$ - 10$_{ 1, 9}$ & \hspace*{-2ex}  101899.795 & \hspace*{-3ex}    8 & \hspace*{-3ex}   26 &          72 & \hspace*{-2ex}    34 &   18 & 0.18 &        0.56(14) & \hspace*{-2ex}        0.81 & \hspace*{-2ex}        0.88 &  uncertain baseline \\ 
  51 & 12$_{ 1,12}$ - 11$_{ 1,11}$ & \hspace*{-2ex}  105777.991 & \hspace*{-3ex}    8 & \hspace*{-3ex}   29 &          79 & \hspace*{-2ex}    43 &   19 & 0.20 &        1.98(18) & \hspace*{-2ex}        0.95 & \hspace*{-2ex}        2.88 &  blend with c-C$_2$H$_4$O and \\ 
 & & & & & & & & & & & &  C$_2$H$_5$CN, v=0 \\ 
  52 & 12$_{ 0,12}$ - 11$_{ 0,11}$ & \hspace*{-2ex}  107283.142 & \hspace*{-3ex}    8 & \hspace*{-3ex}   29 &          80 & \hspace*{-2ex}    24 &   20 & 0.21 &        2.67(10) & \hspace*{-2ex}        1.00 & \hspace*{-2ex}        2.01 &  blend with C$_2$H$_5$OH and U-line \\ 
  53 & 12$_{ 2,11}$ - 11$_{ 2,10}$ & \hspace*{-2ex}  108581.408 & \hspace*{-3ex}    7 & \hspace*{-3ex}   34 &          77 & \hspace*{-2ex}    20 &   21 & 0.19 &        1.49(08) & \hspace*{-2ex}        0.97 & \hspace*{-2ex}        1.94 &  weak blend with C$_2$H$_5$OH \\ 
  58 & 12$_{ 5, 8}$ - 11$_{ 5, 7}$ & \hspace*{-2ex}  108956.206 & \hspace*{-3ex}    6 & \hspace*{-3ex}   60 &          66 & \hspace*{-2ex}    29 &   22 & 0.26 &        2.19(11) & \hspace*{-2ex}        1.34 & \hspace*{-2ex}        3.44 &  blend with C$_2$H$_5$OH \\ 
  59 & 12$_{ 5, 7}$ - 11$_{ 5, 6}$ & \hspace*{-2ex}  108956.229 & \hspace*{-3ex}    6 & \hspace*{-3ex}   60 &          66 & \hspace*{-2ex}    29 &   22 & - & - & \hspace*{-2ex} - & \hspace*{-2ex} - & - \\ 
  68 & 12$_{ 3,10}$ - 11$_{ 3, 9}$ & \hspace*{-2ex}  109030.225 & \hspace*{-3ex}    6 & \hspace*{-3ex}   40 &          75 & \hspace*{-2ex}    29 &   23 & 0.18 &        1.67(11) & \hspace*{-2ex}        0.89 & \hspace*{-2ex}        1.24 &  partial blend with HC$_3$N, v$_4$=1, \\ 
 & & & & & & & & & & & &  C$_2$H$_5$OH, and U-line \\ 
  71 & 12$_{ 1,11}$ - 11$_{ 1,10}$ & \hspace*{-2ex}  111076.901 & \hspace*{-3ex}    8 & \hspace*{-3ex}   31 &          79 & \hspace*{-2ex}    25 &   24 & 0.21 &        1.16(10) & \hspace*{-2ex}        1.08 & \hspace*{-2ex}        1.39 &  slightly shifted? \\ 
  72 & 13$_{ 1,13}$ - 12$_{ 1,12}$ & \hspace*{-2ex}  114528.654 & \hspace*{-3ex}    8 & \hspace*{-3ex}   34 &          86 & \hspace*{-2ex}    37 &   25 & 0.23 &        2.49(15) & \hspace*{-2ex}        1.23 & \hspace*{-2ex}        1.42 &  partial blend with U-line \\ 
  84 & 15$_{10, 5}$ - 14$_{10, 4}$ & \hspace*{-2ex}  136248.969 & \hspace*{-3ex}   10 & \hspace*{-3ex}  169 &          55 & \hspace*{-2ex}    28 &   26 & 0.09 &        2.10(10) & \hspace*{-2ex}        0.72 & \hspace*{-2ex}        1.03 &  blend with U-line \\ 
  85 & 15$_{10, 6}$ - 14$_{10, 5}$ & \hspace*{-2ex}  136248.969 & \hspace*{-3ex}   10 & \hspace*{-3ex}  169 &          55 & \hspace*{-2ex}    28 &   26 & - & - & \hspace*{-2ex} - & \hspace*{-2ex} - & - \\ 
  89 & 15$_{ 4,11}$ - 14$_{ 4,10}$ & \hspace*{-2ex}  136303.599 & \hspace*{-3ex}    6 & \hspace*{-3ex}   65 &          93 & \hspace*{-2ex}    28 &   27 & 0.21 &        3.99(09) & \hspace*{-2ex}        1.62 & \hspace*{-2ex}        4.02 &  blend with a(CH$_2$OH)$_2$ and CH$_3$C$_3$N \\ 
  92 & 15$_{ 3,13}$ - 14$_{ 3,12}$ & \hspace*{-2ex}  136341.155 & \hspace*{-3ex}    6 & \hspace*{-3ex}   57 &          96 & \hspace*{-2ex}    28 &   28 & 0.24 &        2.92(10) & \hspace*{-2ex}        1.81 & \hspace*{-2ex}        2.22 &  partial blend with U-line also in M \\ 
 103 & 16$_{ 5,12}$ - 15$_{ 5,11}$ & \hspace*{-2ex}  145325.871 & \hspace*{-3ex}   30 & \hspace*{-3ex}   83 &          96 & \hspace*{-2ex}    25 &   29 & 0.39 &        2.89(08) & \hspace*{-2ex}        3.30 & \hspace*{-2ex}        4.80 &  uncertain baseline, partial blend  \\ 
 & & & & & & & & & & & & with C$_2$H$_5$CN, v$_{13}$=1/v$_{21}$=1 \\ 
 104 & 16$_{ 5,11}$ - 15$_{ 5,10}$ & \hspace*{-2ex}  145326.209 & \hspace*{-3ex}   30 & \hspace*{-3ex}   83 &          96 & \hspace*{-2ex}    25 &   29 & - & - & \hspace*{-2ex} - & \hspace*{-2ex} - & - \\ 
 \hline
 \end{tabular}
 }
 Notes:
 $(a)$ Numbering of the observed transitions with S$\mu^2$ $>$ 20 D$^2$ (see Table~\ref{t:aan30m}).
 $(b)$ Frequency uncertainty.
 $(c)$ Lower energy level in temperature units (E$_\mathrm{l}/$k$_\mathrm{B}$).
 $(d)$ Calculated rms noise level in T$_{\mathrm{mb}}$ scale.
 $(e)$ Numbering of the observed features.
 $(f)$ Peak opacity of the amino acetonitrile modeled feature.
 $(g)$ Integrated intensity in T$_{\mathrm{mb}}$ scale for the observed spectrum (col. 10), the amino acetonitrile model (col. 11), and the model including all molecules (col. 12). The uncertainty in col. 10 is given in parentheses in units of the last digit.
 \end{table*}
\begin{table*}
 {\centering
 \addtocounter{table}{-1}
 \caption{
(continued)
 }
 \label{t:detectaan30m}
 \vspace*{-1.0ex}
 \begin{tabular}{rlrcrrrcrrrrl}
 \hline\hline
 \multicolumn{1}{c}{N$^a$} & \multicolumn{1}{c}{Transition} & \multicolumn{1}{c}{\hspace*{-2ex} Frequency} & \multicolumn{1}{c}{\hspace*{-3ex} Unc.$^b$} & \multicolumn{1}{c}{\hspace*{-3ex} E$_\mathrm{l}^c$} & \multicolumn{1}{c}{S$\mu^2$} & \multicolumn{1}{c}{\hspace*{-2ex} $\sigma^d$} & \multicolumn{1}{c}{F$^e$} & \multicolumn{1}{c}{$\tau^f$} & \multicolumn{1}{c}{I$_{\mathrm{obs}}^g$} & \multicolumn{1}{c}{\hspace*{-2ex} I$_{\mathrm{AAN}}^g$} & \multicolumn{1}{c}{\hspace*{-2ex} I$_{\mathrm{all}}^g$} & \multicolumn{1}{c}{Comments} \\ 
  & & \multicolumn{1}{c}{\hspace*{-2ex} \scriptsize (MHz)} & \multicolumn{1}{c}{\hspace*{-3ex} \scriptsize (kHz)} & \multicolumn{1}{c}{\hspace*{-3ex} \scriptsize (K)} & \multicolumn{1}{c}{\scriptsize (D$^2$)} & \multicolumn{1}{c}{\hspace*{-2ex} \scriptsize (mK)} & & & \multicolumn{1}{c}{\scriptsize (K~km$/$s)} & \multicolumn{1}{c}{\hspace*{-2ex} \scriptsize (K~km$/$s)} & \multicolumn{1}{c}{\hspace*{-2ex} \scriptsize (K~km$/$s)} & \\ 
 \multicolumn{1}{c}{(1)} & \multicolumn{1}{c}{(2)} & \multicolumn{1}{c}{\hspace*{-2ex} (3)} & \multicolumn{1}{c}{\hspace*{-3ex} (4)} & \multicolumn{1}{c}{\hspace*{-3ex} (5)} & \multicolumn{1}{c}{(6)} & \multicolumn{1}{c}{\hspace*{-2ex} (7)} & \multicolumn{1}{c}{(8)} & \multicolumn{1}{c}{(9)} & \multicolumn{1}{c}{(10)} & \multicolumn{1}{c}{(11)} & \multicolumn{1}{c}{\hspace*{-2ex} (12)} & \multicolumn{1}{c}{\hspace*{-2ex} (13)} \\ 
 \hline
 105 & 16$_{10, 6}$ - 15$_{10, 5}$ & \hspace*{-2ex}  145330.985 & \hspace*{-3ex}   40 & \hspace*{-3ex}  175 &          65 & \hspace*{-2ex}    25 &   30 & 0.11 &        0.97(07) & \hspace*{-2ex}        0.92 & \hspace*{-2ex}        1.02 &  uncertain baseline \\ 
 106 & 16$_{10, 7}$ - 15$_{10, 6}$ & \hspace*{-2ex}  145330.985 & \hspace*{-3ex}   40 & \hspace*{-3ex}  175 &          65 & \hspace*{-2ex}    25 &   30 & - & - & \hspace*{-2ex} - & \hspace*{-2ex} - & - \\ 
 115 & 16$_{ 3,14}$ - 15$_{ 3,13}$ & \hspace*{-2ex}  145443.850 & \hspace*{-3ex}   30 & \hspace*{-3ex}   63 &         103 & \hspace*{-2ex}    25 &   31 & 0.25 &        4.33(08) & \hspace*{-2ex}        2.18 & \hspace*{-2ex}        4.68 &  blend with C$_2$H$_5$CN, v=0 and \\ 
 & & & & & & & & & & & &  U-line \\ 
 118 & 16$_{ 1,15}$ - 15$_{ 1,14}$ & \hspace*{-2ex}  147495.789 & \hspace*{-3ex}    6 & \hspace*{-3ex}   55 &         106 & \hspace*{-2ex}    31 &   32 & 0.29 &        3.27(11) & \hspace*{-2ex}        2.54 & \hspace*{-2ex}       11.47 &  partial blend with H$_3$C$^{13}$CN, v$_8$=1 \\ 
 139 & 17$_{ 4,13}$ - 16$_{ 4,12}$ & \hspace*{-2ex}  154542.406 & \hspace*{-3ex}    5 & \hspace*{-3ex}   79 &         107 & \hspace*{-2ex}   112 &   33 & 0.44 &       13.25(42) & \hspace*{-2ex}        4.63 & \hspace*{-2ex}        5.52 &  blend with U-line \\ 
 140 & 17$_{ 3,15}$ - 16$_{ 3,14}$ & \hspace*{-2ex}  154544.046 & \hspace*{-3ex}    5 & \hspace*{-3ex}   70 &         109 & \hspace*{-2ex}   112 &   33 & - & - & \hspace*{-2ex} - & \hspace*{-2ex} - & - \\ 
 145 & 18$_{ 7,12}$ - 17$_{ 7,11}$ & \hspace*{-2ex}  163454.794 & \hspace*{-3ex}    5 & \hspace*{-3ex}  127 &         101 & \hspace*{-2ex}    38 &   34 & 0.49 &       10.38(13) & \hspace*{-2ex}        5.32 & \hspace*{-2ex}       16.48 &  partial blend with HC$^{13}$CCN,  \\ 
 & & & & & & & & & & & & v$_6$=1 and HCC$^{13}$CN, v$_6$=1 \\ 
 146 & 18$_{ 7,11}$ - 17$_{ 7,10}$ & \hspace*{-2ex}  163454.794 & \hspace*{-3ex}    5 & \hspace*{-3ex}  127 &         101 & \hspace*{-2ex}    38 &   34 & - & - & \hspace*{-2ex} - & \hspace*{-2ex} - & - \\ 
 147 & 18$_{ 8,10}$ - 17$_{ 8, 9}$ & \hspace*{-2ex}  163456.136 & \hspace*{-3ex}    6 & \hspace*{-3ex}  146 &          96 & \hspace*{-2ex}    38 &   34 & - & - & \hspace*{-2ex} - & \hspace*{-2ex} - & - \\ 
 148 & 18$_{ 8,11}$ - 17$_{ 8,10}$ & \hspace*{-2ex}  163456.136 & \hspace*{-3ex}    6 & \hspace*{-3ex}  146 &          96 & \hspace*{-2ex}    38 &   34 & - & - & \hspace*{-2ex} - & \hspace*{-2ex} - & - \\ 
 149 & 18$_{ 9, 9}$ - 17$_{ 9, 8}$ & \hspace*{-2ex}  163470.472 & \hspace*{-3ex}    8 & \hspace*{-3ex}  166 &          90 & \hspace*{-2ex}    38 &   35 & 0.41 &       15.17(14) & \hspace*{-2ex}        5.57 & \hspace*{-2ex}       21.97 &  partial blend with HCC$^{13}$CN,v$_7$=1 \\ 
 150 & 18$_{ 9,10}$ - 17$_{ 9, 9}$ & \hspace*{-2ex}  163470.472 & \hspace*{-3ex}    8 & \hspace*{-3ex}  166 &          90 & \hspace*{-2ex}    38 &   35 & - & - & \hspace*{-2ex} - & \hspace*{-2ex} - & - \\ 
 151 & 18$_{ 6,13}$ - 17$_{ 6,12}$ & \hspace*{-2ex}  163473.305 & \hspace*{-3ex}    5 & \hspace*{-3ex}  111 &         106 & \hspace*{-2ex}    38 &   35 & - & - & \hspace*{-2ex} - & \hspace*{-2ex} - & - \\ 
 152 & 18$_{ 6,12}$ - 17$_{ 6,11}$ & \hspace*{-2ex}  163473.321 & \hspace*{-3ex}    5 & \hspace*{-3ex}  111 &         106 & \hspace*{-2ex}    38 &   35 & - & - & \hspace*{-2ex} - & \hspace*{-2ex} - & - \\ 
 155 & 18$_{11, 7}$ - 17$_{11, 6}$ & \hspace*{-2ex}  163525.533 & \hspace*{-3ex}   11 & \hspace*{-3ex}  216 &          75 & \hspace*{-2ex}    38 &   36 & 0.49 &       10.26(13) & \hspace*{-2ex}        5.27 & \hspace*{-2ex}       17.96 &  blend with HC$_3$N, v$_4$=1 \\ 
 156 & 18$_{11, 8}$ - 17$_{11, 7}$ & \hspace*{-2ex}  163525.533 & \hspace*{-3ex}   11 & \hspace*{-3ex}  216 &          75 & \hspace*{-2ex}    38 &   36 & - & - & \hspace*{-2ex} - & \hspace*{-2ex} - & - \\ 
 157 & 18$_{ 5,14}$ - 17$_{ 5,13}$ & \hspace*{-2ex}  163526.183 & \hspace*{-3ex}    4 & \hspace*{-3ex}   97 &         110 & \hspace*{-2ex}    38 &   36 & - & - & \hspace*{-2ex} - & \hspace*{-2ex} - & - \\ 
 158 & 18$_{ 5,13}$ - 17$_{ 5,12}$ & \hspace*{-2ex}  163527.171 & \hspace*{-3ex}    4 & \hspace*{-3ex}   97 &         110 & \hspace*{-2ex}    38 &   36 & - & - & \hspace*{-2ex} - & \hspace*{-2ex} - & - \\ 
 163 & 18$_{ 4,15}$ - 17$_{ 4,14}$ & \hspace*{-2ex}  163635.326 & \hspace*{-3ex}    5 & \hspace*{-3ex}   86 &         114 & \hspace*{-2ex}    38 &   37 & 0.25 &        4.08(11) & \hspace*{-2ex}        2.82 & \hspace*{-2ex}        5.01 &  partial blend with C$_3$H$_7$CN \\ 
 164 & 18$_{ 3,16}$ - 17$_{ 3,15}$ & \hspace*{-2ex}  163640.468 & \hspace*{-3ex}    5 & \hspace*{-3ex}   78 &         116 & \hspace*{-2ex}    38 &   38 & 0.28 &        4.65(11) & \hspace*{-2ex}        2.99 & \hspace*{-2ex}        6.77 &  partial blend with C$_3$H$_7$CN \\ 
 177 & 19$_{ 6,14}$ - 18$_{ 6,13}$ & \hspace*{-2ex}  172566.092 & \hspace*{-3ex}   50 & \hspace*{-3ex}  119 &         114 & \hspace*{-2ex}    44 &   39 & 0.38 &       10.01(14) & \hspace*{-2ex}        4.39 & \hspace*{-2ex}        6.43 &  partial blend with U-line and \\ 
 & & & & & & & & & & & &  HCC$^{13}$CN, v$_7$=1 \\ 
 178 & 19$_{ 6,13}$ - 18$_{ 6,12}$ & \hspace*{-2ex}  172566.092 & \hspace*{-3ex}   50 & \hspace*{-3ex}  119 &         114 & \hspace*{-2ex}    44 &   39 & - & - & \hspace*{-2ex} - & \hspace*{-2ex} - & - \\ 
 227 & 23$_{ 4,20}$ - 22$_{ 4,19}$ & \hspace*{-2ex}  209272.189 & \hspace*{-3ex}    6 & \hspace*{-3ex}  130 &         148 & \hspace*{-2ex}    58 &   40 & 0.26 &        7.29(29) & \hspace*{-2ex}        4.62 & \hspace*{-2ex}       14.85 &  blend CH$_3$CH$_3$CO, v=0 \\ 
 237 & 23$_{ 1,22}$ - 22$_{ 1,21}$ & \hspace*{-2ex}  209629.913 & \hspace*{-3ex}    9 & \hspace*{-3ex}  113 &         152 & \hspace*{-2ex}    45 &   41 & 0.32 &        9.03(24) & \hspace*{-2ex}        5.54 & \hspace*{-2ex}       30.88 &  blend with HC$^{13}$CCN, v$_7$=2 and \\ 
 & & & & & & & & & & & &  HCC$^{13}$CN, v$_7$=2 \\ 
 247 & 25$_{ 9,16}$ - 24$_{ 9,15}$ & \hspace*{-2ex}  227040.487 & \hspace*{-3ex}   50 & \hspace*{-3ex}  230 &         145 & \hspace*{-2ex}    96 &   42 & 0.29 &        9.58(55) & \hspace*{-2ex}        9.45 & \hspace*{-2ex}       35.33 &  partial blend with CN absorption \\ 
 & & & & & & & & & & & &  and CH$_3$CH$_3$CO, v$_{\mathrm{t}}$=1 \\ 
 248 & 25$_{ 9,17}$ - 24$_{ 9,16}$ & \hspace*{-2ex}  227040.487 & \hspace*{-3ex}   50 & \hspace*{-3ex}  230 &         145 & \hspace*{-2ex}    96 &   42 & - & - & \hspace*{-2ex} - & \hspace*{-2ex} - & - \\ 
 249 & 25$_{ 8,18}$ - 24$_{ 8,17}$ & \hspace*{-2ex}  227045.287 & \hspace*{-3ex}   50 & \hspace*{-3ex}  210 &         149 & \hspace*{-2ex}    96 &   42 & - & - & \hspace*{-2ex} - & \hspace*{-2ex} - & - \\ 
 250 & 25$_{ 8,17}$ - 24$_{ 8,16}$ & \hspace*{-2ex}  227045.287 & \hspace*{-3ex}   50 & \hspace*{-3ex}  210 &         149 & \hspace*{-2ex}    96 &   42 & - & - & \hspace*{-2ex} - & \hspace*{-2ex} - & - \\ 
 251 & 25$_{10,15}$ - 24$_{10,14}$ & \hspace*{-2ex}  227055.944 & \hspace*{-3ex}   50 & \hspace*{-3ex}  254 &         139 & \hspace*{-2ex}    96 &   43 & 0.15 &       -0.64(44) & \hspace*{-2ex}        3.29 & \hspace*{-2ex}        3.62 &  partial blend with CN absorption \\ 
 252 & 25$_{10,16}$ - 24$_{10,15}$ & \hspace*{-2ex}  227055.944 & \hspace*{-3ex}   50 & \hspace*{-3ex}  254 &         139 & \hspace*{-2ex}    96 &   43 & - & - & \hspace*{-2ex} - & \hspace*{-2ex} - & - \\ 
 253 & 25$_{ 7,19}$ - 24$_{ 7,18}$ & \hspace*{-2ex}  227079.847 & \hspace*{-3ex}   50 & \hspace*{-3ex}  191 &         153 & \hspace*{-2ex}    96 &   44 & 0.32 &       10.94(44) & \hspace*{-2ex}        7.16 & \hspace*{-2ex}       57.69 &  blend with CH$_2$CH$^{13}$CN and \\ 
 & & & & & & & & & & & &  CH$_3$OH, v=0 \\ 
 254 & 25$_{ 7,18}$ - 24$_{ 7,17}$ & \hspace*{-2ex}  227079.847 & \hspace*{-3ex}   50 & \hspace*{-3ex}  191 &         153 & \hspace*{-2ex}    96 &   44 & - & - & \hspace*{-2ex} - & \hspace*{-2ex} - & - \\ 
 273 & 25$_{ 2,23}$ - 24$_{ 2,22}$ & \hspace*{-2ex}  231485.527 & \hspace*{-3ex}   50 & \hspace*{-3ex}  138 &         165 & \hspace*{-2ex}    40 &   45 & 0.30 &       12.73(19) & \hspace*{-2ex}        6.27 & \hspace*{-2ex}        6.60 &  blend with U-line? \\ 
 292 & 26$_{ 6,21}$ - 25$_{ 6,20}$ & \hspace*{-2ex}  236269.491 & \hspace*{-3ex}   60 & \hspace*{-3ex}  186 &         163 & \hspace*{-2ex}    37 &   46 & 0.36 &       15.53(18) & \hspace*{-2ex}        8.02 & \hspace*{-2ex}       14.17 &  partial blend with t-C$_2$H$_5$OCHO \\ 
 & & & & & & & & & & & &  and U-line \\ 
 293 & 26$_{ 6,20}$ - 25$_{ 6,19}$ & \hspace*{-2ex}  236270.459 & \hspace*{-3ex}   60 & \hspace*{-3ex}  186 &         163 & \hspace*{-2ex}    37 &   46 & - & - & \hspace*{-2ex} - & \hspace*{-2ex} - & - \\ 
 306 & 28$_{ 0,28}$ - 27$_{ 0,27}$ & \hspace*{-2ex}  244765.968 & \hspace*{-3ex}   21 & \hspace*{-3ex}  160 &         186 & \hspace*{-2ex}    39 &   47 & 0.28 &        9.56(19) & \hspace*{-2ex}        6.62 & \hspace*{-2ex}       10.35 &  blend with CH$_3$$^{13}$CH$_2$CN, v=0 \\ 
 & & & & & & & & & & & &  and U-line \\ 
 322 & 27$_{ 6,22}$ - 26$_{ 6,21}$ & \hspace*{-2ex}  245378.722 & \hspace*{-3ex}   10 & \hspace*{-3ex}  197 &         170 & \hspace*{-2ex}    72 &   48 & 0.35 &       16.69(36) & \hspace*{-2ex}        8.29 & \hspace*{-2ex}       22.21 &  blend with $^{13}$CH$_3$CH$_2$CN, v=0? \\ 
 323 & 27$_{ 6,21}$ - 26$_{ 6,20}$ & \hspace*{-2ex}  245380.146 & \hspace*{-3ex}   10 & \hspace*{-3ex}  197 &         170 & \hspace*{-2ex}    72 &   48 & - & - & \hspace*{-2ex} - & \hspace*{-2ex} - & - \\ 
 368 & 29$_{ 9,20}$ - 28$_{ 9,19}$ & \hspace*{-2ex}  263364.923 & \hspace*{-3ex}   22 & \hspace*{-3ex}  277 &         174 & \hspace*{-2ex}    74 &   49 & 0.26 &        6.72(37) & \hspace*{-2ex}        8.51 & \hspace*{-2ex}        9.17 &  baseline problem?, blend with  \\ 
 & & & & & & & & & & & & U-line \\ 
 369 & 29$_{ 9,21}$ - 28$_{ 9,20}$ & \hspace*{-2ex}  263364.923 & \hspace*{-3ex}   22 & \hspace*{-3ex}  277 &         174 & \hspace*{-2ex}    74 &   49 & - & - & \hspace*{-2ex} - & \hspace*{-2ex} - & - \\ 
 370 & 29$_{10,19}$ - 28$_{10,18}$ & \hspace*{-2ex}  263368.355 & \hspace*{-3ex}   26 & \hspace*{-3ex}  300 &         170 & \hspace*{-2ex}    74 &   49 & - & - & \hspace*{-2ex} - & \hspace*{-2ex} - & - \\ 
 371 & 29$_{10,20}$ - 28$_{10,19}$ & \hspace*{-2ex}  263368.355 & \hspace*{-3ex}   26 & \hspace*{-3ex}  300 &         170 & \hspace*{-2ex}    74 &   49 & - & - & \hspace*{-2ex} - & \hspace*{-2ex} - & - \\ 
 384 & 29$_{ 6,24}$ - 28$_{ 6,23}$ & \hspace*{-2ex}  263604.573 & \hspace*{-3ex}   12 & \hspace*{-3ex}  221 &         184 & \hspace*{-2ex}    74 &   50 & 0.28 &       10.29(36) & \hspace*{-2ex}        8.81 & \hspace*{-2ex}       14.22 &  baseline problem?, partial blend  \\ 
 & & & & & & & & & & & & with CH$_3$CH$_3$CO, v$_{\mathrm{t}}$=1 and \\ 
 & & & & & & & & & & & &  CH$_3$OCH$_3$ \\ 
 385 & 29$_{ 6,23}$ - 28$_{ 6,22}$ & \hspace*{-2ex}  263607.689 & \hspace*{-3ex}   12 & \hspace*{-3ex}  221 &         184 & \hspace*{-2ex}    74 &   50 & - & - & \hspace*{-2ex} - & \hspace*{-2ex} - & - \\ 
 398 & 29$_{ 4,26}$ - 28$_{ 4,25}$ & \hspace*{-2ex}  264055.836 & \hspace*{-3ex}   13 & \hspace*{-3ex}  197 &         189 & \hspace*{-2ex}   108 &   51 & 0.22 &       18.36(49) & \hspace*{-2ex}        5.92 & \hspace*{-2ex}       14.22 &  partial blend with C$_2$H$_5$CN, v=0 \\ 
 & & & & & & & & & & & &  and CH$_3$CH$_3$CO, v=0 \\ 
 \hline
 \end{tabular}
 }
 \end{table*}

We consider it essential for claiming a detection of a new molecule that all 
lines of this molecule in our observed bands are consistent with this claim, 
i.e. are either detected or blended with lines of other species. Therefore, in
the following, we inspect all transitions of amino acetonitrile in our 
frequency range.
Our line survey at 3, 2, and 1.3\,mm covers 596 transitions of our 
amino acetonitrile catalog (v=0 only). Our LTE modeling shows, however, that 
the transitions with the line strength times the appropriate ($a-$ or 
$b-$type) dipole moment squared, S$\mu^2$, smaller than 20 D$^2$ are much too 
weak to be detectable with the sensitivity we achieved. 
Therefore, we list in Table~\ref{t:aan30m} (\textit{online material}) only 
the 398 transitions above this threshold. To save some space, when two 
transitions have a frequency difference smaller than 0.1 MHz which cannot be 
resolved, we list only the first one. We number the transitions in Col.~1 and 
give their quantum numbers in Col.~2. The frequencies, the frequency 
uncertainties, the energies of the lower levels in temperature units, and the
S$\mu^2$ values are listed in Col.~3, 4, 5, and 6, respectively.
Since the spectra are in most cases close to the line confusion limit and it
is difficult to measure the noise level, we give in Col.~7 the rms sensitivity
computed from the system temperature and the integration time: 
$\sigma = \frac{F_{\mathrm{eff}}}{B_{\mathrm{eff}}} \times 
\frac{2\,T_{\mathrm{sys}}}{\sqrt{\delta f \, t}}$, with $F_{\mathrm{eff}}$ and 
$B_{\mathrm{eff}}$ the forward and beam efficiencies, $T_{\mathrm{sys}}$ the 
system temperature, $\delta f$ the spectral resolution, and $t$ the total 
integration time (on-source plus off-source).

\begin{table}
 \centering
 \caption{
 Parameters of our best-fit LTE model of amino acetonitrile.
}
 \label{t:aan30mmodel}
 \vspace*{0.0ex}
 \begin{tabular}{ccccc}
 \hline\hline
 \multicolumn{1}{c}{Size$^{a}$} & \multicolumn{1}{c}{T$_{\mathrm{rot}}$} & \multicolumn{1}{c}{N$_{\mathrm{AAN}}$$^{b}$} & \multicolumn{1}{c}{FWHM} & \multicolumn{1}{c}{V$_{\mathrm{off}}$$^{c}$} \\ 
  \multicolumn{1}{c}{\scriptsize ($''$)} & \multicolumn{1}{c}{\scriptsize (K)} & \multicolumn{1}{c}{\scriptsize (cm$^{-2}$)} & \multicolumn{1}{c}{\scriptsize (km~s$^{-1}$)} & \multicolumn{1}{c}{\scriptsize (km~s$^{-1}$)} \\ 
 \multicolumn{1}{c}{(1)} & \multicolumn{1}{c}{(2)} & \multicolumn{1}{c}{(3)} & \multicolumn{1}{c}{(4)} & \multicolumn{1}{c}{(5)} \\ 
 \hline
 2.0 &  100 & $ 2.80 \times 10^{16}$ & 7.0 & 0.0 \\  \hline
 \end{tabular}
 \begin{list}{}{}
 \item[$(a)$]{Source diameter (FWHM).}
 \item[$(b)$]{Column density of amino acetonitrile.}
 \item[$(c)$]{Velocity offset with respect to the systemic velocity of Sgr~B2(N) V$_{\mathrm{lsr}} = 64$ km~s$^{-1}$.}
 \end{list}
 \end{table}

We list in Col.~8 of Table~\ref{t:aan30m} comments about the blends affecting 
the amino acetonitrile transitions. As can be seen in this table, most of the 
amino acetonitrile lines covered by our survey of Sgr~B2(N) are heavily 
blended with lines of other molecules and therefore cannot be identified in 
this source. 
Only 88 of the 398 transitions are relatively free of contamination from other 
molecules, known or still unidentified according to our modeling. They are 
marked ``Detected'' or ``Group detected'' 
in Col.~8 of Table~\ref{t:aan30m}, and are listed with more information in 
Table~\ref{t:detectaan30m}. They correspond to 51 observed features which are 
shown in Fig.~\ref{f:detectaan30m} (\textit{online material}) and labeled in 
Col.~8 of Table~\ref{t:detectaan30m}. For reference, we show the spectrum 
observed toward Sgr~B2(M) in this figure also. We identified the 
amino acetonitrile lines and the blends affecting them with the LTE model of 
this molecule and the LTE model including all molecules (see 
Sect.~\ref{ss:modeling30m}). The parameters 
of our best-fit LTE model of amino acetonitrile are listed in 
Table~\ref{t:aan30mmodel}, and the model is overlaid in red on the spectrum 
observed toward Sgr~B2(N) in Fig.~\ref{f:detectaan30m}.
The best-fit LTE model including all molecules is shown in green in the same 
figure. The source size we used to model the amino acetonitrile emission was 
derived from our interferometric measurements (see Sect.~\ref{ss:aanpdbi} 
below).

\begin{figure}
\centerline{\resizebox{1.0\hsize}{!}{\includegraphics[angle=270]{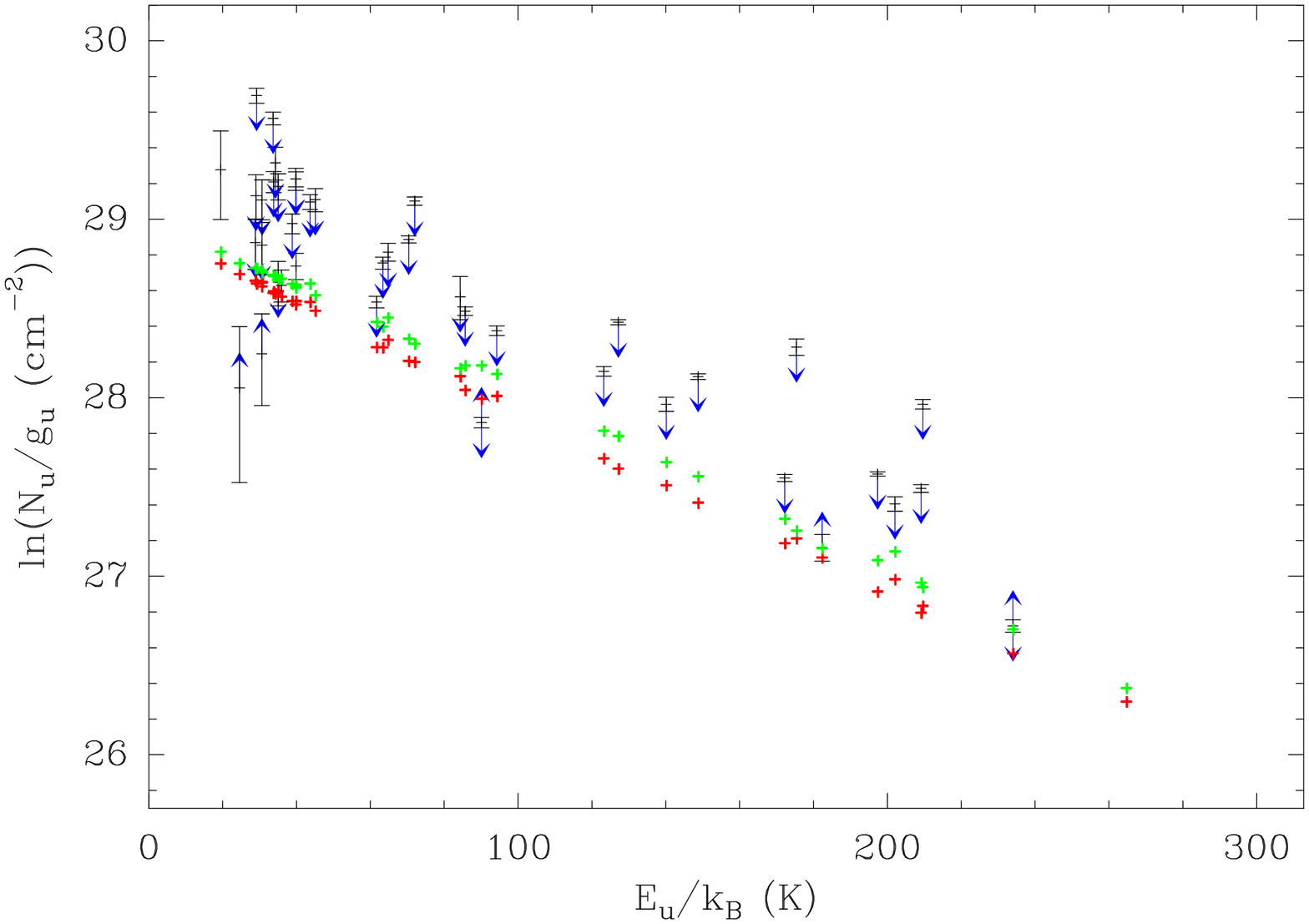}}}
\caption{Population diagram of amino acetonitrile in Sgr~B2(N). The red points 
were computed in the optically thin approximation using the integrated 
intensities of our best-fit \textit{model} of amino acetonitrile, while the 
green points were corrected for the opacity. The black points were computed in 
the optically thin approximation using the integrated intensities of the 
spectrum \textit{observed} with the IRAM 30m telescope. 
The error bars are $1 \sigma$ uncertainties on $N_u/g_u$. Blue arrows pointing
downwards mark the transitions blended with transitions from other molecules, 
while blue arrows pointing upwards indicate that the baseline removed in the 
observed spectrum is uncertain. The arrow length is arbitrary. The measurement 
corresponding to feature 43 (at $E_u/k_B$ = 265 K) is not shown since the 
integrated intensity measured toward Sgr~B2(N) is negative, due to the blend 
with CN absorption lines.}
\label{f:popdiag}
\end{figure}

For the frequency range corresponding to each observed amino acetonitrile 
feature, we list in Table~\ref{t:detectaan30m} the integrated intensities of 
the observed spectrum
(Col.~10), of the best-fit model of amino acetonitrile (Col.~11), and of the 
best-fit model including all molecules (Col.~12). In these columns, the dash 
symbol indicates transitions belonging to the same feature. Columns 1 to 7 are
the same as in Table~\ref{t:aan30m}. The $1\sigma$ uncertainty given in Col.~10
was computed using the estimated noise level of Col.~7. These measurements are 
plotted in the form of a population diagram in Fig.~\ref{f:popdiag}, which 
plots upper level column density divided by statistical weight, $N_u/g_u$, 
versus the upper level energy in Kelvins \citep[see][]{Goldsmith99}.
The data are shown in black and our best-fit model of amino acetonitrile in 
red. Out of 21 features encompassing several transitions, 
10 contain transitions with different energy levels and were ignored in the 
population diagram (features 2, 3, 8, 33, 34, 35, 36, 42, and 49).
We used equation A5 of \citet{Snyder05} to compute the ordinate values. This 
equation assumes optically thin emission. To estimate by how much
line opacities affect this diagram, we applied the opacity correction factor 
$C_\tau = \frac{\tau}{1-e^{-\tau}}$ \citep[see][]{Goldsmith99,Snyder05} to the 
modeled intensities, using the opacities from our radiative transfer 
calculations (Col.~9 of Table~\ref{t:detectaan30m}); the result is shown in 
green in 
Fig.~\ref{f:popdiag}. The population diagram derived from the modeled spectrum 
is slightly shifted upwards but its shape, in particular its slope (the 
inverse of which \textit{approximately} determines the rotation temperature), 
is not significantly changed, since $ln\, C_{\tau}$ does not vary much (from 
0.04 to 0.24). The populations derived from the \textit{observed}
spectrum in the optically thin approximation are therefore not significantly 
affected by the optical depth of the amino acetonitrile 
transitions\footnote{Note that our modeled spectrum is anyway calculated with 
the full LTE radiative transfer which takes into account the optical depth 
effects (see Sect.~\ref{ss:modeling30m}).}. 
The scatter of the black crosses in Fig.~\ref{f:popdiag} is  therefore
dominated by the blends with other molecules and uncertainties in the baseline 
removal (indicated by the downwards and upwards blue arrows, respectively). 
From this analysis, we conclude that our best-fit model for amino acetonitrile 
is fully consistent with our 30m data of Sgr~B2(N).

\begin{figure}
\centerline{\resizebox{1.0\hsize}{!}{\includegraphics[angle=270]{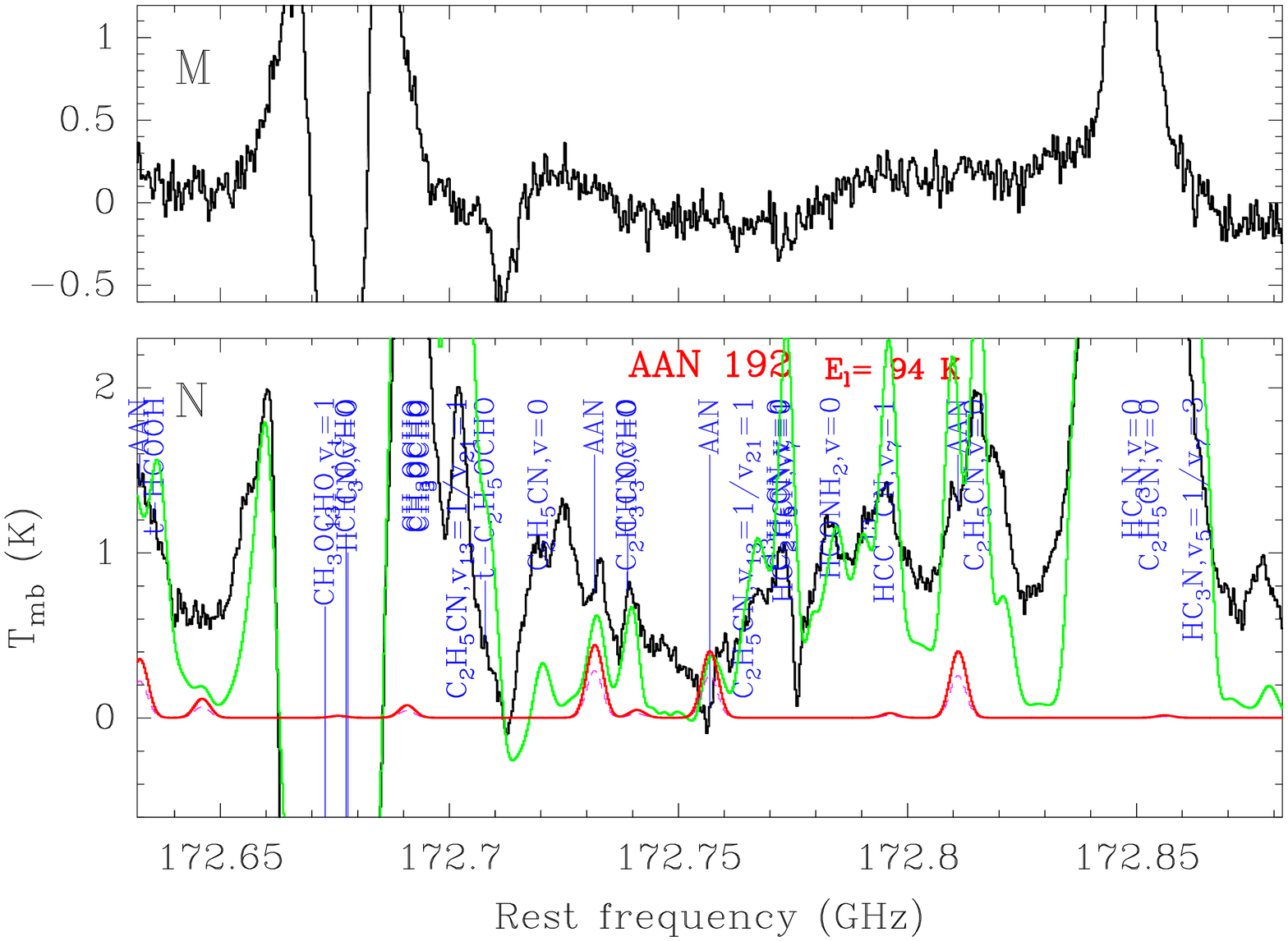}}}
\caption{Spectrum obtained toward Sgr~B2(N) (bottom) and Sgr~B2(M) (top) with 
the IRAM 30m telescope at the frequency of amino acetonitrile (AAN) transition 
192 (see caption of Fig.~\ref{f:detectaan30m} for more details about the color 
coding). There are too many blended lines in the spectrum of Sgr~B2(N) to 
properly remove the baseline, which is very uncertain and most likely at a 
lower level
than could be computed here. This is the only discrepancy concerning the
amino acetonitrile lines in the whole survey. The absorption lines, 
particularly strong in the spectrum of Sgr~B2(M), are velocity components of 
H$^{13}$CN 2$-$1.}
\label{f:badline30m}
\end{figure}

Finally, as mentioned above, the 310 transitions of Table~\ref{t:aan30m} not 
shown in Fig.~\ref{f:detectaan30m} are all but one heavily blended with 
transitions of other molecules and cannot be clearly identified in Sgr~B2(N). 
The single exception is amino acetonitrile transition 192 shown in 
Fig.~\ref{f:badline30m}. There are too many blended lines in this frequency 
range to properly remove the baseline. It is very uncertain and the true 
baseline is most likely at a lower level than computed here. The presence of 
several H$^{13}$CN 2$-$1 velocity components in absorption also complicates 
the analysis. Therefore it 
is very likely that the (single) apparent disagreement concerning
transition 192 between our best-fit model and the 30m spectrum observed toward 
Sgr~B2(N) is not real and does not invalidate our claim of detection of 
amino acetonitrile.

\subsection{Mapping amino acetonitrile with the PdBI}
\label{ss:aanpdbi}

\begin{figure}
\centerline{\resizebox{1.0\hsize}{!}{\includegraphics[angle=270]{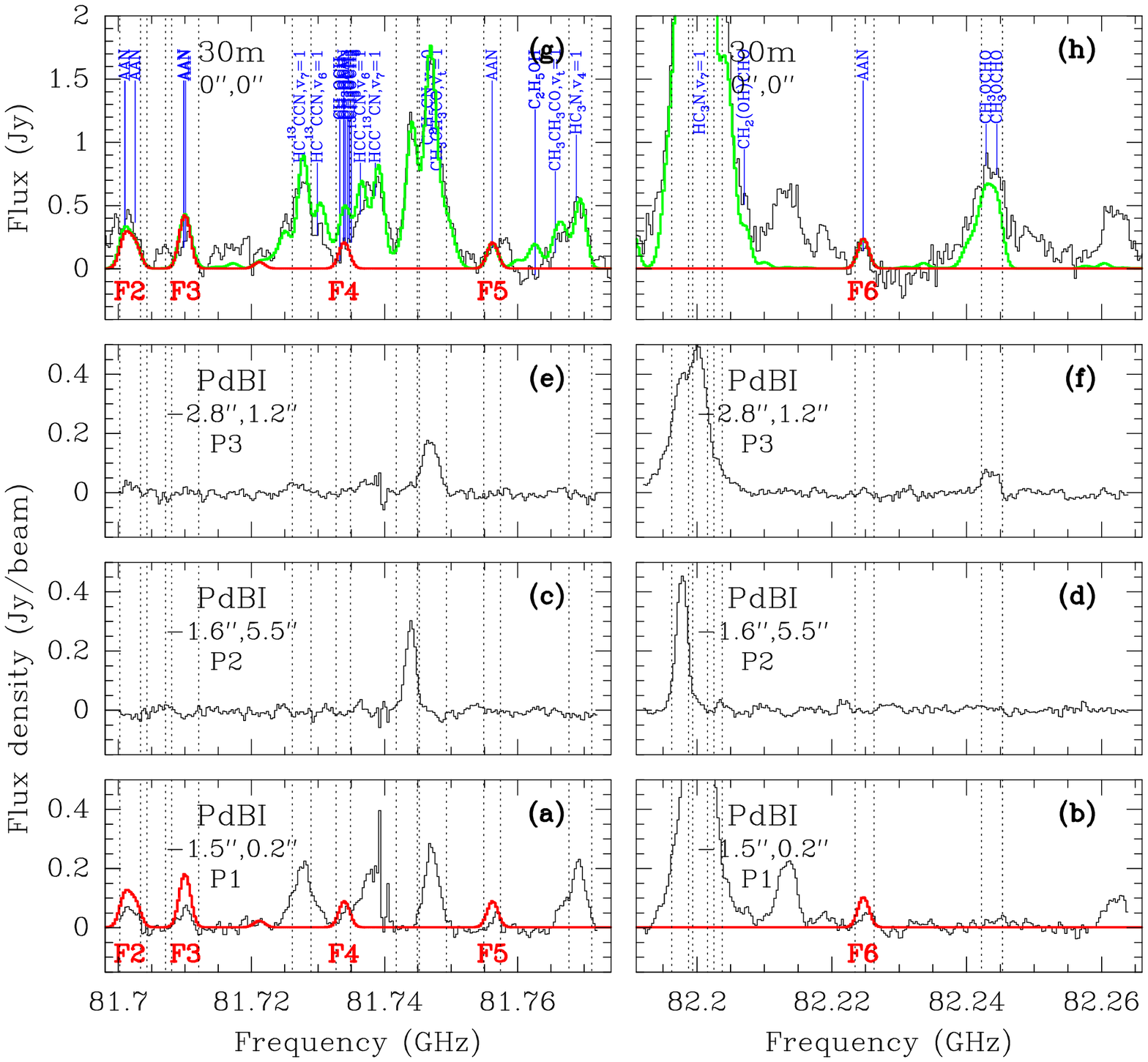}}}
\caption{Spectra obtained with the Plateau de Bure interferometer (\textbf{a} 
to \textbf{f}) and the 30m telescope (\textbf{g} and \textbf{h}) toward 
Sgr~B2(N) (in black). The dotted lines show the frequency ranges listed 
in Table~\ref{t:aanpdbi}. The offset position with respect to the reference 
position of Fig.~\ref{f:pdbimap} is given in each panel, along with a label
(P1 to P3, see their definition in Table~\ref{t:positions}). The 
lines identified in our 30m survey are labeled in blue. The red spectra show 
our best-fit model for amino acetonitrile (AAN) while the green spectrum 
corresponds to the 30m model including all molecules. The observed lines which 
have no counterpart in the green spectrum are still unidentified.}
\label{f:pdbispec}
\end{figure}

The two 3\,mm spectral windows of the PdBI were chosen to cover the five
amino acetonitrile features F2 to F6. The spectra toward Sgr~B2(N) are shown 
for both windows toward 3 positions P1, P2, and P3 in Fig.~\ref{f:pdbispec}a 
to f. Many lines are detected, the strongest one being a line from within the 
vibrationally excited state v$_7$=1 of cyanoacetylene (HC$_3$N) at 82.2 GHz. 
We also easily detect lines from within its vibrationally excited state 
v$_4$=1, from its isotopologues 
HC$^{13}$CCN and HCC$^{13}$CN in the v$_7$=1 state, from ethyl cyanide 
(C$_2$H$_5$CN), as well as two unidentified lines at 82.213 and 82.262 GHz. At 
a lower level, we find emission for all the amino acetonitrile features F2 to 
F6, and we also detect methylformate (CH$_3$OCHO).

\begin{figure*}
\centerline{\resizebox{1.0\hsize}{!}{\includegraphics[angle=270]{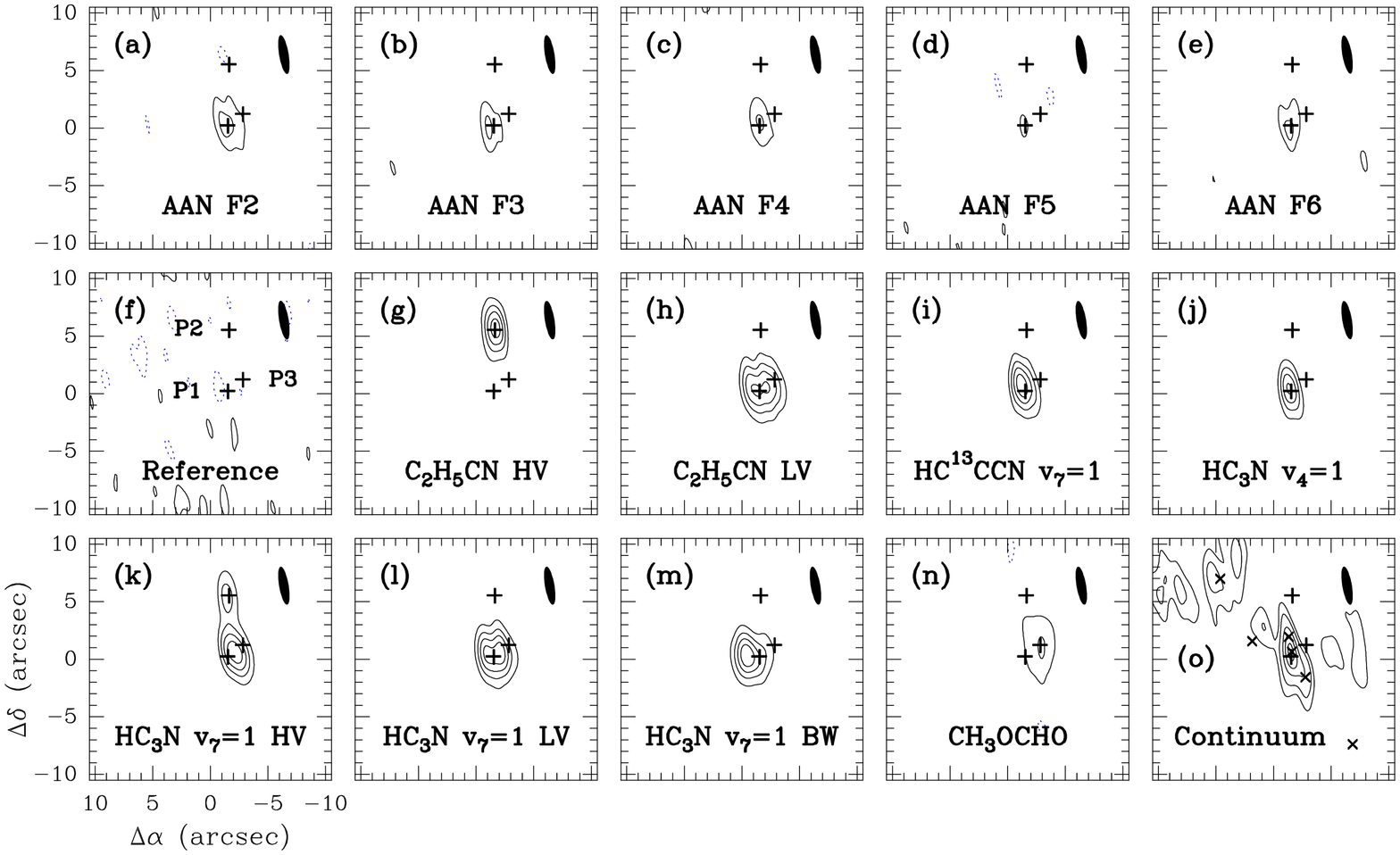}}}
\caption{Integrated intensity maps (panels \textbf{a} to \textbf{n}) and 
continuum map (panel \textbf{o}) obtained toward Sgr~B2(N) with the Plateau 
de Bure interferometer at 82 GHz. Panels \textbf{a} to \textbf{e} show 
the amino acetonitrile (AAN) features F2 to F6 (see Fig.~\ref{f:pdbispec} and 
Table~\ref{t:detectaan30m}). Panel \textbf{f} is a reference map integrated 
on the emission-free frequency range between F2 and F3. Panels \textbf{g} 
to \textbf{n} show the other molecules listed in Table~\ref{t:aanpdbi}. The 
lowest contour (positive in black solid line and negative in blue dotted line) 
and the contour step are 
2$\sigma$ for panel \textbf{f},
3$\sigma$ for panels \textbf{a} to \textbf{e} and panel \textbf{n}, 
4$\sigma$ for panels \textbf{g} and \textbf{h},
5$\sigma$ for panels \textbf{i} to \textbf{l},
and 6$\sigma$ for panel \textbf{m}
(with $\sigma$ given in Col.~5 of Table~\ref{t:aanpdbi}).
For panel \textbf{o}, the first contours are 5$\sigma$ and 10$\sigma$, and the
contour step is 10$\sigma$ for the other contours (with $\sigma$ = 8.5 
mJy/beam). In each panel, the 0,0 position is 
$\alpha_{\mathrm{J2000}} = 17^{\mathrm{h}}47^{\mathrm{m}}20\fs00$, 
$\delta_{\mathrm{J2000}} = -28^\circ22\arcmin19.0\arcsec$, the three plus 
symbols mark the positions P1, P2, and P3 (labeled in panel \textbf{f}), and 
the filled ellipse in the top right corner shows the clean beam 
(HPBW $= 3.35\arcsec \times 0.81\arcsec$ at P.A. = 9.7$^\circ$). 
The cross symbols in panel \textbf{o} are the peak positions of the 
(ultracompact) H{\sc ii} regions detected by \citet{Gaume95} at 1.3~cm (K9.69, 
K1, K2, K3, K5, and K6, from right to left).
The spectral integration was done on the frequency ranges 
given in Table~\ref{t:aanpdbi}. The continuum map was computed on line-free 
frequency ranges. The maps are not corrected for primary beam
attenuation. The amino acetonitrile features emit at the same position as the 
vibrationally excited state v$_4$ = 1 of cyanoacetylene. Feature F4 is 
partially blended with a transition from HCC$^{13}$CN, v$_6$ = 1.}
\label{f:pdbimap}
\end{figure*}

The integrated intensity maps of the amino acetonitrile features F2 to 
F6 are presented in Fig.~\ref{f:pdbimap}a to e, along with two maps of
ethyl cyanide (Fig.~\ref{f:pdbimap}g and h), four maps of cyanoacetylene in the
vibrationally excited states v$_4$=1 and v$_7$=1 (Fig.~\ref{f:pdbimap}j to m),
one map of its isotopologue HC$^{13}$CCN in the state v$_7$=1 
(Fig.~\ref{f:pdbimap}i), one map of methylformate (Fig.~\ref{f:pdbimap}n),
and a reference map computed on the PdBI line-free frequency range between F2 
and F3 (Fig.~\ref{f:pdbimap}f). The frequency intervals used to compute the 
integrated intensities are given in Col.~3 and 4 of Table~\ref{t:aanpdbi} and 
shown with dotted lines in Fig.~\ref{f:pdbispec}. 
We used the fitting routine GAUSS$\_$2D of the GILDAS software to measure the 
position, size, and peak flux of each integrated emission. The results are
listed in Col.~6 to 11 in Table~\ref{t:aanpdbi}.
We label P1 the mean peak position of features F2 to F6, P2 the northern peak 
position of ethyl cyanide, and P3 the peak position of methylformate (see 
Table~\ref{t:positions} and the plus symbols in Fig.~\ref{f:pdbimap}).
Finally, the PdBI velocity-integrated flux spatially integrated over the 
emitting region is listed in Col.~12 and the 30m velocity-integrated intensity 
is given in Col.~13.

\begin{table*}
 \centering
 \caption{
 Measurements obtained toward Sgr~B2(N) with the IRAM Plateau de Bure interferometer at 82 GHz.
}
 \label{t:aanpdbi}
 \vspace*{0.0ex}
 \begin{tabular}{lcccrrrrrrrrr}
 \hline\hline
 \multicolumn{1}{c}{Molecule} & \multicolumn{1}{c}{\hspace*{-3.5ex} F$^a$} & \multicolumn{1}{c}{\hspace*{-2ex} f$_{\mathrm{min}}$$^b$} & \multicolumn{1}{c}{\hspace*{-2ex} f$_{\mathrm{max}}$$^b$} & \multicolumn{1}{c}{$\sigma^c$} & \multicolumn{1}{c}{\hspace*{-2.ex} F$_{\mathrm{peak}}$$^d$} & \multicolumn{1}{c}{$\Delta\alpha$$^d$} & \multicolumn{1}{c}{\hspace*{-1.5ex} $\Delta\delta$$^d$} & \multicolumn{1}{c}{$\theta_{\mathrm{maj}}^{\mathrm{fwhm}}$ $^d$} & \multicolumn{1}{c}{\hspace*{-1.5ex} $\theta_{\mathrm{min}}^{\mathrm{fwhm}}$ $^d$} & \multicolumn{1}{c}{\hspace*{-1ex} P.A.$^d$} & \multicolumn{1}{c}{\hspace*{-2ex} $\Phi_{\mathrm{PdBI}}^e$} & \multicolumn{1}{c}{\hspace*{-2ex} $\Phi_{\mathrm{30m}}^f$} \\ 
  \multicolumn{1}{c}{} & \multicolumn{1}{c}{} & \multicolumn{1}{c}{\hspace*{-2ex} \scriptsize (MHz))} & \multicolumn{1}{c}{\hspace*{-2ex} \scriptsize (MHz)} & \multicolumn{2}{c}{ \hspace*{-3ex} \scriptsize (Jy/beam km/s)} & \multicolumn{1}{c}{\scriptsize ($ '' $)} & \multicolumn{1}{c}{\hspace*{-1.5ex} \scriptsize ($ '' $)} & \multicolumn{1}{c}{\scriptsize ($ '' $)} & \multicolumn{1}{c}{\hspace*{-1.5ex} \scriptsize ($ '' $)} & \multicolumn{1}{c}{\hspace*{-1ex} \scriptsize ($^\circ$)} & \multicolumn{1}{c}{\hspace*{-2ex} \scriptsize (Jy km/s)} & \multicolumn{1}{c}{\hspace*{-2ex} \scriptsize (Jy km/s)} \\ 
 \multicolumn{1}{c}{(1)} & \multicolumn{1}{c}{\hspace*{-3.5ex} (2)} & \multicolumn{1}{c}{\hspace*{-2ex} (3)} & \multicolumn{1}{c}{\hspace*{-2ex} (4)} & \multicolumn{1}{c}{(5)} & \multicolumn{1}{c}{\hspace*{-2.ex} (6)} & \multicolumn{1}{c}{(7)} & \multicolumn{1}{c}{\hspace*{-1.5ex} (8)} & \multicolumn{1}{c}{(9)} & \multicolumn{1}{c}{\hspace*{-1.5ex} (10)} & \multicolumn{1}{c}{\hspace*{-1ex} (11)} & \multicolumn{1}{c}{\hspace*{-2ex} (12)} & \multicolumn{1}{c}{\hspace*{-2ex} (13)} \\ 
 \hline
AAN & \hspace*{-3.5ex} F2 &\hspace*{-2ex}  81700.21 & \hspace*{-2ex}  81703.33 &  0.09 & \hspace*{-2ex}      0.68 &     -1.60 $\pm$      0.05 & \hspace*{-1.5ex}      0.30 $\pm$      0.22 &       3.9 $\pm$       0.4 & \hspace*{-1.5ex}      2.00 $\pm$      0.10 & \hspace*{-1ex}      20.5 $\pm$       0.1 & \hspace*{-2ex}      1.76 & \hspace*{-2ex}      2.89 \\ AAN & \hspace*{-3.5ex} F3 &\hspace*{-2ex}  81708.02 & \hspace*{-2ex}  81712.08 &  0.10 & \hspace*{-2ex}      0.68 &     -1.25 $\pm$      0.06 & \hspace*{-1.5ex}      0.02 $\pm$      0.24 &       3.8 $\pm$       0.5 & \hspace*{-1.5ex}      1.39 $\pm$      0.12 & \hspace*{-1ex}      10.1 $\pm$       0.0 & \hspace*{-2ex}      1.24 & \hspace*{-2ex}      1.75 \\ AAN & \hspace*{-3.5ex} F4 &\hspace*{-2ex}  81732.71 & \hspace*{-2ex}  81734.90 &  0.06 & \hspace*{-2ex}      0.44 &     -1.70 $\pm$      0.06 & \hspace*{-1.5ex}      0.35 $\pm$      0.24 &       3.6 $\pm$       0.5 & \hspace*{-1.5ex}      1.54 $\pm$      0.12 & \hspace*{-1ex}      14.0 $\pm$       0.0 & \hspace*{-2ex}      0.86 & \hspace*{-2ex}      0.98 \\ AAN & \hspace*{-3.5ex} F5 &\hspace*{-2ex}  81754.90 & \hspace*{-2ex}  81757.40 &  0.06 & \hspace*{-2ex}      0.24 &     -1.52 $\pm$      0.10 & \hspace*{-1.5ex}      0.03 $\pm$      0.44 &       3.2 $\pm$       0.9 & \hspace*{-1.5ex}      1.20 $\pm$      0.21 & \hspace*{-1ex}      12.5 $\pm$       1.1 & \hspace*{-2ex}      0.30 & \hspace*{-2ex}      1.15 \\ AAN & \hspace*{-3.5ex} F6 &\hspace*{-2ex}  82223.46 & \hspace*{-2ex}  82226.27 &  0.06 & \hspace*{-2ex}      0.43 &     -1.43 $\pm$      0.06 & \hspace*{-1.5ex}      0.28 $\pm$      0.24 &       3.5 $\pm$       0.5 & \hspace*{-1.5ex}      1.54 $\pm$      0.11 & \hspace*{-1ex}       6.0 $\pm$       0.4 & \hspace*{-2ex}      0.79 & \hspace*{-2ex}      0.97 \\ Reference & \hspace*{-3.5ex}  &\hspace*{-2ex}  81704.27 & \hspace*{-2ex}  81707.08 &  0.07 & \multicolumn{1}{c}{...} & \multicolumn{1}{c}{...} & \multicolumn{1}{c}{...} & \multicolumn{1}{c}{...} & \multicolumn{1}{c}{...} & \multicolumn{1}{c}{...} & ... & ... \\ C$_2$H$_5$CN & \hspace*{-3.5ex} HV &\hspace*{-2ex}  81741.77 & \hspace*{-2ex}  81744.90 &  0.11 & \hspace*{-2ex}      2.05 &     -1.64 $\pm$      0.02 & \hspace*{-1.5ex}      5.58 $\pm$      0.09 &       3.8 $\pm$       0.2 & \hspace*{-1.5ex}      1.50 $\pm$      0.04 & \hspace*{-1ex}       5.7 $\pm$       0.0 & \hspace*{-2ex}      4.07 & \hspace*{-2ex}      6.38 \\ C$_2$H$_5$CN & \hspace*{-3.5ex} LV &\hspace*{-2ex}  81745.21 & \hspace*{-2ex}  81749.27 &  0.15 & \hspace*{-2ex}      2.82 &     -1.74 $\pm$      0.02 & \hspace*{-1.5ex}      0.46 $\pm$      0.09 &       3.8 $\pm$       0.2 & \hspace*{-1.5ex}      2.87 $\pm$      0.04 & \hspace*{-1ex}      13.7 $\pm$       0.0 & \hspace*{-2ex}     10.43 & \hspace*{-2ex}     12.94 \\ {\scriptsize HC$^{13}$CCN v$_7$=1} & \hspace*{-3.5ex}  &\hspace*{-2ex}  81726.15 & \hspace*{-2ex}  81728.96 &  0.09 & \hspace*{-2ex}      2.20 &     -1.35 $\pm$      0.02 & \hspace*{-1.5ex}      0.60 $\pm$      0.07 &       3.7 $\pm$       0.1 & \hspace*{-1.5ex}      1.68 $\pm$      0.03 & \hspace*{-1ex}      12.6 $\pm$       0.0 & \hspace*{-2ex}      4.98 & \hspace*{-2ex}      4.81 \\ HC$_3$N {\scriptsize v$_4$=1} & \hspace*{-3.5ex}  &\hspace*{-2ex}  81767.71 & \hspace*{-2ex}  81771.15 &  0.10 & \hspace*{-2ex}      2.14 &     -1.43 $\pm$      0.02 & \hspace*{-1.5ex}      0.28 $\pm$      0.08 &       3.6 $\pm$       0.2 & \hspace*{-1.5ex}      1.35 $\pm$      0.04 & \hspace*{-1ex}       9.9 $\pm$       0.0 & \hspace*{-2ex}      3.78 & \hspace*{-2ex}      3.85 \\ HC$_3$N {\scriptsize v$_7$=1}$^{g}$ & \hspace*{-3.5ex} HV &\hspace*{-2ex}  82196.27 & \hspace*{-2ex}  82198.77 &  0.25 & \hspace*{-2ex}      6.17 &     -2.16 $\pm$      0.02 & \hspace*{-1.5ex}      0.69 $\pm$      0.07 &       4.0 $\pm$       0.1 & \hspace*{-1.5ex}      1.84 $\pm$      0.03 & \hspace*{-1ex}      16.2 $\pm$      22.5 & \hspace*{-2ex}     16.05 & \hspace*{-2ex}     23.88 \\  & & & &  & \hspace*{-2ex}      3.36 &     -1.50 $\pm$      0.03 & \hspace*{-1.5ex}      5.25 $\pm$      0.12 &       4.0 $\pm$       0.2 & \hspace*{-1.5ex}      1.36 $\pm$      0.06 & \hspace*{-1ex}       5.5 $\pm$      22.5 & \hspace*{-2ex}      5.35 & ... \\ HC$_3$N {\scriptsize v$_7$=1} & \hspace*{-3.5ex} LV &\hspace*{-2ex}  82199.40 & \hspace*{-2ex}  82201.58 &  0.36 & \hspace*{-2ex}      9.06 &     -1.67 $\pm$      0.02 & \hspace*{-1.5ex}      0.42 $\pm$      0.07 &       3.7 $\pm$       0.1 & \hspace*{-1.5ex}      2.50 $\pm$      0.03 & \hspace*{-1ex}      10.2 $\pm$      22.5 & \hspace*{-2ex}     31.04 & \hspace*{-2ex}     33.48 \\ HC$_3$N {\scriptsize v$_7$=1} & \hspace*{-3.5ex} BW &\hspace*{-2ex}  82202.52 & \hspace*{-2ex}  82203.77 &  0.12 & \hspace*{-2ex}      3.37 &     -0.71 $\pm$      0.01 & \hspace*{-1.5ex}      0.24 $\pm$      0.06 &       3.1 $\pm$       0.1 & \hspace*{-1.5ex}      2.77 $\pm$      0.03 & \hspace*{-1ex}      45.0 $\pm$       0.0 & \hspace*{-2ex}     11.75 & \hspace*{-2ex}     12.39 \\ CH$_3$OCHO & \hspace*{-3.5ex}  &\hspace*{-2ex}  82242.21 & \hspace*{-2ex}  82245.33 &  0.10 & \hspace*{-2ex}      0.67 &     -2.83 $\pm$      0.06 & \hspace*{-1.5ex}      1.23 $\pm$      0.26 &       4.8 $\pm$       0.5 & \hspace*{-1.5ex}      2.58 $\pm$      0.12 & \hspace*{-1ex}       9.5 $\pm$      22.5 & \hspace*{-2ex}      2.83 & \hspace*{-2ex}      6.62 \\  \hline
 \end{tabular}
 \begin{list}{}{}
 \item[$(a)$]{Feature numbered like in Col.~8 of Table~\ref{t:detectaan30m} for amino acetonitrile (AAN). HV and LV mean "high" and "low" velocity components, respectively, and BW means blueshifted linewing.}
 \item[$(b)$]{Frequency range over which the intensity was integrated.}
 \item[$(c)$]{Noise level in the integrated intensity map shown in Fig.~\ref{f:pdbimap}.}
 \item[$(d)$]{Peak flux, offsets in right ascension and declination with respect to the reference position of Fig.~\ref{f:pdbimap}, major and minor diameters (FWHM), and position angle (East from North) derived by fitting an elliptical 2D Gaussian to the integrated intensity map shown in Fig.~\ref{f:pdbimap}. The uncertainty in Col.~11 is the formal uncertainty given by the fitting routine GAUSS$\_$2D, while the uncertainties correspond to the beam size divided by two times the signal-to-noise ratio in Col.~7 and 8 and by the signal-to-noise ratio in Col.~9 and 10.}
 \item[$(e)$]{Flux spatially integrated over the region showing emission in the integrated intensity map of Fig.~\ref{f:pdbimap}.}
 \item[$(f)$]{Integrated flux of the 30m spectrum computed over the frequency range given in Col.~3 and 4.}
 \item[$(g)$]{The two emission peaks of Fig.~\ref{f:pdbimap}k were fitted separately.}
 \end{list}
 \end{table*}

We present in Fig.~\ref{f:pdbimap}o the map of continuum emission at 
82.0~GHz, integrated over line-free frequency ranges. The continuum emission
has a complex structure. The main region peaks at
$\alpha_{\mathrm{J2000}} = 17^{\mathrm{h}}47^{\mathrm{m}}19\fs886 \pm 0\fs005$,
$\delta_{\mathrm{J2000}} = -28^\circ22\arcmin18.4\arcsec \pm 0.1\arcsec$, i.e.
within 0.1$\arcsec$ of the position of the ultracompact H{\sc ii} region K2.
It also shows hints of emission at the position of the ultracompact H{\sc ii} 
regions K1 and K3, although the spatial resolution is too poor to resolve them 
\citep[see, e.g.,][]{Gaume95}. There are other secondary peaks. One of them 
coincides with the peak of the shell-like H{\sc ii} region K6 while another 
one is located close ($< 2\arcsec$) to the peak of the shell-like H{\sc ii} 
region K5 and traces most likely the same shell. On the other hand, we detect 
no 3.7\,mm emission at the position of the weak ultracompact H{\sc ii} region 
K9.69 \citep[][]{Gaume95}.

The strong lines detected with the PdBI (Fig.~\ref{f:pdbimap}g to n) allow us 
to gain insight into the distribution of molecular line emission in Sgr~B2(N).
The double-peaked profile of ethyl cyanide seen with the 30m telescope (see 
Fig.~\ref{f:pdbispec}g) is resolved with the PdBI into two sources P1 and P2
separated by about 5.3$\arcsec$ (see Fig.~\ref{f:pdbispec}a, c,
Fig.~\ref{f:pdbimap}g, h, and Table~\ref{t:positions}). P1 and P2 are 
spatially and kinematically 
coincident with the quasi-thermal methanol emission cores ``i'' and ``h''
within $0.6\arcsec$ and $0.2\arcsec$, respectively \citep[][]{Mehringer97a}. 
Cores ``i'' and ``h'' were both 
previously detected in ethyl cyanide \citep[][]{Liu99,Hollis03,Jones07}. Many 
molecules (but not amino acetonitrile within the limits of our sensitivity) 
actually show this double-peaked profile in our 30m survey of Sgr~B2(N) and 
are most likely emitted by these two sources.
P1 and P2 are also detected in our PdBI data in the vibrationally excited 
state v$_7$=1 of cyanoacetylene (see Fig.~\ref{f:pdbispec}b, d, and 
Fig.~\ref{f:pdbimap}k, l), and there is a hint of emission toward P2
in the isotopologue HC$^{13}$CCN while P1 is easily detected (see 
Fig.~\ref{f:pdbispec}c and a). In addition, the wings of the main component of 
cyanoacetylene v$_7$=1 are spatially shifted: the redshifted wing peaks about 
$1\arcsec$ North-West of P1 while the blueshifted wing peaks 
about $1\arcsec$ East of P1 (Fig.~\ref{f:pdbimap}k and m). This 
East-West velocity gradient was previously reported by several authors 
\citep[e.g.][]{Lis93,deVicente00,Hollis03}. Although it could result from 
cloud rotation, it is most likely a sign of outflow activity 
\citep[see, e.g.,][]{Liu99}.
The transition with highest energy in our PdBI sample is a 
transition of cyanoacetylene in the vibrationally excited state v$_4$=1
($E_u/k_B = 1283$ K). Within the limits of our sensitivity, we detect emission 
only toward P1 in this highly excited transition (Fig.~\ref{f:pdbimap}j). 
Finally, methylformate 
peaks at a position significantly offset from P1, at $1.7\arcsec$ to the 
North-West (Fig.~\ref{f:pdbimap}n). It has no counterpart in the continuum map 
of Fig.~\ref{f:pdbimap}o. To sum up, our PdBI data reveal three main positions
of molecular line emission (P1 and P2 corresponding to the methanol cores 
``i'' and ``h'', and P3 the peak position of methylformate), and an East-West 
velocity gradient around P1.

Within the limits of our sensitivity, the amino acetonitrile features F2 to F6 
detected with the PdBI show only one peak, and they all peak at the same 
position
(Fig.~\ref{f:pdbimap}a to e). We are confident that the emission detected in 
features F2 to F6 is not contaminated by the continuum since no significant 
signal is detected in the reference map (Fig.~\ref{f:pdbimap}f). 
Their weighted-mean peak position was labeled P1 above
(offset $-1.5 \pm 0.2\arcsec$, $0.2 \pm 0.2\arcsec$, see 
Table~\ref{t:positions}). 
\textit{The fact that all features are detected at the same position is 
consistent with their assignment to the same molecule} (see above the shifted 
position of methylformate for instance). The \textit{deconvolved} major and 
minor axes of the emission detected in features F2 to F6 are in
the range 0--2.2$\arcsec$ and 1.0--1.9$\arcsec$, respectively. The 
amino acetonitrile emission is therefore slightly resolved and has a size of 
roughly 2$\arcsec$ FWHM, which we used for the LTE modeling.
The spatially integrated fluxes of F4 and F6
agree within 20$\%$ with the fluxes measured with the 30m telescope (see 
Col.~12 and 13 of Table~\ref{t:aanpdbi}). The emission detected with the 
30m telescope in these two features of amino acetonitrile is therefore compact 
(2$\arcsec$) and was not filtered out by the interferometer. The other 
features F2, F3, and F5 have 30m fluxes 
1.6, 1.4, and 3.8 times larger than the PdBI fluxes, respectively: the 
emission filtered out by the interferometer 
most likely corresponds to the unidentified transitions blended with 
these amino acetonitrile features (see Fig.~\ref{f:pdbispec}a). In addition, 
the low signal-to-noise ratio of feature F5 detected with the PdBI may 
significantly affect the flux measurement.

We used the parameters of the 30m model (see Table~\ref{t:aan30mmodel}) to
compute a model spectrum of amino acetonitrile with the spatial resolution of 
the PdBI (using the geometrical mean of the elliptical beam). The agreement 
with the peak spectrum is good, within a factor of 2 
(see Fig.~\ref{f:pdbispec}a and b). The small discrepancy may come from the 
somewhat uncertain source size and from our approximate modeling of the 
interferometric beam pattern: spherical beam and full \textit{uv} coverage 
for the model versus elliptical beam and partially sampled \textit{uv} 
coverage for the observations.
Overall, our LTE model of amino acetonitrile is therefore well consistent with 
the compact emission detected with the PdBI.

\subsection{Mapping amino acetonitrile with the ATCA}
\label{ss:aanatca}

\begin{figure}
\centerline{\resizebox{1.0\hsize}{!}{\includegraphics[angle=0]{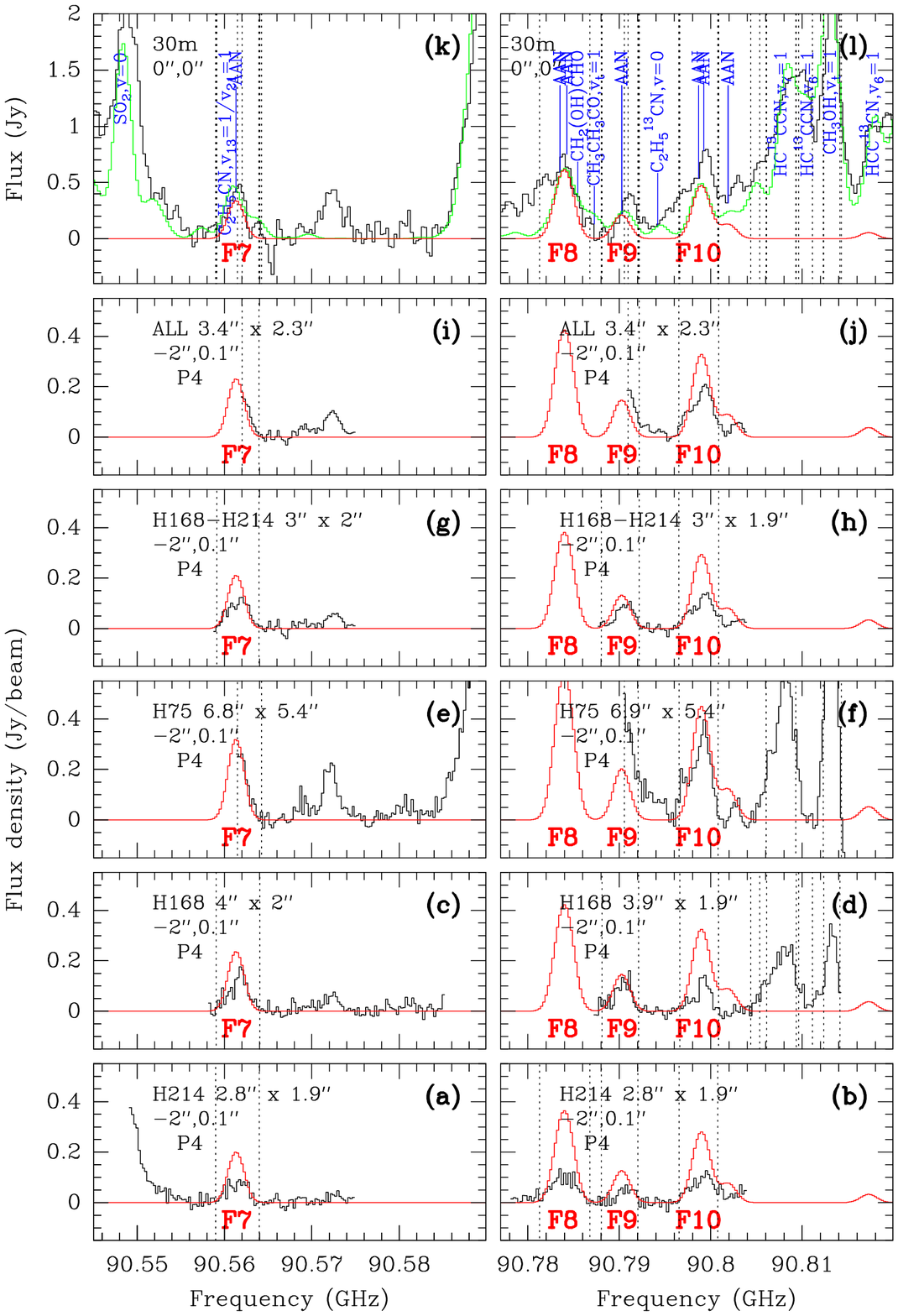}}}
\caption{Spectra obtained with the Australia Telescope Compact Array 
(\textbf{a} to \textbf{j}) and the 30m telescope (\textbf{k} and \textbf{l})
toward Sgr~B2(N) (black histogram). The dotted lines show the frequency ranges 
listed in Table~\ref{t:aanatca}. The offset position with respect to the 
reference position of Fig.~\ref{f:atcamap} is given in each panel. The lines 
identified in our 30m survey are labeled in blue. The red spectrum shows our
best-fit model for amino acetonitrile (AAN) while the green spectrum 
corresponds to the 30m model including all molecules. The observed lines which 
have no counterpart in the green spectrum are still unidentified.
Figures \textbf{a} and \textbf{b} show
the extended configuration H\,214, \textbf{c} and \textbf{d} the intermediate
configuration H\,168, \textbf{e} and \textbf{f} the compact configuration 
H\,75, \textbf{g} and \textbf{h} the combination of H\,214 and H\,168, and 
\textbf{i} and \textbf{j} the combination of all three configurations. The 
spectral coverage is not the same for all configurations because the sky 
tuning frequency for H\,168 and H\,75 was not corrected for the observatory 
velocity variations. The clean beam size (HPBW) is given in each panel.}
\label{f:atcaspec}
\end{figure}

The two 3mm spectral windows of the ATCA were chosen to cover the four
amino acetonitrile features F7 to F10. The spectra toward position P4 of 
Sgr~B2(N) (offset $-2\arcsec$,0.1$\arcsec$) are presented for both windows in 
Fig.~\ref{f:atcaspec}. Since the spectral windows were not exactly the same in 
each configuration (see Sect.~\ref{ss:atca}), we show the spectra for each 
configuration (Fig.~\ref{f:atcaspec}a to f), plus the combination of the two 
broadest ones (Fig.~\ref{f:atcaspec}g and h), and the combination of all 
three configurations (Fig.~\ref{f:atcaspec}i and j). Nearly all the lines seen 
with the 30m telescope are detected with the ATCA toward P4. In the 90.6 GHz 
band,
we detect the blue wing of an SO$_2$ transition (Fig.~\ref{f:atcaspec}a), the 
red wing of an HC$^{13}$CCN ground-state transition (Fig.~\ref{f:atcaspec}e),
an unidentified line, and feature F7. In the 90.8 GHz band, we detect the 
low-velocity component of a v$_{\mathrm{t}}$=1 transition of methanol, the 
low-velocity component of a v$_7$=1 transition of HC$^{13}$CCN 
(Fig.~\ref{f:atcaspec}d and f), and the three amino acetonitrile features F8, 
F9, and F10. Toward the northern position P5, we detect a second velocity 
component of methanol v$_{\mathrm{t}}$=1 and HC$^{13}$CCN v$_7$=1 (not shown 
in Fig.~\ref{f:atcaspec}, see below).

\begin{table}
 \centering
 \caption{Peak positions of continuum and molecular line emission detected 
 with the PdBI and the ATCA toward Sgr~B2(N).}
\label{t:positions}
 \vspace*{0.0ex}
 \begin{tabular}{cccl}
  \hline\hline
  & RA (J2000) & Dec (J2000) & Comments\\
  & $17^{\mathrm{h}}47^{\mathrm{m}}$ & $-28^\circ22\arcmin$ & \\
  \hline
  & \multicolumn{2}{c}{PdBI} & \\
  82 & $19\fs886 \pm 0\fs005$ & $18.4\arcsec \pm 0.1\arcsec$ & continuum 82.0 GHz\\
  P1 & $19\fs89 \pm 0\fs01$ & $18.8\arcsec \pm 0.2\arcsec$ & mean AAN$^a$ F2 to F6\\
  P2 & $19\fs88 \pm 0\fs01$ & $13.5\arcsec \pm 0.2\arcsec$ & mean HC$_3$N v$_7$=1,  C$_2$H$_5$CN\\
  P3 & $19\fs79 \pm 0\fs01$ & $17.8\arcsec \pm 0.3\arcsec$ & peak CH$_3$OCHO\\
  \hline
  & \multicolumn{2}{c}{ATCA} & \\
  95 & $19\fs87 \pm 0\fs01$ & $18.7\arcsec \pm 0.1\arcsec$ & mean continuum\\
  & & & 93.2 and 97.4 GHz\\
  P4 & $19\fs85 \pm 0\fs02$ & $18.9\arcsec \pm 0.2\arcsec$ & mean AAN F7 to F10\\
  P5 & $19\fs86 \pm 0\fs01$ & $13.8\arcsec \pm 0.1\arcsec$ & mean HC$^{13}$CCN v$_7$=1,\\
  & & & CH$_3$OH v$_{\mathrm{t}}$=1\\
  \hline
 \end{tabular}
 \begin{list}{}{}
  \item[$(a)$] AAN stands for amino acetonitrile. 
 \end{list}
\end{table}

\begin{figure*}
\centerline{\resizebox{1.0\hsize}{!}{\includegraphics[angle=270]{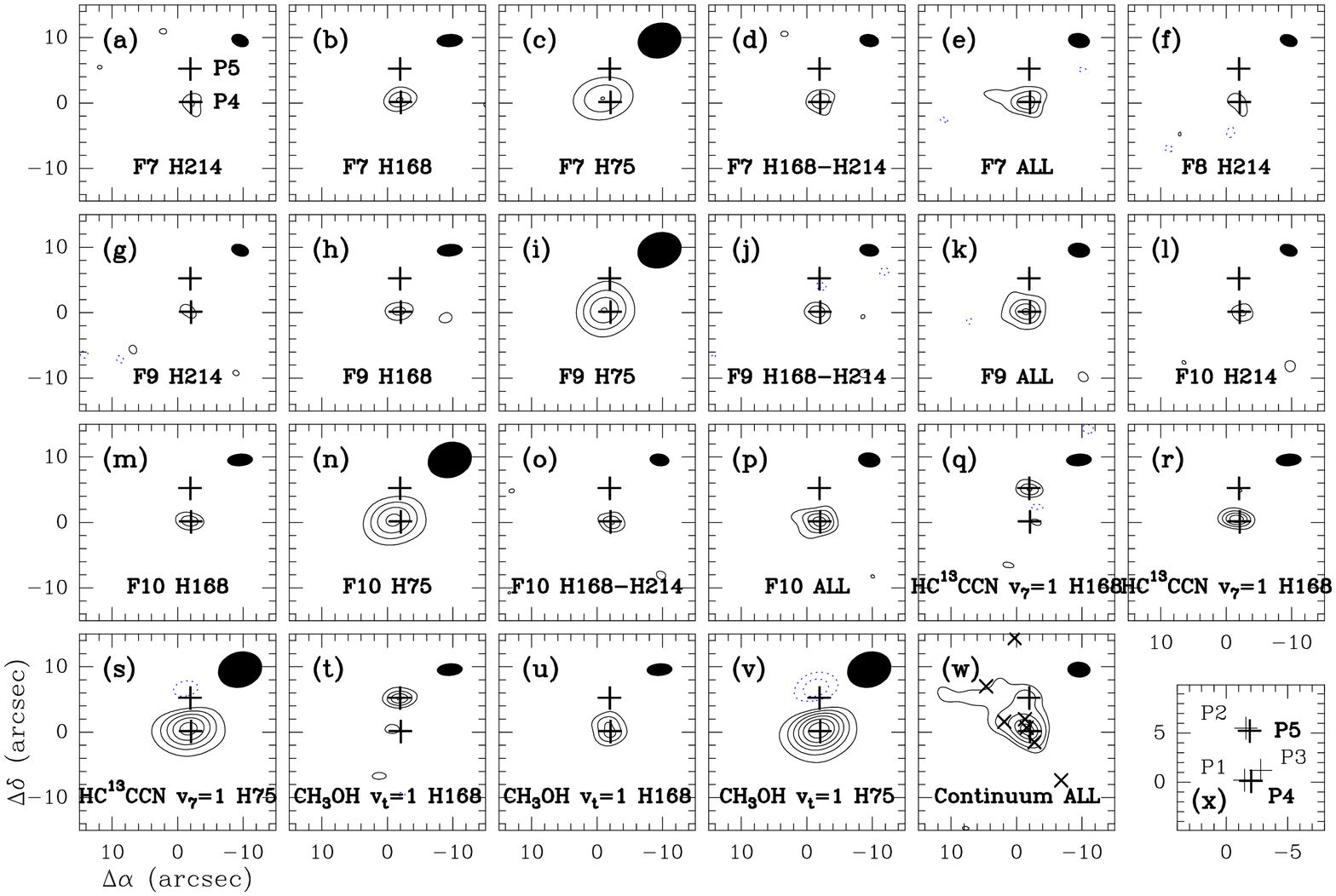}}}
\caption{Integrated intensity maps (panels \textbf{a} to \textbf{v}) and
continuum map (panel \textbf{w}) obtained toward Sgr~B2(N) with the Australia 
Telescope Compact Array at 3\,mm. Panels \textbf{(a)} to \textbf{(p)} 
show the amino acetonitrile features F7 to F10 in the different 
configurations (see Fig.~\ref{f:atcaspec}). 
Panels \textbf{q} to \textbf{v} show the other molecules listed in 
Table~\ref{t:aanatca}. The lower contour (positive in black solid line 
and negative in blue dotted line) and the contour step are 
3$\sigma$ for panels \textbf{a}, \textbf{b}, \textbf{d} to \textbf{h}, 
\textbf{j} to \textbf{m}, and \textbf{o} to \textbf{q},
4$\sigma$ for panels \textbf{r}, \textbf{t}, and \textbf{u},
5$\sigma$ for panels \textbf{c}, \textbf{i} and \textbf{n},
and 6$\sigma$ for panels \textbf{s} and \textbf{v}
(with $\sigma$ given in Col.~5 of Table~\ref{t:aanatca}).
For panel \textbf{w}, the first contours are 3 and 6$\sigma$, and the contour
step is 6$\sigma$ for the other contours (with $\sigma = 37$ mJy/beam).
In each panel, the 0,0 position is 
$\alpha_{\mathrm{J2000}} = 17^{\mathrm{h}}47^{\mathrm{m}}20\fs00$, 
$\delta_{\mathrm{J2000}} = -28^\circ22\arcmin19.0\arcsec$, the two thick plus 
symbols mark positions P4 and P5 (labeled in panel \textbf{a}), and the 
filled ellipse in the top right corner shows the clean beam. 
The cross symbols in panel \textbf{w} are the peak positions of the 
(ultracompact) H{\sc ii} regions detected by \citet{Gaume95} at 1.3~cm (K9.69, 
and K1 to K6, from right to left). The spectral integration was done on 
the frequency ranges given in Table~\ref{t:aanatca}. The maps are not 
corrected for primary beam attenuation. Panel \textbf{x} displays the ATCA 
positions P4 and P5 with thick plus symbols and the PdBI positions P1 to P3 
with thin plus symbols, for visual comparison.}
\label{f:atcamap}
\end{figure*}

The integrated intensity maps of the amino acetonitrile features F7 to F10 in 
the different configurations are presented in Fig.~\ref{f:atcamap}a to p, 
along with maps of the excited states of methanol and HC$^{13}$CCN 
(Fig.~\ref{f:atcamap}q to v). The frequency intervals used to compute the 
integrated intensities are given in Col.~3 and 4 of Table~\ref{t:aanatca} and
drawn in dotted lines in Fig.~\ref{f:atcaspec}. We used the fitting routine 
GAUSS$\_$2D of the GILDAS software to measure the position, size, and peak 
flux of each integrated emission. The results are listed in Col.~6 to 11 of 
Table~\ref{t:aanatca}. We label P4 the weighted-mean peak position of features
F7 to F10, computed using the combined configuration H\,214\,+\,H\,168 (only 
H\,214 for F8), and P5 the average northern peak position of methanol and 
HC$^{13}$CCN. 
The mean peak position P4 is at the
offset ($-2.0 \pm 0.3\arcsec$, $0.1 \pm 0.2\arcsec$), and the average position
P5 is at ($-1.9 \pm 0.1\arcsec$, $5.2 \pm 0.1\arcsec$) (see coordinates in 
Table~\ref{t:positions} and positions in Fig.~\ref{f:atcamap}x).
Finally, the ATCA velocity-integrated flux spatially integrated over the 
emitting region is listed in Col.~12 of Table~\ref{t:aanatca} and the 30m 
velocity-integrated intensity is given in Col.~13.

\begin{table*}
 \centering
 \caption{
 Measurements obtained toward Sgr~B2(N) with the Australia Telescope Compact Array at 91 GHz.
}
 \label{t:aanatca}
 \vspace*{0.0ex}
 \begin{tabular}{lcccrrrrrrrrr}
 \hline\hline
 \multicolumn{1}{c}{\hspace*{-2ex} Molecule$^a$} & \multicolumn{1}{c}{\hspace*{-5ex} Conf.$^b$} & \multicolumn{1}{c}{\hspace*{-3ex} f$_{\mathrm{min}}$$^c$} & \multicolumn{1}{c}{\hspace*{-1ex} f$_{\mathrm{max}}$$^c$} & \multicolumn{1}{c}{$\sigma^d$} & \multicolumn{1}{c}{\hspace*{-2ex} F$_{\mathrm{peak}}$$^e$} & \multicolumn{1}{c}{\hspace*{-1ex} $\Delta\alpha$$^e$} & \multicolumn{1}{c}{\hspace*{-2ex} $\Delta\delta$$^e$} & \multicolumn{1}{c}{ \hspace*{-1ex} $\theta_{\mathrm{maj}}^{\mathrm{fwhm}}$ $^e$} & \multicolumn{1}{c}{ \hspace*{-2ex} $\theta_{\mathrm{min}}^{\mathrm{fwhm}}$ $^e$} & \multicolumn{1}{c}{\hspace*{-2ex} P.A.$^e$} & \multicolumn{1}{c}{\hspace*{-3ex} $\Phi_{\mathrm{ATCA}}^f$} & \multicolumn{1}{c}{\hspace*{-2ex} $\Phi_{\mathrm{30m}}^g$} \\ 
  \multicolumn{1}{c}{} & \multicolumn{1}{c}{} & \multicolumn{1}{c}{\hspace*{-3ex} \scriptsize (MHz))} & \multicolumn{1}{c}{\hspace*{-1ex} \scriptsize (MHz)} & \multicolumn{2}{c}{ \hspace*{-2ex} \scriptsize (Jy/beam km/s)} & \multicolumn{1}{c}{\hspace*{-1ex} \scriptsize ($ '' $)} & \multicolumn{1}{c}{\hspace*{-2ex} \scriptsize ($ '' $)} & \multicolumn{1}{c}{\scriptsize ($ '' $)} & \multicolumn{1}{c}{\scriptsize ($ '' $)} & \multicolumn{1}{c}{\hspace*{-2ex} \scriptsize ($^\circ$)} & \multicolumn{1}{c}{\hspace*{-3ex} \scriptsize (Jy km/s)} & \multicolumn{1}{c}{\hspace*{-2ex} \scriptsize (Jy km/s)} \\ 
 \multicolumn{1}{c}{\hspace*{-2ex} (1)} & \multicolumn{1}{c}{\hspace*{-5ex} (2)} & \multicolumn{1}{c}{\hspace*{-3ex} (3)} & \multicolumn{1}{c}{\hspace*{-1ex} (4)} & \multicolumn{1}{c}{(5)} & \multicolumn{1}{c}{\hspace*{-2ex} (6)} & \multicolumn{1}{c}{\hspace*{-1ex} (7)} & \multicolumn{1}{c}{\hspace*{-2ex} (8)} & \multicolumn{1}{c}{ \hspace*{-1ex} (9)} & \multicolumn{1}{c}{ \hspace*{-2ex} (10)} & \multicolumn{1}{c}{\hspace*{-2ex} (11)} & \multicolumn{1}{c}{\hspace*{-3ex} (12)} & \multicolumn{1}{c}{\hspace*{-2ex} (13)} \\ 
 \hline
AAN F7 & \hspace*{-5ex} E &\hspace*{-3ex}  90558.99 & \hspace*{-1ex}  90563.99 &  0.12 & \hspace*{-2ex}      0.87 & \hspace*{-1ex}     -2.32 $\pm$      0.20 & \hspace*{-2ex}     -0.22 $\pm$      0.13 & \hspace*{-1ex}       2.9 $\pm$       0.4 & \hspace*{-2ex}       2.2 $\pm$       0.3 & \hspace*{-2ex}      29.2 $\pm$       0.1 & \hspace*{-3ex}      1.12 & \hspace*{-2ex}      3.06 \\ AAN F7 & \hspace*{-5ex} I &\hspace*{-3ex}  90559.05 & \hspace*{-1ex}  90564.05 &  0.15 & \hspace*{-2ex}      1.47 & \hspace*{-1ex}     -1.94 $\pm$      0.20 & \hspace*{-2ex}      0.58 $\pm$      0.10 & \hspace*{-1ex}       3.6 $\pm$       0.4 & \hspace*{-2ex}       2.7 $\pm$       0.2 & \hspace*{-2ex}     -79.8 $\pm$       1.5 & \hspace*{-3ex}      1.73 & \hspace*{-2ex}      3.06 \\ AAN F7 & \hspace*{-5ex} C &\hspace*{-3ex}  90561.48 & \hspace*{-1ex}  90564.23 &  0.10 & \hspace*{-2ex}      1.53 & \hspace*{-1ex}     -0.96 $\pm$      0.21 & \hspace*{-2ex}      0.71 $\pm$      0.17 & \hspace*{-1ex}       7.0 $\pm$       0.4 & \hspace*{-2ex}       4.8 $\pm$       0.3 & \hspace*{-2ex}     -80.3 $\pm$       0.0 & \hspace*{-3ex}      1.37 & \hspace*{-2ex}      1.79 \\ AAN F7 & \hspace*{-5ex} M &\hspace*{-3ex}  90559.13 & \hspace*{-1ex}  90563.96 &  0.12 & \hspace*{-2ex}      1.19 & \hspace*{-1ex}     -2.10 $\pm$      0.16 & \hspace*{-2ex}      0.25 $\pm$      0.10 & \hspace*{-1ex}       3.1 $\pm$       0.3 & \hspace*{-2ex}       2.7 $\pm$       0.2 & \hspace*{-2ex}      83.5 $\pm$       0.0 & \hspace*{-3ex}      1.71 & \hspace*{-2ex}      2.96 \\ AAN F7 & \hspace*{-5ex} A &\hspace*{-3ex}  90562.03 & \hspace*{-1ex}  90563.96 &  0.06 & \hspace*{-2ex}      0.61 & \hspace*{-1ex}     -1.23 $\pm$      0.15 & \hspace*{-2ex}      0.26 $\pm$      0.10 & \hspace*{-1ex}       5.3 $\pm$       0.3 & \hspace*{-2ex}       3.3 $\pm$       0.2 & \hspace*{-2ex}      87.1 $\pm$      22.5 & \hspace*{-3ex}      1.37 & \hspace*{-2ex}      1.00 \\ AAN F8 & \hspace*{-5ex} E &\hspace*{-3ex}  90781.27 & \hspace*{-1ex}  90786.77 &  0.21 & \hspace*{-2ex}      1.22 & \hspace*{-1ex}     -1.94 $\pm$      0.24 & \hspace*{-2ex}     -0.18 $\pm$      0.16 & \hspace*{-1ex}       3.8 $\pm$       0.5 & \hspace*{-2ex}       2.1 $\pm$       0.3 & \hspace*{-2ex}      45.0 $\pm$       0.0 & \hspace*{-3ex}      1.70 & \hspace*{-2ex}      7.23 \\ AAN F9 & \hspace*{-5ex} E &\hspace*{-3ex}  90788.02 & \hspace*{-1ex}  90792.02 &  0.09 & \hspace*{-2ex}      0.45 & \hspace*{-1ex}     -1.70 $\pm$      0.29 & \hspace*{-2ex}     -0.06 $\pm$      0.19 & \hspace*{-1ex}       3.4 $\pm$       0.6 & \hspace*{-2ex}       1.9 $\pm$       0.4 & \hspace*{-2ex}      45.0 $\pm$       0.1 & \hspace*{-3ex}      0.49 & \hspace*{-2ex}      2.14 \\ AAN F9 & \hspace*{-5ex} I &\hspace*{-3ex}  90788.11 & \hspace*{-1ex}  90792.11 &  0.13 & \hspace*{-2ex}      0.97 & \hspace*{-1ex}     -1.77 $\pm$      0.26 & \hspace*{-2ex}      0.23 $\pm$      0.13 & \hspace*{-1ex}       3.6 $\pm$       0.5 & \hspace*{-2ex}       2.2 $\pm$       0.3 & \hspace*{-2ex}     -84.0 $\pm$       0.0 & \hspace*{-3ex}      1.01 & \hspace*{-2ex}      2.14 \\ AAN F9 & \hspace*{-5ex} C &\hspace*{-3ex}  90790.54 & \hspace*{-1ex}  90792.04 &  0.10 & \hspace*{-2ex}      2.00 & \hspace*{-1ex}     -1.25 $\pm$      0.16 & \hspace*{-2ex}      0.49 $\pm$      0.13 & \hspace*{-1ex}       6.1 $\pm$       0.3 & \hspace*{-2ex}       5.8 $\pm$       0.3 & \hspace*{-2ex}     -45.0 $\pm$       0.0 & \hspace*{-3ex}      1.86 & \hspace*{-2ex}      1.59 \\ AAN F9 & \hspace*{-5ex} M &\hspace*{-3ex}  90788.04 & \hspace*{-1ex}  90792.16 &  0.10 & \hspace*{-2ex}      0.85 & \hspace*{-1ex}     -1.67 $\pm$      0.17 & \hspace*{-2ex}      0.02 $\pm$      0.11 & \hspace*{-1ex}       3.1 $\pm$       0.3 & \hspace*{-2ex}       2.4 $\pm$       0.2 & \hspace*{-2ex}      75.9 $\pm$       0.1 & \hspace*{-3ex}      1.07 & \hspace*{-2ex}      2.14 \\ AAN F9 & \hspace*{-5ex} A &\hspace*{-3ex}  90790.95 & \hspace*{-1ex}  90792.16 &  0.05 & \hspace*{-2ex}      0.63 & \hspace*{-1ex}     -1.26 $\pm$      0.14 & \hspace*{-2ex}      0.23 $\pm$      0.09 & \hspace*{-1ex}       4.8 $\pm$       0.3 & \hspace*{-2ex}       3.9 $\pm$       0.2 & \hspace*{-2ex}      84.5 $\pm$       0.0 & \hspace*{-3ex}      1.51 & \hspace*{-2ex}      1.31 \\ AAN F10 & \hspace*{-5ex} E &\hspace*{-3ex}  90796.52 & \hspace*{-1ex}  90800.77 &  0.16 & \hspace*{-2ex}      1.17 & \hspace*{-1ex}     -2.51 $\pm$      0.20 & \hspace*{-2ex}      0.02 $\pm$      0.13 & \hspace*{-1ex}       2.7 $\pm$       0.4 & \hspace*{-2ex}       2.2 $\pm$       0.3 & \hspace*{-2ex}     -81.2 $\pm$       0.2 & \hspace*{-3ex}      1.38 & \hspace*{-2ex}      6.48 \\ AAN F10 & \hspace*{-5ex} I &\hspace*{-3ex}  90796.61 & \hspace*{-1ex}  90800.86 &  0.10 & \hspace*{-2ex}      0.93 & \hspace*{-1ex}     -1.81 $\pm$      0.21 & \hspace*{-2ex}      0.20 $\pm$      0.11 & \hspace*{-1ex}       3.3 $\pm$       0.4 & \hspace*{-2ex}       2.0 $\pm$       0.2 & \hspace*{-2ex}      82.3 $\pm$       0.0 & \hspace*{-3ex}      0.79 & \hspace*{-2ex}      6.19 \\ AAN F10 & \hspace*{-5ex} C &\hspace*{-3ex}  90796.54 & \hspace*{-1ex}  90800.79 &  0.15 & \hspace*{-2ex}      3.36 & \hspace*{-1ex}     -1.08 $\pm$      0.15 & \hspace*{-2ex}      0.28 $\pm$      0.12 & \hspace*{-1ex}       6.6 $\pm$       0.3 & \hspace*{-2ex}       4.9 $\pm$       0.2 & \hspace*{-2ex}     -77.7 $\pm$       0.0 & \hspace*{-3ex}      2.78 & \hspace*{-2ex}      6.48 \\ AAN F10 & \hspace*{-5ex} M &\hspace*{-3ex}  90796.52 & \hspace*{-1ex}  90800.88 &  0.13 & \hspace*{-2ex}      1.32 & \hspace*{-1ex}     -2.28 $\pm$      0.15 & \hspace*{-2ex}      0.03 $\pm$      0.09 & \hspace*{-1ex}       2.9 $\pm$       0.3 & \hspace*{-2ex}       2.2 $\pm$       0.2 & \hspace*{-2ex}      83.7 $\pm$      22.5 & \hspace*{-3ex}      1.42 & \hspace*{-2ex}      6.48 \\ AAN F10 & \hspace*{-5ex} A &\hspace*{-3ex}  90796.52 & \hspace*{-1ex}  90800.88 &  0.13 & \hspace*{-2ex}      1.82 & \hspace*{-1ex}     -1.66 $\pm$      0.12 & \hspace*{-2ex}      0.12 $\pm$      0.08 & \hspace*{-1ex}       4.5 $\pm$       0.2 & \hspace*{-2ex}       3.1 $\pm$       0.2 & \hspace*{-2ex}     -89.3 $\pm$      22.5 & \hspace*{-3ex}      3.24 & \hspace*{-2ex}      6.48 \\ {\scriptsize HC$^{13}$CCN v$_7$=1 HV} & \hspace*{-5ex} I &\hspace*{-3ex}  90804.36 & \hspace*{-1ex}  90805.36 &  0.05 & \hspace*{-2ex}      0.49 & \hspace*{-1ex}     -1.94 $\pm$      0.20 & \hspace*{-2ex}      5.16 $\pm$      0.10 & \hspace*{-1ex}       3.0 $\pm$       0.4 & \hspace*{-2ex}       2.0 $\pm$       0.2 & \hspace*{-2ex}      84.4 $\pm$       0.0 & \hspace*{-3ex}      0.41 & \hspace*{-2ex}      2.39 \\ {\scriptsize HC$^{13}$CCN v$_7$=1 LV} & \hspace*{-5ex} I &\hspace*{-3ex}  90806.11 & \hspace*{-1ex}  90809.36 &  0.14 & \hspace*{-2ex}      2.74 & \hspace*{-1ex}     -1.56 $\pm$      0.10 & \hspace*{-2ex}      0.55 $\pm$      0.05 & \hspace*{-1ex}       3.7 $\pm$       0.2 & \hspace*{-2ex}       2.1 $\pm$       0.1 & \hspace*{-2ex}      83.6 $\pm$       0.0 & \hspace*{-3ex}      2.69 & \hspace*{-2ex}     10.30 \\ {\scriptsize HC$^{13}$CCN v$_7$=1 LV} & \hspace*{-5ex} C &\hspace*{-3ex}  90806.04 & \hspace*{-1ex}  90809.29 &  0.15 & \hspace*{-2ex}      4.98 & \hspace*{-1ex}     -1.71 $\pm$      0.10 & \hspace*{-2ex}      0.25 $\pm$      0.08 & \hspace*{-1ex}       7.1 $\pm$       0.2 & \hspace*{-2ex}       4.4 $\pm$       0.2 & \hspace*{-2ex}     -83.8 $\pm$       0.0 & \hspace*{-3ex}      4.12 & \hspace*{-2ex}     10.30 \\ {\scriptsize CH$_3$OH v$_{\mathrm{t}}$=1 HV} & \hspace*{-5ex} I &\hspace*{-3ex}  90809.61 & \hspace*{-1ex}  90811.11 &  0.09 & \hspace*{-2ex}      1.52 & \hspace*{-1ex}     -1.85 $\pm$      0.12 & \hspace*{-2ex}      5.23 $\pm$      0.06 & \hspace*{-1ex}       3.7 $\pm$       0.2 & \hspace*{-2ex}       2.1 $\pm$       0.1 & \hspace*{-2ex}     -85.7 $\pm$       0.1 & \hspace*{-3ex}      1.54 & \hspace*{-2ex}      6.55 \\ {\scriptsize CH$_3$OH v$_{\mathrm{t}}$=1 LV} & \hspace*{-5ex} I &\hspace*{-3ex}  90812.36 & \hspace*{-1ex}  90814.11 &  0.11 & \hspace*{-2ex}      1.60 & \hspace*{-1ex}     -1.84 $\pm$      0.13 & \hspace*{-2ex}      0.44 $\pm$      0.06 & \hspace*{-1ex}       3.8 $\pm$       0.3 & \hspace*{-2ex}       3.6 $\pm$       0.1 & \hspace*{-2ex}      23.7 $\pm$      22.5 & \hspace*{-3ex}      2.72 & \hspace*{-2ex}     10.40 \\ {\scriptsize CH$_3$OH v$_{\mathrm{t}}$=1 LV} & \hspace*{-5ex} C &\hspace*{-3ex}  90812.29 & \hspace*{-1ex}  90814.29 &  0.12 & \hspace*{-2ex}      5.22 & \hspace*{-1ex}     -1.86 $\pm$      0.08 & \hspace*{-2ex}     -0.03 $\pm$      0.06 & \hspace*{-1ex}       7.1 $\pm$       0.2 & \hspace*{-2ex}       4.5 $\pm$       0.1 & \hspace*{-2ex}     -77.9 $\pm$       0.0 & \hspace*{-3ex}      4.29 & \hspace*{-2ex}     10.40 \\  \hline
 \end{tabular}
 \begin{list}{}{}
 \item[$(a)$]{For amino acetonitrile (AAN), we give the feature number like in Col.~8 of Table~\ref{t:detectaan30m}. For the other molecules, HV and LV mean high and low velocity component, respectively.}
 \item[$(b)$]{Interferometer configuration: E: extended (H\,214), I: intermediate (H\,168), C: compact (H\,75), M: mixed (H\,214 + H\,168), A: all (H\,214 + H\,168 + H\,75).}
 \item[$(c)$]{Frequency range over which the intensity was integrated.}
 \item[$(d)$]{Noise level in the integrated intensity map shown in Fig.~\ref{f:atcamap}.}
 \item[$(e)$]{Peak flux, offsets in right ascension and declination with respect to the reference position of Fig.~\ref{f:atcamap}, major and minor diameters (FWHM), and position angle (East from North) derived by fitting an elliptical 2D Gaussian to the integrated intensity map shown in Fig.~\ref{f:atcamap}. The uncertainty in Col.~11 is the formal uncertainty given by the fitting routine GAUSS$\_$2D, while the uncertainties correspond to the beam size divided by two times the signal-to-noise ratio in Col.~7 and 8 and by the signal-to-noise ratio in Col.~9 and 10.}
 \item[$(f)$]{Flux spatially integrated over the region showing emission in the integrated intensity map of Fig.~\ref{f:atcamap}.}
 \item[$(g)$]{Integrated flux of the 30m spectrum computed over the frequency range given in Col.~3 and 4.}
 \end{list}
 \end{table*}

Since the line-free frequency ranges in the two spectral windows are too
small to compute a reliable continuum map, we present in Fig.~\ref{f:atcamap}w
a map combining the emission obtained at 93.2 and 97.4 GHz in the continuum 
mode. The contaminating lines were avoided in the integration. The continuum 
emission detected with the ATCA has properties very similar to the emission 
detected with the PdBI (see Sect.~\ref{ss:aanpdbi}). The emission peaks at 
$\alpha_{\mathrm{J2000}} = 17^{\mathrm{h}}47^{\mathrm{m}}19\fs87 \pm 0\fs01$, 
$\delta_{\mathrm{J2000}} = -28^\circ22\arcmin18.7\arcsec \pm 0.1\arcsec$, i.e.
within 0.3$\arcsec$ of the ultracompact H{\sc ii} region K2 
\citep[][]{Gaume95}. The continuum emission is somewhat extended around the 
main peak: it extends toward the other ultracompact H{\sc ii} regions K1, K3, 
and the shell-like H{\sc ii} region K5, but these features are not resolved 
with the ATCA. We also detect a shell-like emission close to K6, but its 
signal-to-noise ratio is lower than in the PdBI data. We do not detect any 
emission toward the weak ultracompact H{\sc ii} region K9.69 
\citep[][]{Gaume95}.

The strong methanol v$_{\mathrm{t}}$=1 and HC$^{13}$CCN v$_7$=1 lines show the
same structure as the strong lines detected with the PdBI, namely emission from
two different positions separated by about 5.1$\arcsec$ in declination, each 
with a distinct velocity (Fig.~\ref{f:atcamap}q, r, t, and u). These two 
positions P4 and P5 coincide with the PdBI positions P1 and P2, respectively, 
within 0.5$\arcsec$ which is about one fifth of the ATCA synthesized beam (see 
Fig.~\ref{f:atcamap}x for a visual comparison). Therefore we are very 
confident that they correspond to the same regions. The third position P3 
detected with the PdBI in methylformate is not detected in the small number of 
transitions observed with the ATCA.

Within the limits of our sensitivity, the amino acetonitrile features F7 to F10
show only one peak, and they all peak at the same position P4 coincident with 
the PdBI position P1. \textit{Therefore, all amino acetonitrile features 
detected with both the ATCA and the PdBI peak at the same position, which is 
consistent with their assignment to the same molecule}.
The emission detected in features F7 to F10 is barely resolved with the ATCA. 
Given the uncertainties, it is consistent with the source size of 2$\arcsec$ 
suggested by our measurements with the PdBI. The ATCA spatially integrated 
flux of 
feature F7 in the mixed configuration H\,214\,+\,H168 agrees with the 
30m telescope flux within 40$\%$. Given the calibration uncertainties of 
both instruments, and the somewhat larger noise in the 30m spectrum, most of 
the flux of feature F7 is recovered with the ATCA. On the other hand, the ATCA
spatially integrated fluxes of features  F8 to F10 are significantly weaker 
than the 30m telescope fluxes. The main reason for this disagreement may be 
the contamination of the 30m spectrum by emission from transitions of 
other molecules which are still unidentified in our survey (see 
Fig.~\ref{f:atcaspec}l). This contaminating emission is missed by the 
interferometer either because it peaks at an offset position or because it is 
extended and filtered out.

The model spectrum of amino acetonitrile computed with the same parameters as
the 30m model (see Table~\ref{t:aan30mmodel}) but with the spatial resolution
of the ATCA is shown in red in Fig.~\ref{f:atcaspec}a to j. The agreement with 
the spectra obtained with the ATCA is good, within a factor of 2 (see
comment about the interferometric modeling in Sect.~\ref{ss:aanpdbi}).
Overall, our LTE model of amino acetonitrile is therefore well consistent with
the compact emission detected with the ATCA.

\subsection{Abundance of amino acetonitrile in Sgr~B2(N)}
\label{ss:abundance}

The continuum emission detected with the PdBI (see Fig.~\ref{f:pdbimap}o) has 
a peak intensity of $0.459 \pm 0.009$ Jy/$3.3\arcsec \times 0.8\arcsec$-beam.
In the 2''-diameter region convolved by the PdBI beam centered on the AAN peak 
position P1 (convolved FWHM: $3.9\arcsec \times 2.2\arcsec$), we measure a 
mean intensity of 0.257 Jy/beam and a flux of 0.556~Jy. The continuum emission 
from Sgr~B2(N) at 3.7\,mm is dominated by thermal dust emission, at most 
30$\%$ come from free-free emission \citep[][]{Kuan96b}, probably even less 
toward the ultracompact H{\sc ii} region K2 since its flux is only 20 and 
60~mJy at 2 and 1.3\,cm, respectively \citep[][]{Gaume90,Gaume95}. In 
addition, the thermal dust emission is optically thin \citep[][]{Carlstrom89}, 
so we can estimate the H$_2$ column density using the equation:
\begin{equation}
N_{\mathrm{H}_2} = \frac{S_\nu^{\mathrm{beam}}}{\Omega_{\mathrm{beam}} \mu m_{\mathrm{H}} \kappa_\nu B_\nu(T_{\mathrm{dust}})},
\end{equation}
with $S_\nu^{\mathrm{beam}}$ the intensity at $\nu = 82.0$ GHz, 
$\Omega_{\mathrm{beam}} = 7.19 \times 10^{-11}$ rad$^2$ the solid angle of the 
synthesized beam, $\mu = 2.33$ the mean molecular weight, $m_{\mathrm{H}}$ the 
mass of atomic hydrogen, $\kappa_\nu$ the dust mass opacity (for a standard 
gas-to-dust ratio of 100 by mass), and $B_\nu(T_{\mathrm{dust}})$ the Planck 
function at the dust temperature $T_{\mathrm{dust}}$. We assume a dust mass 
opacity \hbox{$\kappa_\nu$(230 GHz) = 0.01 cm$^2$~g$^{-1}$} valid for dust 
grains that have coagulated at high density and acquired ice mantles 
\citep[][]{Ossenkopf94} and a dust emissivity exponent $\beta = 1.5$, which 
yield a dust mass opacity $\kappa_\nu$(82 GHz) = 0.0036 cm$^2$~g$^{-1}$. Since 
gas and dust are well thermally coupled via collisions at densities above 
$\sim 10^5$ cm$^{-3}$ \citep[see, e.g.,][]{Lesaffre05}, we assume a dust 
temperature equal to the excitation temperature derived for amino acetonitrile 
(100~K, see Table~\ref{t:aan30mmodel} and Fig.~\ref{f:popdiag}). We obtain a 
mean H$_2$ column density $N_{\mathrm{H}_2} = 1.3 \times 10^{25}$ cm$^{-2}$ in 
the inner region where we detect the amino acetonitrile emission with the 
PdBI. Assuming a distance of 8 kpc (see Sect.~\ref{ss:sgrb2intro}), the 
central region of 
deconvolved FWHM diameter 2$\arcsec$ has a linear diameter of 16000 AU. We 
derive a total mass of 2340 M$_\odot$ from its integrated flux (0.556 Jy). 
Assuming spherical symmetry, this translates into a mean density 
$n_{\mathrm{H}_2} = 1.7 \times 10^8$ cm$^{-3}$.

Using the column density derived from our modeling (see 
Table~\ref{t:aan30mmodel}), we find an amino acetonitrile abundance relative to
H$_2$ of about \hbox{$2.2 \times 10^{-9}$} in the inner region of deconvolved 
FWHM diameter 2$\arcsec$, with an uncertainty on the abundance of at least a 
factor of 2 given the uncertainties on the dust mass opacity and the dust 
emissivity exponent.

The continuum emission detected with the ATCA at a mean frequency of 95.3 GHz 
has a peak intensity of 1.00 Jy/$3.5\arcsec\times2.4\arcsec$-beam. This 
intensity is 1.8 times larger than the PdBI continuum flux measured over an 
area $3.9\arcsec\times2.2\arcsec$ (see above). If this flux difference comes 
from the frequency dependence of the continuum emission, then we derive a 
frequency exponent of 3.9 which is, within the calibration uncertainties, 
consistent with the exponent of 3.5 expected for thermal dust emission with a 
dust emissivity exponent $\beta = 1.5$ assumed above. We are therefore 
confident that the continuum emission detected toward K2 with the PdBI and the 
ATCA is largely dominated by thermal dust emission.

\subsection{Limits on possible extended emission of cold amino acetronitrile 
with the VLA}
\label{ss:aanvla}

\begin{figure}
\centerline{\resizebox{1.0\hsize}{!}{\includegraphics[angle=0]{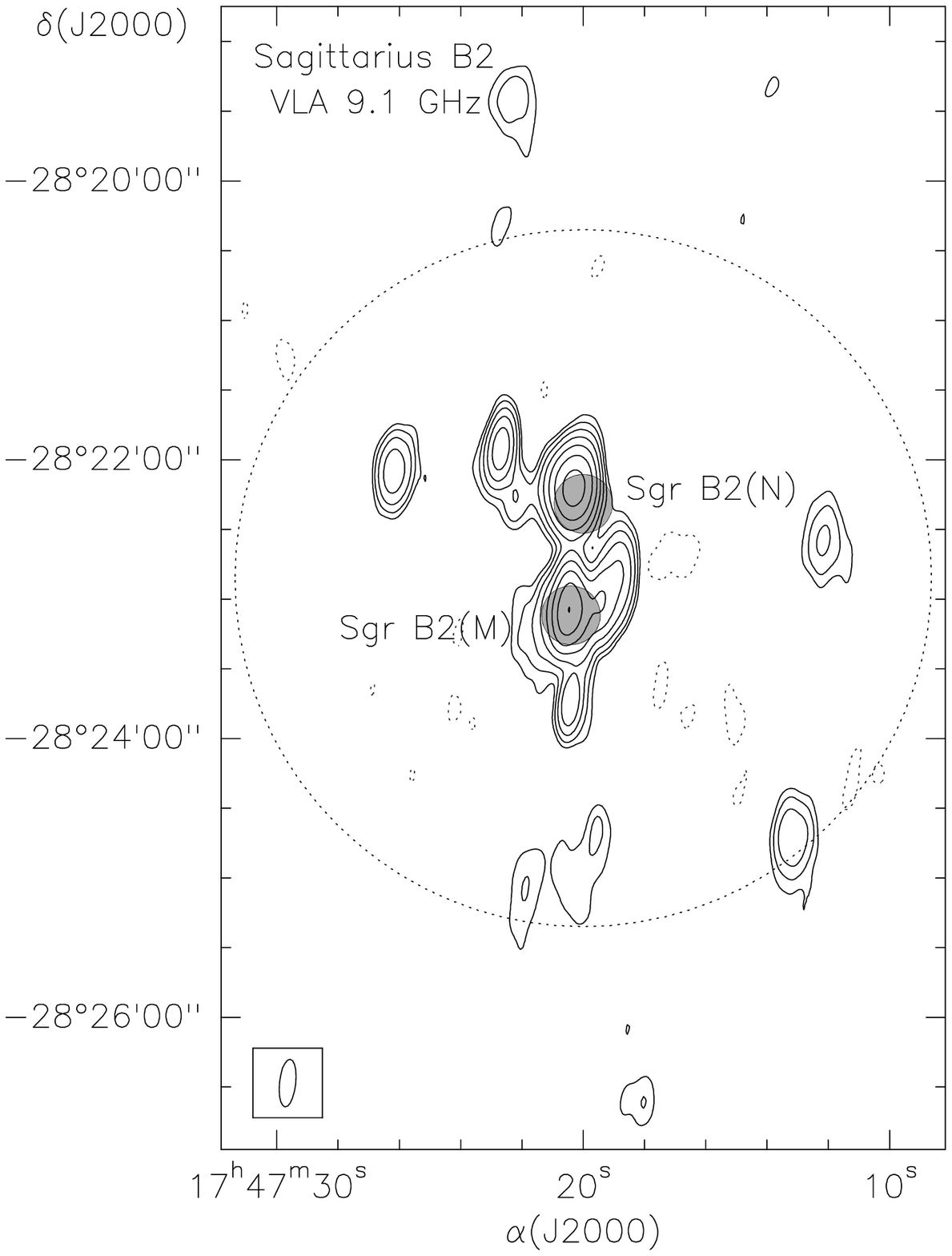}}}
\caption{Continuum image of the Sgr~B2 region obtained with the Very Large 
Array at 9.1 GHz . The lowest contour 
level is 4 times the rms noise level of 8 mJy~beam$^{-1}$ and contours double 
in value until they reach 512 times that level. The dotted circle represents 
the FWHM of the VLA antennas' primary beam at 9.1 GHz. The image is not 
corrected for attenuation due to the primary beam's response. Note that it is 
produced from data taken over only a 1 hour period and has limited dynamic 
range and sensitivity. It is, however, consistent with the higher 
sensitivity three-pointing-mosaic 4.8 GHz image presented by 
\citet{Mehringer95}, which has a similar resolution. The synthesized beam is 
represented in the lower left corner. The upper and lower shaded circles are 
centered at the Sgr~B2(N) and (M) pointing positions of our 30m telescope 
spectral line survey, respectively. Their size corresponds to the $25\arcsec$ 
FWHM of the 30m telescope at 100 GHz.
}
\label{f:vlamap}
\end{figure}

Figure~\ref{f:vlamap} shows the continuum image of the Sgr~B2 region we 
obtained at 9.1 GHz with the VLA. At this frequency, the primary beam has a 
FWHM of $\sim 5\arcmin$. Our $uv$-data should adequately sample structures 
with sizes of up to $1.5\arcmin$. There are two major 
continuum sources, Sgr~B2(N) and (M), whose emission is
a blend of the contributions of many separate sources that are resolved in 
higher resolution images \citep[e.g.,][]{Gaume95}. A number of secondary 
sources are evident. The peak brightness, $S_{\rm p}$, toward Sgr B2(N) is 
2.85 Jy~beam$^{-1}$. Toward Sgr B2(M) we measure $S_{\rm p} = 4.12$ 
Jy~beam$^{-1}$. The image has a relatively high 1$\sigma$ rms noise level of
8.4 mJy~beam$^{-1}$ that is determined by dynamic range limitations. Nowhere 
in the whole line datacube do we find any significant line emission or 
absorption above 3.4~mJy~beam$^{-1}$ (5 times the 1$\sigma$ rms noise level in 
the images of the spectral line emission). This 
corresponds to a brightness temperature upper limit of 0.35~K. With this, and
assuming a width of 30 km~s$^{-1}$, similar to other lines with extended 
emission \citep[e.g., ][]{Hunt99,Hollis04} we can use the standard relation to 
calculate an upper limit to the column densities of the upper
levels of the strongest of the $J_{K_a,K_c} = 1_{01}-0_{00}$ hfs components
of $\sim 8 \times 10^{12}$~cm$^{-2}$. \citet{Hollis04} use a two temperature 
component picture to explain their multitransition glycolaldehyde study, 
invoking components with temperatures of $T = 8$ and 50~K. For 8~K, our limit 
on the total column density of extended amino acetonitrile would be
$\sim 3 \times 10^{14}$~cm$^{-2}$ and for 50~K it would be 
$\sim 3 \times 10^{16}$~cm$^{-2}$. The low-$T$ value is on the same order as 
the total column density of glycolaldehyde that \citet{Hollis04} calculate for 
this temperature.

Using the ATCA, \citet{Hunt99} imaged Sgr~B2 in the $J=1-0$ line of HC$_3$N  
at a frequency near 9.1 GHz, very close to our amino acetonitrile frequency. 
In their 
$4.4\arcsec \times 9.9\arcsec$ resolution images they found the spatial 
distribution of the line emission to resemble that of the continuum emission 
very closely. Assuming that possible extended amino acetonitrile had a similar
distribution, the high continuum flux densities allow us to determine very 
sensitive limits on possible absorption (or weakly inverted emission) 
\textit{toward the continuum emission}. We calculate very low
$5\sigma$ limits on the absolute value of the optical depth of 
$8.3 \times 10^{-4}$ and $1.2 \times 10^{-3}$ toward Sgr~B2(N) and (M). This 
implies $\sim 125$ and 20 times lower column density limits for $T = 8$ and 
50~K, respectively, compared to limits on possible thermal emission quoted 
above. 

\section{Discussion}
\label{s:discussion}

\subsection{Amino acetonitrile in Sgr~B2(N)}
\label{ss:sgrb2n}

We detected compact emission from amino acetonitrile in Sgr~B2(N) 
with a source size of 2$\arcsec$ FWHM, a column density of 
$2.8 \times 10^{16}$ cm$^{-2}$, an excitation temperature of 100 K, a 
linewidth of 7 km~s$^{-1}$, and a centroid velocity of 64 km~s$^{-1}$ (see 
Table~\ref{t:aan30mmodel}). We estimated the abundance of amino acetonitrile 
to be $2.2 \times 10^{-9}$ in this compact region. We found no evidence for 
a possible colder, more extended emission. The compact emission peaks at 
position P1 (see Sect.~\ref{ss:aanpdbi} and Table~\ref{t:positions}), which is 
located 0.4$\arcsec$ South of the ultracompact H{\sc ii} region K2 where the 
3.7\,mm thermal dust continuum emission detected with the PdBI also peaks. 
This angular separation is at a level of 2$\sigma$ only, so it may not be 
significant. Our PdBI and ATCA data show that ethyl cyanide C$_2$H$_5$CN, 
cyanoacetylene HC$_3$N in its excited states v$_7 = 1$ and v$_4 = 1$, 
HC$^{13}$CCN in its excited state v$_7 = 1$, and methanol CH$_3$OH in its 
excited state v$_t = 1$ also peak at this position P1. The amino acetonitrile 
emission arises therefore from the hot core region called the 
``Large Molecule Heimat'' (see Sect.~\ref{sss:introlmh}). 
Our PdBI continuum data show that this compact region is extremely 
dense ($1.7 \times 10^8$ cm$^{-3}$) and massive (2340~M$_\odot$). The 
ultracompact H{\sc ii} region K2 
is most likely still embedded in the dense, hot core traced by the thermal 
dust emission and seems therefore to be the youngest source among the numerous 
ultracompact H{\sc ii} regions populating the Sgr~B2 molecular cloud.
In addition, the LMH hosts the powerful Sgr B2(N) H$_2$O maser region. The 
distribution of the maser emission over $4\arcsec \times 2\arcsec$ was mapped 
using Very Long Baseline Interferometry by \citet{Reid88}, who also fitted a 
kinematical model invoking expansion and rotation. The best fit center of 
expansion is displaced from our interferometric position P1 by
($\Delta \alpha, \Delta \delta$) = ($-0.49\arcsec,-0.66\arcsec$), which is 
less than the combined positional uncertainty of the VLBI and PdBI data. The 
best fit radial component of the expansion velocity, $63 \pm 3$~km~s$^{-1}$, 
is also in excellent agreement with the 64~km~s$^{-1}$ that we obtain for the 
LMH (see Table~\ref{t:aan30mmodel}). H$_2$O masers are associated with young 
stellar objects in their earliest stages when they drive powerful outflows, 
such as the one found in Sgr~B2(N) which has a total velocity extent of 
$\sim \pm 50$~km~s$^{-1}$ \citep[see, e.g.,][]{Reid88}. Thus, the 
water vapor maser provides evidence for the youth of the LMH, the very compact 
region where the amino acetonitrile emission originates from.

\citet{Bisschop07} measured the abundances of various complex molecules in
massive hot core regions and classified these molecules as ``cold'' ($<100$~K) 
or ``hot'' ($>100$~K). Based on the high abundances, the similar high rotation 
temperatures, and the relative constant abundance ratios of the oxygen-bearing 
species and two nitrogen-bearing species, they concluded that the 
``hot'' molecules share a common solid state formation scheme. From an
analysis of the emission of complex organic molecules in molecular clouds in
the Galactic center region and a comparison to results previously obtained in 
hot cores, \citet{RequenaTorres06} support also the scenario in which complex 
organic molecules are formed on the grain surfaces. The high temperature and 
abundance we measured for amino acetonitrile suggest it shares the same 
properties as the ``hot'' molecules found by \citet{Bisschop07}, which favors 
its formation on the grain surfaces, although its detection in other hot cores 
to check if it follows the abundance correlations found by \citet{Bisschop07} 
and \citet{RequenaTorres06} is needed to prove this conclusion. 

\citet{Wirstroem07} failed to detect amino acetonitrile in the hot cores 
\object{Orion~KL}, \object{W51~e1$/$e2}, \object{S140}, and \object{W3(OH)} 
with the Onsala 20m telescope. They 
found \textit{beam-averaged} column density upper limits of 
$1.1-3.5 \times 10^{13}$ cm$^{-2}$ for amino acetonitrile, while they detected 
vinylcyanide C$_2$H$_3$CN with a column density of $\sim 2 \times 10^{14}$ 
cm$^{-2}$ in the first two sources. Our 30m observations of Sgr~B2(N) imply a 
column density of $8 \times 10^{17}$~cm$^{-2}$ for vinylcyanide with a source 
size of 2.3$\arcsec$ (Belloche et al., \textit{in prep.}), i.e. about 30 times 
our amino acetonitrile column density. If the column density ratio of these 
two species in Sgr~B2(N) holds for other hot cores, then the observations of 
\citet{Wirstroem07} were not sensitive enough to detect amino acetonitrile in 
their sources.

Our PdBI and ATCA data show that the double peak structure seen in many 
transitions detected with the 30m telescope are produced by two sources
separated by about 5.3$\arcsec$ in the North-South direction (positions P1 and 
P2, see Sect.~\ref{ss:aanpdbi}). The centroid velocity difference between these
two positions is about 9~km~s$^{-1}$. The northern and more redshifted source
(P2) is about twice weaker in the molecular emission detected in our 30m data 
(Belloche et al., \textit{in prep.}). These two sources were already detected 
in ethylcyanide with high-resolution observations 
\citep[see, e.g.,][]{Liu99,Hollis03,Jones07}. Our interferometric data 
show that cyanoacetylene HC$_3$N and its isotopologue HC$^{13}$CCN in their 
excited state v$_7 = 1$, and methanol in its excited state v$_t = 1$ are also 
detected toward both sources. In the PdBI spectra shown in 
Fig.~\ref{f:pdbispec}c and d, we do not find a clear evidence for 
amino acetonitrile at a velocity of $\sim 73$~km~s$^{-1}$. There may be a hint 
of emission at a level about a factor of 2 lower than the emission toward P1 
(see also Feature F3 in the 30m spectrum in Fig.~\ref{f:pdbispec}g), but it is 
below our 3$\sigma$ detection limit. Therefore we \textit{cannot rule out} 
that amino acetonitrile shares the same property as, e.g., ethylcyanide, 
cyanoacetylene, and methanol and is also present in the northern source P2 at 
a level about twice lower than in P1. 

The molecular source P2 is not detected at a 3$\sigma$ level of 
26~mJy/$3.4\arcsec\times0.81\arcsec$-beam in the continuum map we obtained 
with the PdBI at 82.0 GHz (see Fig.~\ref{f:pdbimap}o). On the other hand, we 
detect some emission with an intensity of 
280~mJy/$3.5\arcsec\times2.4\arcsec$-beam 
in the continuum map obtained with the ATCA at a mean frequency of 95.3 GHz 
(see Fig.~\ref{f:atcamap}w). If this
emission is more extended than the PdBI beam, then the PdBI upper limit 
translates into 79 mJy in the ATCA beam, which yields an unphysical frequency
exponent of $\sim 8$ for the continuum emission toward P2 (see 
Sect.~\ref{ss:abundance}). We suspect that the ATCA continuum toward P2 is 
contaminated by a low-level line emission. In any case, since
the continuum emission toward P2 is much weaker than toward P1 at 3mm, P2
must be less dense and/or less hot than the hot core P1. P2 coincides with
a weak blob of emission in the 1.3\,cm VLA map of \citet{Gaume95} (see their
Fig.~6 and 7). However, this blob is located within the shell-like, weak, 
extended emission associated with K5 and it is difficult to know if it is 
compact or not from the 1.3 cm map published by these authors. Therefore the 
hot core P2 traced by the molecular emission may also be associated with an 
ultracompact H{\sc ii} region, weaker than K2.
Alternatively, if it is not directely associated with a compact source of 
free-free emission, it may have been formed by the interaction of the 
shell-like structure K5 with the ambient medium, and could be in an earlier
stage of evolution than the hot core P1 (LMH) associated with the ultracompact 
H{\sc ii} region K2.

\subsection{Amino acetonitrile in Sgr~B2(M)}
\label{ss:sgrb2m}

We do not detect amino acetonitrile in our 30m survey toward Sgr~B2(M). Using 
the same parameters as for Sgr~B2(N) (100~K and a FWHM source size of 
$2\arcsec$), we find a $\sim3\sigma$ column density upper limit of 
$6 \times 10^{15}$ cm$^{-2}$ in the LTE approximation. The column density of 
amino acetonitrile is thus at least a factor $\sim$5 weaker toward Sgr~B2(M) 
than toward Sgr~B2(N). This is not surprising since, e.g., \citet{Nummelin00} 
found that hot-core-type molecules are more abundant in Sgr~B2(N) by factors 
3--8 as compared to Sgr~B2(M).

\subsection{Amino acetonitrile, a precursor of glycine?}
\label{ss:precursor}

Amino acids, building blocks of proteins and therefore key ingredients to 
explain the origin of life, have been found in meteorites on Earth. Their 
deuterium isotopic composition suggests that they, or at 
least their direct precursors, were formed in the cold interstellar medium 
\citep[e.g.][]{Pizzarello05}. Looking for amino acids in the interstellar 
medium is therefore appealing. However, the simplest amino acid glycine has 
been intensively searched for in the past 30 years, but has unfortunately not 
been discovered yet \citep[e.g.][]{Brown79,Snyder05,Cunningham07}.

Amino acetonitrile was proposed early on as a possible direct precursor of 
glycine in the interstellar medium \citep[e.g.][]{Brown77}. The formation of 
glycine via a Strecker-cyanohydrin synthesis has long been favored
\citep[][]{Peltzer84,Ehrenfreund01,Bernstein04}. This pathway involves a 
carbonyl compound (such as an aldehyde or a ketone), hydrogen cyanide, and 
ammonia, and produces the amino nitrile which, after hydrolysis, yields the 
amino acid. However, the Strecker synthesis cannot explain the 
higher deuterium fractionation of amino acids compared to hydroxy acids 
which was measured in meteorites 
\citep[see][ and references therein]{Elsila07}.

Amino acids were successfully produced in the laboratory by UV-photolysis of 
ice mixtures mimicking the mantles of insterstellar grains 
\citep[][]{Bernstein02,Munoz02}. For an ice mixture composed of H$_2$O, 
CH$_3$OH, NH$_3$, and HCN, \citet{Woon02} proposed theoretically a pathway of 
radical-radical reactions involving the radicals t-HOCO and CH$_2$NH$_2$ 
produced by UV irradiation. This hypothesis was tested and verified 
experimentally by \citet{Holtom05} with an ice mixture of CH$_3$NH$_2$ and 
CO$_2$ bombarded by energetic electrons mimicking the impact of cosmic rays in 
the interstellar medium. This pathway leading to glycine does not involve the
formation of amino acetonitrile. On the other hand, \citet{Elsila07} 
experimented the UV-photolysis of an ice mixture of H$_2$O, CH$_3$OH, HCN, and 
NH$_3$, and found with isotopic labeling techniques multiple pathways leading 
to the formation of amino acids. The main pathway involves the formation of 
the amino nitrile and they proposed ``a modified radical-radical mechanism 
that takes into account the formation of nitriles as amino acid precursor 
molecules''. They also noticed that a Strecker-type synthesis may be at most a 
minor contributor to the formation of glycine.

The formation of glycine in the gas phase was also investigated.
\citet{Blagojevic03} synthesized ionized glycine via the reaction of the 
hydroxylamine ion NH$_2$OH$^+$ with acetic acid CH$_3$COOH. They proposed the 
formation of the precursor hydroxylamine NH$_2$OH in the grain  mantles and
the formation of acetic acid via ion-molecule reactions in the gas phase.
Based on quantum chemical calculations, \citet{Maeda06} found another pathway 
to form glycine in the gas phase involving barrierless reactions between 
closed-shell species. Their pathway starts from CO$_2$, NH$_3$, and CH$_2$, 
and leads to glycine via the reaction of CO$_2$ with the closed-shell molecule 
CH$_2$NH$_3$, a higher energy isomer of methylamine CH$_3$NH$_2$. However, 
they mentioned that CH$_2$NH$_3$ should be efficiently destroyed by 
H$_2$O, so this pathway may be unlikely in the interstellar gas phase where 
water can be very abundant. Both gas phase formation routes do not involve 
amino acetonitrile as a direct precursor of glycine.

This brief overview of the experimental and theoretical work on the formation
of amino acids in the interstellar medium shows that there is no consensus 
about the chemical precursors of amino acids. It is however important
to note that the amino acids produced in the ice experiments mentioned above 
\citep[except][]{Holtom05} are found experimentally \textit{after} the 
hydrolysis of the ice residues. It is possible that only their precursors 
(e.g. amino nitriles) are synthesized by the ice photochemistry, and that the 
amino acids are formed only later, e.g. on the comet/asteroids surfaces, by 
hydrolysis \citep[][]{Elsila07}. Therefore amino acetonitrile may well be a 
direct precursor of glycine.

The formation of amino acetonitrile itself was also investigated theoretically 
by \citet{Koch08}. They found that water can efficiently catalyze a reaction 
between methylenimine CH$_2$NH and hydrogen isocyanide HNC to form 
amino acetonitrile in the grain mantles at a temperature of 50 K. 
Methylenimine 
was detected in the gas phase toward Sgr~B2(N) by, e.g., \citet{Nummelin00}. 
They found evidence for both hot, compact and cold, extended components and 
derived a column density of $3.3 \times 10^{17}$ cm$^{-2}$ for the compact 
component, with a source size of 2.7$\arcsec$ and a temperature of 
$210^{+400}_{-80}$ K, which is consistent with our own analysis 
(Belloche et al., \textit{in prep.}). This 
column density is an order of magnitude larger than 
the column density we derived for amino acetonitrile, which does not rule out
methylenimine as a precursor of amino acetonitrile.

\subsection{Glycine in Sgr~B2(N)}
\label{ss:glycine}

The frequency coverage of our 30m survey of Sgr~B2(N) includes many transitions
of glycine as listed in the CDMS catalog (entries 75511 and 75512), but we do 
not detect this molecule within the limits of our LTE 
analysis. Using the same parameters as for amino acetonitrile (100 K and a 
source size of 2$\arcsec$, see Table~\ref{t:aan30mmodel}), we derive a column 
density upper limit of $2.0 \times 10^{17}$~cm$^{-2}$ for conformer I and 
$5.0 \times 10^{15}$ cm$^{-2}$ for conformer II. Alternatively, the upper 
limit on emission from glycine more extended than the 30m beam at 3mm
($\sim 26\arcsec$) is $1.2 \times 10^{15}$~cm$^{-2}$ for conformer I and 
$3.0 \times 10^{13}$ cm$^{-2}$ for conformer II. For a temperature of 75 K,
we find column density upper limits for conformer I of $1.5 \times 10^{17}$ 
cm$^{-2}$ for a source size of 2$\arcsec$ and $8.9 \times 10^{14}$~cm$^{-2}$
for emission more extended than the 30m beam, and for conformer II  
$3.7 \times 10^{15}$ cm$^{-2}$ and $2.2 \times 10^{13}$ cm$^{-2}$, 
respectively.

\citet{Jones07} did not detect glycine conformer I in Sgr~B2(N) with the ATCA 
and derived a 3$\sigma$ upper limit of $1.4\times10^{15}$ cm$^{-2}$ for the 
beam-averaged column density at 75 K, which translates into an upper 
limit of $2.0 \times 10^{16}$ cm$^{-2}$ for a source size of 2$\arcsec$ after 
correction for beam dilution ($17.0\arcsec \times 3.4\arcsec$). This upper 
limit on any compact emission from glycine is at a level 8 times lower than
the one we derive with the 30m telescope. As \citet{Jones07} mentioned, the
tentative detection of \citet{Kuan03} is \textit{inconsistent} with this upper 
limit in the case of compact emission.
On the other hand, the ATCA non-detection does not exclude extended emission 
at the level found by \citet{Kuan03} who reported a beam-averaged column 
density of $4.2 \times 10^{14}$ cm$^{-2}$ with the NRAO 12m telescope and
a rotational temperature of 75 K. However \citet{Cunningham07} did not detect 
glycine conformer I in Sgr~B2(N) with the 22m Mopra telescope 
and derived a 3$\sigma$ upper limit of $3.7 \times 10^{14}$ cm$^{-2}$ for the
beam-averaged column density at 75 K. This upper limit is at a level 
2.4 times lower than our beam-averaged upper limit. However, their upper 
limit was derived for a position offset by 26$\arcsec$ from the hot core 
position, so it only rules out emission from Sgr~B2(N) more extended than 
$\sim 50\arcsec$ in diameter. If glycine's emission were centered on the hot 
core position with a diameter of $\sim 25\arcsec$, Mopra would have missed 
$\sim 80\%$ of the flux,
and their upper limit would be $1.9 \times 10^{15}$ cm$^{-2}$. In that case,
our upper limit is more significant, but not low enough to rule out the
tentative detection reported by \citet{Kuan03} if glycine is confined to a 
source size of $\sim 30\arcsec$. However, the arguments presented by
\citet{Snyder05} do rule out this case. As a conclusion, the upper limits of
\citet{Cunningham07}, \citet{Jones07}, and \citet{Snyder05} rule out emission
of glycine conformer I at the level reported by \citet{Kuan03} in Sgr~B2(N) 
for any source size.

\citet{Cunningham07} found an upper limit of $7.7 \times 10^{12}$~cm$^{-2}$
for the beam-averaged column density of glycine conformer II with Mopra toward
the central position of Sgr~B2(N). This upper limit is nearly a factor of 3 
lower than the upper limit we derived above with the 30m telescope for 
extended emission. With the ATCA, \citet{Jones07} found an upper limit of 
$8.6 \times 10^{13}$ cm$^{-2}$ for the beam-averaged column density, which
translates into $1.2 \times 10^{15}$ cm$^{-2}$ for a source size of 
2$\arcsec$. This upper limit is again a factor of 3 lower than the upper limit 
we derived above with the 30m telescope for compact emission.

\citet{Bernstein04} found experimentally that organic acids are less stable 
than organic nitriles against UV photodestruction but they concluded that
in dense molecular clouds, the ratio nitrile to acid should 
be affected by less than a factor of 2 over the lifetime of the cloud. 
Therefore it could be instructive to compare the pairs 
methylcyanide/acetic acid (CH$_3$CN/CH$_3$COOH) and 
amino acetonitrile/glycine (NH$_2$CH$_2$CN/NH$_2$CH$_2$COOH). In our 30m line
survey (Belloche et al., \textit{in prep.}), we derive a column density ratio 
on the order of 200 for CH$_3$CN/CH$_3$COOH toward Sgr~B2(N). If the two pairs
are produced by similar chemical pathways yielding similar column density 
ratios, then we expect the glycine column density to be two orders of 
magnitude smaller than the amino acetonitrile column density, i.e. about 
$2 \times 10^{14}$ cm$^{-2}$ for a compact source of 2$\arcsec$ diameter, which
is nearly two orders of magnitude smaller than the upper limit derived above 
for glycine conformer I from the ATCA measurements of \citet{Jones07}, and a 
factor 5 smaller for conformer II. Therefore glycine emission may be well 
below the confusion limit in Sgr~B2(N).

\section{Conclusions}
\label{s:concl}

We used the complete 3\,mm and partial 2 and 1.3\,mm line surveys obtained 
with the IRAM 30m telescope toward the hot cores Sgr~B2(N) and (M) 
to search for emission from the complex molecule 
amino acetonitrile. We carried out follow-up observations with the IRAM 
Plateau de Bure and ATCA interferometers at selected frequencies. We also 
looked for extended emission from cold amino acetonitrile with the VLA.
We report the detection of amino acetonitrile toward the hot
core Sgr~B2(N)-LMH, which is the first detection of this molecule in the 
interstellar medium. Our main results and conclusions are the following:

 \begin{enumerate}
   \item In the course of this work, we prepared an amino acetonitrile 
   entry (56507) for the catalog of the Cologne Database for Molecular 
   Spectroscopy (CDMS) using the laboratory transition frequencies reported
   by \citet*{Bogey90}.
   \item 88 of the 398 significant transitions of amino acetonitrile covered by
   our 30m line survey are relatively free of contamination from other 
   molecules and are detected in the form of 51 observed features toward 
   Sgr~B2(N).
   \item Nine features out of 51 were followed-up upon and detected with the 
   IRAM PdB and ATCA interferometers. The amino acetonitrile emission looks 
   compact and we derive a source size of about 2$\arcsec$ in diameter (FWHM).
   \item With a source size of 2$\arcsec$ and an LTE analysis, we derive an 
   amino acetonitrile column density of $2.8 \times 10^{16}$ cm$^{-2}$ for a 
   temperature of 100~K and a linewidth of 7 km~s$^{-1}$.
   \item The compact continuum emission detected with the PdB interferometer
   yields a mean H$_2$ column density $N_{\mathrm{H}_2} = 1.3 \times 10^{25}$ 
   cm$^{-2}$ in the central region of diameter 2$\arcsec$ for a temperature of
   100 K, which implies a mean density $n_{\mathrm{H}_2} = 1.7 \times 10^8$
   cm$^{-3}$, a mass of 2340~M$_\odot$, and an amino acetonitrile fractional 
   abundance of $2.2 \times 10^{-9}$.
   \item The high abundance and temperature may indicate that amino 
   acetonitrile is formed by grain surface chemistry.
   \item We detected emission from ethylcyanide C$_2$H$_5$CN, cyanoacetylene 
   HC$_3$N and its isotopologue HC$^{13}$CCN in their v$_7$=1 excited states,
   and methanol CH$_3$OH in its excited state v$_{\mathrm{t}}$=1 in two
   compact sources toward Sgr~B2(N). The two sources are separated by about 
   $5.3\arcsec$ in the North-South direction and by 9 km~s$^{-1}$ in velocity.
   They produce double peaked line shapes for many molecules detected in our
   30m line survey. Only the southern source is detected with the PdBI in 
   continuum emission. It is associated with the ultracompact H{\sc ii} region 
   K2 and a powerful H$_2$O maser region, and must be very young. The 
   northern source is weaker in molecular emission and must be less 
   dense and/or less hot. It may be associated with a weaker ultracompact 
   H{\sc ii} region and may be in an even earlier stage of evolution. 
   The sensitivity of our observations was not good enough to detect 
   amino acetonitrile toward the northern source. It is at least a factor of
   2 weaker than toward the southern source.
   \item We did not detect amino acetonitrile toward Sgr~B2(M) and derived a
   column density upper limit of $6 \times 10^{15}$~cm$^{-2}$.
   \item We failed to detect any extended emission from cold amino acetonitrile
   with the VLA and derived a column density upper limit of 
   $3 \times 10^{12-14}$~cm$^{-2}$ at 8~K.
   \item Amino acetonitrile may be a chemical precursor of glycine. 
   We did not detect the glycine conformers I and II in our 30m line 
   survey. The column density upper limits we derive are less constraining 
   than upper limits previously published by other authors. Based on 
   our detection of amino acetonitrile and a comparison to the pair 
   methylcyanide/acetic acid (CH$_3$CN/CH$_3$COOH) both of which are
   detected in our survey, we conclude that the column density of glycine 
   conformers I and II in Sgr~B2(N) may be two orders of magnitude and a 
   factor 5 below the current best upper limits, respectively, which would be 
   below the confusion limit of Sgr~B2(N) in the 1--3\,mm range.
 \end{enumerate}

\begin{acknowledgements}
We thank the IRAM staff in Grenoble for observing at the PdBI and for their 
help with the data reduction, the IRAM staff in Granada for service observing 
in January 2005, and Sergio Martin for providing the reference (off) position 
for our 30m observations. We thank John Pearson for his predictions of the 
first excited state of ethylcyanide, Claus Nielsen for providing transition 
frequencies of formamide isotopologues, Isabelle Kleiner, Vadim Ilyushin, 
and Frank Lovas for acetic acid frequencies, and Brian Drouin for his 
predictions of acetone.
H.~S.~P.~M. and the CDMS had been supported initially through the Deutsche
Forschungsgemeinschaft (DFG) via the collaborative research grant SFB 494.
Recent support is provided by the Bundesministerium f\"ur Bildung und Forschung
administered through Deutsches Zentrum f\"ur Luft- und Raumfahrt (DLR; the
German space agency).
J.~O. is a Jansky Fellow of the National Radio Astronomy Observatory.
C.~H. is a fellow of the Studienstiftung des deutschen Volkes and member of the
International Max-Planck Research School for Radio and Infrared Astronomy.

\end{acknowledgements}


\listofobjects


\begin{thebibliography}{}
\bibitem[Bernstein et al.(2002)]{Bernstein02} Bernstein, M.~P., 
Dworkin, J.~P., Sandford, S.~A., Cooper, G.~W., \& Allamandola, L.~J.\ 
2002, \nat, 416, 401
\bibitem[Bernstein et al.(2004)]{Bernstein04} Bernstein, M.~P.,
Ashbourn, S.~F.~M., Sandford, S.~A., \& Allamandola, L.~J.\ 2004, \apj,
601, 365
\bibitem[Berulis et al.(1985)]{Berulis85} Berulis, I.~I.,
Winnewisser, G., Krasnov, V.~V., \& Sorochenko, R.~L.\ 1985, Soviet
Astronomy Letters, 11, 251
\bibitem[Bisschop et al.(2007)]{Bisschop07} Bisschop, S.~E., J{\"o}rgensen, 
J.~K., van Dishoeck, E.~F., \& de Wachter, E.~B.~M.\ 2007, \aap, 465, 913
\bibitem[Blagojevic et al.(2003)]{Blagojevic03} Blagojevic, V.,
Petrie, S., \& Bohme, D.~K.\ 2003, \mnras, 339, L7
\bibitem[Bogey, Dubus, \& Guillemin(1990)Bogey et al.]{Bogey90} Bogey, M., 
Dubus, H., \& Guillemin, J.~C.\ 1990, J. Mol. Spectrosc., 143, 180 
\bibitem[Brown et al.(1977)]{Brown77} Brown, R.~D., Godfrey, P.~D., Ottrey, 
A.~L., \& Storey, J.~W.~V.\ 1977, J. Mol. Spectrosc., 68 (3), 359 
\bibitem[Brown et al.(1979)]{Brown79} Brown, R.~D., Godfrey, P.~D., Storey, 
J.~W.~V., Bassez, M.-P., et al.\ 1979, \mnras, 186, 5P
\bibitem[Carlstrom \& Vogel(1989)]{Carlstrom89} Carlstrom, J.~E., \& Vogel, 
S.~N.\ 1989, \apj, 337, 408
\bibitem[Ceccarelli et al.(2000)]{Ceccarelli00} Ceccarelli, C.,
Loinard, L., Castets, A., Faure, A., \& Lefloch, B.\ 2000, \aap, 362, 1122
\bibitem[Clark(1980)]{Clark80} Clark, B.~G.\ 1980, \aap, 89, 377 
\bibitem[Combes et al(1996)Combes et al.]{Combes96} Combes, 
F., Q-Rieu, N., \& Wlodarczak, G.\ 1996, \aap, 308, 618
\bibitem[Comito et al.(2005)]{Comito05} Comito, C., Schilke, P., 
Phillips, T.~G., Lis, D.~C., Motte, F., \& Mehringer, D.\ 2005, \apjs, 156, 
127
\bibitem[Cummins, Linke, \& Thaddeus(1986)Cummins et al.]{Cummins86} Cummins, 
S.~E., Linke, R.~A., \& Thaddeus, P.\ 1986, \apjs, 60, 819
\bibitem[Cunningham et al.(2007)]{Cunningham07} Cunningham, M.~R., 
et al.\ 2007, \mnras, 376, 1201
\bibitem[Dahmen et al.(1997)]{Dahmen97} Dahmen, G., et al.\ 1997, \aaps, 126, 
197
\bibitem[de Vicente et al.(2000)]{deVicente00} de Vicente, P.,
Mart{\'{\i}}n-Pintado, J., Neri, R., \& Colom, P.\ 2000, \aap, 361, 1058
\bibitem[Ehrenfreund et al.(2001)]{Ehrenfreund01} Ehrenfreund, P.,
Glavin, D.~P., Botta, O., Cooper, G., \& Bada, J.~L.\ 2001, Proceedings of
the National Academy of Science, 98, 2138
\bibitem[Eisenhauer et al.(2003)]{Eisenhauer03} Eisenhauer, F., 
Sch{\"o}del, R., Genzel, R., Ott, T., Tecza, M., Abuter, R., Eckart, A., \& 
Alexander, T.\ 2003, \apjl, 597, L121 
\bibitem[Elsila et al.(2007)]{Elsila07} Elsila, J.~E., Dworkin,
J.~P., Bernstein, M.~P., Martin, M.~P., \& Sandford, S.~A.\ 2007, \apj,
660, 911
\bibitem[Gaume \& Claussen(1990)]{Gaume90} Gaume, R.~A., \&
Claussen, M.~J.\ 1990, \apj, 351, 538
\bibitem[Gaume et al.(1995)]{Gaume95} Gaume, R.~A., Claussen, M.~J., de Pree, 
C.~G., Goss, W.~M., \& Mehringer, D.~M.\ 1995, \apj, 449, 663
\bibitem[Goldsmith \& Langer(1999)]{Goldsmith99} Goldsmith, P.~F.,
\& Langer, W.~D.\ 1999, \apj, 517, 209
\bibitem[Gordy \& Cook(1984)]{Gordy84} Gordy, W., \& Cook, R.~L.\ 1984, 
Microwave Molecular Spectra, 3$^{\mathrm{rd}}$ edition, Wiley (New York)
\bibitem[Hollis et al.(1980)]{Hollis80} Hollis, J.~M., Snyder, 
L.~E., Suenram, R.~D., \& Lovas, F.~J.\ 1980, \apj, 241, 1001
\bibitem[Hollis et al.(2000)]{Hollis00} Hollis, J.~M., Lovas, 
F.~J., \& Jewell, P.~R.\ 2000, \apjl, 540, L107
\bibitem[Hollis et al.(2001)]{Hollis01} Hollis, J.~M., Vogel,
S.~N., Snyder, L.~E., Jewell, P.~R., \& Lovas, F.~J.\ 2001, \apjl, 554, L81
\bibitem[Hollis et al.(2002)]{Hollis02} Hollis, J.~M., Lovas,
F.~J., Jewell, P.~R., \& Coudert, L.~H.\ 2002, \apjl, 571, L59
\bibitem[Hollis et al.(2003)]{Hollis03} Hollis, J.~M., Pedelty,
J.~A., Boboltz, D.~A., Liu, S.-Y., Snyder, L.~E., Palmer, P., Lovas, F.~J.,
\& Jewell, P.~R.\ 2003, \apjl, 596, L235
\bibitem[Hollis et al.(2004)]{Hollis04} Hollis, J.~M., Jewell, P.~R., Lovas, 
F.~J., Remijan, A., \& M{\o}llendal, H.\ 2004, \apjl, 610, L21
\bibitem[Hollis et al.(2007)]{Hollis07} Hollis, J.~M., Jewell,
P.~R., Remijan, A.~J., \& Lovas, F.~J.\ 2007, \apjl, 660, L125
\bibitem[Holtom et al.(2005)]{Holtom05} Holtom, P.~D., Bennett,
C.~J., Osamura, Y., Mason, N.~J., \& Kaiser, R.~I.\ 2005, \apj, 626, 940
\bibitem[Hunt et al.(1999)]{Hunt99} Hunt, M.~R., Whiteoak, J.~B., Cragg, 
D.~M., White, G.~L., \& Jones, P.~A.\ 1999, \mnras, 302, 1
\bibitem[Jones et al.(2007)]{Jones07} Jones, P.~A., Cunningham, M.~R., 
Godfrey, P.~D., \& Cragg, D.~M.\ 2007, \mnras, 374, 579
\bibitem[Jones et al.(2008)]{Jones08} Jones, P.~A., Burton, M.~G., Cunningham, 
M.~R., Menten, K.~M., Schilke, P., Belloche, A., Leurini, S., Ott, J., Walsh, 
A.~J.\ 2008, \mnras, \textit{in press}
\bibitem[Koch et al.(2008)]{Koch08} Koch, D.~M., Toubin, C., Peslherbe, 
G.~H., \& Hynes, J.~T.\ 2008, J. Phys. Chem. C, 112(8), 2972
\bibitem[Kuan \& Snyder(1996)]{Kuan96a} Kuan, Y.-J., \& Snyder, 
L.~E.\ 1996, \apj, 470, 981 
\bibitem[Kuan et al.(1996)]{Kuan96b} Kuan, Y.-J., Mehringer, D.~M., \& Snyder, 
L.~E.\ 1996, \apj, 459, 619
\bibitem[Kuan et al.(2003)]{Kuan03} Kuan, Y.-J., Charnley,
S.~B., Huang, H.-C., Tseng, W.-L., \& Kisiel, Z.\ 2003, \apj, 593, 848
\bibitem[Lesaffre et al.(2005)]{Lesaffre05} Lesaffre, P., Belloche, A., 
Chi{\`e}ze, J.-P., \& Andr{\'e}, P.\ 2005, \aap, 443, 961
\bibitem[Lis et al.(1991)]{Lis91} Lis, D.~C., Carlstrom,
J.~E., \& Keene, J.\ 1991, \apj, 380, 429
\bibitem[Lis et al.(1993)]{Lis93} Lis, D.~C., Goldsmith,
P.~F., Carlstrom, J.~E., \& Scoville, N.~Z.\ 1993, \apj, 402, 238
\bibitem[Liu \& Snyder(1999)]{Liu99} Liu, S.-Y., \& Snyder, L.~E.\ 1999, \apj, 
523, 683
\bibitem[Liu et al.(2001)]{Liu01} Liu, S.-Y., Mehringer, 
D.~M., \& Snyder, L.~E.\ 2001, \apj, 552, 654
\bibitem[MacDonald \& Tyler(1972)]{MacDonald72} MacDonald, J.~N. \& Tyler, 
J.~K.\ 1972, J. Chem. Soc. Chem. Comm., 995
\bibitem[Maeda \& Ohno(2006)]{Maeda06} Maeda, S., \& Ohno, K.\ 2006, \apj, 
640, 823
\bibitem[Mehringer et al.(1995)]{Mehringer95} Mehringer, D.~M.,
Palmer, P., \& Goss, W.~M.\ 1995, \apjs, 97, 497
\bibitem[Mehringer \& Menten(1997)]{Mehringer97a} Mehringer, D.~M.,
\& Menten, K.~M.\ 1997, \apj, 474, 346
\bibitem[Mehringer et al.(1997)]{Mehringer97b} Mehringer, D.~M.,
Snyder, L.~E., Miao, Y., \& Lovas, F.~J.\ 1997, \apjl, 480, L71
\bibitem[Menten(2004)]{Menten04} Menten, K.~M.\ 2004, in The Dense 
Interstellar Medium in Galaxies, Eds. Pfalzner, Kramer, Staubmeier, \& 
Heithausen, Springer proceedings in physics, Vol. 91 (Berlin, Heidelberg: 
Springer), 69 
\bibitem[Miao et al.(1995)]{Miao95} Miao, Y., Mehringer, D.~M., Kuan, Y.-J., 
\& Snyder, L.~E.\ 1995, \apjl, 445, L59 
\bibitem[Miao \& Snyder(1997)]{Miao97} Miao, Y., \& Snyder, L.~E.\ 1997, 
\apjl, 480, L67 
\bibitem[Minh et al.(1992)]{Minh92} Minh, Y.~C., Irvine,
W.~M., \& Friberg, P.\ 1992, \aap, 258, 489
\bibitem[Morris \& Serabyn(1996)]{Morris96} Morris, M., \&
Serabyn, E.\ 1996, \araa, 34, 645
\bibitem[M{\"u}ller et al.(2001)]{Mueller01} M{\"u}ller, H.~S.~P., Thorwirth, 
S., Roth, D.~A., \& Winnewisser, G.\ 2001, \aap, 370, L49
\bibitem[M{\"u}ller et al.(2005)]{Mueller05} M{\" u}ller, H.~S.~P.,
Schl{\" o}der, F., Stutzki, J., \& Winnewisser, G.\ 2005, J. Mol. Struct.,
742, 215
\bibitem[Mu{\~n}oz Caro et al.(2002)]{Munoz02} Mu{\~n}oz Caro, G.~M., 
Meierhenrich, U.~J., Schutte, W.~A., Barbier, B., et al.\ 2002, \nat, 416, 403
\bibitem[Nummelin et al.(2000)]{Nummelin00} Nummelin, A., Bergman, P., 
Hjalmarson, {\AA}., Friberg, P., Irvine, W.~M., Millar, T.~J., Ohishi, M., \& 
Saito, S.\ 2000, \apjs, 128, 213
\bibitem[Ossenkopf \& Henning(1994)]{Ossenkopf94} Ossenkopf, V., \& 
Henning, T.\ 1994, \aap, 291, 943
\bibitem[Peltzer et al.(1984)]{Peltzer84} Peltzer, E.~T., Bada, 
J.~L., Schlesinger, G., \& Miller, S.~L.\ 1984, Advances in Space Research, 
4, 69
\bibitem[Pickett(1973)]{Pickett73} Pickett, H.~M.\ 1973, J. Mol. Spectrosc., 
46, 335
\bibitem[Pickett(1991)]{Pickett91} Pickett, H.~M.\ 1991, J. Mol. Spectrosc., 
148, 371
\bibitem[Pickett et al.(1998)]{Pickett98} Pickett, H.~M., Poynter, R.~L., 
Cohen, E.~A., Delitsky, M.~L., Pearson, J.~C., \& M{\"u}ller, H.~S.~P.\ 1998
, J. Quant. Spectrosc. \& Rad. Transfer, 60, 883
\bibitem[Pizzarello \& Huang(2005)]{Pizzarello05} Pizzarello, S., \& Huang, 
Y.\ 2005, Geochim. Cosmoschim. Acta, 69, 599
\bibitem[Reid(1993)]{Reid93} Reid, M.~J.\ 1993, \araa, 31, 345
\bibitem[Reid et al.(1988)]{Reid88} Reid, M.~J., Schneps, M.~H., Moran, 
J.~M., Gwinn, C.~R., Genzel, R., Downes, D., \& Roennaeng, B.\ 1988, \apj, 
330, 809
\bibitem[Remijan et al.(2002)]{Remijan02} Remijan, A., Snyder,
L.~E., Liu, S.-Y., Mehringer, D., \& Kuan, Y.-J.\ 2002, \apj, 576, 264
\bibitem[Requena-Torres et al.(2006)]{RequenaTorres06} Requena-Torres,
M.~A., Mart{\'{\i}}n-Pintado, J., Rodr{\'{\i}}guez-Franco, A.,
Mart{\'{\i}}n, S., Rodr{\'{\i}}guez-Fern{\'a}ndez, N.~J., \& de Vicente,
P.\ 2006, \aap, 455, 971
\bibitem[Sault, Teuben, \& Wright(1995)Sault et al.]{Sault95} Sault, R.~J., 
Teuben, P.~J., \& Wright, M.~C.~H.\ 1995, Astronomical Data Analysis Software 
and Systems IV, 77, 433
\bibitem[Snyder, Kuan, \& Miao(1994)Snyder et al.]{Snyder94} 
Snyder, L.~E., Kuan, Y.-J., \& Miao, Y.\ 1994, Lecture Notes in Physics 439: 
The Structure and Content of Molecular Clouds, Eds. T.~L. Wilson and K.~J. '
Johnston, Springer-Verlag, 439, 187
\bibitem[Snyder et al.(2002)]{Snyder02} Snyder, L.~E., Lovas,
F.~J., Mehringer, D.~M., Miao, N.~Y., Kuan, Y., Hollis, J.~M., \& Jewell,
P.~R.\ 2002, \apj, 578, 245
\bibitem[Snyder et al.(2005)]{Snyder05} Snyder, L.~E., Lovas, F.~J., Hollis, 
J.~M., Friedel, D.~N., et al.\ 2005, \apj, 619, 914
\bibitem[Storey(1976)]{Storey76} Storey, J.~W.~V.\ 1976, Ph.D.~Thesis, Monash 
University
\bibitem[Turner \& Apponi(2001)]{Turner01} Turner, B.~E., \& Apponi, A.~J.\ 
2001, \apjl, 561, L207
\bibitem[Vogel et al.(1987)]{Vogel87} Vogel, S.~N., Genzel, R.,
\& Palmer, P.\ 1987, \apj, 316, 243
\bibitem[Wirstr{\" o}m et al.(2007)]{Wirstroem07} Wirstr{\" o}m, E.~S., 
Bergman, P., Hjalmarson, {\AA}., \& Nummelin, A.\ 2007, \aap, 473, 177
\bibitem[Woon(2002)]{Woon02} Woon, D.~E.\ 2002, \apjl, 571, L177
\end{thebibliography}
\end{document}